\documentclass[11pt]{article}

\usepackage[latin1]{inputenc}
\usepackage[abbr]{harvard}
\usepackage{times}

\usepackage{psfrag, amsmath, amscd, amsfonts, latexsym, epsf}
\usepackage{graphicx}
\usepackage[small]{caption}
\usepackage{color}
\usepackage{url}
\usepackage{tikz}
\newcommand{\argmax}{\operatorname{argmax}}

\def\bbbr{{\mathbb R}}

\newcommand{\htransf}{H}

\def\urltilda{\kern -.15em\lower .7ex\hbox{\~{}}\kern .04em}
\def\urldot{\kern -.10em.\kern -.10em}
\def\urlhttp{http\kern -.10em\lower -.1ex\hbox{:}\kern -.12em\lower
  0ex\hbox{/}\kern -.18em\lower 0ex\hbox{/}}

\begin{document}

\pagenumbering{roman}

\title{\bf Idealized computational models for\\ auditory receptive fields%
\thanks{The support from the Swedish Research Council (contract 2010-4766) is gratefully acknowledged. }}

\author{\em Tony Lindeberg$^1$ and Anders Friberg$^2$\\
        \\
        $^1$Department of Computational Biology\\
        $^2$Department of Speech, Music and Hearing\\
        School of Computer Science and Communication\\
        KTH Royal Institute of Technology\\
        SE-100 44 Stockholm, Sweden}

\date{}

\maketitle

\begin{abstract}
\noindent
This paper presents a theory by which idealized models of auditory
receptive fields can be derived in a principled axiomatic manner, from a set of
structural properties to (i)~enable invariance of receptive field responses under natural
sound transformations and (ii)~ensure internal consistency between
spectro-temporal receptive fields at different temporal and spectral scales.

For defining a time-frequency transformation
of a purely temporal sound signal, it is shown that the framework allows for a new way
of deriving the Gabor and Gammatone filters as well as a novel family
of generalized Gammatone filters,
with additional degrees of freedom to obtain different trade-offs
between the spectral selectivity and the temporal delay of time-causal
temporal window functions.

When applied to the definition of a second-layer of receptive fields
from a spectrogram,
it is shown that the framework leads to two canonical families of
spectro-temporal receptive fields, in terms of spectro-temporal
derivatives of either spectro-temporal Gaussian kernels for non-causal
time or the
combination of a time-causal generalized Gammatone filter over the temporal domain
and a Gaussian filter over the logspectral domain.
For each filter family, the spectro-temporal receptive fields can be
either separable over the time-frequency domain or be adapted to local
glissando transformations that represent variations in logarithmic
frequencies over time.
Within each domain of either non-causal or time-causal time, these
receptive field families are derived by uniqueness from the assumptions.

It is demonstrated how the presented framework allows for 
computation of basic auditory features for audio processing
and that it leads to predictions about auditory receptive fields
with good qualitative similarity to biological receptive fields measured in the
inferior colliculus (ICC) and primary auditory cortex (A1) of mammals.

\medskip
\noindent
{\em Keywords:\/} 
receptive field, auditory, temporal, spectro-temporal, scale space, spectrogram,
Gabor filter, Gammatone filter, Gaussian derivative,
feature detection, onset detection, partial tone detection, glissando,
inferior colliculus, primary auditory cortex,
auditory perception

\medskip
\noindent
Copyright: The Authors
\end{abstract}

\newpage

\tableofcontents

\newpage

\pagenumbering{arabic}

\section{Introduction}

The information in sound is based on variations in the air pressure
over time, which for many sound sources can be modelled as a
superposition of sine wave oscillations of different frequencies.
To capture this information by auditory perception or signal
processing, the sound signal has to be processed over some
non-infinitesimal amount of time and in the case of a spectral
analysis also over some range of frequencies. 
Such a region over time or over the spectro-temporal domain is
referred to as a temporal or spectro-temporal receptive field \cite{AerJoh81-BICY,TheSenDou-JNeuroSci,MilEscReaSch01-JNeuroPhys,FriShaElhKle03-NatureNeuSci}.

If one considers the theoretical or algorithmic problem of designing
an auditory system that is going to analyse the variations in air
pressure over time, one may ask what types of auditory operations
should be performed on the sound signal. Would any operation be
reasonable? Specifically, regarding the notion of receptive fields,
what types of temporal or spectro-temporal receptive field profiles
would be reasonable. Is it possible to derive a theoretical model of
how receptive fields ``ought to'' respond to auditory data?

In vision, the corresponding problem of formulating a theoretical
model for visual receptive fields \cite{Lin13-BICY} can be addressed in detail
based on a framework developed in the area of
computer vision known as {\em scale-space theory\/}
\cite{Iij62,Wit83,Koe84,KoeDoo92-PAMI,Lin93-Dis,Lin94-SI,SpoNieFloJoh96-SCSPTH,Flo97-book,Haa04-book,Lin08-EncCompSci,Lin10-JMIV}.
A paradigm that has been developed in this field is to impose
{\em structural constraints\/} on the first stages of visual
processing that reflect {\em symmetry properties\/} of the
environment.
Interestingly, it turns out to be possible to substantially reduce the
class of permissible image operations from such arguments, and it has
been shown that biological receptive fields as measured in the lateral
geniculate nucleus (LGN) and the primary visual cortex (V1) of higher
mammals
\cite{HubWie59-Phys,HubWie62-Phys,DeAngOhzFre95-TINS,deAngAnz04-VisNeuroSci,ConLiv06-JNeurSci}
can be well modelled by idealized scale-space operations
\cite{You87-SV,YouLesMey01-SV,Lin13-BICY,Lin13-PONE}.

The subject of this article is to show how a corresponding 
theory for receptive fields can
be developed for auditory stimuli, and how idealized models of
auditory receptive fields can be derived by applying 
scale-space theory to auditory signals.
Our aim is to express auditory operations that are well localized over time and
frequencies and which allow for well-founded handling of temporal
phenomena that occur at different temporal scales as well as receptive
fields that operate over different ranges of frequencies in such a way
that operations over different ranges of frequencies can be related 
in a well-defined manner.

It will be shown that when applied to the definition of spectrograms 
alternatively the formulation of an idealized cochlea model, the scale-space
approach can be used for deriving the Gabor \cite{Gab46,WolGodDor01-ASAA,LobLoi03-ICASSP,QiuSchEsc03-JNeuroPhys,WuZhaShi11-ASLP} and Gamma-tone \cite{Joh72-HearTheory,PatNimHolRic87-GammaTone,HewMed94-JASA,PatAllGig95-JASA}
approaches for computing local windowed Fourier transforms as specific
cases of a complex-valued scale-space transform over different
frequencies.
In addition, the scale-space approach to defining
spectrograms leads to a new family of generalized Gamma-tone filters
where the time constants of the individual first-order integrators
that are coupled in cascade are not equal as for regular Gamma-tone
filters but instead distributed logarithmically over temporal scales
and thereby allowing for different trade-offs in
terms of {\em e.g.\/} the frequency selectivity of the spectrogram and the
temporal delay of time-causal receptive fields.

When applied to a logarithmic transformation of the spectrogram, as
motivated from the desire of handling sound signals of different
strength (sound pressure) in
an invariant manner and with a logarithmic transformation of the
frequencies as motivated by the desire of enabling invariance
properties under a frequency shift, such as transposing a
musical piece by one octave, we will show how this theory also allows
for the formulation of spectro-temporal receptive fields at higher
levels in the auditory hierarchy in terms of spectro-temporal derivatives
of spectro-temporal smoothing operations as obtained from scale-space theory.

It will be demonstrated how such second-layer receptive fields can be
used for computing basic auditory features such as onset detection,
partial tone enhancement and formants, and specifically how
different types of features can be obtained at different temporal
scales $\tau$, spectral scales $s$ and how this theory naturally 
also leads to a glissando parameter $v$ that represents how
logarithmic frequencies $\nu$ may vary over time $t$ according
to a local linear approximation $\nu' = \nu + v t$.

Compared to the more common approach of computing auditory features in
digital signal processing by local windowed fast Fourier transforms (FFT),
we argue that proposed theory provides a way to avoid artifacts of
performing the computation in temporal blocks that later have to be
combined again.
Furthermore, by the built-in covariance properties of the model 
under temporal shifts, variations in sound pressure and frequency shifts,
the proposed approach allows for provable invariance properties under
such transformations of sound signals.

It will also be shown how idealized models of spectro-temporal
receptive fields as obtained from the presented theory in terms of
spectro-temporal derivatives of spectro-temporal scale-space kernels
can be used for generating predictions of auditory receptive fields
that are qualitatively similar to biological receptive fields
as measured by cell recordings in the inferior colliculus (ICC) and the
primary auditory cortex (A1)
\cite{MilEscReaSch01-JNeuroPhys,QiuSchEsc03-JNeuroPhys,MacWehZad04-JNeuroSci,AndLiPol07-JNeuroSci,ElhFriChiSha07-JNeuroSci,AteSch12-PONE}.

\subsection{Outline of the presentation}

The presentation is organized as follows:
Section~\ref{sec-struct-req-temp-rec-fields} starts by describing basic
constraints on temporal receptive fields as motivated by the desire of
capturing temporal structures at different temporal scales in a
theoretically well-defined manner.
Section~\ref{sec-spat-temp-scsp-concepts} then describes the temporal
scale-space concepts that satisfy these properties, with a distinction
on whether the auditory processing operations are required to be time-causal or not.
For off-line processing of pre-recorded sound signals, we may take
the liberty of accessing the virtual future in relation to any pre-recorded time
moment, whereas one in a real-time situation has to take explicit
account to the fact that the future cannot be accessed.
Thereby, we obtain different theories depending on whether time is
treated in a non-causal or a time-causal manner.

In section~\ref{sec-spectr} we apply these temporal scale-space
theories to the definition of multi-scale spectrograms by the
formulation of local windowed Fourier transforms of different temporal extent to
be able to capture temporal phenomena at different temporal scales.
Section~\ref{sec-2nd-layer-rec-fields} develops a corresponding theory
for spectro-temporal receptive fields applied to the spectrogram, and
it is shown how auditory receptive fields over the spectro-temporal
domain can be expressed in an analogous way to how visual
receptive fields are defined over space-time, with the conceptual
difference that the two spatial dimensions in vision are
replaced by a logarithmic frequency dimension.
Specifically, we demonstrate how basic auditory features can
computed in this way from spectro-temporal derivatives of idealized
receptive fields as obtained from the auditory scale-space theory.

Section~\ref{sec-biol-aud-rec-field} gives examples of how biological auditory receptive fields can be modelled
by the proposed theory.
Section~\ref{sec-rel-audio-proc} relates the presented theory to
previous approaches in audio processing, and
section~\ref{sec-summ-disc} concludes with a summary and discussion.

Appendix~\ref{app-freq-anal} complements the above 
treatment by an in-depth analysis of the frequency selectivity
properties of the temporal scale-space kernels.
Appendix~\ref{app-temp-dyn} gives a corresponding analysis of the
temporal delays of the time-causal receptive fields.
Finally, Appendix~\ref{sec-comp-impl} shows how the presented continuous theory
can be transferred to a discrete implementation while still preserving
the theoretical scale-space properties, and thereby allowing for 
theoretically well-founded digital implementation.

\section{Structural requirements on temporal receptive fields}
\label{sec-struct-req-temp-rec-fields}

In the following, we will describe a set of structural requirements that
can be stated concerning the temporal receptive fields for a general sensory system
that processes a scalar time-dependent signal regarding 
(i)~the measurement of sensory data with its close relationship to the
notion of temporal scale,
(ii)~internal derived representations of the signal that are to be
computed by a general sensory system, and
(iii)~the special nature of time.

If we regard the sensory signal $f$ as defined on a
one-dimensional continuous temporal axis $f \colon \bbbr \rightarrow \bbbr$, 
then the problem of defining a set of early sensory
operations can be formulated as finding a family of operators 
${\cal T}_{\tau}$ that are to act on $f$ to produce a family of new
intermediate representations of the signal%
\footnote{In equation~(\ref{eq-def-visual-op-T-s}), the symbol
                ``$\cdot$'' at the position of the first argument of
                $L$ is a place holder to emphasize that 
                in this relation, $L$ is regarded as a function and
                not evaluated with respect to its first argument $t$.
                The following semi-colon emphasizes the different
                natures of the temporal coordinate $t$ and the filter parameter $\tau$.}
\begin{equation}
  \label{eq-def-visual-op-T-s}
  L(\cdot;\; \tau) = {\cal T}_{\tau} \, f(\cdot)
\end{equation}
which are also to be defined as functions on $\bbbr$, {\em i.e.\/}, 
$L(\cdot;\; \tau) \colon \bbbr \rightarrow \bbbr$.

\paragraph{Linearity.}

If we want these initial visual processing stages to make as few irreversible 
decisions as possible, it is natural to initially require ${\cal T}_{\tau}$ to be a 
{\em linear operator\/} such that
 \begin{equation}
   {\cal T}_{\tau}(a_1 f_1 + a_2 f_2) = a_1 {\cal T}_{\tau} f_1 + a_2 {\cal T}_{\tau} f_2
\end{equation}
holds for all functions $f_1, f_2 \colon \bbbr \rightarrow \bbbr$ 
and all scalar constants $a_1, a_2 \in \bbbr$.

Linearity also implies that a number of special properties of
receptive fields (to be developed below) will transfer to 
temporal derivatives of these and do therefore
imply that different types of time-dependent structures in the signal will be treated in a
similar manner irrespective of what types of linear filters they are
captured by.

Concerning the assumption of linearity, it should be noted that there
is an implicit degree of freedom in this formulation concerning 
the parameterization of the units by
which the input signal $f$ is measured.
Given an underlying measurement signal $I(t)$ in units of energy from
a sensor, one could for a positive
signal also consider defining the input signal $f$ in terms of a 
reparameterization of the sensor signal $I$ according to a
self-similar power law
\begin{equation}
  f(t) = \left( I(t) \right)^{\alpha}
\end{equation}
for some $\alpha > 0$ or using a self-similar logarithmic transformation
\begin{equation}
  f(t) = \log \left( \frac{I(t)}{I_0} \right)
\end{equation}
defined relative to some reference level $I_0$.
Both of these transformations are self-similar in the sense that 
(i)~they
are well-behaved under rescalings of the measurement domain
$I(t) \mapsto a \, I(t)$ for $a > 0$ and 
(ii)~the local magnification/compression around any measurement value
as defined from the derivative will also follow a self-similar power law.

\paragraph{Temporal shift invariance.}

Let us require ${\cal T}_{\tau}$ to be a 
{\em shift-invariant operator\/} in the sense that it commutes with
the temporal shift operator ${\cal S}_{\Delta t}$ defined by
$({\cal S}_{\Delta t} f)(t) = f(t - \Delta t)$,
such that
\begin{equation}
   {\cal T}_{\tau} \left( {\cal S}_{\Delta t} f \right) 
   = {\cal S}_{\Delta t} \left( {\cal T}_{\tau} f \right)
  \end{equation}
holds for all $\Delta t \in \bbbr$.
The motivation behind this assumption is the basic requirement
that the representation of a sensory event should be similar
irrespective of the time at which it occurs.

\paragraph{Convolution structure.}

Together, the assumptions of linearity and shift-invariance imply that
the internal representations $L(\cdot;\; \tau)$ are given by 
{\em convolution transformations\/} \cite{HirWid55}
\begin{equation}
   L(t;\; \tau)  = (T(\cdot;\; \tau) * f)(t) =
   \int_{\xi \in \bbbr} 
      T(\xi;\; \tau) \, f(t - \xi) \, d\xi 
\end{equation}
where $T(\cdot;\; \tau)$ denotes some family of convolution kernels.
These convolution kernels and their temporal derivatives can
also be referred to as temporal receptive fields.

\paragraph{Regularity.}

To be able to use tools from functional analysis,
we will initially assume that both the original signal $f$ and 
the family of convolution kernels $T(\cdot;\; \tau)$
are in the Banach space $L^2(\bbbr)$,
{\em i.e.\/} that $f \in L^2(\bbbr)$ and 
$T(\cdot;\; \tau) \in L^2(\bbbr)$ 
with the norm
\begin{equation}
 \| f \|_2^2 
  = \int_{t \in \bbbr}  |f(t)|^2 \, dx \, dt.
\end{equation}
Then also the intermediate representations $L(\cdot;\; \tau)$ will 
be in the same Banach space, and the operators ${\cal T}_{\tau}$ 
can be regarded as well-defined.

\paragraph{Positivity (non-negativity).}

Concerning the convolution kernels $T$, one may require these to be
non-negative to constitute smoothing transformations
  \begin{equation}
    T(t;\; \tau) \geq 0.
  \end{equation}

\paragraph{Normalization.}

Furthermore, it is natural to require the convolution kernels to be
normalized to unit mass
  \begin{equation}
  \label{eq-norm-mass-of-kernel-spat-temp}
    \/ T(\cdot;\; \tau) \|_1 = \int_{t \in \bbbr} T(t;\; \tau) \, dt = 1
  \end{equation}
to leave a constant signal unaffected by the temporal smoothing transformation.

\paragraph{Quantitative measurement of the temporal extent and 
  the temporal offset of non-negative
  scale-space kernels.}

For a non-negative convolution kernel, we can measure the temporal
offset $\bar{t}$ by the mean operator 
\begin{equation}
  m = {\bar t} 
  = M(T(\cdot;\; \tau))  
  = \frac{\int_{t \in \bbbr} t \, T(t;\; \tau) \, dt}
             {\int_{t \in \bbbr} T(t;\; \tau) \, dt}
\end{equation}
and the temporal extent by the temporal variance 
\begin{equation}
  \Sigma 
  = V(T(\cdot;\; \tau))
  = \frac{\int_{t \in \bbbr} (t - \bar{t})^2 \, T(t;\; \tau) \, dt}
              {\int_{t \in \bbbr} T(t;\; \tau) \, dt}.
\end{equation}
Using the additive properties of mean values and variances
under convolution, which hold for non-negative distributions, it
follows that
\begin{align}
   \begin{split}
       \label{eq-add-prop-mean-val-spat-temp}
       m = M(T(\cdot;\; \tau_1) * T(\cdot;\; \tau_2)) =
       M(T(\cdot;\; \tau_1)) + M(T(\cdot;\; \tau_2)) = m_1 + m_2,
   \end{split}\\
   \begin{split}
       \label{eq-add-prop-cov-mat-spat-temp}
       \Sigma = V(T(\cdot;\; \tau_1) * T(\cdot;\; \tau_2)) 
        = V(T(\cdot;\; \tau_1)) + V(T(\cdot;\; \tau_2)) = \Sigma_1 + \Sigma_2.
   \end{split}
\end{align}

\paragraph{Identity operation with continuity.}

To guarantee that the limit case of the internal
scale-space representations when the scale parameter $\tau$
tends to zero should correspond to the original
image data $f$, we will assume that
\begin{equation}
   \label{eq-init-cond-scsp-zero-order-spat-temp}
    \lim_{\tau \downarrow 0} L(\cdot;\; \tau) 
   = \lim_{\tau \downarrow 0} {\cal T}_{\tau} f = f.
\end{equation}
Hence, the intermediate signal representations $L(\cdot;\, \tau)$ can be
regarded as a family of derived representations parameterized by
a temporal scale parameter $\tau$. 

\paragraph{Semi-group alternatively Markov structure over scale.}

For such sensory measurements to be properly related {\em between\/} different
temporal scales, it is natural to require the operators ${\cal T}_{\tau}$ with their
associated convolution kernels $T(\cdot;\; \tau)$ to form a
{\em semi-group\/} over $\tau$
\begin{equation}
  \label{eq-semi-group-spat-temp}
    {\cal T}_{\tau_1}  {\cal T}_{\tau_2} 
    =  {\cal T}_{\tau_1 + \tau_2} 
\end{equation}
which means that the composition of two convolution kernels from the
semi-group should also be a member of the same family of kernels 
and with added parameters values
\begin{equation}
    T(\cdot;\; \tau_1) * T(\cdot;\; \tau_2) 
    = T(\cdot;\; \tau_1 + \tau_2).
\end{equation}
Then, the transformation between any different and ordered
scale levels $\tau_1$ and $\tau_2$ with 
$\tau_2 \geq \tau_1$ will obey the 
{\em cascade property\/}
\begin{equation}
  \label{eq-casc-prop-spat-temp}
  L(\cdot;\; \tau_2) 
  = T(\cdot;\; \tau_2 - \tau_1) * T(\cdot;\; \tau_1) * f 
  = T(\cdot;\; \tau_2 - \tau_1) * L(\cdot;\; \tau_1)
\end{equation}
implying that we can compute the representation $L(\cdot;\; \tau_2)$ 
at a coarser scale from the representation $L(\cdot;\; \tau_1)$ 
at any finer scale using a similar type of transformation as when
computing the representation at any scale from the original
data $f$. 

For a temporal scale-space representation based on a discrete set of
temporal scale levels $\tau_{k}$ with $k = 0 \dots K$, we can
alternatively require a Markov property of the form
\begin{equation}
  \label{eq-Markov-prop}
  T(\cdot;\; \tau_{k+1})  = (\Delta T)(\cdot;\; k) \, T(\cdot;\; \tau_{k})
\end{equation}
where $(\Delta T)(\cdot;\; k)$ represents the transformation between
the adjacent scale levels $\tau_k$ and $\tau_{ki+1}$.
Then, the representation between any pair of temporal scale levels
$m < n$ 
\begin{equation}
  \label{eq-Markov-cascade}
  T(\cdot;\; \tau_n)  = (\Delta T)(\cdot;\; m \mapsto n) \, T(\cdot;\; \tau_m)
\end{equation}
will be given by convolution with the kernel
\begin{equation}
  (\Delta T)(\cdot;\; m \mapsto n) = *_{k = m}^{n-1} (\Delta T)(\cdot;\; k).
\end{equation}
The reason for relaxing the semi-group structure to a Markov structure
is to make it possible to take larger temporal scale steps $\delta \tau_k$ 
at coarser temporal scales, and thereby not requiring the
transformations between adjacent temporal scale levels to be equal.

A representation of a signal that possesses these properties is referred to as a
{\em temporal multi-scale representation\/}.

\paragraph{Self-similarity over scale.}

Regarding the family of convolution kernels used for computing a
multi-scale representation, one may require them to 
{\em self-similar over temporal scales\/}, 
such that all the kernels correspond to rescaled copies
\begin{equation}
  \label{eq-self-sim-req}
   T(t;\; \tau) 
  = \frac{1}{\varphi(\tau)} \bar{T} \left(\frac{t}{\varphi(\tau)} \right)
\end{equation}
of some prototype kernel $\bar{T}$ for 
some transformation%
\footnote{The reason for introducing a function $\varphi$ for
  transforming the scale parameter $s$ into a scaling factor
  $\varphi(\tau)$ over time, is that the requirement of a
  semi-group structure (\ref{eq-semi-group-spat-temp}) does not imply any restriction on how 
  the parameter $\tau$ should be related to sound measurements in dimensions
  of time --- the semi-group structure only implies an abstract
  ordering relation between coarser and finer scales $\tau_2 > \tau_1$ that
  could also be satisfied for any monotonously increasing
  transformation of the parameter $\tau$. 
  For the Gaussian temporal scale-space concept 
  according to equations  (\ref{eq-nD-scsp-undet-par-spat-temp})--(\ref{eq-gauss-gen-spattemp-2+1-D}) 
  this transformation is given by $\sigma = \varphi(\tau) = \sqrt{\tau}$.}
of $\varphi(\tau)$ of the temporal scale parameter $\tau$.

\paragraph{Temporal covariance.}

If the same sensory stimulus is recorded by two sensors that sample the
variations in the signal with different temporal sampling rates, or
if similar temporal events occur at a somewhat
different speed, it seems natural that the auditory system should be
able to relate the temporal scale-space
representations that are computed from the data.
Therefore, one may require that if the temporal dimension is rescaled
by a uniform scaling factor
\begin{equation}
  \label{eq-temp-sc-model}
   f' = {\cal S} \, f
 \quad \mbox{corresponding to} \quad
  f'(t') = f(t) 
   \quad \mbox{with} \quad
   t' = S \, t,
\end{equation}
then there should exist some transformation of the temporal scale
parameter $\tau' = B(\tau)$ such that the corresponding
temporal scale-space representations are equal:
\begin{equation}
   L'(t';\; \tau') = L(t;\; \tau) 
   \quad \mbox{corresponding to} \quad
   {\cal T}_{B(\tau)} \, {\cal B} \, f = {\cal B} \, {\cal T}_{\tau} \, f.
\end{equation}
In the case of a discrete set of temporal scale levels, we cannot
however require self-similarity or temporal covariance to hold exactly.
At best, we can aim at approximate transformation properties 
{\em e.g.\/} in terms
of the temporal variance of the temporal scale-space kernels.

\paragraph{Non-creation of new structures with increasing scale.} 

A necessary requirement on a scale-space representation is
that convolution with the scale-space kernel $T(\cdot;\; \tau)$
should correspond to {\em smoothing transformation\/}
in the sense that coarser scale representations should be guaranteed
to constitute {\em simplifications\/} of corresponding finer scale representations.
 
\paragraph{Non-creation of local extrema (zero-crossings).}

One way of formalizing such a requirement for a one-dimensional signal $f \colon \bbbr \rightarrow \bbbr$, 
is by the requirement that
the number of local extrema in the data must not increase with scale for any signal
and is referred to as {\em non-creation of local extrema\/}.
Formally, a one-dimensional kernel $T$ is a scale-space kernel if for
any signal $f$, the number of local extrema in $T * f$ is guaranteed
to not exceed the number of local extrema in $f$ \cite{Lin90-PAMI}.
It can be shown that for a one-dimensional signal, this condition can
also be equivalently expressed in terms of zero-crossings.

For one-dimensional signals, it can be shown that the requirement of
non-creation of local extrema implies that a scale-space kernel must
be positive and unimodal (having one peak only) both in the spatial 
domain and the Fourier domain \cite{Lin90-PAMI}.

\subsection{Specific scale-space axioms for a non-causal temporal domain}
\label{sec-non-caus-scsp-axioms}

Depending on the conditions under which the sensory data
is processed, we can consider two types of cases.
For pre-recorded signals, we may in
principle assume access to the data at all temporal moments
simultaneously and thereby apply operations to the signal that would
correspond to access to virtual future.
For real-time signal processing or when modelling biological perception, there is,
however, no way of having access to the future, which imposes
fundamental additional structural requirements on a temporal visual front-end.
For pre-recorded temporal signals, we require the following:

\paragraph{Non-enhancement of local extrema.} 

In the case of a continuous scale parameter, one way of formalizing the requirement of
non-creation of new image structures with increasing scale is that
{\em local extrema must not be enhanced with increasing scale\/}.
In other words, if a point $(t_0;\; \tau_0)$ is a local (spatial) maximum of the mapping
$t \mapsto L(t;\; \tau_0)$ then the value must not increase with scale.
Similarly, if a point $(t;\; \tau_0)$ is a local (spatial) minimum of the mapping
$t \mapsto L(t;\; \tau_0)$, then the value must not decrease with scale.
Given the above mentioned differentiability property with respect to
scale, we say that the multi-scale representation constitutes a
{\em scale-space representation\/} if it for a scalar scale parameter
satisfies the following conditions \cite{Lin96-ScSp}:
 \begin{align}
    \label{eq-non-enh-loc-extr}
    \begin{split}
      \partial_{\tau} L(t_0;\; \tau_0) \leq 0 \quad\quad \mbox{at any non-degenerate local maximum},
    \end{split}\\
    \begin{split}
      \partial_{\tau} L(t_0;\; t\tau_0) \geq 0 \quad\quad \mbox{at any non-degenerate local minimum}.
    \end{split}
  \end{align}

\noindent
By considering the response to a constant signal, it 
follows from the requirement of non-enhancement of local extrema that
a scale-space kernel should be normalized to unit $L_1$-norm,
corresponding to the normalization requirement in equation~(\ref{eq-norm-mass-of-kernel-spat-temp}).

\subsection{Special scale-space axioms for a time-causal temporal
  domain}
\label{sec-time-caus-scsp-axioms}

When processing sensory data in a real-time scenario,
the following additional temporal requirements are instead needed:

\paragraph{Temporal causality.}

For a sensory system that interacts in with the environment in a
real-time setting, a fundamental constraint on the convolution kernels
(the temporal receptive fields) is that there is no way of having access to future
information, which implies that the temporal smoothing kernels
must be {\em time-causal\/} in the sense that the convolution
kernel must be zero for any relative time moment that would imply
access to the future:
\begin{equation}
  \label{eq-def-time-caus}
    T(t;\; \tau) = 0 \quad \mbox{if} \quad t < 0.
\end{equation}
Note that the possibly pragmatic solution of using a truncated
symmetric filter of finite support in combination with a temporal
delay may not be appropriate for a time-critical real-time system,
since it would need to unnecessarily long time delays in particular at
coarser temporal scales. 
Therefore, a dedicated theory for truly time-causal spatio-temporal
scale-space concepts is needed.

\paragraph{Time-recursivity.}

Another fundamental constraint on a real-time system is that it cannot
be expected to keep a full record of everything that has happened in the past.
To keep down memory requirements it is therefore desirable that the
computations can be based on a limited internal 
{\em temporal buffer\/} $M(t)$, which should provide:
\begin{itemize}
\item
    a sufficient record of past information and
\item
   sufficient information to update its internal state in a recursive
   manner over time as new information arrives.
\end{itemize}
A particularly useful solution in this context is to use the internal 
temporal representations $L$ at different temporal scales
as a sufficient memory buffer of the past.

\paragraph{Non-creation of structure in the context of discrete
  temporal scale levels.}

For a temporal scale-space representation involving a discrete
set of scale levels only, we build on the requirement of
non-creation of local extrema as expressed for a one-dimensional
temporal signal depending on time $t$ only.
Let us therefore regard a one-dimensional temporal smoothing kernel $T_{time}$ 
as a {\em temporal scale-space kernel\/} if and only if the kernel is
(i)~time-causal and in addition (ii)~for any purely temporal signal $f$, 
the number of local extrema in $T _{time}* f$ is guaranteed
to not exceed the number of local extrema in $f$ \cite{LF96-ECCV}.

\section{Scale-space concepts for purely temporal domains}
\label{sec-spat-temp-scsp-concepts}

In this section we will describe how the structural requirements
listed in
section~\ref{sec-struct-req-temp-rec-fields} delimit the class of
temporal scale-space kernels and thus the class of possible temporal
receptive fields.

\subsection{Non-causal Gaussian temporal scale-space}
\label{sec-gauss-spat-temp-scsp}

If we for the purpose of analyzing pre-recorded auditory data allow for
unlimited freedom of accessing the sensory data at all temporal moments
simultaneously, we can apply a similar way of reasoning as has been
used for deriving scale-space concepts for image data over a spatial
domain \cite{Iij62,Wit83,Koe84,Lin93-Dis,Lin96-ScSp,Flo97-book,WeiIshImi99-JMIV,Haa04-book,Lin08-EncCompSci,Lin10-JMIV}.

Given time-dependent sensory data $f \colon \bbbr \rightarrow \bbbr$
defined over a one-dimensional temporal domain, let us
assume that the first stage of sensory processing as represented
by the operator ${\cal T}_{\tau}$ should satisfy the following
structural requirements:
(i)~{\em linearity\/},
(ii)~{\em shift invariance\/} and
(iii)~obey a {\em semi-group structure over temporal scales\/} $\tau$,
where we also have to assume 
(iv)~certain {\em regularity properties\/} of the semi-group 
${\cal T}_{\tau}$ {\em over scale\/} $\tau$
to guarantee sufficient differentiability properties with respect
  to time $t$ and temporal scales $\tau$.
Let us furthermore require
(iv)~{\em non-enhancement of local extrema\/} to hold for
{\em any\/} smooth function 
$f \in C^{\infty}(\bbbr) \cap L^1(\bbbr)$
and for any positive scale direction $s$.

Then, it follows from \cite[theorem~5, page 42]{Lin10-JMIV} that these conditions together
imply that the scale-space family $L$ must satisfy a diffusion
equation of the form
  \begin{equation}
    \label{eq-nD-scsp-undet-par-spat-temp}
    \partial_{\tau} L 
    = \frac{1}{2} \, \Sigma_0 \, \partial_{tt} L - \delta_0 \, \partial_t L
  \end{equation}
with initial condition $L(t;\; 0) = f(t)$ 
for some positive constant $\Sigma_0$ and some constant $\delta_0$.
Equivalently, this spatio-temporal scale-space representation at
scale $\tau$ can be obtained by convolution with 
{\em temporal Gaussian kernels\/} of the form
\begin{equation}
  \label{eq-gauss-gen-spattemp-2+1-D}
  g(t;\; \tau) 
   = \frac{1}{\sqrt{2 \pi \Sigma_{\tau}}} \,
      e^{- (t - \delta_{\tau})^2/2\tau}
\end{equation}
with $\Sigma_{\tau} = \tau \, \Sigma_0$ and time and $\delta_{\tau} =
\tau \, \delta_0$.
Since the parameter $\Sigma_0$ only corresponds to an unessential
rescaling of the temporal scale parameter $\tau$, we will in the
following set $\Sigma_0 = 1$.

\begin{figure}[!hbt]
  \begin{center}
    \begin{tabular}{ccc}
      $g(t;\; \tau, \delta)$ & $g_t(t;\; \tau, \delta)$ & $g_{tt}(t;\; \tau, \delta)$ \\
      \includegraphics[width=0.30\textwidth]{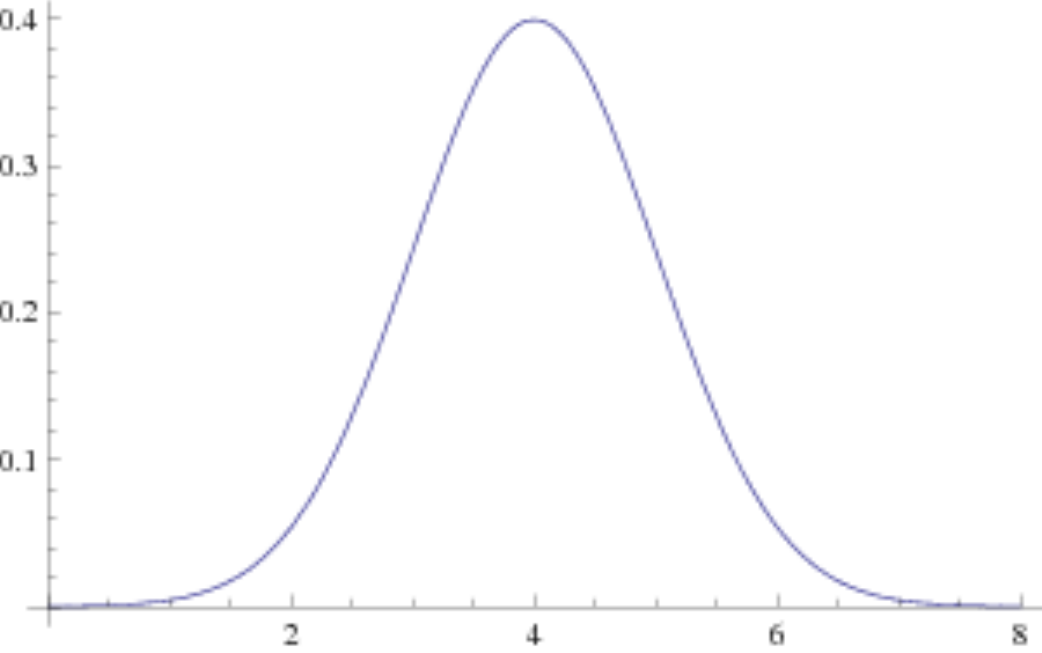} \hspace{-2mm} &
      \includegraphics[width=0.30\textwidth]{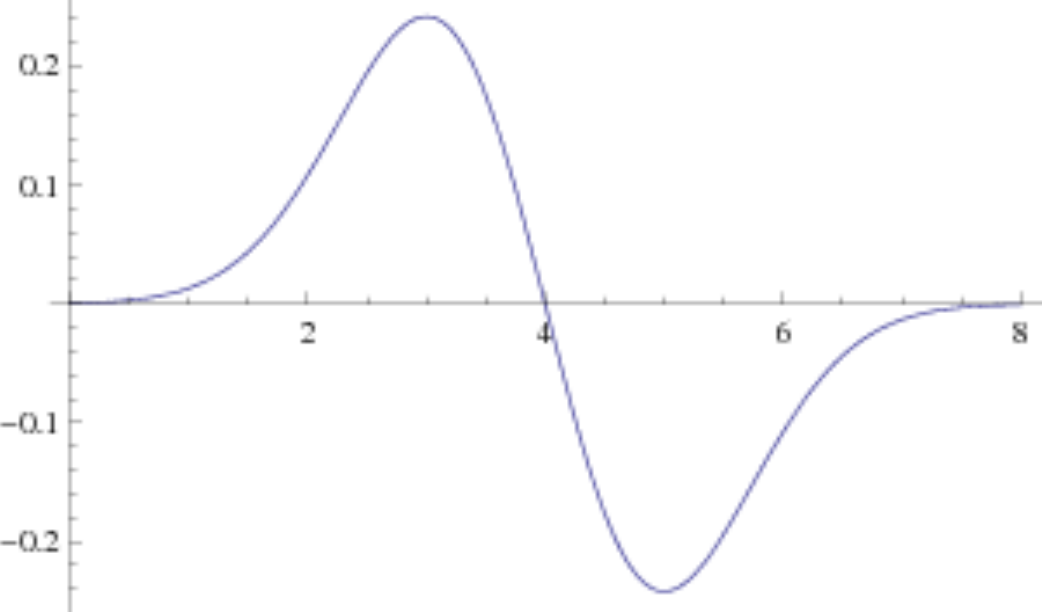} \hspace{-2mm} &
      \includegraphics[width=0.30\textwidth]{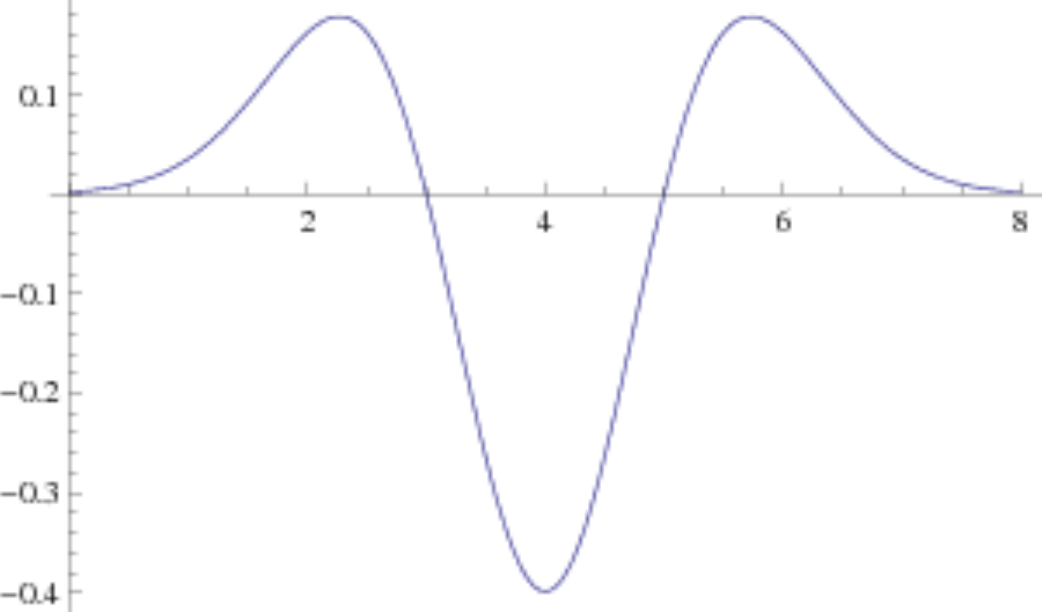} \\
    \end{tabular} 
  \end{center}
  \vspace{-3mm}
  \caption{The time-shifted Gaussian kernel 
           $g(t;\; \tau, \delta) = 
              1/\sqrt{2 \pi \tau} \exp(-(t-\delta)^2/2 \tau)$
           for $\tau = 1$ and $\delta = 4$ with its first- and
           second-order temporal derivatives.}
  \label{fig-time-shift-gauss-ders-1D}
\end{figure}

Graphs of these kernels are shown in 
  figure~\ref{fig-time-shift-gauss-ders-1D}.
  Notably, these kernels are not strictly time causal.
  To arbitrary degree of accuracy, however, they can 
  be approximated
  by truncated time-causal kernels, provided that the time delay $\delta$ is
  chosen sufficiently long in relation to the temporal scale $\tau$.
  Hence, the choice of $\delta$ leads to a trade-off between
  the computational accuracy of the implementation and the
  temporal response properties as delimited by a non-zero time delay.
  This problem, however, arises only for real-time analysis.
  For off-line computations, the time delay can in many cases be set
  to zero, corresponding to kernels that are mirror symmetric 
  $T(-t;\, s) = T(t;\; s)$ through the origin.
  In this respect, the truncated and time-shifted Gaussian kernels
  may serve as a simplest possible model for  a
  temporal scale-space representation, provided that the
  requirements of temporal causality and temporal recursivity can be
  relaxed.

\paragraph{Derived receptive fields in terms of temporal derivatives.}

In addition to the zero-order smoothing kernel $T$, we have in
figure~\ref{fig-time-shift-gauss-ders-1D} also shown its first- and
second-order temporal derivatives $T_t$ and $T_{tt}$.
The reason for this is that derivatives of scale-space kernels do also
obey desirable structural properties in terms of linearity, shift
invariance and nice properties over scale in terms of non-enhancement
of local extrema, with the semi-group property replaced by a
cascade property over scale 
\begin{equation}
  (\partial_{t^{\alpha}} L)(\cdot;\; \tau_2) 
  = T(\cdot;\; \tau_2-\tau_1) *  (\partial_{t^{\alpha}} L) (\cdot;\; \tau_1) 
\end{equation}
and with the limit case when the temporal scale goes to zero
(\ref{eq-init-cond-scsp-zero-order-spat-temp}) replaced by
\begin{equation}
   \label{eq-init-cond-scsp-zero-order-spat-temp-der}
    \lim_{\tau \downarrow 0} (\partial_{t^{\alpha}} L)(\cdot;\; \tau) 
   = \lim_{\tau \downarrow 0} \partial_{t^{\alpha}} ({\cal T}_{\tau} f) = \partial_{t^{\alpha}} f
\end{equation}
provided that the corresponding derivative of $f$ exists.
Regarding temporal receptive fields that are expressed in terms of
derivatives of scale-space kernels, the normalization condition
(\ref{eq-norm-mass-of-kernel-spat-temp}) is replaced by the integral
of the receptive field being zero
  \begin{equation}
    \| (\partial_{t^{\alpha}}  T)(\cdot;\; \tau) \|_1 = \int_{t \in \bbbr} (\partial_{t^{\alpha}}  T)(t;\; \tau) \, dt = 0.
  \end{equation}
In all other major respects, such receptive fields satisfy essential
scale-space properties in terms of non-creation of new structures with
increasing scale in the sense that local extrema in the receptive
field response are not enhanced from a fine to a coarser scale or that
the number of local extrema or zero-crossings in the signal is
guaranteed to not increase from any fine to any coarser scale.

In addition, receptive fields that are expressed in terms of temporal
derivatives are invariant under additive transformations of the signal
\begin{equation}
   f(t) \mapsto f(t) + C
\end{equation}
and thereby providing a mechanism for capturing local variations in
the signal under variabilities of its magnitude.

\subsection{Time-causal temporal scale-space}
\label{sec-time-caus-scale-spaces}

When constructing a system for real-time processing of sensory data,
a fundamental constraint on the temporal smoothing
kernels is that they have to be time-causal. 
As previously mentioned, the ad hoc solution of using a truncated
symmetric filter of finite temporal extent in combination with a
temporal delay is not appropriate in a time-critical context.
Because of computational and memory efficiency, the computations
should furthermore be based on a compact temporal buffer that contains
sufficient information for representing sensory information at multiple
temporal scales and computing features therefrom.
Corresponding requirements are also necessary in computational
modelling of biological perception.

\paragraph{Time-causal scale-space kernels for a purely temporal domain.}

\begin{figure}[!h]
   \begin{center}
      \includegraphics[width=0.60\textwidth]{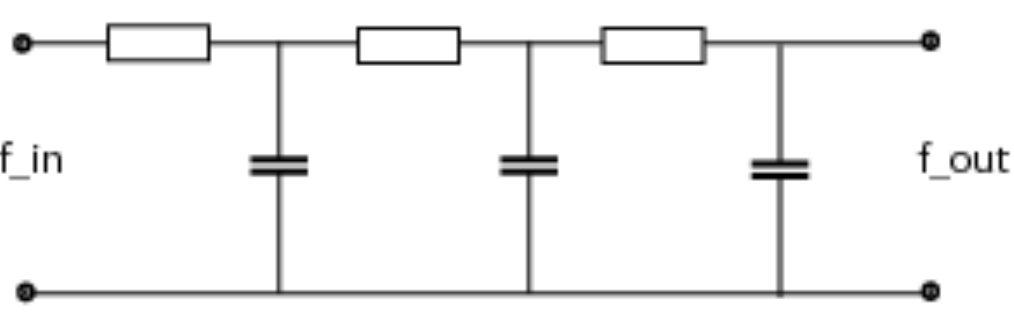}
   \end{center}
\caption{Electric wiring diagram consisting of a set of resistors and
  capacitors that emulate a series of first-order integrators coupled in
  cascade, if we regard the time-varying voltage $f_{in}$ as
  representing the time varying input signal and the resulting output
  voltage $f_{out}$ as representing the time varying output signal at a
  coarser temporal scale.
  According to the theory of temporal scale-space
  kernels for one-dimensional signals \protect\cite{Lin90-PAMI,LF96-ECCV},
  the corresponding equivalent truncated exponential kernels are the only
  primitive temporal smoothing kernels that guarantee both temporal
  causality and non-creation of local extrema
  (alternatively zero-crossings) with increasing temporal scale.
  Such first-order temporal integration can be used as a straightforward
  computational model for temporal processing in biological neurons
  (see also \protect\cite[Chapters~11--12]{Koch99-book} regarding 
  physical modelling of the information transfer in dendrites of neurons).}
  \label{fig-first-order-integrators-electric}
\end{figure}

Given the requirement on a temporal scale-space kernel in terms of
non-creation of local extrema over a purely temporal domain,
truncated exponential kernels 
  \begin{equation}
    h_{exp}(t;\; \mu_i) 
    = \left\{
        \begin{array}{ll}
          \frac{1}{\mu_i} e^{-t/\mu_i} & t \geq 0 \\
          0         & t < 0
        \end{array}
      \right.
  \end{equation}
can be shown to constitute the only
class of time-causal scale-space kernels over a continuous domain \cite{Lin90-PAMI,LF96-ECCV}.
The Laplace transform of such a kernel is given by
\begin{equation}
    H_{exp}(q;\; \mu_i) 
    = \int_{t = - \infty}^{\infty} h_{exp}(t;\; \mu_i) \, e^{-qt} \, dt
    = \frac{1}{1 + \mu_i q}
  \end{equation}
  and by coupling $k$ such kernels in cascade, we obtain a composed filter
  \begin{equation}
    \label{eq-comp-trunc-exp-cascade}
    h_{composed}(t;\; \mu) 
    = *_{i=1}^{k} h_{exp}(t;\; \mu_i)
  \end{equation}
  having a Laplace transform of the form
  \begin{equation}
    \label{eq-expr-comp-kern-trunc-exp-filters}
    H_{composed}(q;\; \mu) 
    = \int_{t = - \infty}^{\infty} (*_{i=1}^{k} h_{exp}(t;\; \mu_i)) \, e^{-qt} \, dt
    =  \prod_{i=1}^{k} \frac{1}{1 + \mu_i q}
  \end{equation}
  The composed filter has temporal mean value
  \begin{equation}
    \label{eq-mean-trunc-exp-filters}
    M(h_{composed}(\cdot;\; \mu)) = \sum_{t=1}^{k} \mu_i
  \end{equation}
  and temporal variance
  \begin{equation}
    \label{eq-var-trunc-exp-filters}
    \tau_k = V(h_{composed}(\cdot;\; \mu)) = \sum_{t=1}^{k} \mu_i^2
  \end{equation}
In terms of physical models, repeated convolution with this set of
truncated exponential kernels corresponds to coupling a series
of first-order integrators with time constants $\mu_k$ in cascade:
\begin{equation}
   \partial_t L(t;\; \tau_k) 
   = \frac{1}{\mu_k} \left( L(t;\; \tau_{k-1}) - L(t;\; \tau_k) \right)
\end{equation}
with $L(t;\; 0) = f(t)$.
These temporal smoothing kernels satisfy scale-space properties in
  the sense that the number of local extrema or the number of zero-crossings
  in the temporal signal are guaranteed to not increase with
  the temporal scale.
  In this respect, these kernels have a desirable and well-founded smoothing
  property that can be used for defining multi-scale observations over time.
  A limitation of this type of temporal scale-space representation,
  however, is that the {\em scale levels are required to be discrete\/}
  and that the scale-space representation does hence not admit
  a continuous scale parameter.
  Computationally, however, the scale-space representation based
  on truncated exponential kernels can be highly efficient and admits
  for direct implementation in terms of hardware (or wetware) that emulates
  first-order integration over time 
  (see figure~\ref{fig-first-order-integrators-electric} for
  illustration of a corresponding electric wiring diagram).

\begin{figure}[p]
  \begin{center}
    \begin{tabular}{ccc}
      $h(t;\; \mu, K=4)$ & $h_{t}(t;\; \mu, K=4)$ & $h_{tt}(t;\; \mu, K=4)$ \\
      \includegraphics[width=0.30\textwidth]{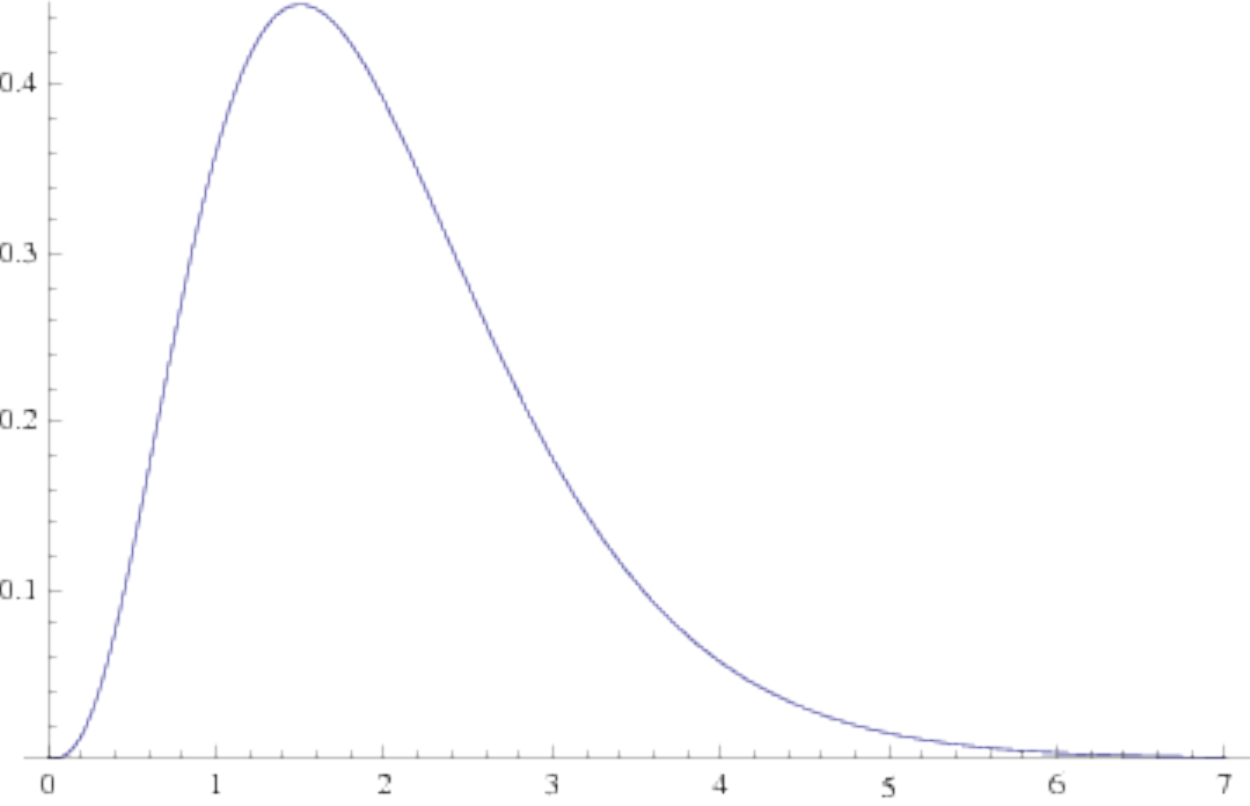} \hspace{-2mm} &
      \includegraphics[width=0.30\textwidth]{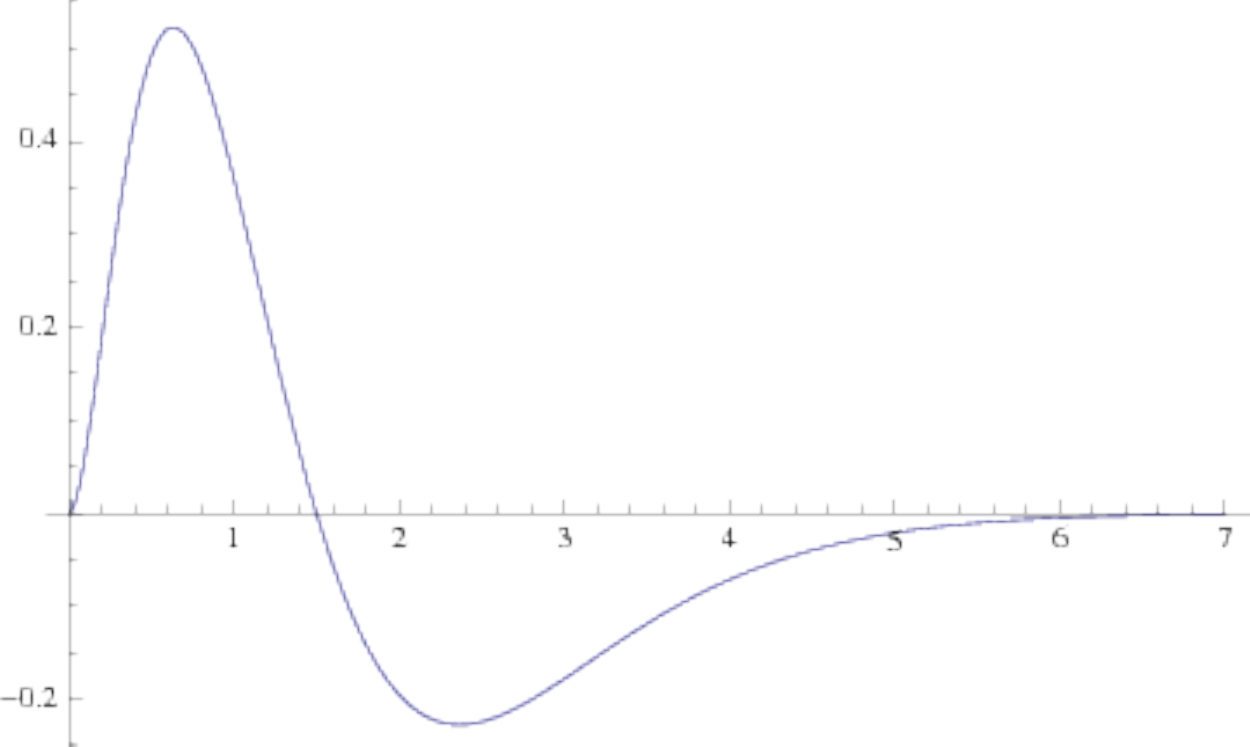} \hspace{-2mm} &
      \includegraphics[width=0.30\textwidth]{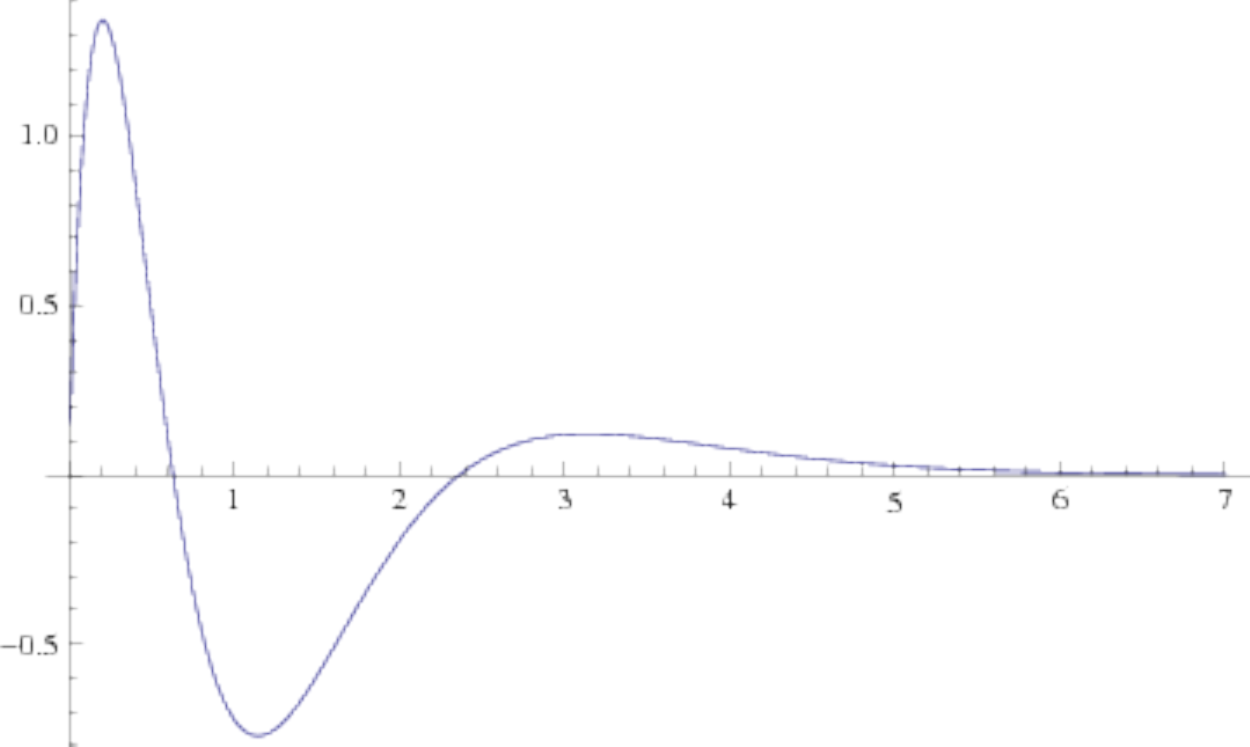} \\
    \end{tabular} 
  \end{center}

  \begin{center}
    \begin{tabular}{ccc}
      $h(t;\; \mu, K=7)$ & $h_{t}(t;\; \mu, K=7)$ & $h_{tt}(t;\; \mu, K=7)$ \\
      \includegraphics[width=0.30\textwidth]{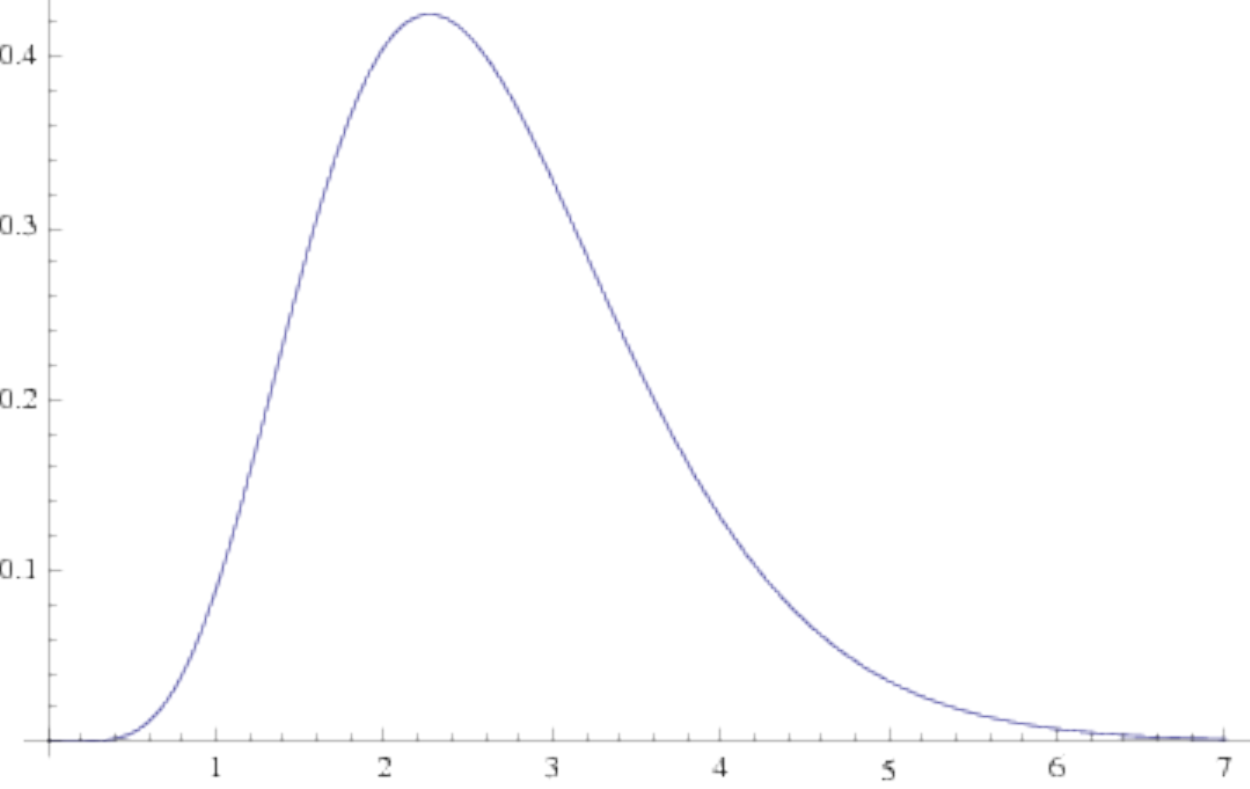} \hspace{-2mm} &
      \includegraphics[width=0.30\textwidth]{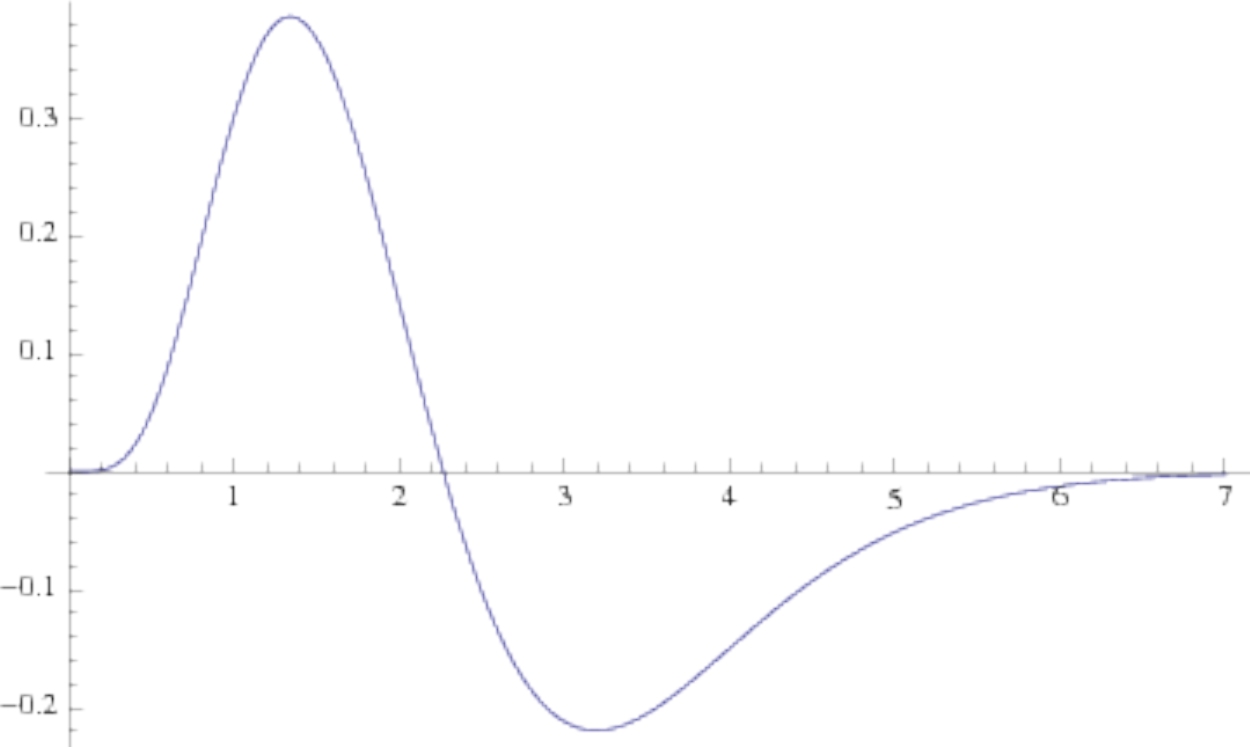} \hspace{-2mm} &
      \includegraphics[width=0.30\textwidth]{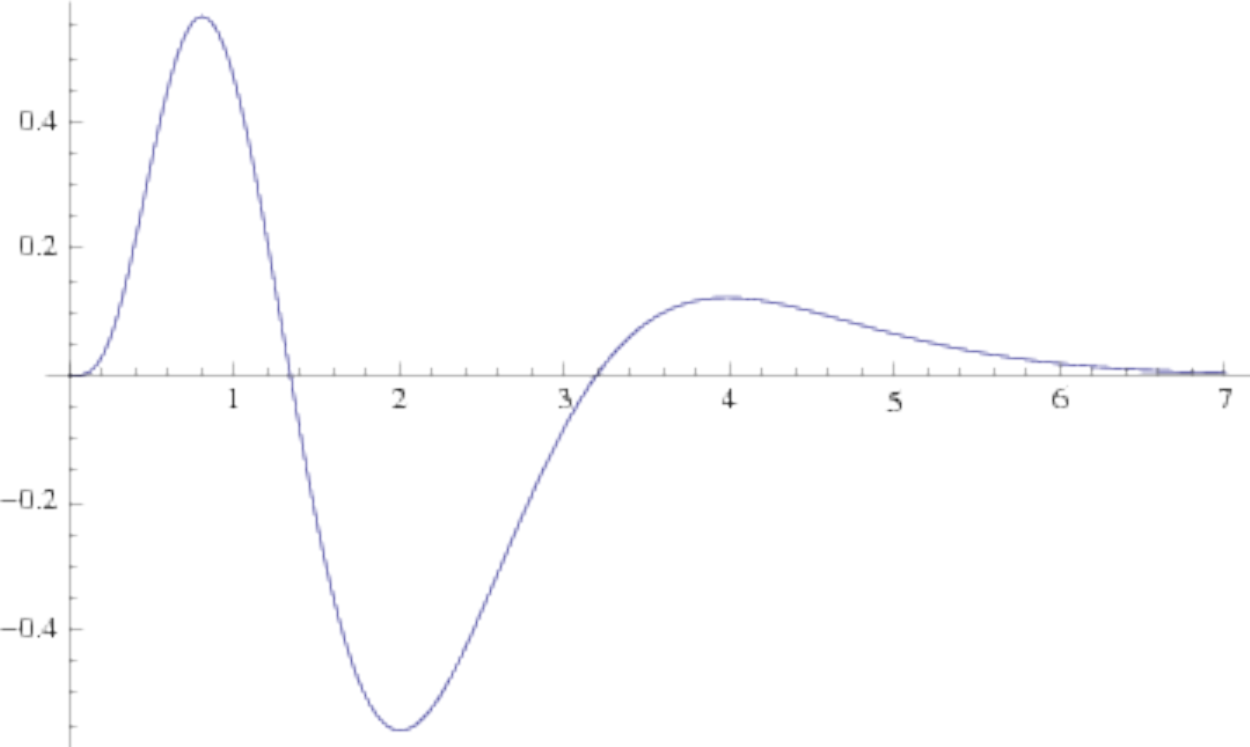} \\
    \end{tabular} 
  \end{center}
  \vspace{-3mm}
  \caption{Equivalent kernels 
           $h_{composed}(t;\; \mu) = *_{i=1}^{k} h_{exp}(t;\; \mu)$
           with temporal variance $\tau = 1$ corresponding to the composition of
           $k$ truncated exponential kernels 
           $h_{exp}(t;\; \mu) = \frac{1}{\mu} \exp{-t/\mu}$ 
           with similar time constants $\mu$ and
           their first- and second-order derivatives.
           (top row) $k = 4$ and $\mu = \sqrt{1/4}$.
           (bottom row) $k = 7$ and $\mu = \sqrt{1/7}$.}
  \label{fig-trunc-exp-kernels-1D}

  \bigskip

 \begin{center}
    \begin{tabular}{ccc}
      {\small $h(t;\; K=4, c = \sqrt{2})$} 
      & {\small $h_{t}(t;\; K=4, c = \sqrt{2})$}
      & {\small $h_{tt}(t;\; K=4, c = \sqrt{2}))$} \\
      \includegraphics[width=0.30\textwidth]{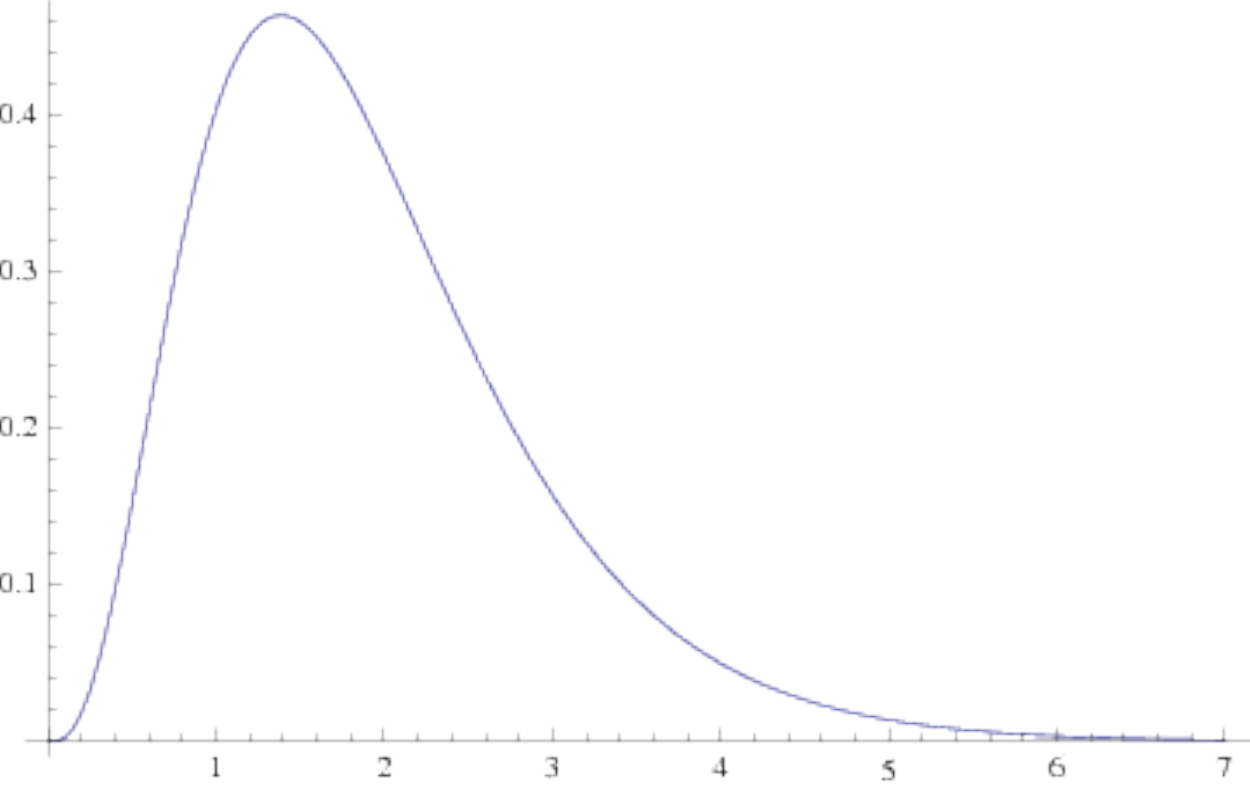} \hspace{-2mm} &
      \includegraphics[width=0.30\textwidth]{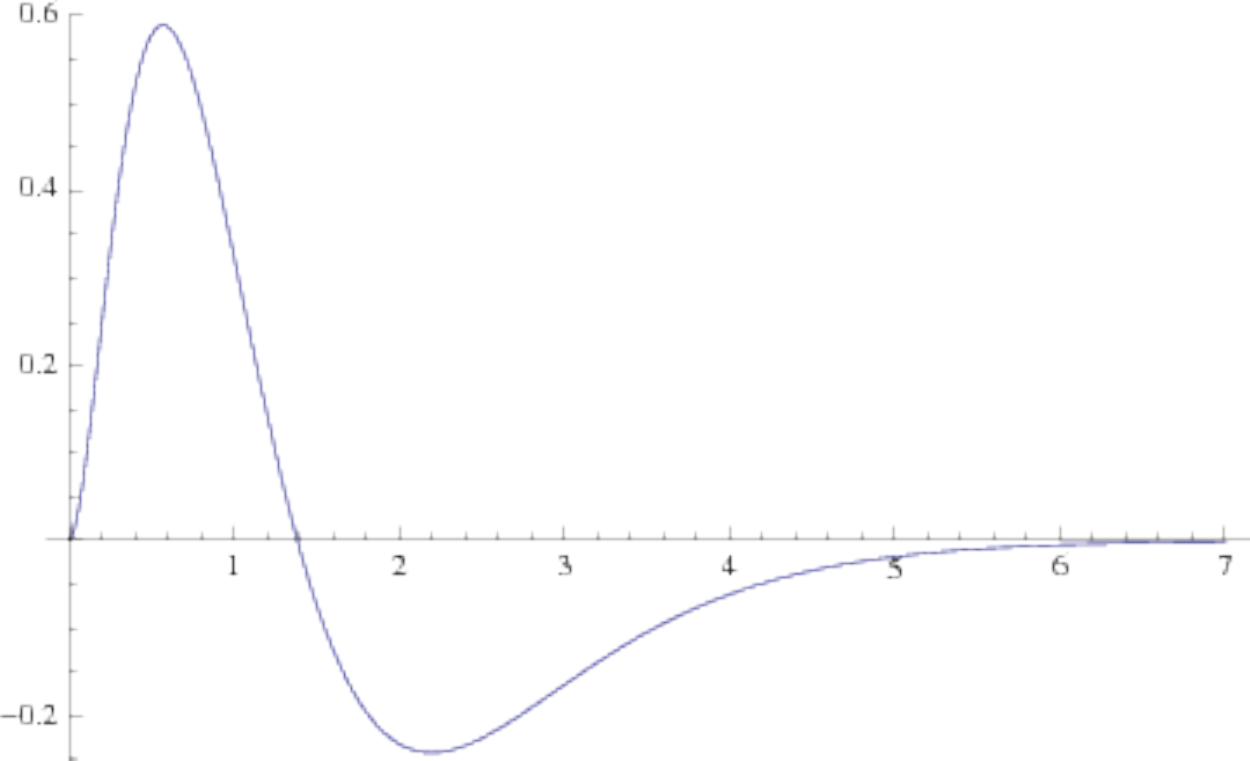} \hspace{-2mm} &
      \includegraphics[width=0.30\textwidth]{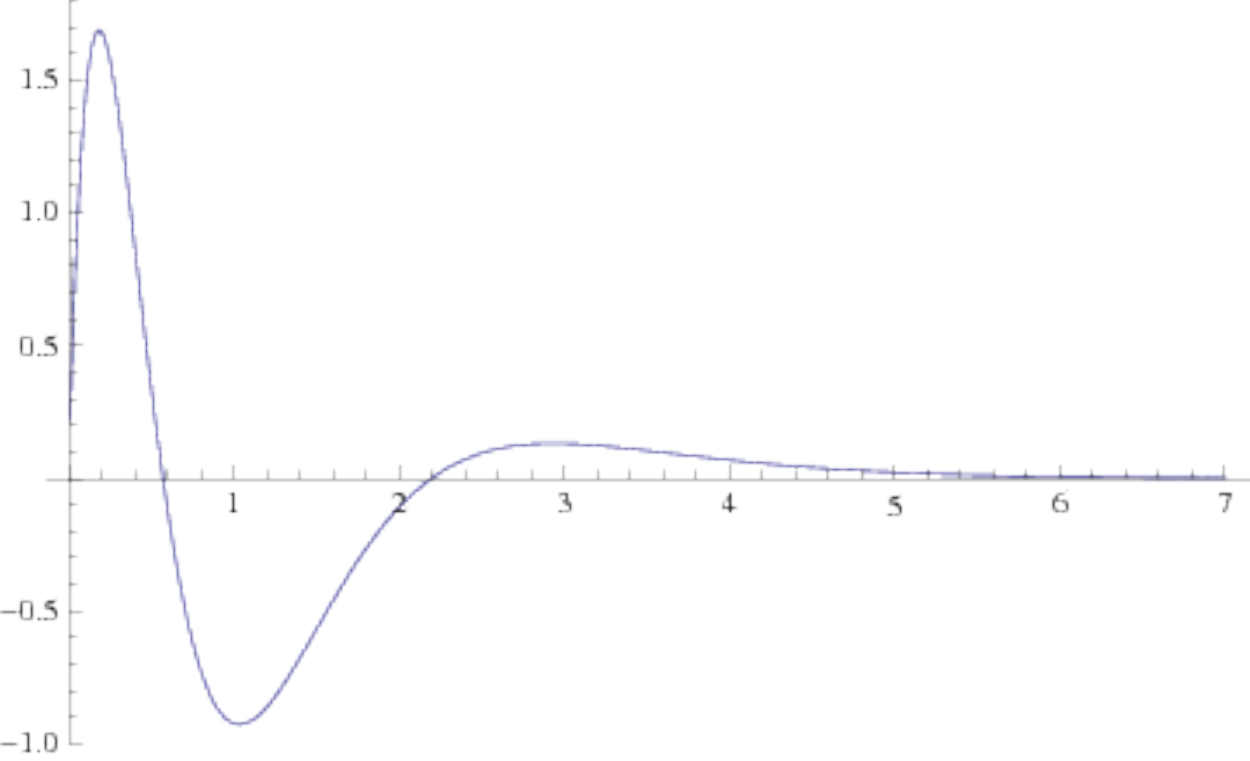} \\
    \end{tabular} 
  \end{center}

  \begin{center}
    \begin{tabular}{ccc}
      {\small $h(t;\; K=7, c = \sqrt{2})$} 
      & {\small $h_{t}(t;\; K=7, c = \sqrt{2}))$} 
      & {\small $h_{tt}(t;\; K=7, c = \sqrt{2}))$} \\
      \includegraphics[width=0.30\textwidth]{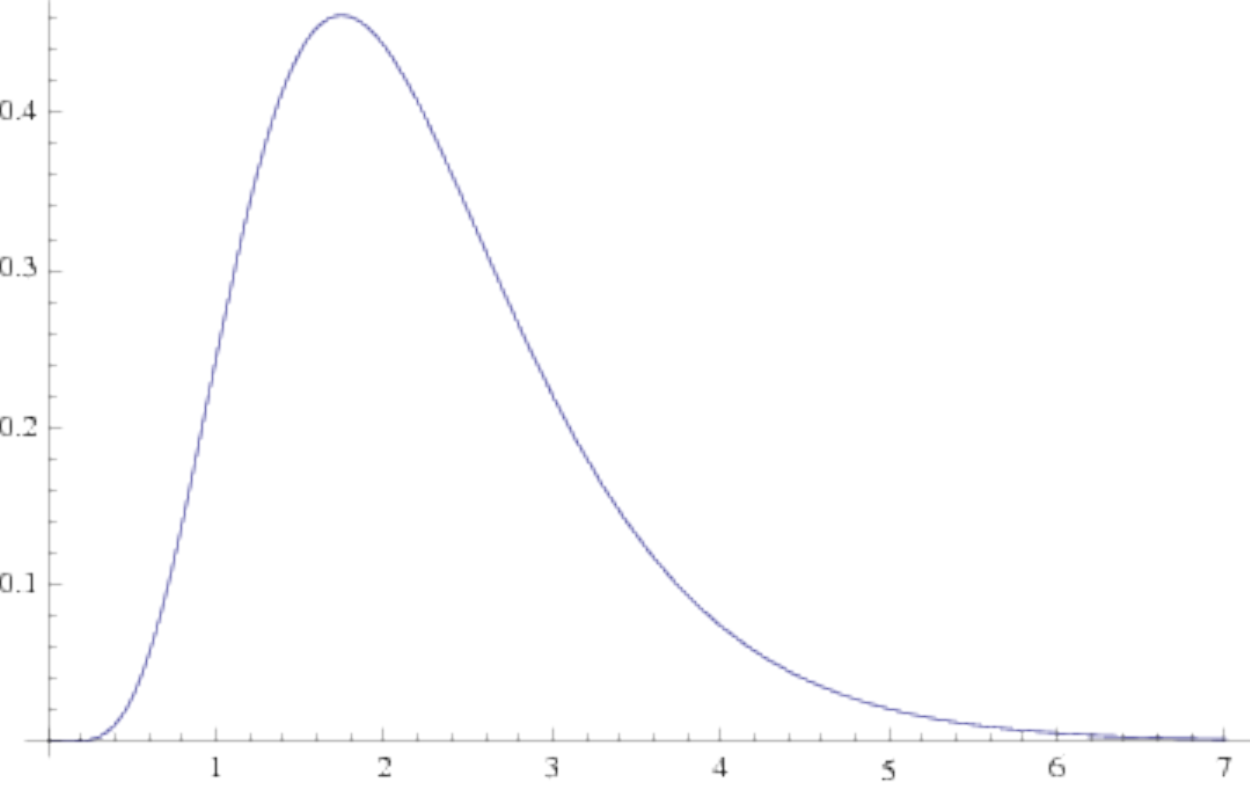} \hspace{-2mm} &
      \includegraphics[width=0.30\textwidth]{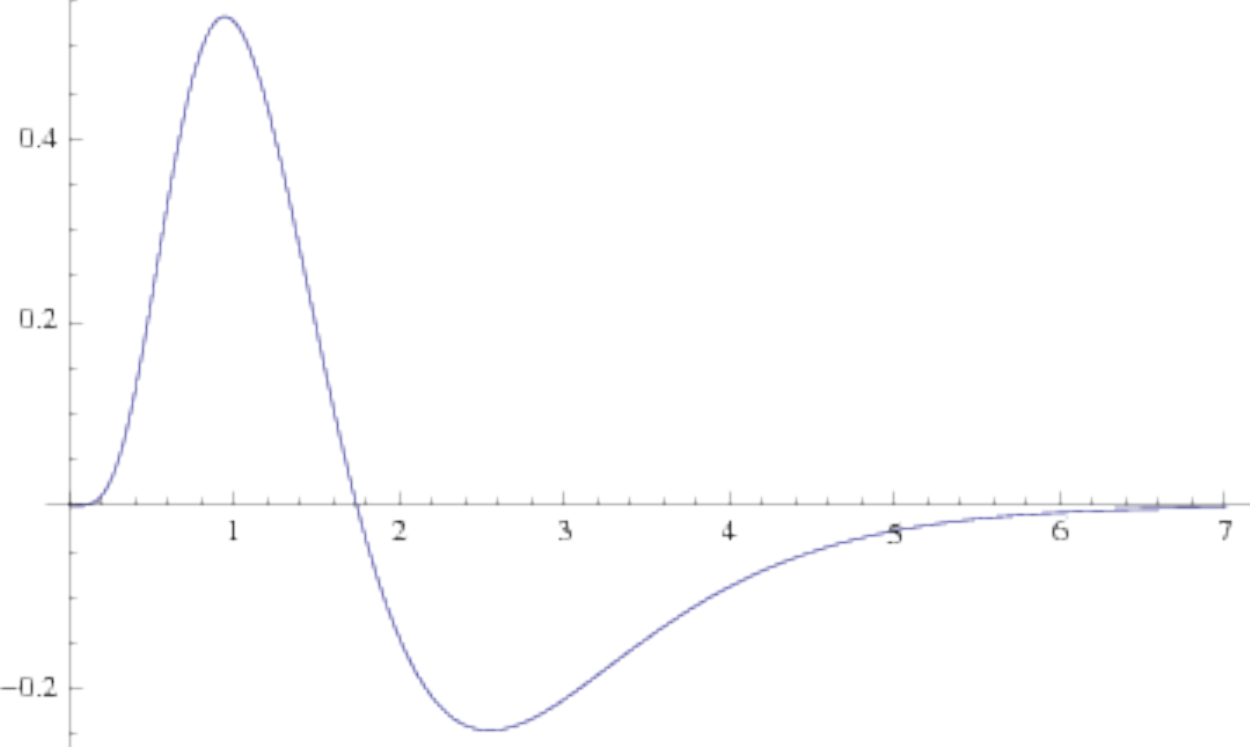} \hspace{-2mm} &
      \includegraphics[width=0.30\textwidth]{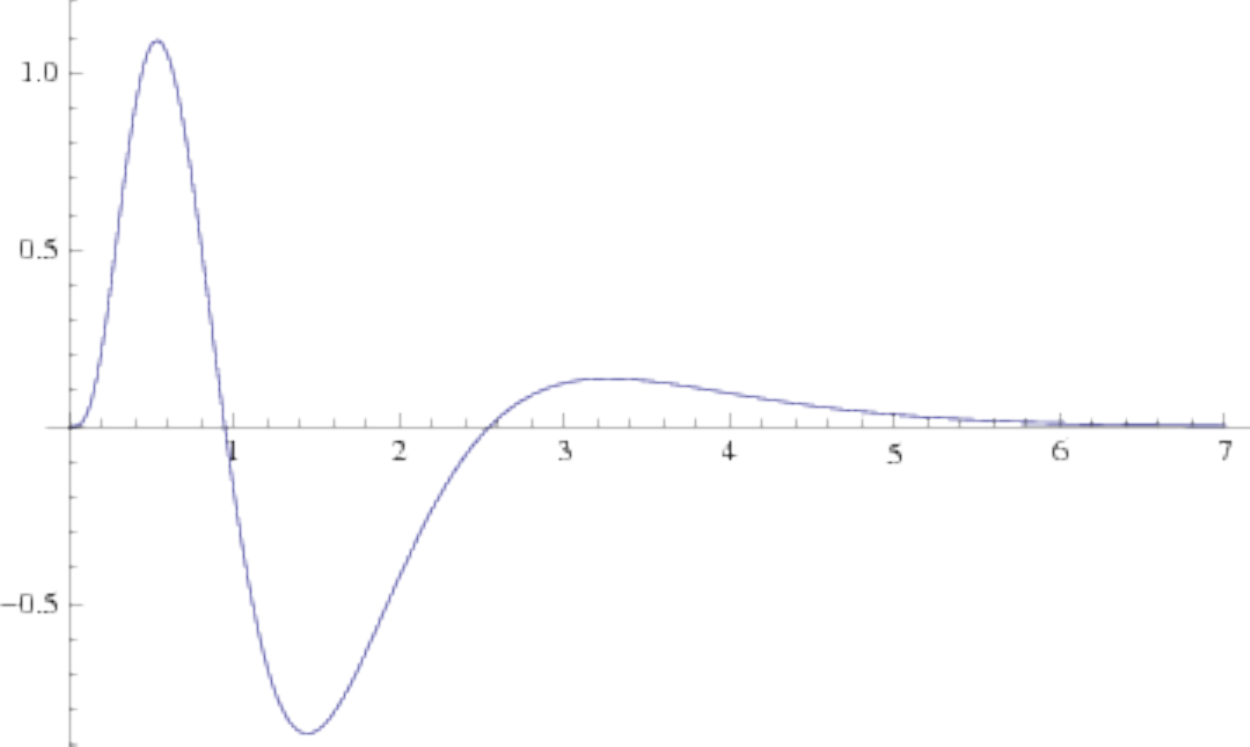} \\
    \end{tabular} 
  \end{center}
  
 \vspace{-3mm}
  \caption{Equivalent kernels 
           $h_{composed}(t;\; \mu) = *_{i=1}^{k} h_{exp}(t;\; \mu_i)$
           with temporal variance $\tau = 1$
           corresponding to the composition of
           $K = 4$ or $K = 7$ truncated exponential kernels with different time constants
           defined from a self-similar distribution of the temporal
           scale levels according to
           equations~(\protect\ref{eq-distr-tau-values}),
          (\protect\ref{eq-mu1-log-distr}) and (\protect\ref{eq-muk-log-distr})
           and corresponding to a
           uniform distribution in terms of effective temporal scale
           $\tau_{eff} = \log \tau$ for $c = \sqrt{2}$ and with their
           first- and second-order derivatives.}
  \label{fig-trunc-exp-kernels-nonequal-mu-1D-csqrt2}
\end{figure}

\begin{figure}[hbtp]
\begin{center}
    \begin{tabular}{ccc}
      {\small $h(t;\; K=4, c = 2^{3/4})$} 
      & {\small $h_{t}(t;\; K=4, c = 2^{3/4})$}
      & {\small $h_{tt}(t;\; K=4, c = 2^{3/4}))$} \\
      \includegraphics[width=0.30\textwidth]{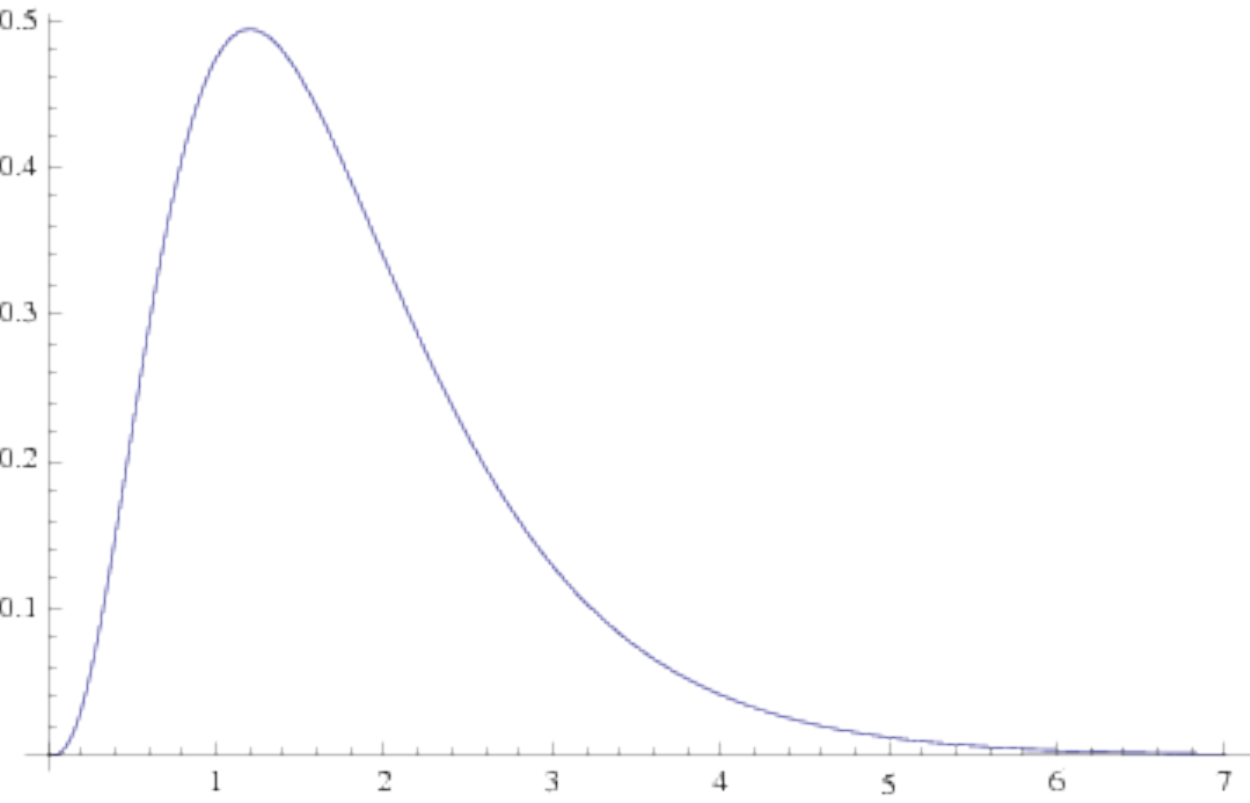} \hspace{-2mm} &
      \includegraphics[width=0.30\textwidth]{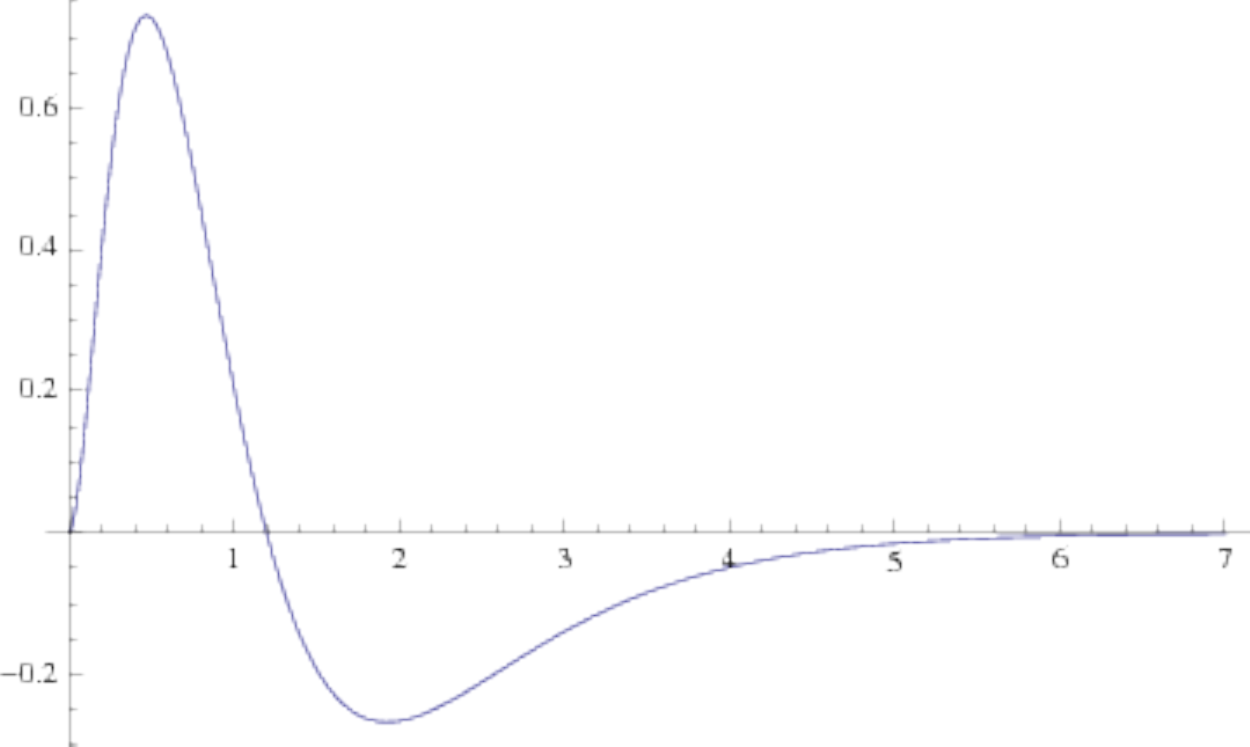} \hspace{-2mm} &
      \includegraphics[width=0.30\textwidth]{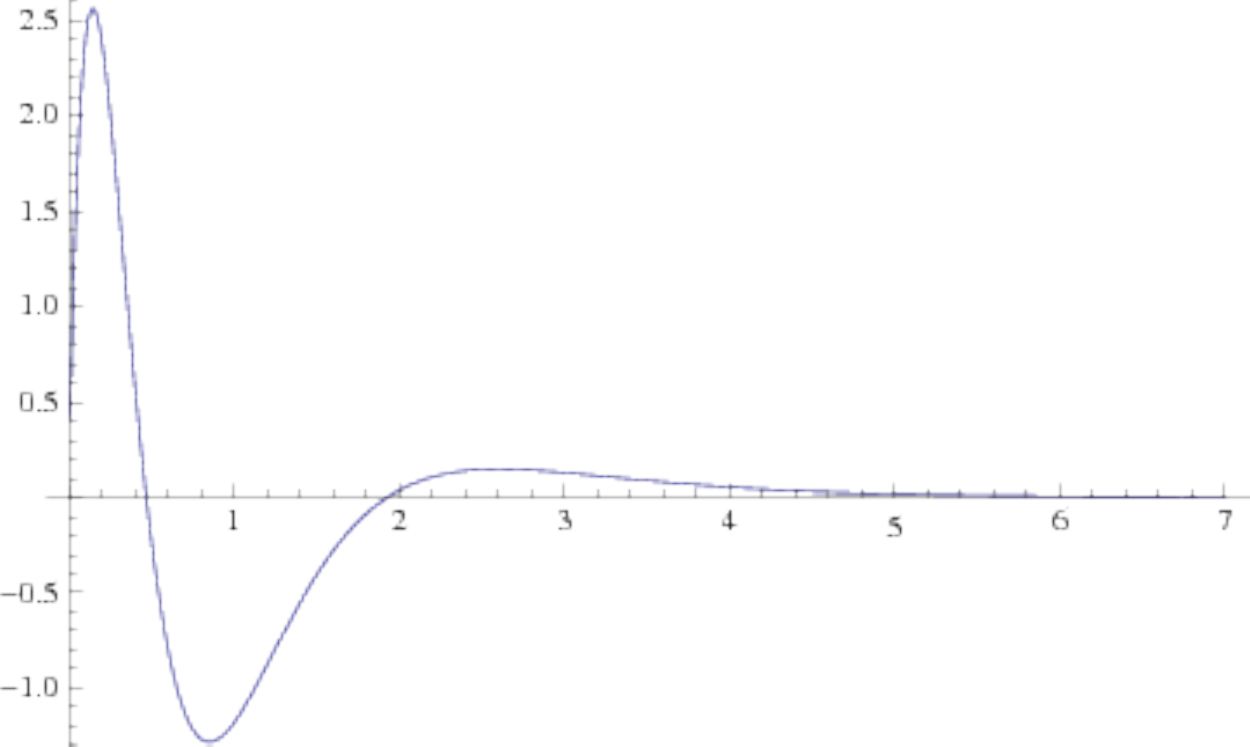} \\
    \end{tabular} 
  \end{center}

  \begin{center}
    \begin{tabular}{ccc}
      {\small $h(t;\; K=7, c = 2^{3/4})$} 
      & {\small $h_{t}(t;\; K=7, c = 2^{3/4}))$} 
      & {\small $h_{tt}(t;\; K=7, c = 2^{3/4}))$} \\
      \includegraphics[width=0.30\textwidth]{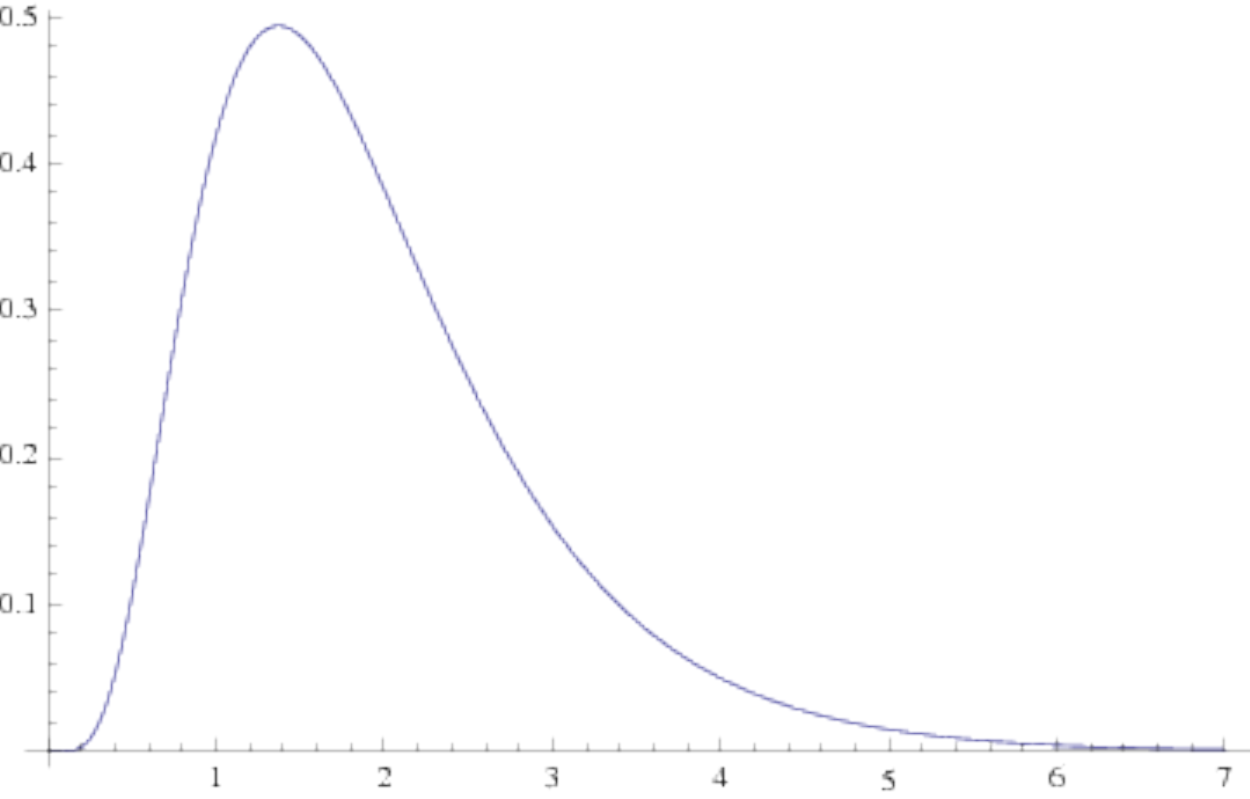} \hspace{-2mm} &
      \includegraphics[width=0.30\textwidth]{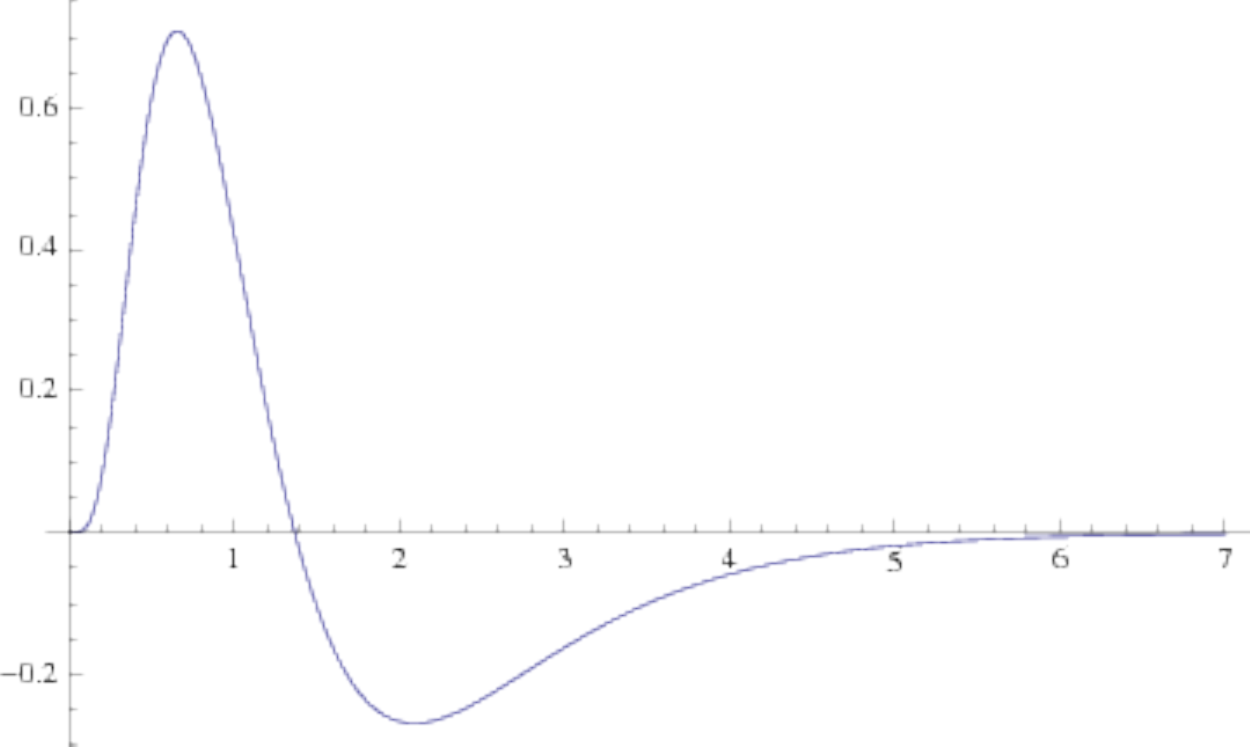} \hspace{-2mm} &
      \includegraphics[width=0.30\textwidth]{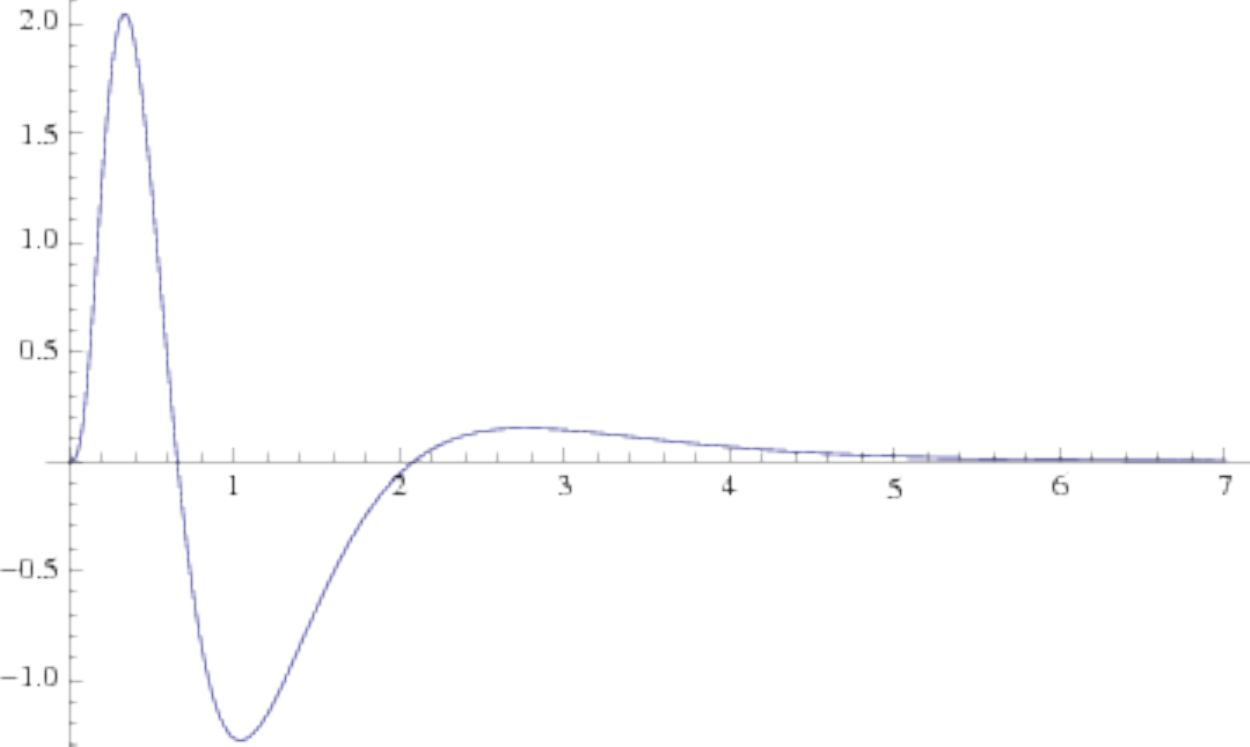} \\
    \end{tabular} 
  \end{center}
  
 \vspace{-3mm}
  \caption{Equivalent kernels 
           $h_{composed}(t;\; \mu) = *_{i=1}^{k} h_{exp}(t;\; \mu_i)$
           with temporal variance $\tau = 1$
           corresponding to the composition of
           $K = 4$ or $K = 7$ truncated exponential kernels with different time constants
           defined from a self-similar distribution of the temporal
           scale levels according to
           equations~(\protect\ref{eq-distr-tau-values}),
          (\protect\ref{eq-mu1-log-distr}) and (\protect\ref{eq-muk-log-distr})
           and corresponding to a
           uniform distribution in terms of effective temporal scale
           $\tau_{eff} = \log \tau$ for $c = 2^{3/4}$ and with their
           first- and second-order derivatives.}
  \label{fig-trunc-exp-kernels-nonequal-mu-1D-c2r0p75}
\end{figure}

When implementing this temporal scale-space concept, a set of
intermediate scale levels has to be distributed between some minimum
scale level $\tau_{min}$ and some maximum scale level $\tau_{max}$.
Assuming that a total number of $K$ scale levels is to be used, it is
natural to distribute the temporal scale levels according to a geometric series,
corresponding to a uniform distribution in units of
{\em effective temporal scale\/} $\tau_{eff} = \log \tau$
\cite{Lin92-PAMI}.
Using such a logarithmic distribution of the temporal scale levels,
the different levels in the temporal scale-space representation at
increasing temporal scales will serve as a logarithmic memory of the
past, with qualitative similarity similarity to the mapping of the
past onto a logarithmic time axis in the scale-time model by \cite{Koe88-BC}.
If we have the freedom of choosing $\tau_{min}$ freely, a natural way
of parameterizing these temporal scale levels is by using a distribution
parameter $c > 1$ such that
\begin{equation}
  \label{eq-distr-tau-values}
  \tau_k = c^{2(k-K)} \tau_{max} \quad\quad (1 \leq k \leq K)
\end{equation}
which by equation~(\ref{eq-var-trunc-exp-filters}) implies that time
constants of the individual first-order integrators will be given by
\begin{align}
  \begin{split}
     \label{eq-mu1-log-distr}
     \mu_1 & = c^{1-K} \sqrt{\tau_{max}}
  \end{split}\\
  \begin{split}
     \label{eq-muk-log-distr}
     \mu_k & = \sqrt{\tau_k - \tau_{k-1}} = c^{k-K-1} \sqrt{c^2-1} \sqrt{\tau_{max}} \quad\quad (2 \leq k \leq K)
  \end{split}
\end{align}
If the temporal signal is on the other hand given at some minimum
temporal scale level $\tau_{min}$, 
we can instead determine $c$ in (\ref{eq-distr-tau-values})
such that $\tau_1 = \tau_{min}$ 
\begin{equation}
  c = \left( \frac{\tau_{max}}{\tau_{min}} \right)^{\frac{1}{2(K-1)}}
\end{equation}
and add $K - 1$ temporal scale levels with $\mu_k$ according to (\ref{eq-muk-log-distr}).
Alternatively, if one chooses a uniform distribution of the
intermediate temporal scale levels 
\begin{equation}
  \label{eq-distr-tau-values-uni}
  \tau_k = \frac{k}{K} \, \tau_{max}
\end{equation}
implying
\begin{equation}
  \mu_k = \mu = \sqrt{\frac{\tau_{max}}{K}}
\end{equation}
then it becomes straightforward to compute the explicit expression
for the composed kernel 
\begin{equation}
  \label{eq-comp-conv-kernel-rep-trunc-exp-same-time-const}
  h_{composed}(t;\; \mu, k) =
  {\cal L}^{-1} 
    \left(
      \frac{1}{(1 + \mu q)^k}
    \right)
  = \frac{t^{k-1} \, e^{-t/\mu}}{\mu^k \, \Gamma(k)}
  \quad\quad (t > 0)
\end{equation}
which has temporal mean value $m_k = k \, \mu$ and variance $\tau = k \, \mu^2$.
Note that in contrast to the primitive truncated exponentials,
which are discontinuous at the origin, these kernels are continuous of
order $k-1$, thus allowing for differentiation up to order $k-1$.
The corresponding expressions for the first- and second-order
derivatives are 
  \begin{align}
    \begin{split}
      h_{composed,t}(t;\; \mu, k) 
        & = \mu^{-k-1} t^{k-2} 
              \frac{((k-1) \mu -t)}{\Gamma(k)}
              e{-t/\mu}
    \end{split}\nonumber\\
    \begin{split}
              & = - \frac{(t - (k-1) \mu)}{\mu  t}
                    \, h_{composed,t}(t;\; \mu, k),
    \end{split}\\
    \begin{split}
      h_{composed,tt}(t;\; \mu, k) 
       & = \mu ^{-k-2} t^{k-3}
            \frac{\left((k^2-3 k+2) \mu^2
                        -2 (k-1) t \mu 
                        +t^2
                  \right) }{\Gamma(k)}
            e^{-t/\mu}
    \end{split}\nonumber\\
    \begin{split}
        & = \frac{\left(
                    \left(k^2-3 k+2\right) \mu ^2-2 (k-1) t \mu +t^2
                  \right)}
                 {\mu ^2 t^2}
            \, h_{composed,t}(t;\; \mu, k).
    \end{split}
   \end{align}
Figure~\ref{fig-trunc-exp-kernels-1D} shows graphs of these kernels 
for two combinations of $\mu$ and $K$ that correspond to the same
value of the composed variance $\tau = K \, \mu^2$.
As can be seen from the graphs, the kernels are highly asymmetric for small values
of $k$, whereas they become gradually more symmetric as $k$ increases.
Figures~\ref{fig-trunc-exp-kernels-nonequal-mu-1D-csqrt2}--\ref{fig-trunc-exp-kernels-nonequal-mu-1D-c2r0p75} 
show corresponding
compositions of truncated exponential kernels for
self-similar distributions of the intermediate time constants
according to equations~(\ref{eq-distr-tau-values}),
(\ref{eq-mu1-log-distr}) and (\protect\ref{eq-muk-log-distr}) 
for $c = \sqrt{2}$ and $c = 2^{3/4}$.
As can be seen from a comparison between 
figure~\ref{fig-trunc-exp-kernels-1D} 
and figures~\ref{fig-trunc-exp-kernels-nonequal-mu-1D-csqrt2}--\ref{fig-trunc-exp-kernels-nonequal-mu-1D-c2r0p75},
the use of a self-similar distribution of the time constants 
(in figures~\ref{fig-trunc-exp-kernels-nonequal-mu-1D-csqrt2}--\ref{fig-trunc-exp-kernels-nonequal-mu-1D-c2r0p75})
allows for smoother behaviour near the origin with increasing $K$ while not
increasing the temporal delay as much as for the kernels corresponding to
a uniform distribution of the intermediate temporal scale levels (in figure~\ref{fig-trunc-exp-kernels-1D}).

\paragraph{Time-recursive computation of temporal derivatives.}

Temporal scale-space derivatives of order $r$ can be defined from 
  this scale-space model according to
  \begin{equation}
    L_{t^r}(\cdot;\; \tau_k)
    = \partial_{t^k} L(\cdot;\; \tau_k)
    = (\partial_{t^k} (*_{i=1}^{k} h_{exp}(t;\; \mu_i)) * f
  \end{equation}
where the Laplace transform of the composed (equivalent) derivative
  kernel is 
  \begin{equation}
  \label{eq-transfer-fcn-der}
    H_{composed}^{(r)}(q;\; \tau_k) 
    =q^r \prod_{i=1}^{k} \frac{1}{1 + \mu_i q}
  \end{equation}
For this kernel to have a net integration effect, 
and to enable well-posed derivative operators, 
an obvious requirement is that the total order of differentiation 
should not exceed the total order of integration.
Thereby, $r < k$ is a necessary requirement.
As a consequence, the composed transfer function must have finite $L_2$-norm.
 
A very useful observation that can be made concerning derivative 
computations is that temporal derivatives can equivalently be computed 
from differences between different temporal channels.
Let us first assume that all time constants $\mu_i$ are different in 
(\ref{eq-transfer-fcn-der}).
Then, a partial fraction division gives 
\begin{equation}
  \htransf_{composed}^{(r)}(q;\; \tau_k) 
  = \sum_{i = 1}^{k} A_i \, \htransf_{prim}(q;\; \mu_i)
\end{equation}
where
\begin{equation}
  A_i = 
  \frac{(-1)^r}{\mu_i^r}
  \prod_{j=1,j \neq i}^{k}
    \frac{1}{(1 - \mu_j/\mu_i)}
 \quad\quad
 (1 \leq i \leq k)
\end{equation}
showing that {\em each temporal derivative can be computed as
a linear combination of the representations at the
different time-scales\/}.

More realistically, however, the channels that we can regard as available 
at a certain temporal scale with index $k$ will not be the results of direct
filtering with different time constants $\mu_i$. 
Rather, we would like to use the intermediate outputs from the cascade coupled
recursive filters $\htransf_{composed}(q;\; \tau_i)$ 
for $k - r \leq i \leq k$.
Decomposition of $\htransf_{composed}^{(r)}$ into a sum of $r$ such transfer 
functions 
\begin{equation}
  \label{eq-decomp-exp-der-filt}
  \htransf_{composed}^{(r)}(q;\; \tau_k) 
  = \sum_{i = k-r}^{k} B_i \, \htransf_{composed}(q;\; \tau_i) 
\end{equation}
shows that the weights $B_i$ are given as the solution
of a triangular system of equations 
provided that the necessary condition $r < k$ is satisfied
\begin{equation}
  \label{eq-coeff-decomp-exp-der-filt}
  \frac{(-1)^r}{\mu_i^r}
  \prod_{j = i+1}^{k}
    \frac{1}{(1 - \mu_j/\mu_i)}
  =  
  B_i 
  +
  \sum_{\nu = i+1}^{k}
    B_{\nu}
    \prod_{j = i+1}^{\nu}
      \frac{1}{(1 - \mu_j/\mu_i)}
 \quad\quad
 (k-r \leq i \leq k).
\end{equation}
After a few more calculations it can be shown
that the Laplace transforms of the equivalent derivative
computation kernels satisfy the recurrence relation \cite{LF96-ECCV}
\begin{equation}
  \label{eq-rec-rel-temp-ders}
  \htransf_{composed}^{(r)}(q;\; \tau_k) 
  = \frac{1}{\mu_k}
      \left(
        \htransf_{composed}^{(r-1)}(q;\; \tau_{k-1}) 
        -
       \htransf_{composed}^{(r-1)}(q;\; \tau_k)
      \right),
\end{equation}
implying that higher-order temporal derivatives can be
computed from small-support finite differences of lower-order
derivatives (analogous to pure finite differences in
the spatial domain) where the temporal scale-space representations at
different temporal scales serve as a sufficient temporal buffer of
what has occurred in the past.
Derivative computations will therefore be highly efficient.
Specifically, it follows that both the temporal smoothing operation
and the computation of temporal derivatives are time-recursive.

\section{Multi-scale spectrograms for auditory signals}
\label{sec-spectr}

The above treatment concerning temporal receptive fields is general
and can be used for modelling desirable properties of receptive fields for
a variety of time-dependent sensory signals.
For auditory signals, an additional structural requirement arises 
from the fact that the auditory information is transferred in terms of
sound waves that travel from the transmitter to the receiver and 
the auditory information can be encoded in terms of oscillation
frequencies of the air pressure that generates the sensory signal.
For this reason and from our knowledge that the variations due to the
geometry and other properties of the cochlea leads to physical resonances whose effect can 
be modelled by as a physical Fourier transform, spectrograms are a common tool for analyzing 
auditory information. 

Note that our primary aim is not to specifically
model, for example, the measured response of the nerves coming from the
cochlea as typically done in previous auditory models
\cite{PatAllGig95-JASA}.
Instead we are following the scale-space theory using the principle of invariance
as outlined in section~\ref{sec-struct-req-temp-rec-fields}
and section~\ref{sec-spat-temp-scsp-concepts}.

\begin{figure}[!p]
  \begin{center}
    \begin{tabular}{cc}
      {\small $g_{cos}(t,\omega;\; \tau)$} & {\small $g_{sin}(t,\omega;\; \tau)$}\\
      \includegraphics[width=0.32\textwidth]{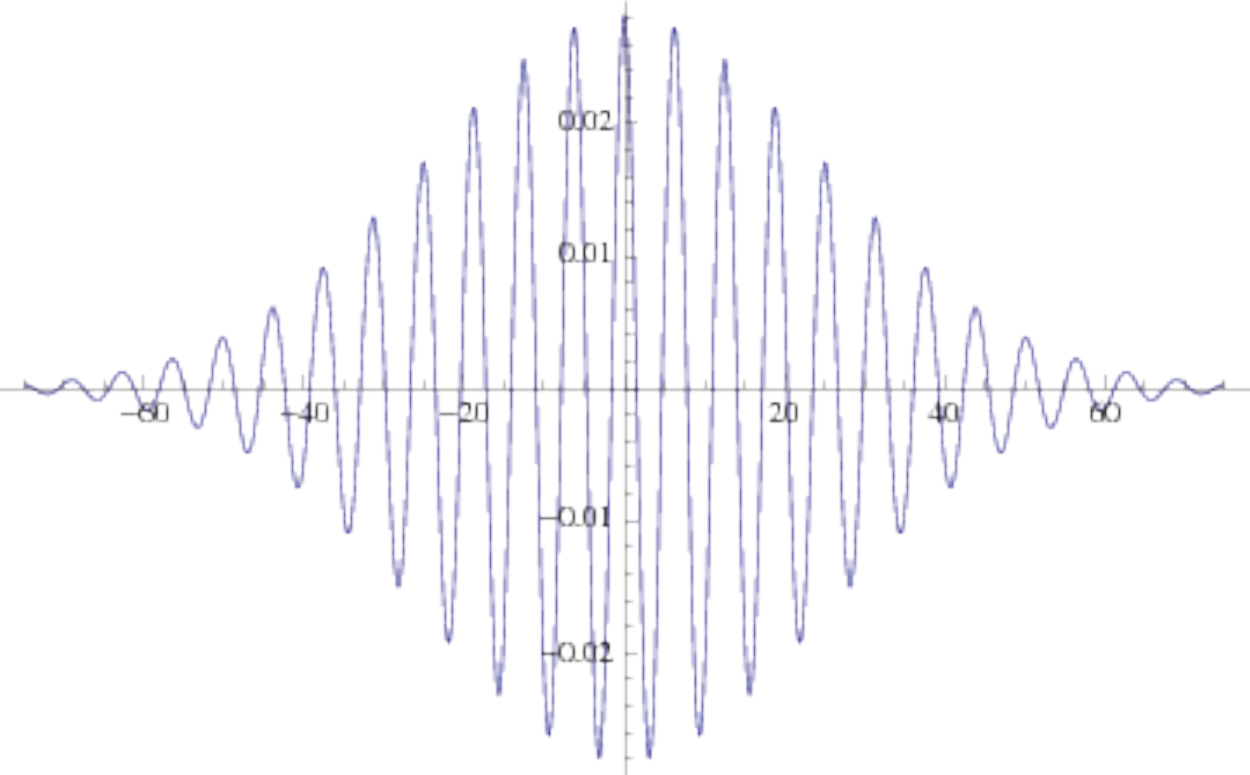} \hspace{-2mm} &
      \includegraphics[width=0.32\textwidth]{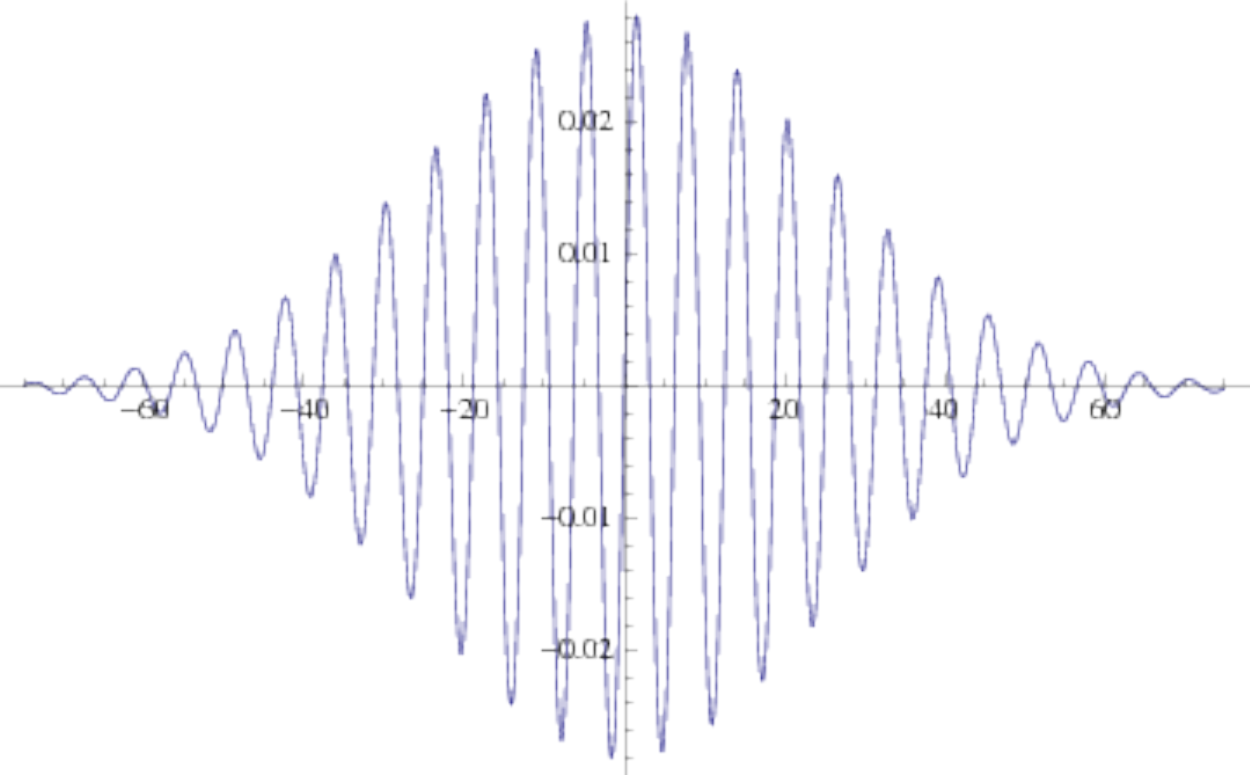} \\
    \end{tabular} 
  \end{center}
 \begin{center}
    \begin{tabular}{cc}
      {\small $h_{cos,log}(t, \omega;\; \mu, K = 4)$} & {\small $h_{sin,log}(t, \omega;\; \mu, K = 4)$} \\
      \includegraphics[width=0.32\textwidth]{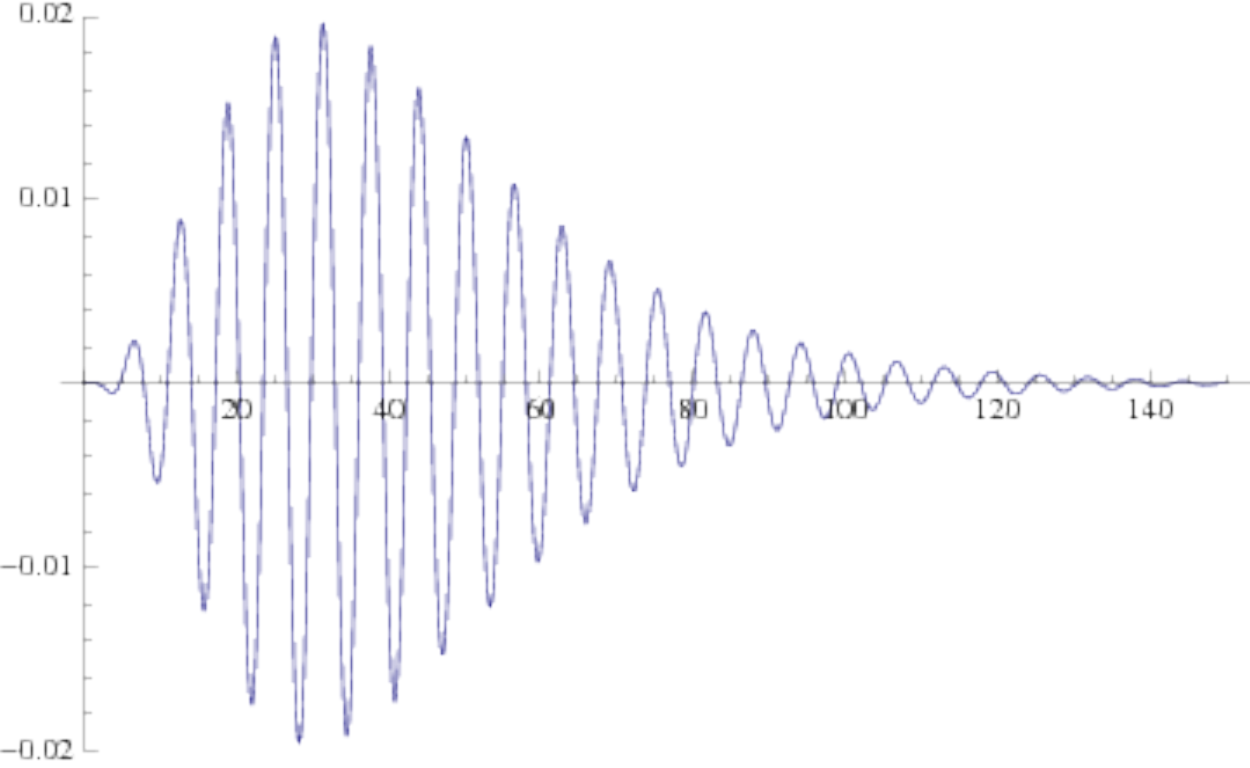} \hspace{-2mm} &
      \includegraphics[width=0.32\textwidth]{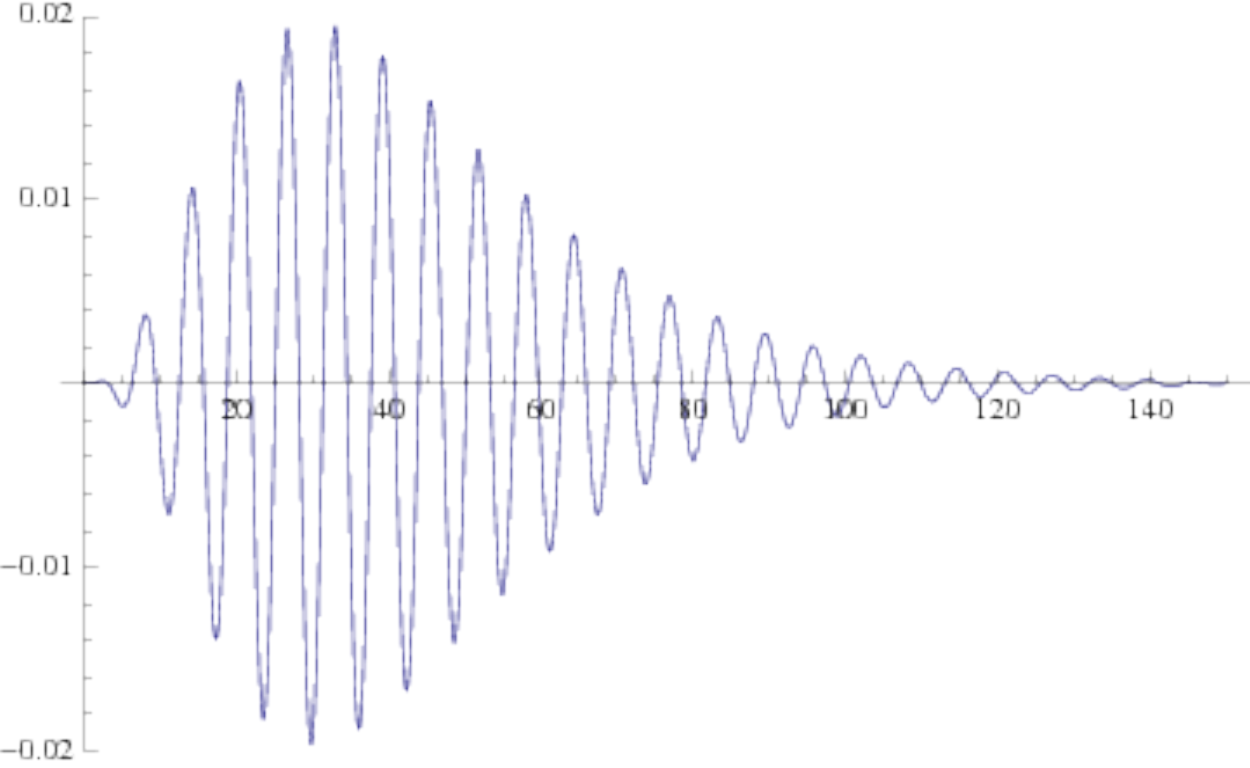} \\
    \end{tabular} 
  \end{center}
  \begin{center}
    \begin{tabular}{cc}
      {\small $h_{cos,uni}(t, \omega;\; \mu, K = 4)$} & {\small $h_{sin,uni}(t, \omega;\; \mu, K = 4)$} \\
      \includegraphics[width=0.32\textwidth]{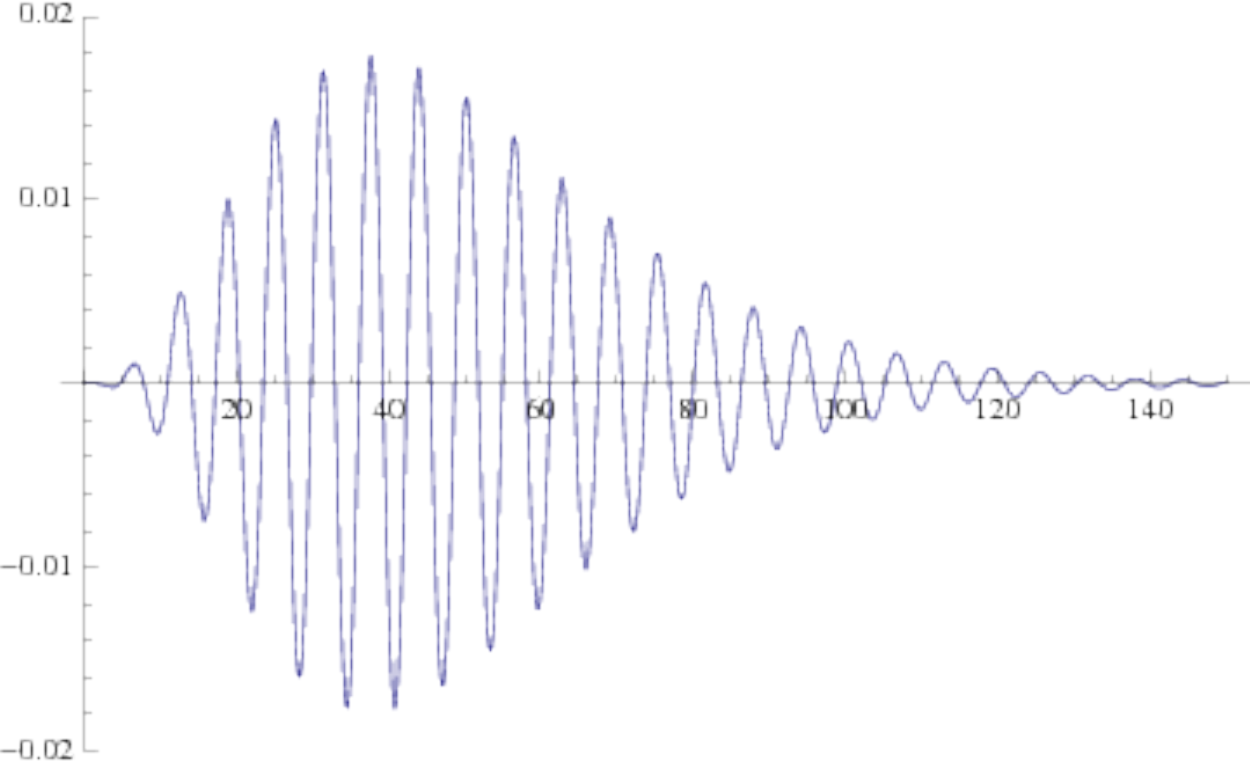} \hspace{-2mm} &
      \includegraphics[width=0.32\textwidth]{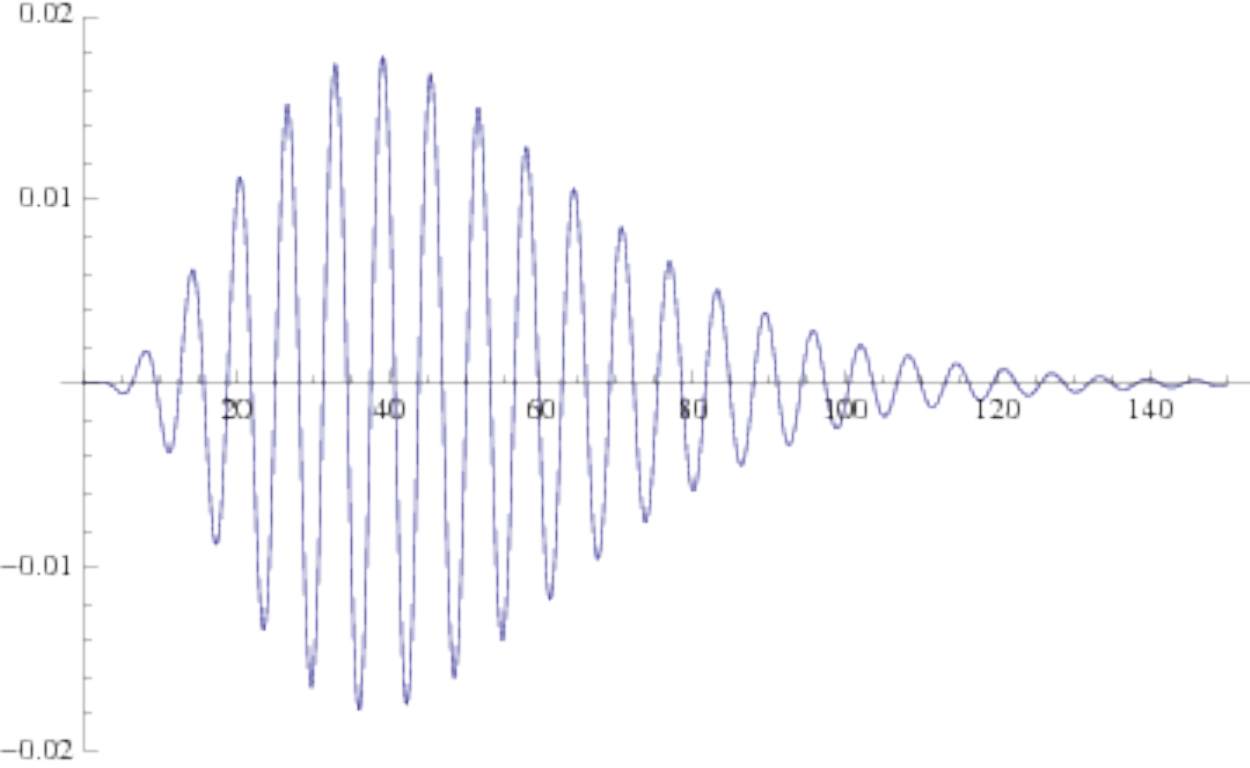} \\
    \end{tabular} 
  \end{center}
 \begin{center}
    \begin{tabular}{cc}
      {\small $h_{cos,log}(t, \omega;\; \mu, K = 7)$} & {\small $h_{sin,log}(t, \omega;\; \mu, K = 7)$} \\
      \includegraphics[width=0.32\textwidth]{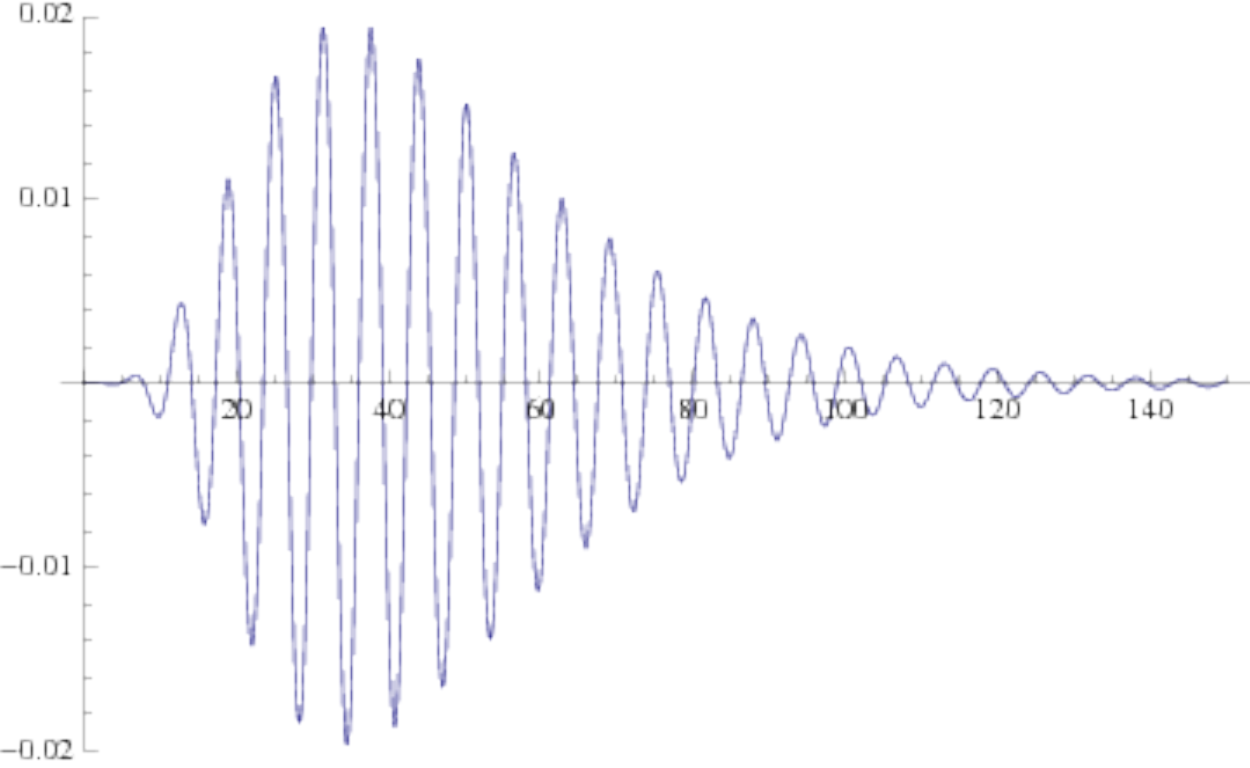} \hspace{-2mm} &
      \includegraphics[width=0.32\textwidth]{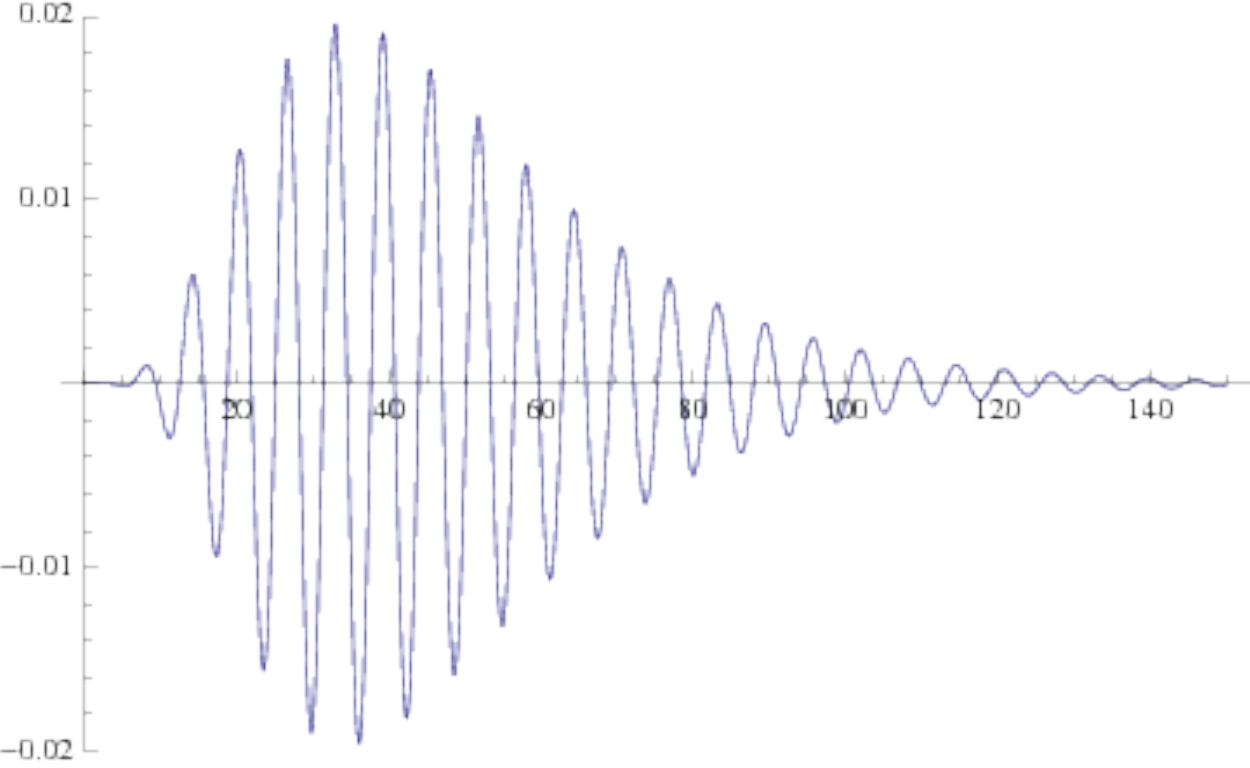} \\
    \end{tabular} 
  \end{center}
  \begin{center}
    \begin{tabular}{cc}
      {\small $h_{cos,uni}(t, \omega;\; \mu, K = 7)$} & {\small $h_{sin,uni}(t, \omega;\; \mu, K = 7)$} \\
      \includegraphics[width=0.32\textwidth]{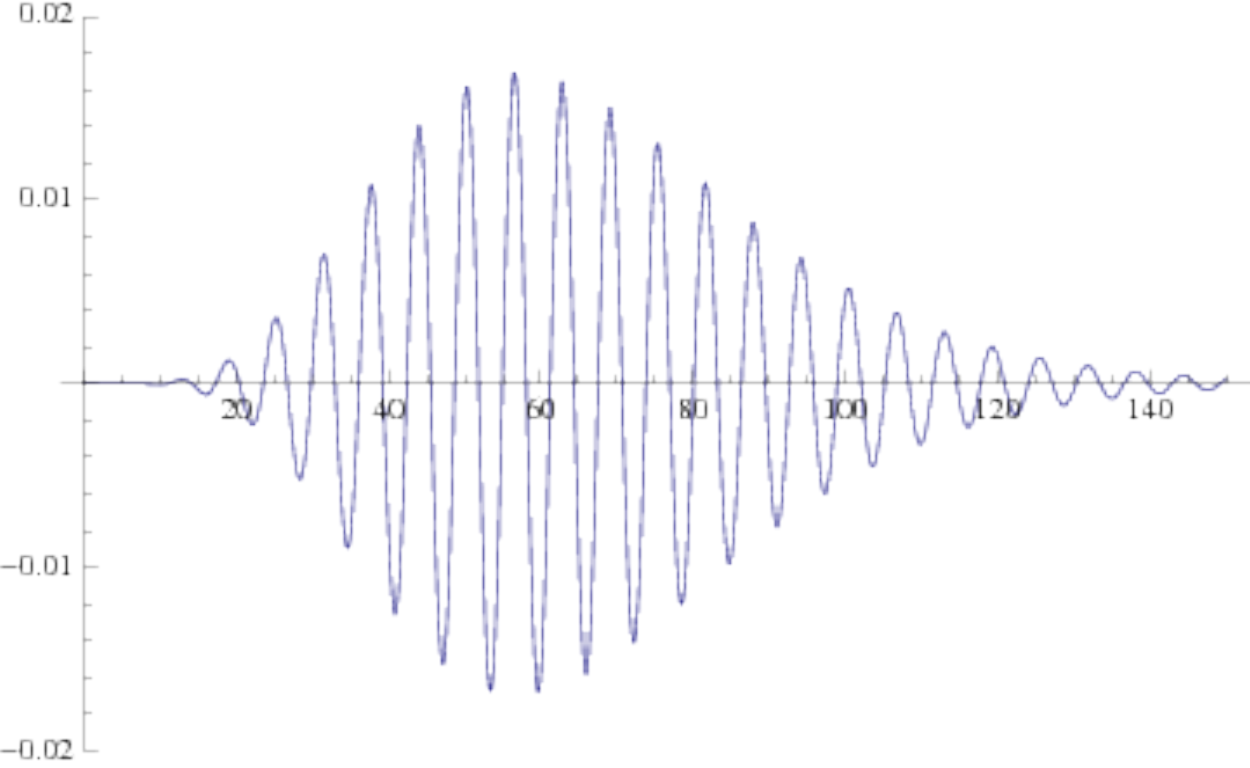} \hspace{-2mm} &
      \includegraphics[width=0.32\textwidth]{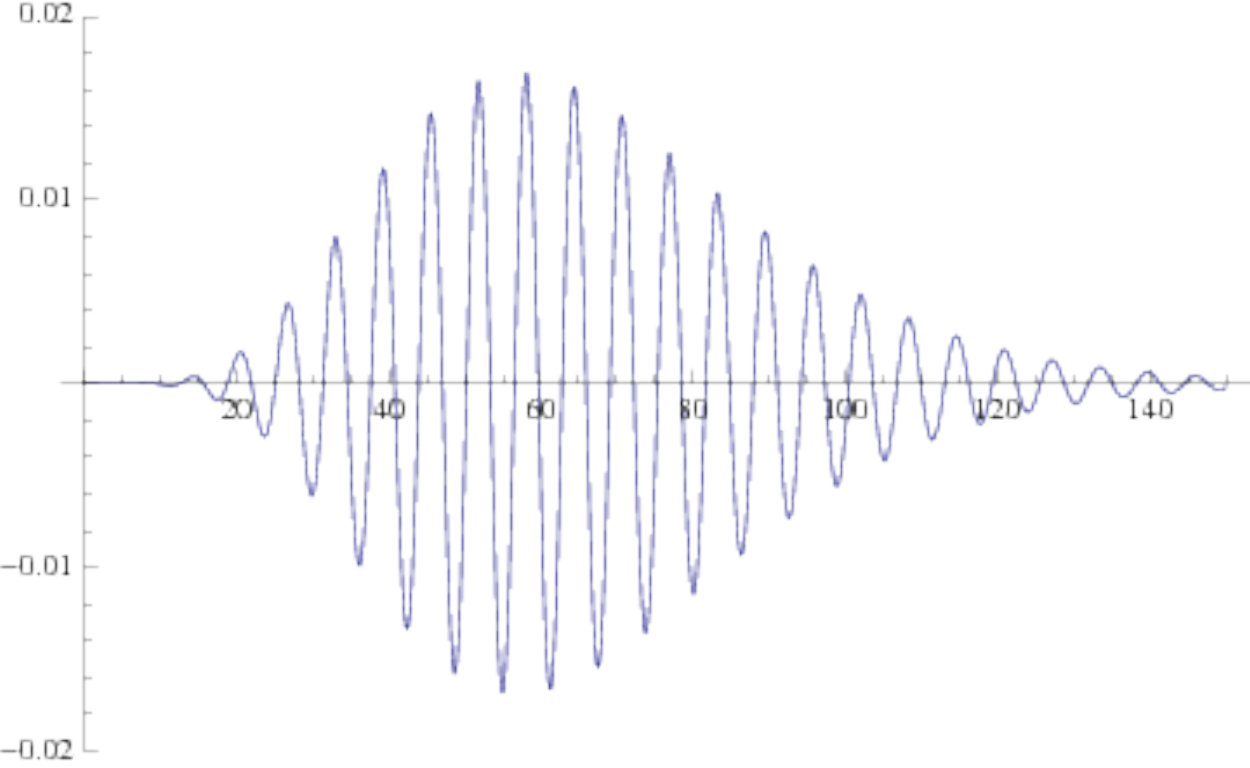} \\
    \end{tabular} 
  \end{center}
  \vspace{-3mm}
  \caption{Examples of underlying frequency selective scale-space filters for computing multi-scale
    spectrograms according to the presented framework with the
    temporal extent proportional to the wavelength according to
    $\sigma = \sqrt{t} = 2 \pi n/\omega$ for $n = 4$.
    (top row) The real and the imaginary parts of a Gabor
    function $g_{cos}(t, \omega;\; \tau) = g(t;\; \tau) \, \cos \omega t$
    and $g_{sin}(t, \omega;\; \tau) = g(t;\; \tau) \, \sin \omega t$.
    (second row) The equivalent kernels corresponding to $K = 4$
    truncated exponential filters coupled in cascade with a logarithmic distribution of
    the time constants according to
    (\protect\ref{eq-distr-tau-values}) for $c = 2^{3/4}$
   and multiplied by the real and the imaginary components of a
    complex sine wave $e^{i \omega t}$.
    (third row) The equivalent kernel corresponding to $K = 4$
    truncated exponential filters with equal time constants coupled in
    cascade and multiplied by the real and the imaginary components of a
    complex sine wave, corresponding to Gammatone filters.
    (fourth and fifth rows) Similar kernels as in the second and third
    rows while using $K = 7$ instead of $K = 4$. (Horizontal axis:
    Time relative to $\omega = 1$.)}
  \label{fig-first-stage-aud-rec-fields}
\end{figure}

\begin{figure}[hbtp]
  \begin{center}
    \begin{tabular}{cc}
      {\small $\log |S_{discgauss}(t,\omega;\;  \tau)|$} 
      & {\small $\log |S_{discgauss}(t,\omega;\;  \tau)|$}  \\
      \includegraphics[width=0.48\textwidth]{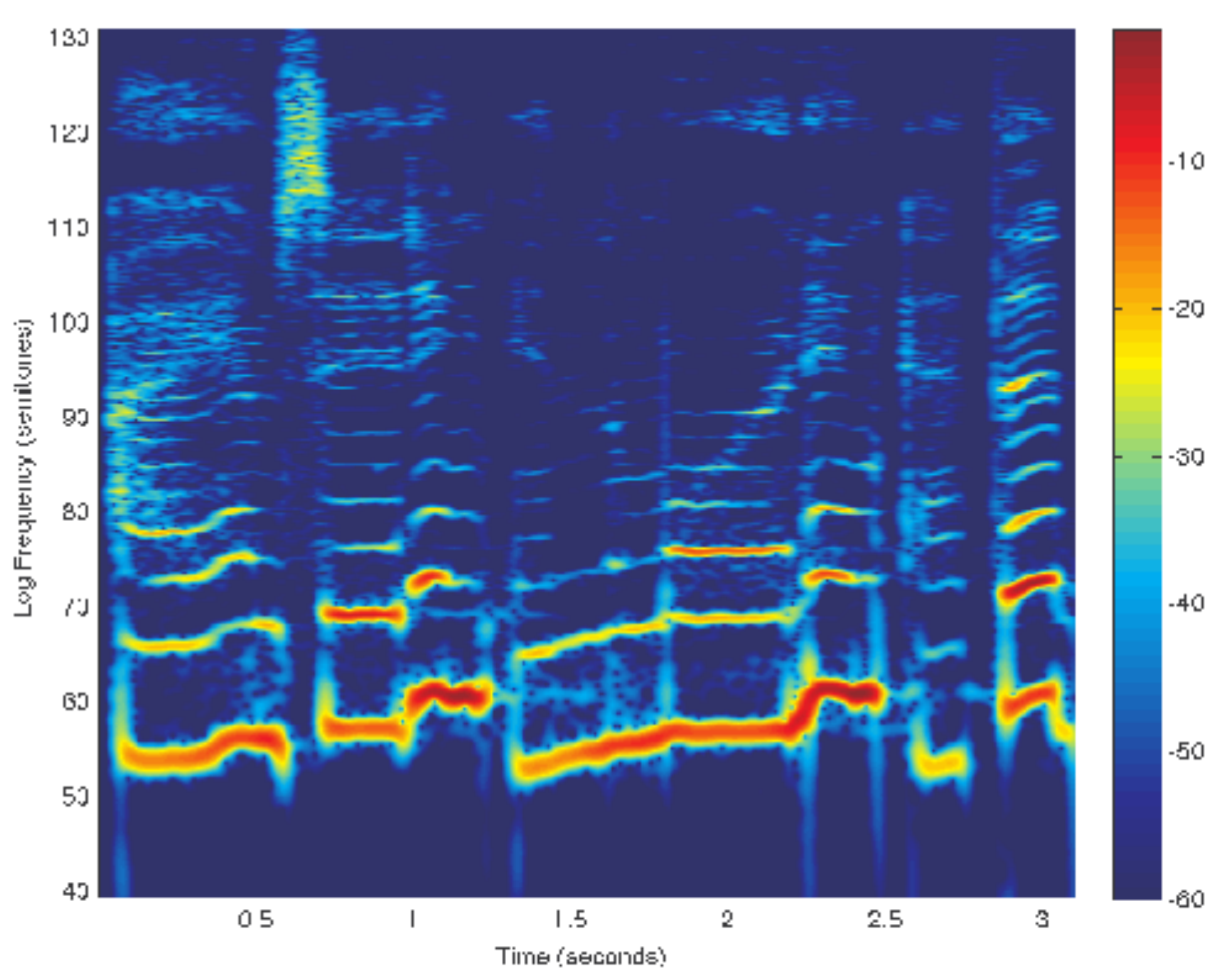} \hspace{-2mm} 
      & \includegraphics[width=0.48\textwidth]{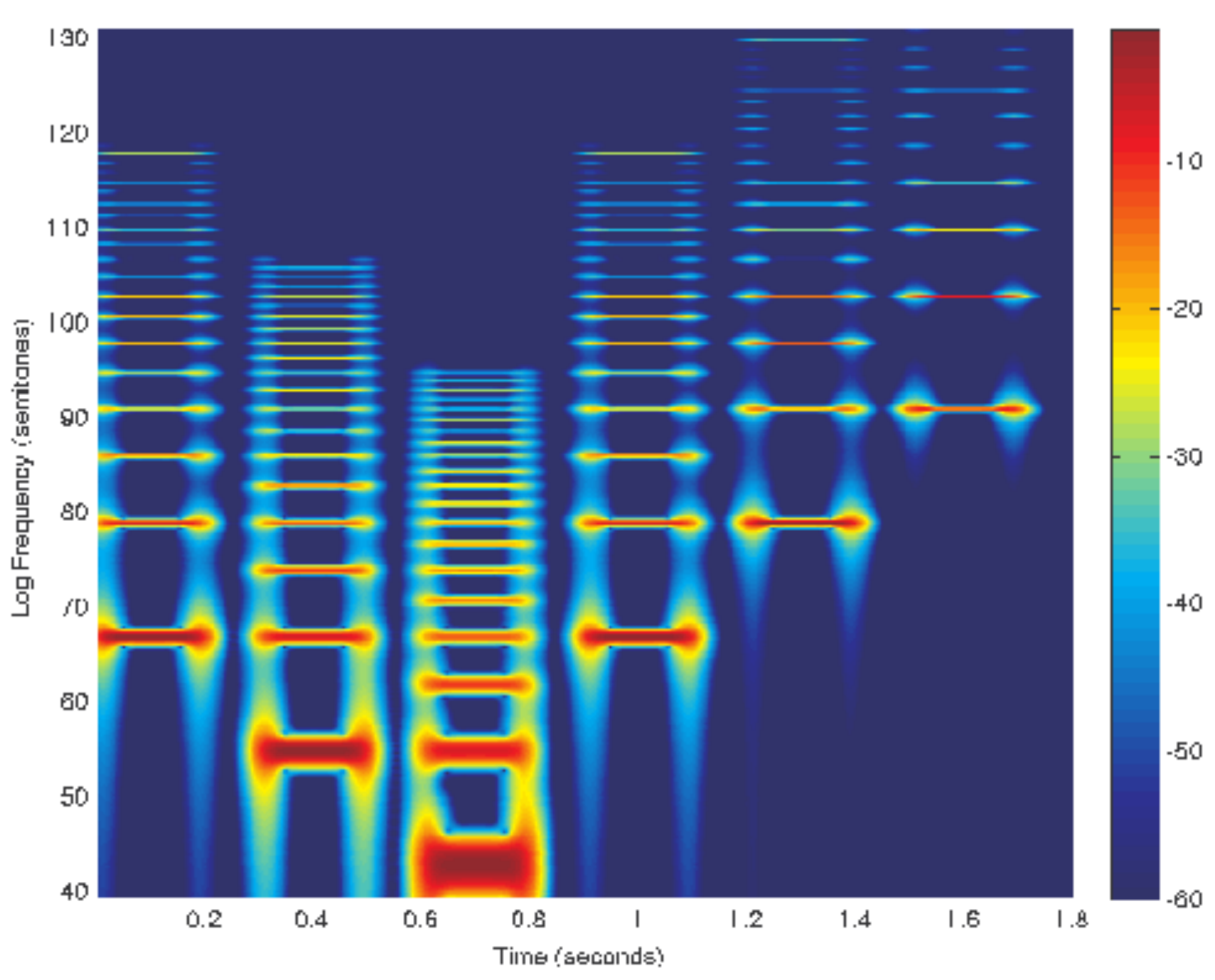}
      \\
      {\small $\log |S_{rec-uni}(t,\omega;\; \tau)|$} 
      & {\small $\log |S_{rec-uni}(t,\omega;\; \tau)|$} \\
      \includegraphics[width=0.48\textwidth]{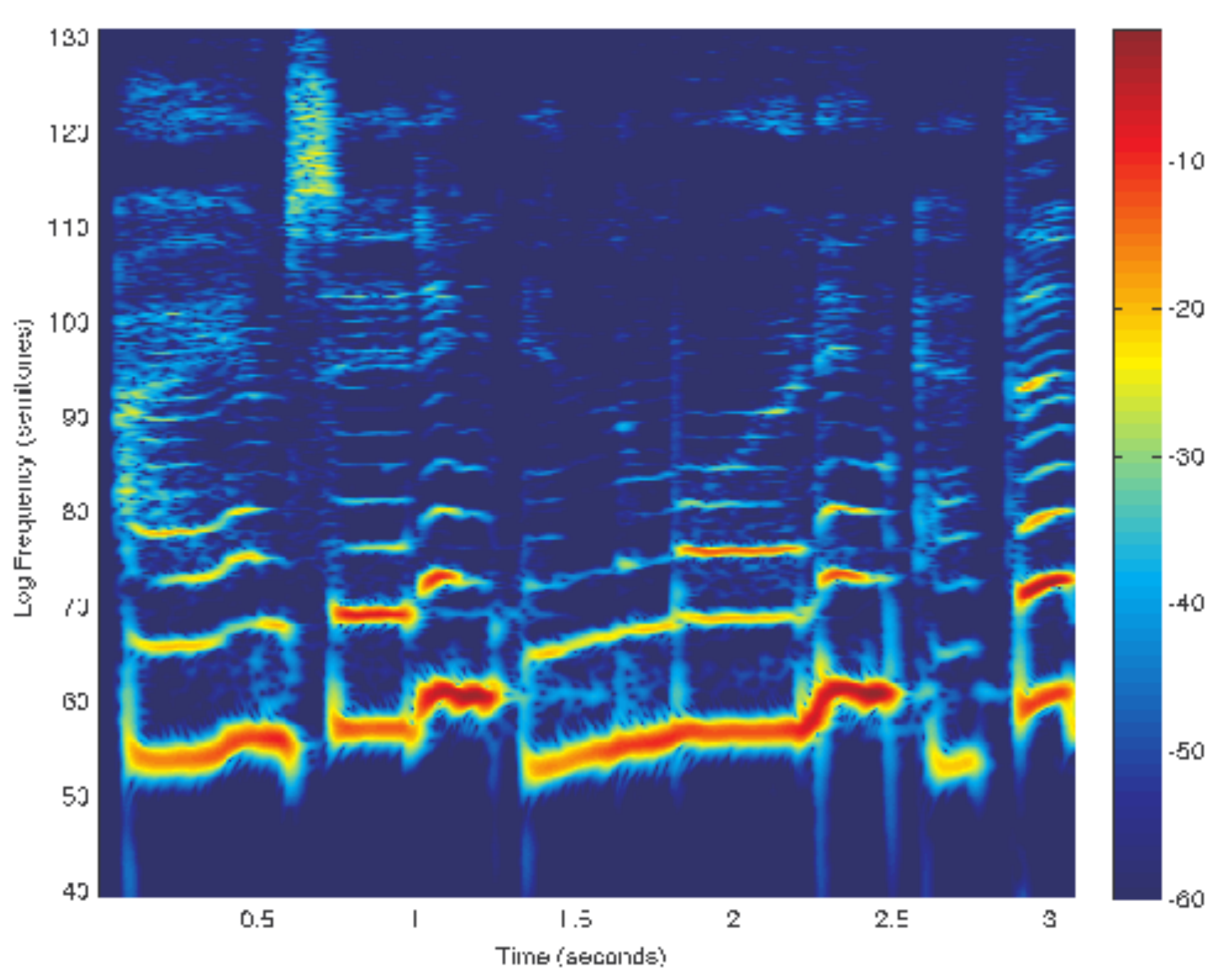} \hspace{-2mm} 
     & \includegraphics[width=0.48\textwidth]{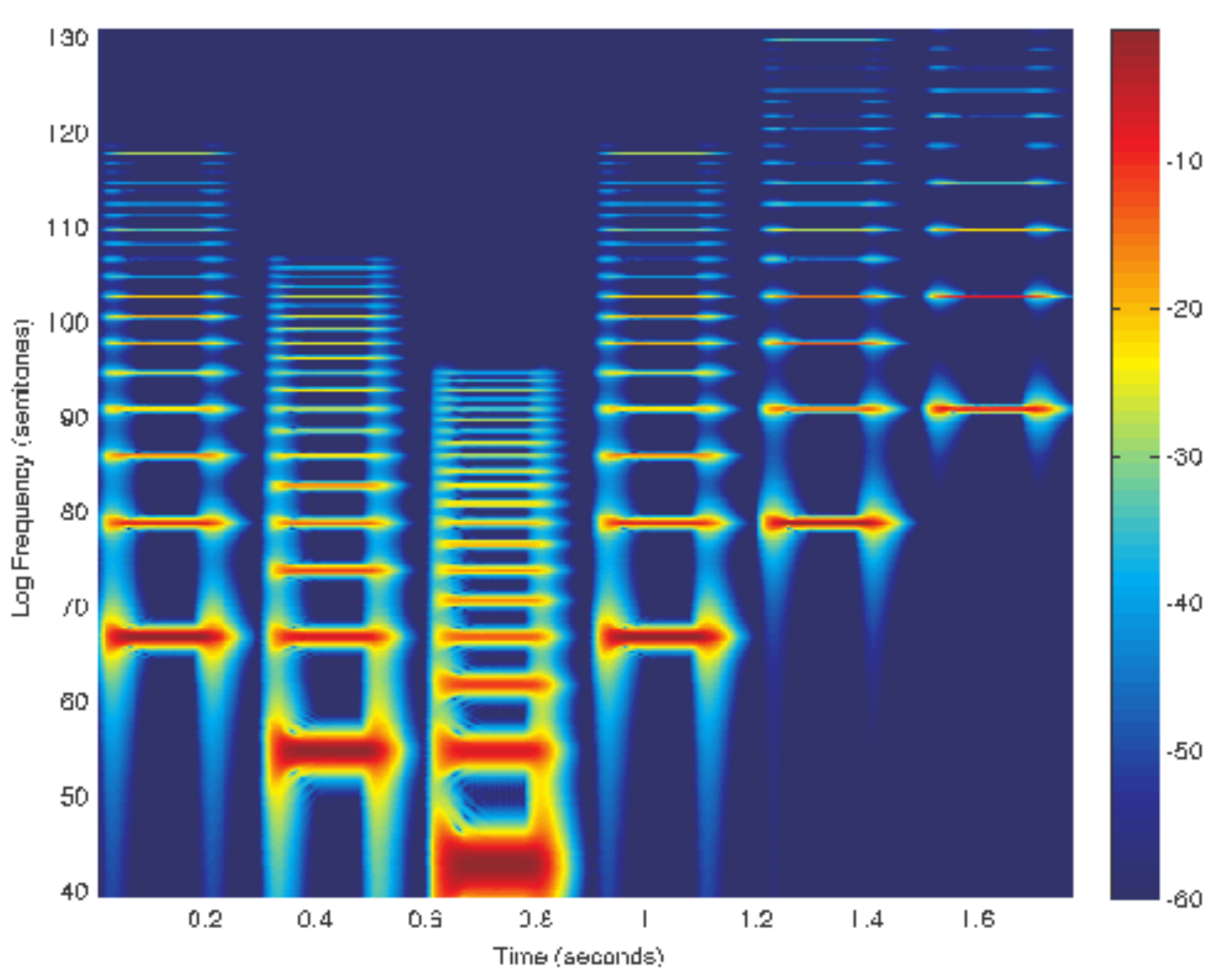}
     \\
      {\small $\log |S_{rec-log}(t,\omega;\;  \tau)|$}  
      & {\small $\log |S_{rec-log}(t,\omega;\;  \tau)|$} \\
      \includegraphics[width=0.48\textwidth]{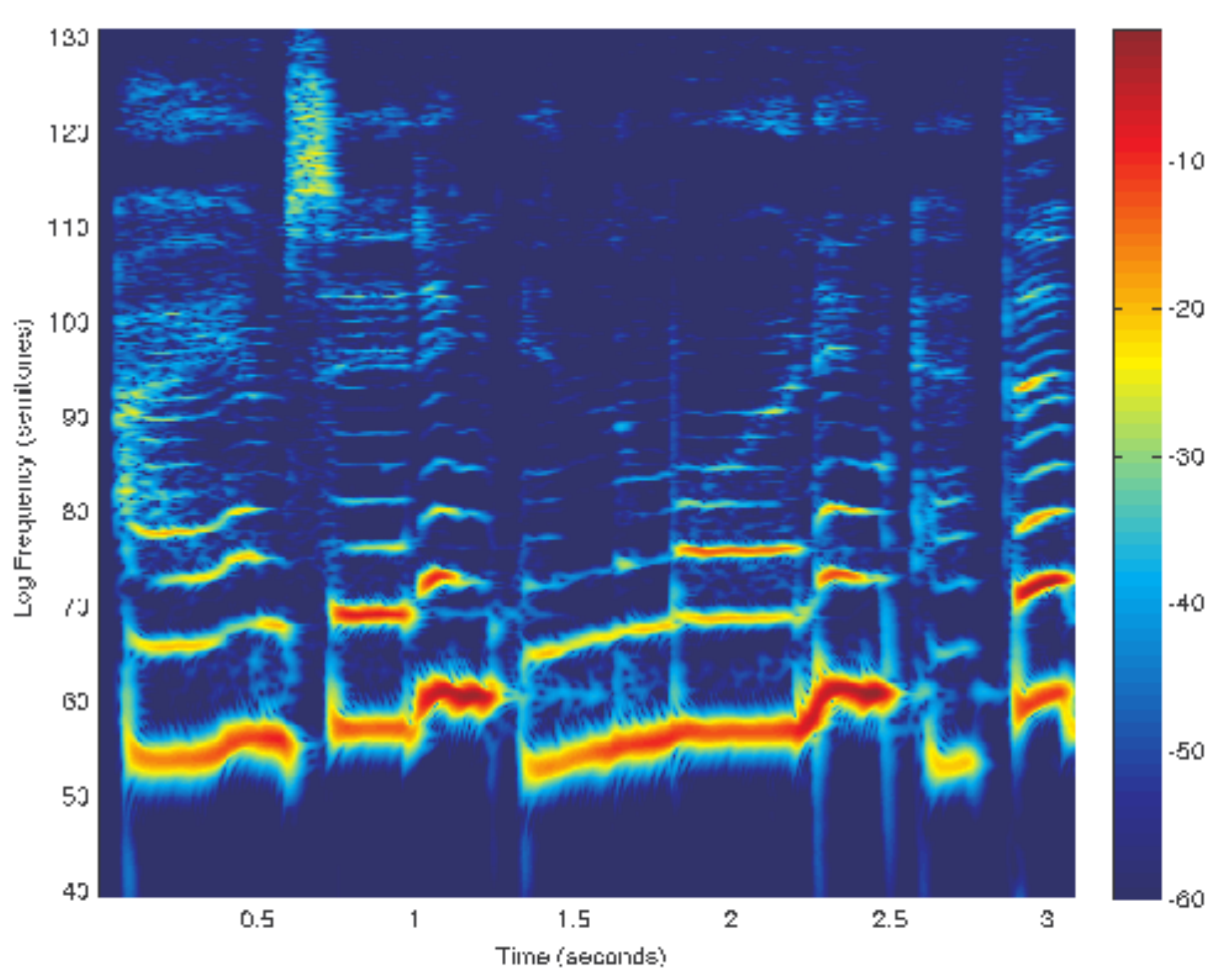} \hspace{-2mm} 
      & \includegraphics[width=0.48\textwidth]{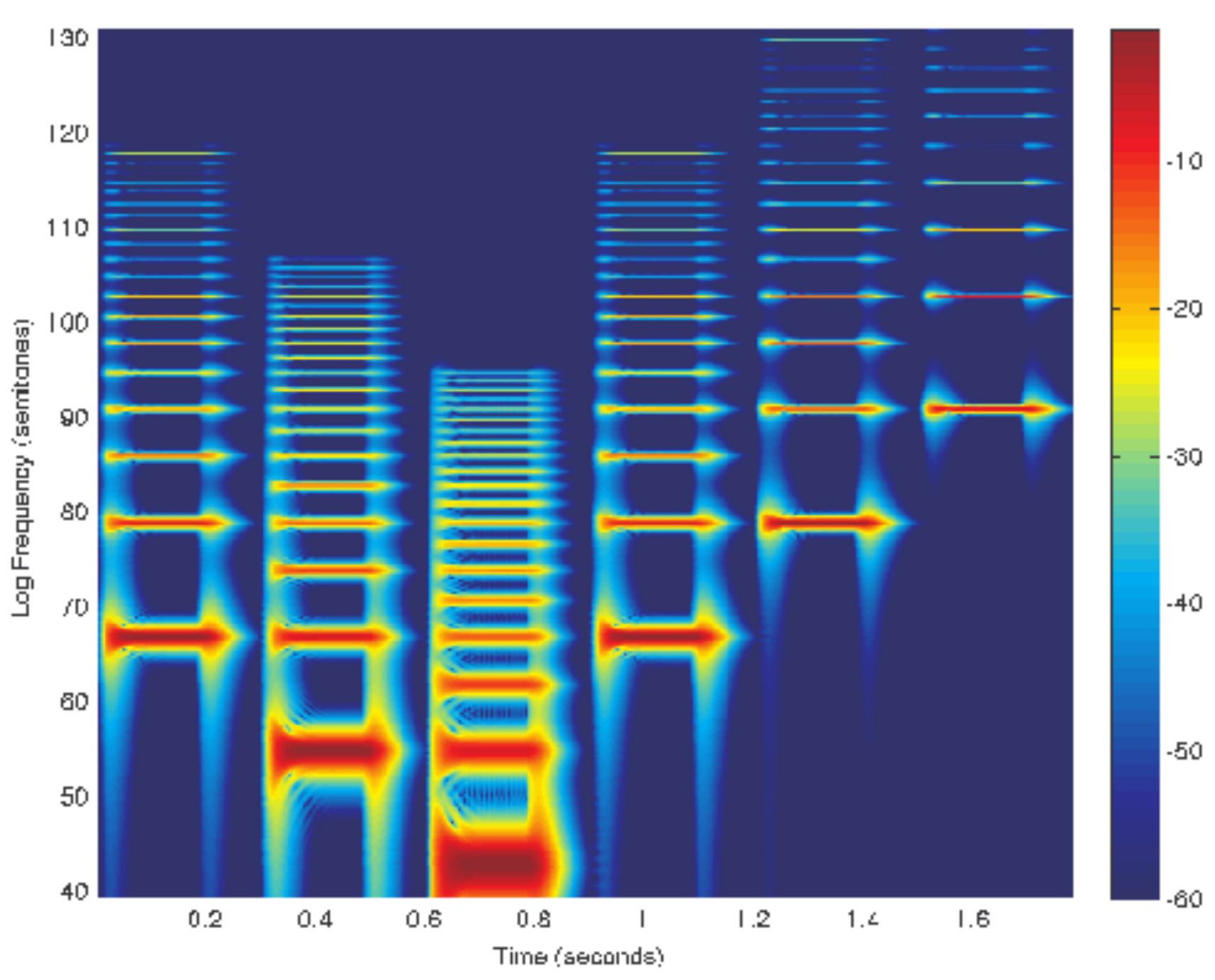} \\
    \end{tabular} 
  \end{center}
 
  \vspace{-3mm}
  \caption{Spectrograms computed using different temporal scale-space
    concepts for (left column) the first 3 seconds of ``Tom's diner'' by Suzanne Vega
    with the lyrics ``I am sitting in the morning at the ...'' and
    (right column) a synthetic signal containing harmonic spectra with
    different fundamental frequencies
    over a logarithmic frequency scale from 80 Hz to 16 kHz using 48
    frequency levels per octave and with a {\em fixed temporal window scale\/}
   $\sigma_t = \sqrt{\tau} = 20~\mbox{ms}$ for all frequencies for
   (top row) the discrete analogue of the Gaussian kernel,
   (middle row)
    a cascade of seven time-causal recursive filters having a uniform
    distribution of the temporal scale levels and
(bottom row)
    a cascade of seven time-causal recursive filters having a
    logarithmic distribution of the temporal scale levels
   with $c = \sqrt{2}$. 
  The vertical axis shows the logarithmic frequency expressed in semitones with 69 corresponding to the tone A4 (440 Hz).}
  \label{fig-first-stage-aud-rec-fields-toms-diner-same-wdw-sc}
\end{figure}

\begin{figure}[hbtp]
  \begin{center}
    \begin{tabular}{cc}
      {\small $\log |S_{discgauss}(t,\omega;\;  \tau)|$} 
      & {\small $\log |S_{discgauss}(t,\omega;\;  \tau)|$}  \\
      \includegraphics[width=0.48\textwidth]{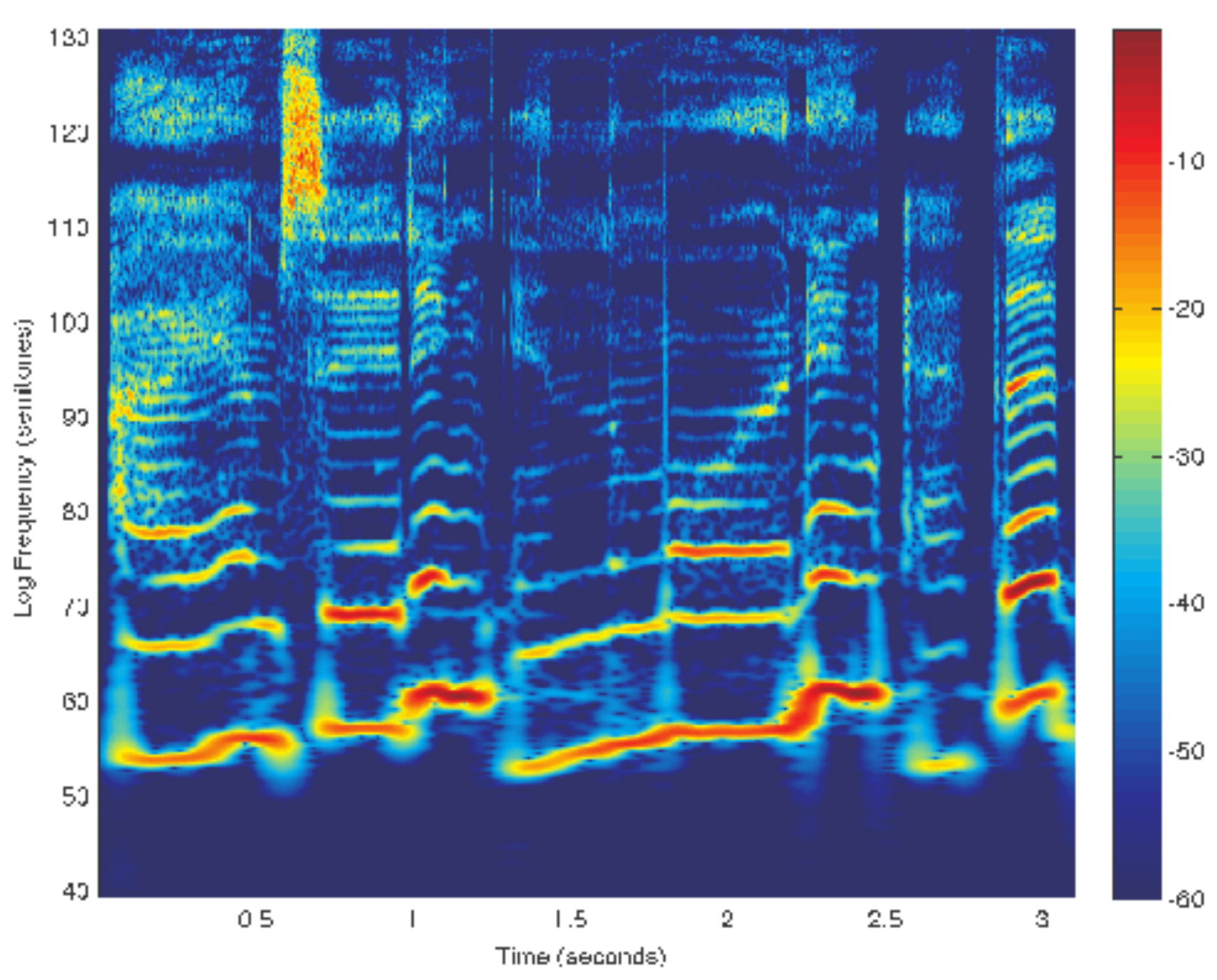} \hspace{-2mm} 
      & \includegraphics[width=0.48\textwidth]{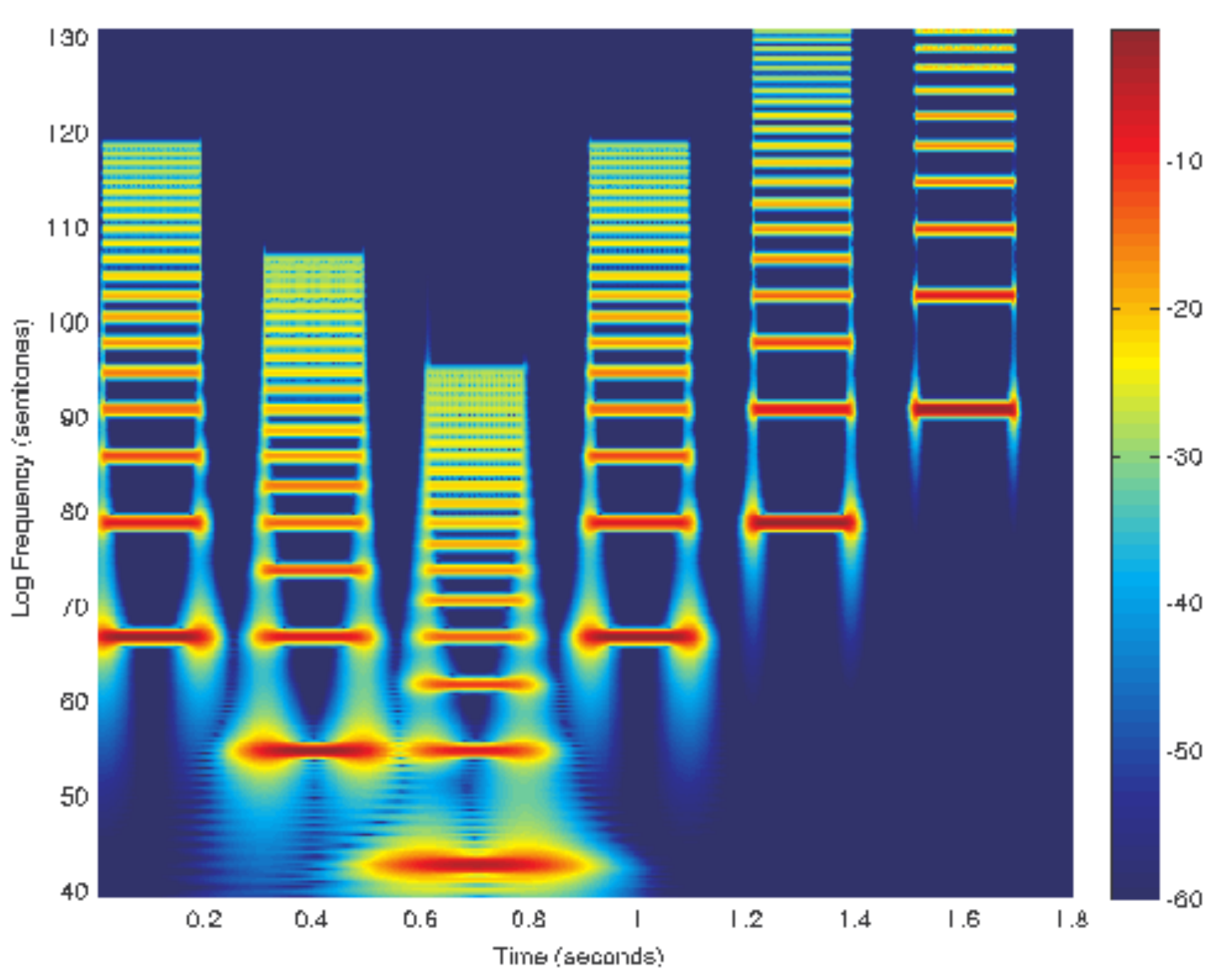}
      \\
      {\small $\log |S_{rec-uni}(t,\omega;\; \tau)|$} 
      & {\small $\log |S_{rec-uni}(t,\omega;\; \tau)|$} \\
      \includegraphics[width=0.48\textwidth]{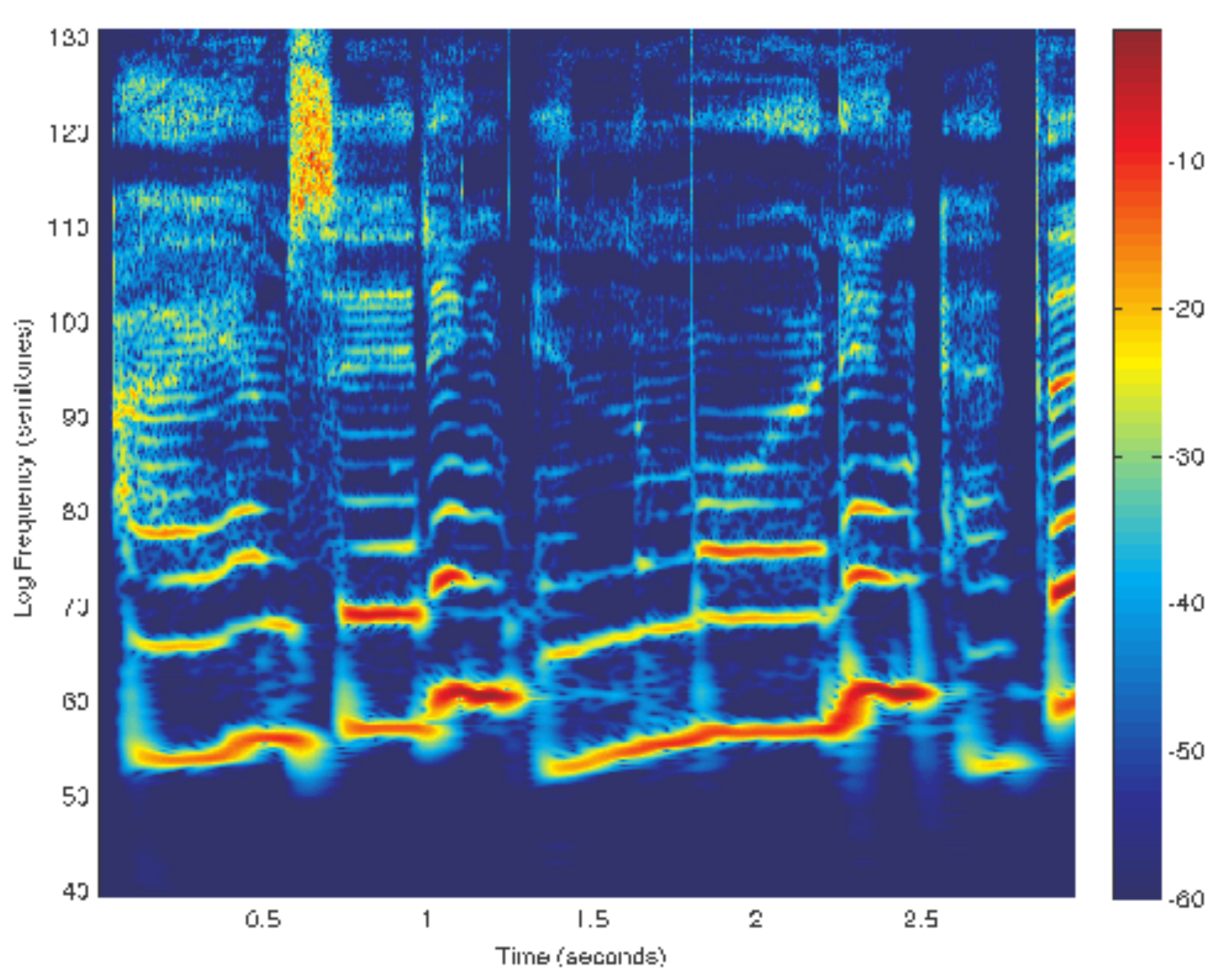} \vspace{-2mm} 
     & \includegraphics[width=0.48\textwidth]{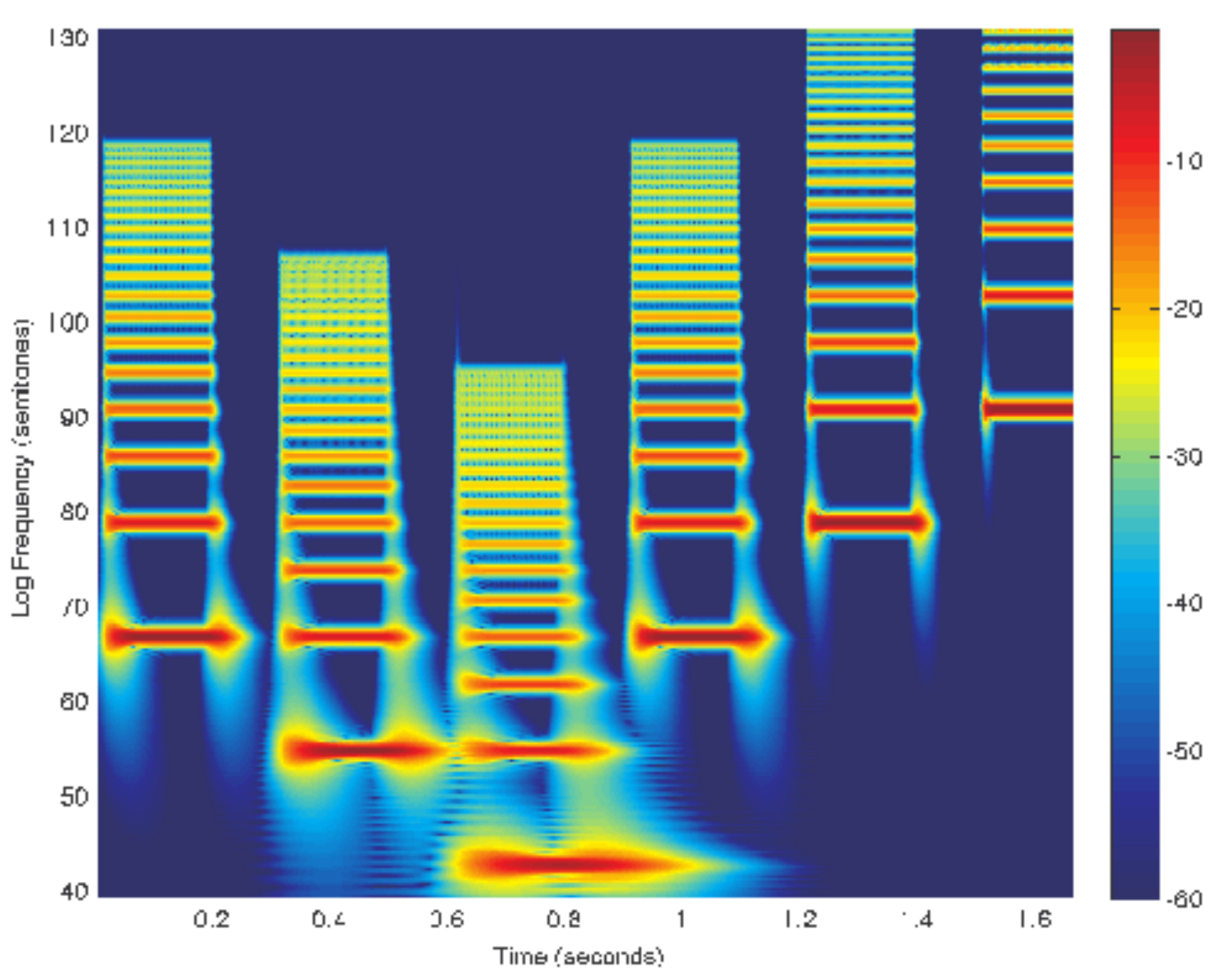}
     \\
      {\small $\log |S_{rec-log}(t,\omega;\;  \tau)|$}  
      & {\small $\log |S_{rec-log}(t,\omega;\;  \tau)|$} \\
      \includegraphics[width=0.48\textwidth]{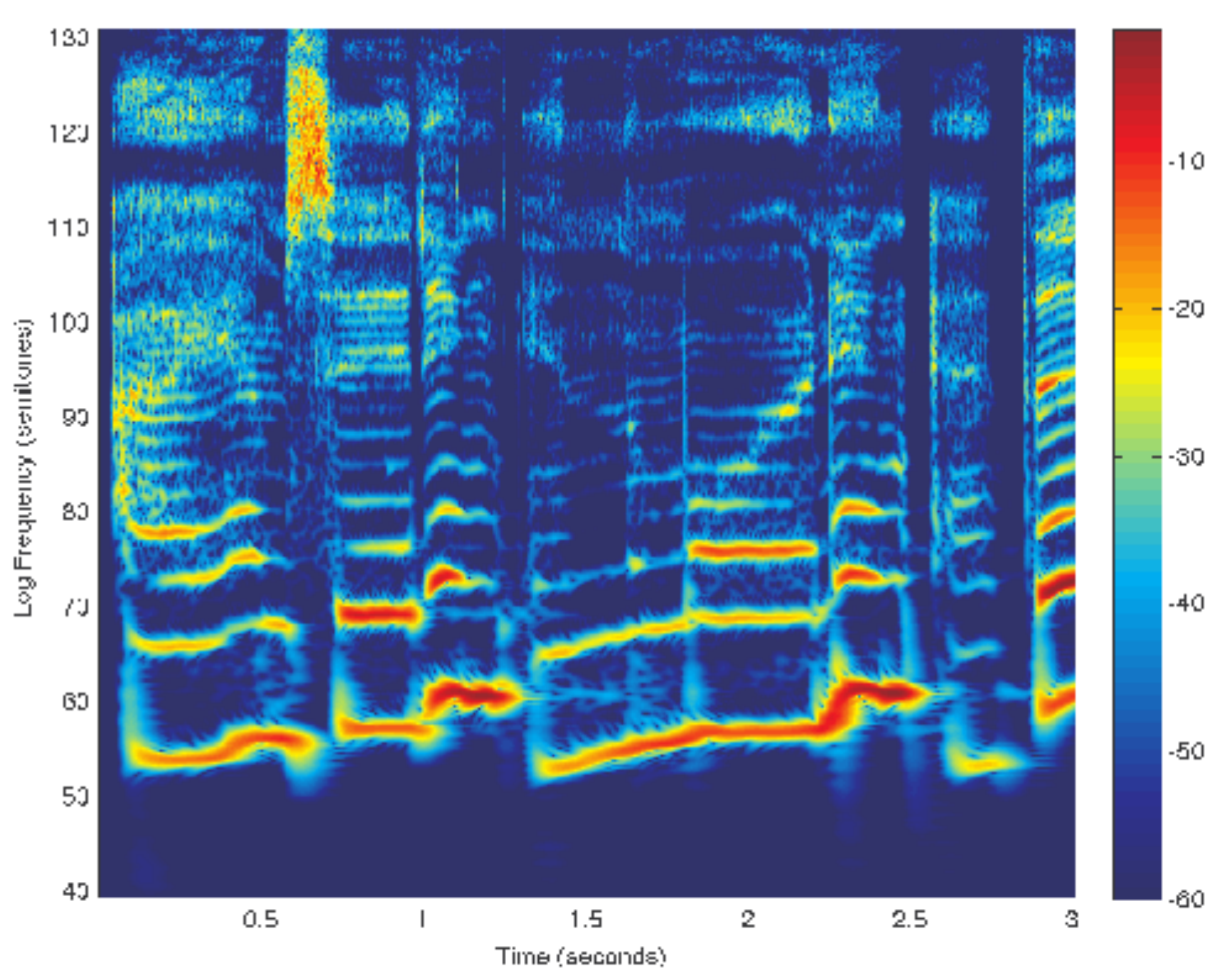} \hspace{-2mm} 
      & \includegraphics[width=0.48\textwidth]{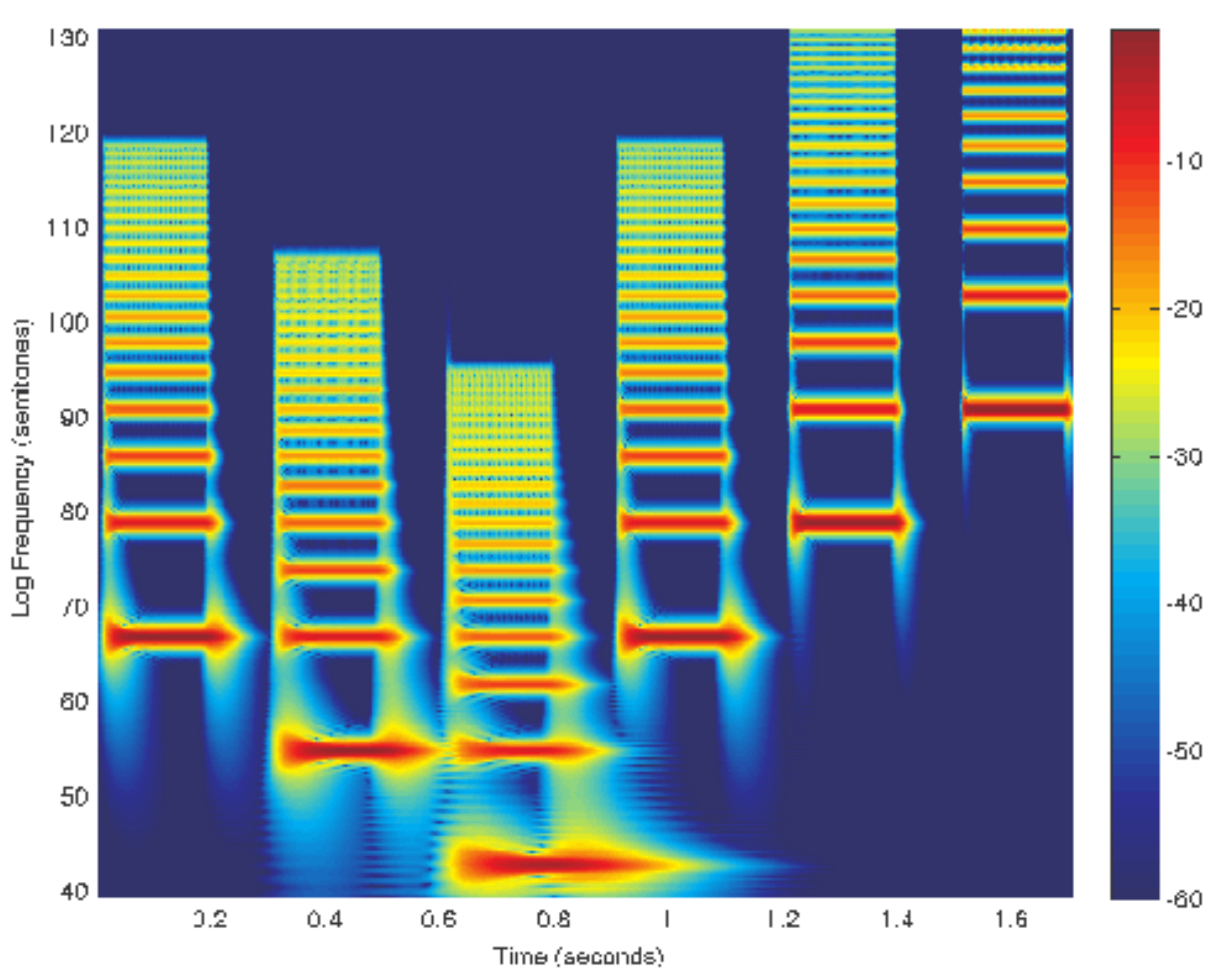}
      \\
    \end{tabular} 
  \end{center}
 
  \vspace{-3mm}
  \caption{Spectrograms computed using different temporal scale-space
    concepts for (left column) the first 3 seconds of ``Tom's diner'' by Suzanne Vega
    with the lyrics ``I am sitting in the morning at the ...'' and
    (right column) a synthetic signal containing harmonic spectra with
    different ground tones
    over a logarithmic frequency range from 80 Hz to 16 kHz using 48
    frequency levels per octave
    and with the {\em temporal window scale proportional to the wavelength\/}
    for $n = 8$ and with a soft lower threshold $\sigma_0 = \sqrt{\tau_0} =  1~\mbox{ms}$ for
   (top row) the discrete analogue of the Gaussian kernel,
   (middle row)
    a cascade of seven time-causal recursive filters having a uniform
    distribution of the temporal scale levels and
(bottom row)
    a cascade of seven time-causal recursive filters having a
    logarithmic distribution of the temporal scale levels with $c =
    \sqrt{2}$. The vertical axis shows the logarithmic frequency expressed in semitones with 69 corresponding to the tone A4 (440 Hz).
   The spectrograms computed with time-causal kernels have been delay
 compensated by a temporal delay defined from the position of the
 first inflection point of the temporal window function.}
  \label{fig-first-stage-aud-rec-fields-toms-diner-prop-wdw-sc}
\end{figure}

Based on the two models for temporal receptive fields
(non-casual in section~\ref{sec-gauss-spat-temp-scsp} and
 time-casual in section~\ref{sec-time-caus-scale-spaces}),
we can use the temporal smoothing functions in these two temporal
scale-space models as scale-dependent
window functions for defining two types of complex-valued
{\em multi-scale spectrograms\/} according to
\begin{align}
   \begin{split}
      \label{eq-complex-spectr-gauss-def}
      S_{Gauss}(t, \omega;\; \tau) 
      & = \int_{t'=-\infty}^{\infty}
                g(t - t';\; \tau) \, f(t') \,
                e^{-i\omega t'} \, dt'
   \end{split}\\
   \begin{split}
      \label{eq-complex-spectr-exp-def}
      S_{exp}(t, \omega;\; \mu) 
      & = \int_{t'=-\infty}^{\infty}
                h_{composed}(t - t';\; \mu) \, f(t') \,
                e^{-i\omega t'} \, dt'
   \end{split}
\end{align}
where 
\begin{itemize}
\item
  $g(t;\; \tau)$ is a temporal Gaussian kernel of the form
(\ref{eq-gauss-gen-spattemp-2+1-D}),
\item
$h_{composed}(t;\; \mu)$ with $\mu = (\mu_1, \dots, \mu_k)$ is the equivalent convolution kernel
corresponding to a cascade of truncated exponential filters of the
form (\ref{eq-comp-trunc-exp-cascade}).
\end{itemize}
These definitions imply that the convolution kernels used for defining
temporal scale-space for a general time-varying signal are here used
as scale-dependent window functions for defining windowed Fourier
transforms of different temporal extent. 

For a given value of $\tau$, the spectrogram becomes a 2-D function.
With the definition extended to all values of $\tau$, the spectrogram
based on Gaussian window functions instead becomes a 3-D volume over all
temporal extents of the window function or alternatively a set of
discrete 2-D slices for the window functions based on truncated exponential
functions coupled in cascade for vectors $\mu = (\mu_1, \dots, \mu_k)$
of different length $k$.

Note that {\em a priori\/}
there may be no principled reason for preferring a particular duration of
the temporal window function for the windowed Fourier transform over some
other temporal duration. Specifically, different temporal durations may be appropriate
for different auditory tasks, such as a preference for a short temporal 
duration for onset detection and a preference for a longer temporal
duration to separate sounds with nearby frequencies.
In this context, the scale-space approach allows for the definition of
windowed Fourier transforms for all temporal extents in such a way that 
any windowed Fourier transforms at a coarse temporal scale 
can be related to a windowed Fourier 
transform at any finer scale using the cascade property
(\ref{eq-casc-prop-spat-temp}) 
derived from the semi-group structure  
(\ref{eq-semi-group-spat-temp}) or the Markov property
(\ref{eq-Markov-prop}) of the underlying scale-space kernels.
In combination with the additional scale-space properties of
non-creation of new structures with increasing scale, this
guarantees well-founded theoretical properties between 
corresponding windowed Fourier transforms computed at different temporal scales. 

In most other work on auditory signal processing, there is often an
implicit assumption that one chooses a scale for computing the auditory
features that seems to work and on which later stage analysis is then based.
By the presented formulation of multi-scale spectrograms, we aim at making the
consequences of such assumptions explicit, and emphasizing the
possibility of computing auditory features at multiple temporal scales
as an integrated part of the analysis.
Compared to the more traditional approach
of computing spectrograms from local fast Fourier transforms combined
with local windowing operations, this formulation of multi-scale
spectrograms avoids the concatenation of such windowing operations
altogether and thereby the artifacts caused by these.

The scale-space approach for defining multi-scale auditory
spectrograms implies that instead of computing a scale-space
representation of the original auditory signal, the auditory signal is
first projected onto the two orthogonal dimensions $\cos \omega t$ and
$ i \sin \omega t$
of a complex sine wave $e^{-i\omega t}$
\begin{equation}
  f_{cos}(t, \omega) = f(t) \, \cos \omega t \quad\quad
  f_{sin}(t, \omega) = f(t) \, \sin \omega t
\end{equation}
for which temporal scale-space representations are then defined,
implying that the multi-scale spectrogram can be interpreted as a
complex-valued scale-space transform.

\paragraph{Invariance and covariance properties.}

Concerning the symmetry requirements of a general temporal sensory
front-end described in section~\ref{sec-struct-req-temp-rec-fields},
the linearity of the scale-space operations is transferred to a
linearity in the complex multi-scale spectrograms (\ref{eq-complex-spectr-gauss-def})--(\ref{eq-complex-spectr-exp-def}).
This implies that multiple sources of sound will combined in
an additive manner in terms of their complex-valued responses and that sound sources of different 
strength (sound pressure) will be handled in a similar manner up to a multiplication of
the strength of the signal.

Regarding temporal shift invariance, the magnitude maps $|S_{Gauss}|$  
of $|S_{exp}|$ are invariant under a shift of the temporal axis, 
whereas the phase of the truly complex spectrograms $S_{Gauss}$ 
and $S_{exp}$ will be transformed in a predictable manner between
similar sound signals that occur at different time moments or at
different distances to the observer.

Under a local rescaling of the temporal axis, the temporal receptive fields
obtained from the Gaussian scale-space model are fully scale covariant.
Under a rescaling of the temporal axis 
\begin{equation}
  t \mapsto \alpha \, t 
\end{equation}
the corresponding complex-valued multi-scale spectrograms are transformed according to
\begin{align}
   \begin{split}
      S(t, \omega;\; \tau) 
      & \mapsto S(\alpha \, t, \frac{\omega}{\alpha};\;
      \alpha^2 \tau) 
   \end{split}
\end{align}
If we let the window scale $\sigma = \sqrt{\tau}$ for any angular frequency $\omega$ be 
proportional to the wavelength $\lambda = 2 \pi/\omega$ corresponding
to that frequency, then the corresponding spectrograms 
within the same 2-D slice of the extended 3-D multi-scale spectrogram
can therefore be perfectly matched under a rescaling of the temporal
axis corresponding to a frequency shift
\begin{equation}
  \omega \mapsto \frac{\omega}{\alpha}
\end{equation}
If the temporal window functions on the other hand do not have the
temporal extent proportional to the wavelength, then temporal covariance
does not hold within the same 2-D slice 
but still holds within the 3-D multi-scale spectrogram based on
Gaussian window functions because of their
self-similarity over scale, whereas the corresponding scaling relations can only
be approximate for the truncated exponential functions coupled in
cascade, because of the temporal scale levels being restricted to
a discrete set of values.

Again there may not be any principled reason for preferring a
particular temporal scale over another. The multi-scale nature of the
corresponding spectrogram makes this aspect explicit and opens up for
using different temporal scales for different auditory tasks, where
different temporal scales may have complementary advantages.

\paragraph{Relations to Gabor functions.}

By rewriting the expression (\ref{eq-complex-spectr-gauss-def})  
for the complex-valued spectrogram based
on the Gaussian temporal scale-space concept as
\begin{equation}
S_{Gauss}(\omega, t;\; \tau) 
      = e^{-i \omega t} \int_{t'=-\infty}^{\infty}
                g(t - t';\; \tau) \,
                e^{i\omega (t-t')} \, f(t') \, dt',
\end{equation}
it can be seen that up to a phase shift by this
multi-scale spectrogram can equivalently be interpreted as the convolution of the
original auditory signal $f$ by Gabor functions \cite{Gab46}  of the form
(see figure~\ref{fig-first-stage-aud-rec-fields}, top row)
\begin{equation}
  G(t, \omega;\; \tau) = g(t;\; \tau) \, e^{i \omega t}
\end{equation}
Such Gabor functions have been previously used for analyzing auditory signals by
several authors, including \cite{WolGodDor01-ASAA,Kle02-ActAcust,KleGel02-InterSpeech,LobLoi03-ICASSP,QiuSchEsc03-JNeuroPhys,BooLie06-ICASSP,EzzBouPog07-InterSpeech,DomHecJouGoe08-ICASSP,HeLecMadAll09-AffComp,HecDomJouGoe11-SpeechComm,WuZhaShi11-ASLP,SchMeyKol12-JASA,SamLac12-SpeechTech}.
Our theory provides a new way of deriving this form of representation
with special emphasis on the multi-scale nature of the Gaussian window
functions and their resulting cascade properties between spectrograms
at different temporal scales.

\paragraph{Relations to Gammatone filters.}

In the special case when the time constants of all the $K$ truncated
exponential filters that are coupled in cascade are all equal $\mu_i = \mu$,
it follows from combination of
equations~(\ref{eq-complex-spectr-exp-def}) and
(\ref{eq-comp-conv-kernel-rep-trunc-exp-same-time-const})
that the multi -scale spectrogram is given by
\begin{equation}
      C_{caus-exp}(t, \omega;\; \mu) 
= e^{-i \omega t} \int_{t'=-\infty}^{\infty}
   \frac{(t-t')^{K-1} \, e^{-(t-t')/\mu}}{\mu^K \, \Gamma(K)}
   e^{i\omega (t-t')} 
\, f(t') \, dt'
\end{equation}
corresponds to convolution of the input signal $f$ by filters of the
form (see figure~\ref{fig-first-stage-aud-rec-fields}, third and fifth
rows)
\begin{align}
  \begin{split}
     h_{cos}(t, \omega;\; \mu, k) 
     & = \frac{t^{K-1} \, e^{-t/\mu}}{\mu^K \, \Gamma(K)} \cos \omega t
  \end{split}\\
  \begin{split}
     h_{sin}(t, \omega;\; \mu, k) 
     & = \frac{t^{K-1} \, e^{-t/\mu}}{\mu^K \, \Gamma(K)} \sin \omega t
  \end{split}
\end{align}
For comparison, the Gammatone filter with parameters $a$ and $b$ and
frequency $\phi$ is defined according to 
\begin{equation}
   \gamma(t) = a \, t^{n-1} e^{-2\pi b t} \cos(2\pi \phi \, t + \alpha), \,
\end{equation}
By identification of the parameters 
\begin{equation}
   a = \frac{1}{\mu^K \, \Gamma(K)} \quad\quad b = \frac{1}{2 \pi \mu}
\end{equation}
and using $\omega = 2 \pi \, \phi$ it follows that we can derive the Gammatone filter as a special case
of applying a time-causal scale-space representation with discrete
scale levels to the projections $f_{cos}(t, \omega)$ and
$f_{sin}(t, \omega)$ of an auditory signal $f(t)$ onto a complex sine
wave $e^{- i \omega t}$.

Gammatone filter banks are also commonly used in audio processing
\cite{Joh72-HearTheory,PatNimHolRic87-GammaTone,HewMed94-JASA,PatAllGig95-JASA,IriPat97-JASA,AmbEppLin01-ICASSP,Hoh02-GammaTone,ImmPee03-AcResLettOnLine,SchBezWagNey07-ICASSP,NgaSawHisSer10-ISCAS}.
The present treatment provides a new way of deriving them in a
principled and
conceptually similar way as the Gabor filters can be derived, with the
differences that the temporal filtering operations are
required to be truly time-causal and that only a discrete set of
temporal scale levels is to be used.

\paragraph{Generalization of Gammatone filters.}

In addition, by allowing for different time
constants in the primitive truncated exponential filters, this
scale-space concept leads to a generalized family of such kernels
\begin{align}
  \begin{split}
     h_{cos}(t, \omega;\; \mu) 
     & = h_{composed}(t;\; \mu) \, \cos \omega t
  \end{split}\\
  \begin{split}
     h_{sin}(t, \omega;\; \mu) 
     & = h_{composed}(t;\; \mu) \,  \sin \omega t
  \end{split}
\end{align}
with $h_{composed}$ according to
(\ref{eq-expr-comp-kern-trunc-exp-filters}).
As can be seen from a comparison between the two classes of
time-causal window functions in
figure~\ref{fig-first-stage-aud-rec-fields}, 
which are shown for the same 
temporal extent as measured in terms of
the variance $\tau = \sum \mu_i^2$ of the underlying filter
$h_{composed}$ without the complex sine wave, the frequency selective
filters based on truncated exponential filters having a logarithmic
distribution of the intermediate temporal scale levels allow for a 
faster response compared to the corresponding filters based 
on a uniform distribution. Thereby, this family of generalized
Gammatone filters allows for additional degrees of freedom
to obtain different trade-offs between {\em e.g.\/} 
the frequency selectivity and the temporal delay of time-causal window
functions by varying the number of levels $K$ and the distribution
parameter $c$ --- see appendix~\ref{app-freq-anal} and
appendix~\ref{app-temp-dyn} for in-depth analysis of the
frequency selectivity and the temporal delay of such kernels.

\paragraph{Frequency-dependent window scale.}

To guarantee basic covariance properties of the spectrogram under a frequency shift
\begin{equation}
  \omega \mapsto \alpha \, \omega
\end{equation}
it is as earlier mentioned natural to let the temporal window scale 
vary with the frequency $\omega$ in such a a way that the temporal
window scale in units of $\sigma = \sqrt{\tau}$ is proportional to 
the wavelength $\lambda = 2 \pi/\omega$
\begin{equation}
  \label{eq-tau-depend-on-omega-gauss-exp-spectr}
   \tau = \left( \frac{2 \pi \, n}{\omega} \right)^2
\end{equation}
where $n$ is a parameter.
Thereby, it follows that {\em e.g.\/}\ a shift by one octave of a
musical piece implies that the corresponding spectrogram will 
also appear similar while shifted by one octave if the frequency
axis of the spectrogram is parameterized on a logarithmic scale.

To prevent the temporal window scale from being too short for high
frequency sounds, we have additionally chosen to add a soft lower threshold
such that the temporal extent is instead chosen according to 
\begin{equation}
  \label{eq-tau-depend-on-omega-gauss-exp-spectr-soft-threshold}
   \tau = \tau_0 + \left( \frac{2 \pi \, n}{\omega} \right)^2
\end{equation}
where $\tau_0 = \sigma_0^2$ denotes a lower bound on the temporal window scale.
Thereby, frequency covariance of a 2-D spectrogram will only be
approximate, while being a good approximation if $\tau \gg \tau_0$.
If we quantify $\tau \gg \tau_0$ as $\tau = \beta^2 \tau_0$,
then the soft threshold corresponds to
\begin{equation}
  \omega = \frac{2 \pi n}{\sqrt{\beta^2-1} \, \sigma_0}
\end{equation}
which with $\sigma_0 = 1~\mbox{ms}$, $n = 8$ and $\beta = 2$
corresponds to a frequency of about 4 600 Hz. 
By varying the parameters $\sigma_0$ and $n$, we can move the
frequency where deviations from true invariance begin to occur
for a given value of the tolerance parameter $\beta$.

In human hearing, there is different evidence that the resolution of pitch perception
decreases in the area around 2-5 kHz (see e.g. \cite{Har96-JASA}).
The frequency difference limen for sine tones decreases \cite{Moo73-JASA}; 
the synchrony in the neural firing in the auditory nerve decreases
\cite{Joh80-JASA}, and the ability to identify the pitch of a mistuned
harmonic decreases \cite{HarMcAdaSmi90-JASA}. All three effects occur
within the same frequency range. Therefore, there should presumable
be an upper limit above which self-similarity will not hold.

To prevent the temporal delay from being too long at low
frequencies, one can also introduce a soft upper bound on the temporal
scale
\begin{equation}
  \tau'
  = \frac{\tau}
             {\left(
                1 + \left( \frac{\tau}{\tau_{\infty}} \right)^p
              \right)^{1/p}}
\end{equation}
for suitable values of the parameters $\tau_{\infty}$ and $p$.
Then, approximate frequency covariance will hold over some subset of
frequencies as defined by the parameters $n$, $\tau_0$,
$\tau_{\infty}$ and $p$.

\section{Receptive fields defined over the spectrogram}
\label{sec-2nd-layer-rec-fields}

\begin{figure}[hbtp]
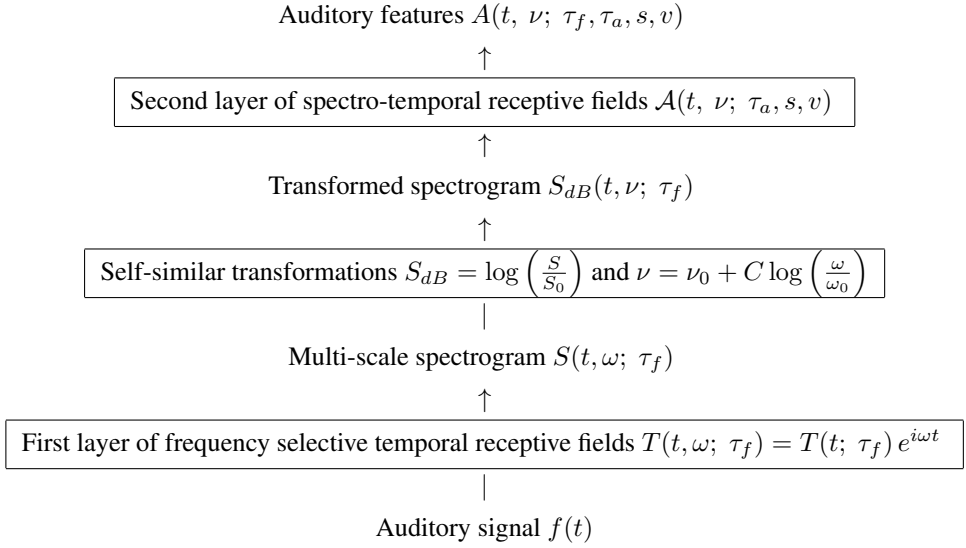

  \begin{center}

  \small

 \vspace{1.4mm}
  \begin{tabular}{c}
    Auditory features $A(t, \nu;\; \tau_f, \tau_a, s, v)$\\
  \end{tabular}
  \vspace{1.4mm}

 \begin{tabular}{c}
    $\uparrow$
  \end{tabular}

 \begin{tabular}{|c|}
  \hline
    Second layer of spectro-temporal receptive fields 
    ${\cal A}(t, \nu;\; \tau_a, s, v)$ 
    \vphantom{$S_{dB} = \log \left( \frac{S}{S_0} \right)$} \\
  \hline
  \end{tabular}

 \begin{tabular}{c}
    $\uparrow$
  \end{tabular}

  \vspace{1.4mm}
  \begin{tabular}{c}
    Transformed spectrogram $S_{dB}(t, \nu;\; \tau_f)$\\
  \end{tabular}
  \vspace{1.4mm}

  \begin{tabular}{c}
    $\uparrow$
  \end{tabular}

  \begin{tabular}{|c|}
    \hline
    Self-similar transformations 
    $S_{dB} = \log \left( \frac{S}{S_0} \right)$  
    and $\nu = \nu_0 + C \log \left( \frac{\omega}{\omega_0} \right)$\\
   \hline
  \end{tabular}

  \begin{tabular}{c}
    $\mid$
  \end{tabular}

  \vspace{1.4mm}
  \begin{tabular}{c}
    Multi-scale spectrogram $S(t, \omega;\; \tau_f)$\\
  \end{tabular}
  \vspace{1.4mm}

  \begin{tabular}{c}
    $\uparrow$
  \end{tabular}

  \begin{tabular}{|c|}
  \hline
    First layer of frequency selective temporal receptive fields $T(t, \omega;\; \tau_f) = T(t;\; \tau_f) \, e^{i \omega  t}$
   \vphantom{$S_{transf} = \log \left( \frac{S}{S_0} \right)$} \\
  \hline
  \end{tabular}

  \begin{tabular}{c}
    $\mid$
  \end{tabular}

  \vspace{1.4mm}
  \begin{tabular}{c}
    Auditory signal $f(t)$\\
  \end{tabular}

\end{center}
\caption{Schematic illustration of the definition of auditory features 
  from a second layer of receptive
  fields over the spectrogram, where we also allow for a
  logarithmic transformation of the magnitude values $|S|$ of the
  spectrogram prior to the application of the second layer of linear
  receptive fields and make use of a logarithmic transformation of
  the frequencies $\nu = \nu_0 + C \log \left( \frac{\omega}{\omega_0} \right)$ 
  before defining the linear receptive fields over the spectro-temporal domain.
  Regarding the scale parameters, the first layer of temporal
  receptive fields depends on a single temporal scale parameter
  $\tau_f$ for the frequency selective temporal filters,
  whereas the second layer of auditory receptive fields also depends
  on an additional temporal scale parameter $\tau_a$, a spectral scale parameter
  $s$ over the logarithmic frequencies $\nu$ and a glissando parameter
  $v$ representing the rate by which the logarithmic frequencies may vary
  over time.}
  \label{fig-sec-layer-rec-fields}
\end{figure}

Given that a spectrogram has been computed by a first
layer of auditory receptive fields, we will define a second layer of
receptive fields by operating on the spectrogram with 2-D
spectro-temporal filters as illustrated in
figure~\ref{fig-sec-layer-rec-fields},
in a structurally similar way as visual
receptive fields are applied to time-varying visual input
\cite{Lin13-BICY,Lin13-PONE}.

\subsection{Logarithmic transformations of the spectrogram}
\label{sec-log-trans-spectr}

Prior to the definition of receptive fields from the spectrogram, it
is natural to allow for a self-similar transformation of
the magnitude values of the spectrogram
\begin{equation}
  \label{eq-transf-spectrogram}
   S_{dB} = 20 \log_{10} \left( \frac{|S|}{S_0} \right).
\end{equation}
A logarithmic transformation of the magnitude values of the spectrogram
implies that a multiplicative transformation of the sound pressure
$f \mapsto a \,  f$, corresponding to $|S| \mapsto a \, |S|$,
or an inversely proportional reduction in the sound pressure of the signal from a
single auditory point source as function of distance 
$f \mapsto f/R$, corresponding to $|S| \mapsto |S|/R$,
are both transformed into a subtraction of the logarithmic 
magnitude values by a constant
\begin{equation}
  \label{eq-basic-transf-spectr-magn}
 |S| \mapsto \frac{a\, |S|}{R} \Rightarrow
  S_{dB} \mapsto S_{dB} + 20 \log_{10} a - 20 \log_{10} R.
\end{equation}
If we operate on the logarithmically transformed spectrogram by a
receptive field ${\cal A}_{\Sigma}$ that is based on a combination of a
spectro-temporal smoothing operation ${\cal T}_{\Sigma}$ 
with spectral and temporal scale parameters as determined by a
spectro-temporal covariance matrix $\Sigma$, 
spectral and/or temporal derivatives 
$\partial_{t^{\alpha}} \partial_{\nu^{\beta}}$ of orders $\alpha$ and
$\beta$ with at least one of $\alpha > 0$ or $\beta > 0$
\begin{equation}
   {\cal R} \, S_{dB} 
  = \partial_{t^{\alpha}} \partial_{\nu^{\beta}} {\cal T}_{\Sigma} \, S_{dB} 
\end{equation}
then it follows that the influence on the receptive field responses of
the constants $a$ and $R$ 
\begin{equation}
   {\cal R} \, S_{dB} 
  = \partial_{t^{\alpha}} \partial_{\nu^{\beta}} {\cal T}_{\Sigma} \, (S_{dB} + 20 \log_{10} a - 20 \log_{10} R)
  = \partial_{t^{\alpha}} \partial_{\nu^{\beta}} {\cal T}_{\Sigma}  \,
  S_{dB} + 0 + 0
\end{equation}
will be eliminated by the derivative operation if the constants $a$ and $R$ do not depend on time $t$ or
logarithmic frequency $\nu$, implying invariance of the second-layer receptive field
responses to variations in the sound pressure or the distance to a sound
source.

A logarithmic transformation of the magnitude values is also
compatible with the Weber-Fechner law, which states that the ratio of
an increment threshold $\Delta I$ of a stimulus for a just noticeable
different in relation to the background intensity $I$ is constant over
large ranges of magnitude variations, which approximately holds in
both visual and auditory perception \cite{Pal99-Book,KanSchJes00-PrincNeurSci}.

Furthermore, since logarithmic frequencies constitutes a natural metric for
relating frequencies of sound
\cite{Fle34-JASA,KanSchJes00-PrincNeurSci,You05-JASA}
and there is an approximately logarithmic distribution of frequencies
both on  the basilar membrane \cite{Gre90-JASA} and in the
organization of the auditory cortex \cite{RomWilKau82-Science}, it is natural to express
these derived receptive fields in terms of logarithmic frequencies parameterized by
\begin{equation}
   \nu = \nu_0 + C \, \log \left( \frac{\omega}{\omega_0} \right)
\end{equation}
for some constants $C$ and $\omega_0$, where specifically $C = 69$ and
$\omega_0 = 2 \pi \cdot 440$ corresponds to logarithmic frequencies
according to the MIDI standard.

\subsection{Structural requirements on second-layer spectro-temporal receptive fields}
\label{sec-struct-req-2nd-layer-RF}

Given a transformed spectrogram defined in this way, let us
define a family of second layer spectro-temporal receptive fields $A(t, \omega;\; \Sigma)$ 
that are to operate on the transformed spectrogram 
$S_{dB}(t, \nu;\, \tau)$ and be parameterized by some
multi-dimensional spectro-temporal scale parameter $\Sigma$
that includes smoothing over time $t$ and logarithmic frequencies
$\nu$, and for which the corresponding operator ${\cal A}_{\Sigma}$ 
is required to obey:
\begin{itemize}
\item[(i)]
  {\em linearity\/} over the logarithmic spectrogram
  \begin{equation}
    {\cal A}_{\Sigma} (a \, S_1 + b \, S_2) 
    = a {\cal A}_{\Sigma} (S_1) + b \, {\cal A}_{\Sigma} (S_2)
  \end{equation}
  to ensure that (a)~the multiplicative relations of the magnitude of
  the spectrogram (\ref{eq-basic-transf-spectr-magn}) that are mapped to linear relations
  by the logarithmic transformation (\ref{eq-transf-spectrogram})
  are preserved as linear relations
  over the receptive field responses and (b)~that scale-space properties that are imposed to ensure non-creation
  of new structures in smoothed spectrograms as defined by
  spectro-temporal smoothing kernels do also transfer to
  spectro-temporal derivatives of these,
\item[(ii)]
  {\em shift-invariance\/} with respect to translations over time 
  $t \mapsto t + \Delta t$ and logarithmic frequencies $\nu \mapsto
  \nu + \Delta \nu$
  \begin{equation}
     {\cal A}_{\Sigma} ({\cal S}_{(\Delta t, \Delta \nu)} S) = {\cal S}_{(\Delta t, \Delta \nu)}  ({\cal A}_{\Sigma} S)
  \end{equation}
  such that all temporal moments and all logarithmic frequencies are treated in a similar manner.
  Temporal shift invariance implies that an auditory stimulus
  should be perceived in a similar manner irrespective of when it
  occurs. 
  Shift-invariance in the logarithmic frequency domain implies that,
  for example,  a piece of music should be perceived in a similar manner if it is transposed by
  {\em e.g\/} one octave.
\end{itemize}
These conditions together imply that the spectro-temporal receptive
fields should be given by convolution with some two-dimensional kernel
over the spectro-temporal domain \cite{HirWid55}
\begin{equation}
  ({\cal A}_{\Sigma} S_{dB})(t, \nu;\; \tau_f, \Sigma)
  = \int_{\xi = -\infty}^{\infty} \int_{\eta = -\infty}^{\infty} 
      T(\xi, \eta;\; \Sigma) \, S_{dB}(t-\xi, \nu-\eta;\; \tau_f) \,
      d\xi \, d\eta.
\end{equation}
To characterize what types of receptive fields are compatible with
scale-space properties, we will next impose additional structural
requirements, which will take different forms depending on time is
treated in a time-causal or non-causal manner.

\paragraph{Relations between receptive fields at different
  spectro-temporal scales.}

For pre-recorded sound signals, for which we can take the freedom of
accessing data from the virtual future in relation to any time moment, we impose a 
\begin{itemize}
\item[(iii.a)]
  continuous semi-group structure over spectro-temporal scales on the second layer of receptive
  fields
  \begin{equation} 
    T(\cdot, \cdot;\; \Sigma_2) = T(\cdot, \cdot;\; \Sigma_2 -
    \Sigma_1) \, T(\cdot, \cdot;\; \Sigma_1)
  \end{equation}
  corresponding to an additive structure over the
  multi-dimensional scale parameter $\Sigma$.
\end{itemize}
For time-causal causal signals, we require:
\begin{itemize}
\item[(iii.b)]
  a continuous semi-group structure over spectral scales $s$
  \begin{equation}
     T(\cdot;\; s_2)  = T(\cdot;\; s_2-s_1) \, T(\cdot;\; s_1)
  \end{equation}
  and a Markov property between adjacent temporal scales:
  \begin{equation}
     T(\cdot;\; \tau_{k+1})  = (\Delta T)(\cdot;\; k) \, T(\cdot;\; \tau_{k}).
  \end{equation}
\end{itemize}
These requirements are analogous to previous treatment in
section~\ref{sec-struct-req-temp-rec-fields}, with extensions from a
purely temporal domain to a spectro-temporal domain.

\paragraph{Non-creation of new spectro-temporal structures with
  increasing scale.}

When processing the spectrogram at different spectro-temporal scales,
we want to ensure that the spectro-temporal receptive fields do
not create new structures at coarser scales that do not correspond to
simplifications of corresponding structures at finer scales.
Depending on whether time is treated in a time-causal or non-causal
manner, we formalize this conditions in different manners:
\begin{itemize}
\item[(iv.a)]
  For the non-causal Gaussian spectrogram
  (\ref{eq-complex-spectr-gauss-def}), for which temporal causality of
  the temporal smoothing kernels is disregarded,
  we require {\em non-enhancement of local extrema\/} in the sense that if for
  some scale $\Sigma_0$ the point $(t_0, \nu_0)$ is a local 
  maximum (minimum) for the mapping 
  $(t, \nu) \mapsto F(t, \nu;\; \Sigma_0)$ then the value at this point
  must not increase (decrease) with increasing scale $\Sigma$.
\item[(iv.b)]
  For the time-causal spectrogram (\ref{eq-complex-spectr-exp-def})
  based on truncated exponential filters coupled in cascade
  (\ref{eq-comp-trunc-exp-cascade}), we require: (iv.b1) the smoothing
  operation over the
  log-spectral domain to satisfy non-enhancement of local extrema in
  the sense that if at some log-spectral scale $s_0$ a point $\nu_0$
  is a local maximum (minimum) of the mapping $\nu \mapsto S_{dB}(\nu;\; s_0)$
  obtained by disregarding the temporal variations, then the value at this point
  must not increase (decrease) with increasing log-spectral scale $s$, 
  and (iv.b2)  the
  smoothing operation over time to be a time-causal scale-space kernel
  in the sense that it is guaranteed to not create new local extrema
  under an increase of the temporal scale parameter $\tau$.
\end{itemize}

\paragraph{Glissando covariance.}

In musical performance, the frequencies may vary continuously over time
in such a way that the fundamental frequency $\omega_1$ and the
harmonics (overtones)
$\omega_j$ are all multiplied by the same time-varying factor 
$\omega_j (t) = \psi(t) \, \omega_j$.
This is in particular prominent in singing, but may occur in all instruments with continuous pitch control.
In terms of logarithmic frequencies, we can model a local
linearization of this temporal variability as a glissando
transformation of the form
\begin{equation}
  \nu(t) = \nu_0 + v \, t
\end{equation}
analogous to the way spatial image data may be subject to local
Galilean transformations over time.
Comparing two spectrograms, one with constant frequencies over time
and one with linearily varying logarithmic frequencies,
the glissando transformation can in operator form be expressed as
\begin{equation}
  S' = {\cal G}_v \, S
  \quad\quad \mbox{corresponding to} \quad\quad
  S'(t, \nu') = S(t, \nu)
\end{equation}
for $\nu' = \nu + v \, t$.
Specifically, in relation to receptive field responses that are
computed over the two domains with
spectro-temporal scale parameters $\Sigma$ and $\Sigma'$, we may require:
\begin{itemize}
\item[(v)]
  If two local patches of two spectrograms are related by a local
  glissando transformation, then it should be possible to relate the
  local spectro-temporal receptive field responses such that
  \begin{equation}
    {\cal A}_{G_v(\Sigma)} \, G_v \, S = G_v \, {\cal A}_{\Sigma} \, S
  \end{equation}
  for some transformation $\Sigma' = G_v(\Sigma)$ of the spectro-temporal scale
  parameters $\Sigma$.
\end{itemize}

\begin{figure}[!p]
  \begin{center}
    \begin{tabular}{cc}
      {\small\em Gaussian} & {\small\em Time-causal} \\
     \\
     {\small\em Separable} & {\small\em Separable} \\
      $T$ & $T$\\
      \includegraphics[width=0.22\textwidth]{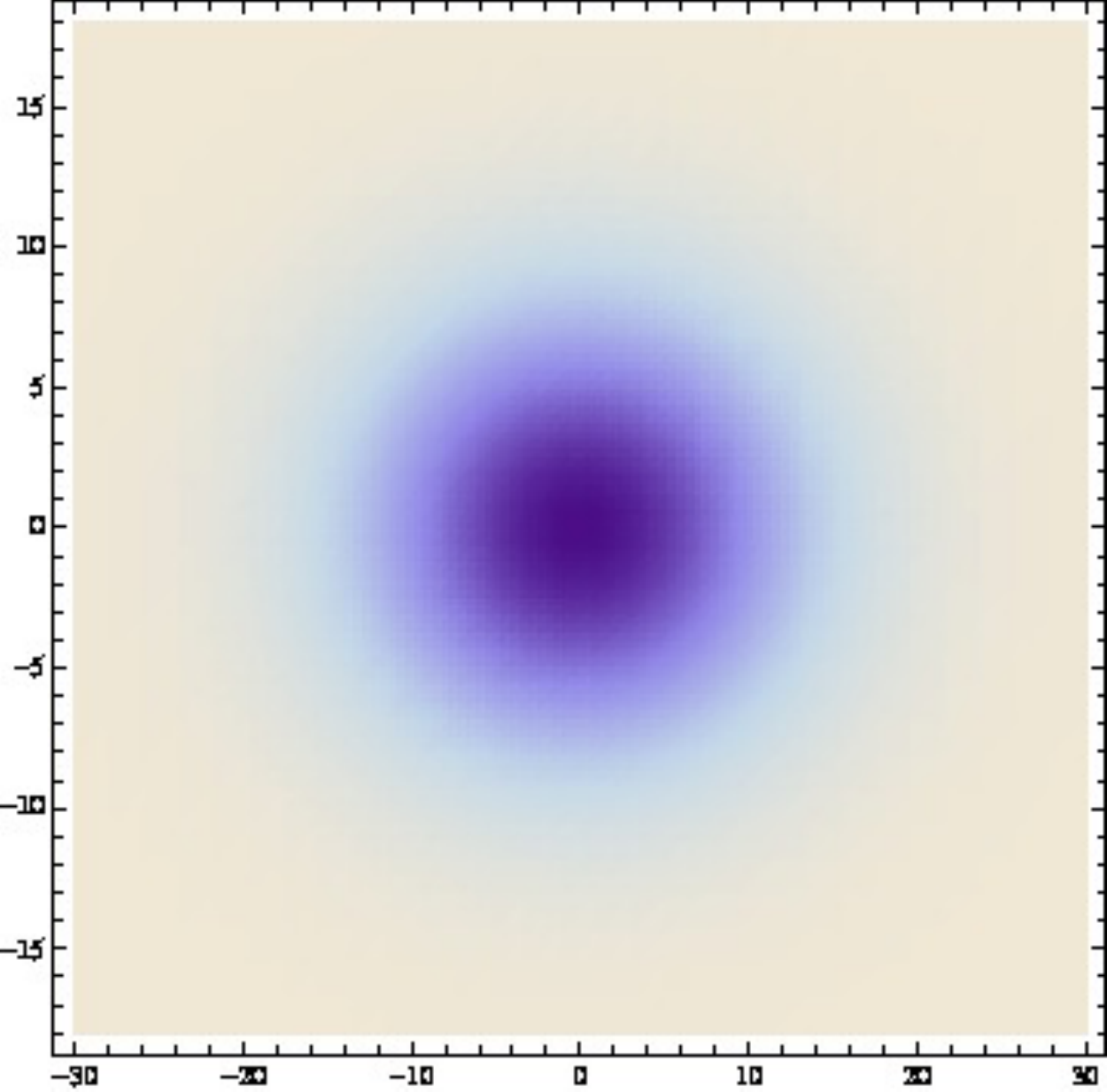} \hspace{-2mm} &
      \includegraphics[width=0.22\textwidth]{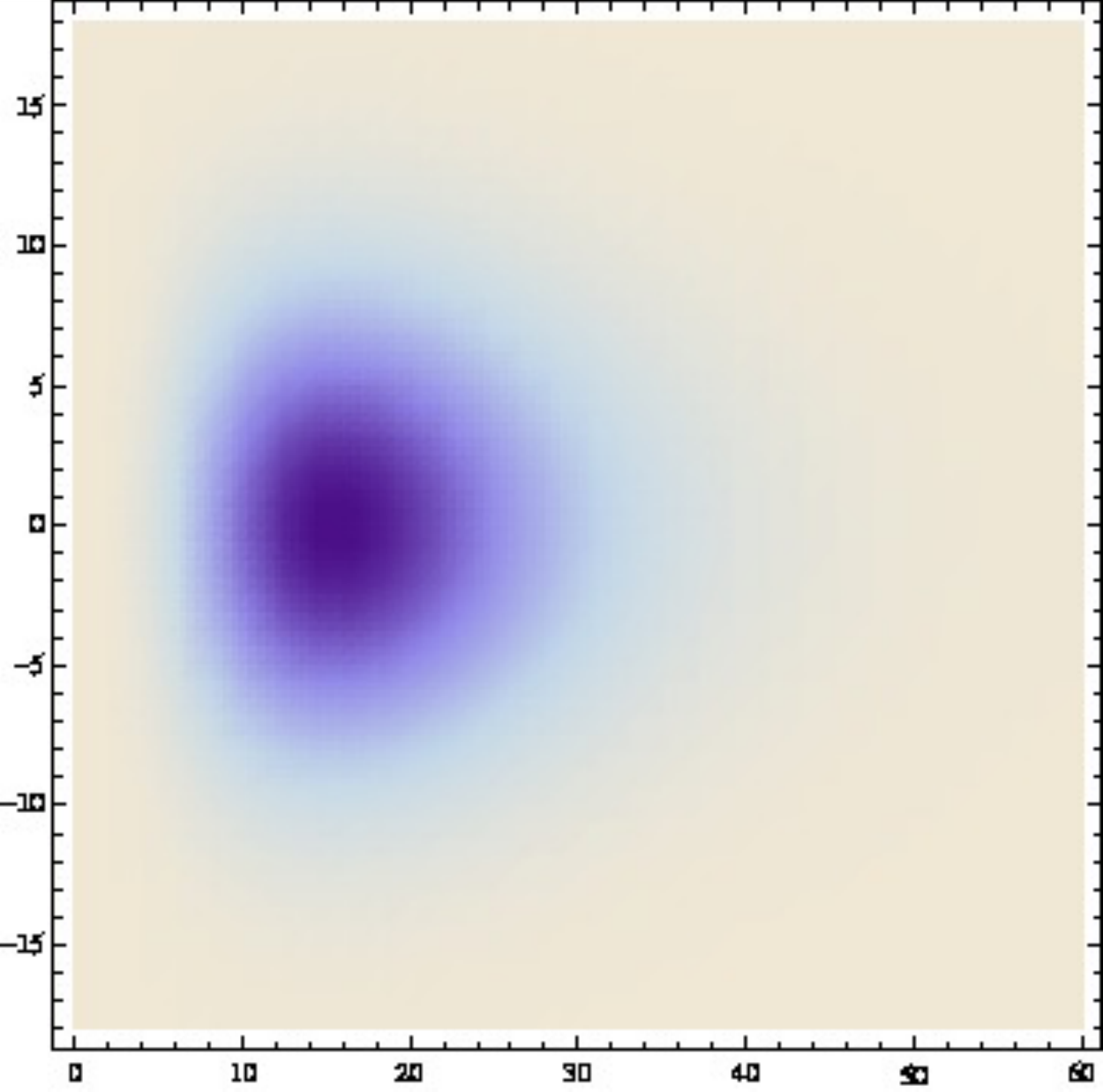} \\
    \end{tabular} 
  \end{center}
  \vspace{-5mm}
  \begin{center}
    \begin{tabular}{cc}
      {\small $\partial_t T$} & {\small $\partial_t T$}\\
      \includegraphics[width=0.22\textwidth]{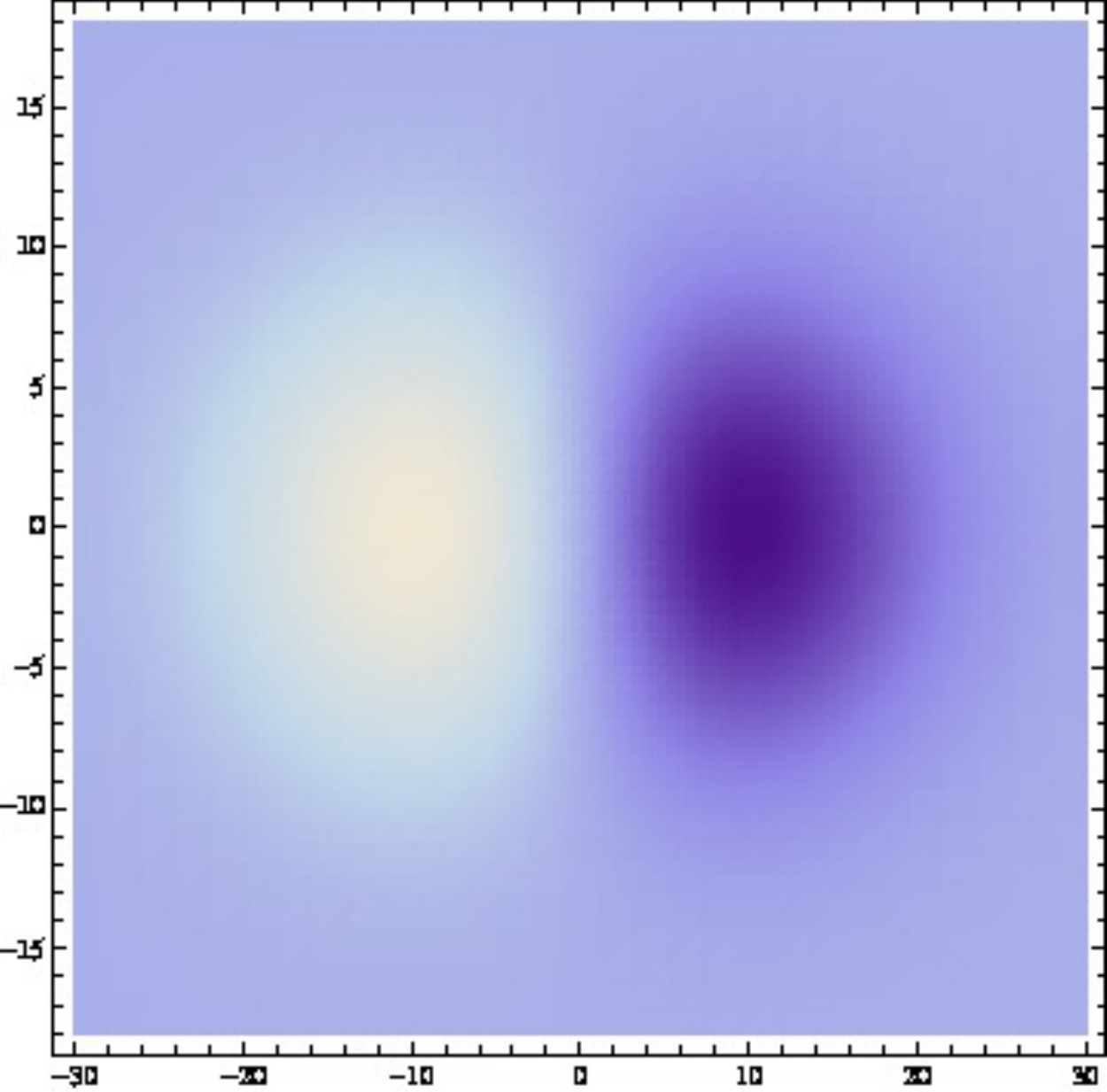} \hspace{-2mm} &
      \includegraphics[width=0.22\textwidth]{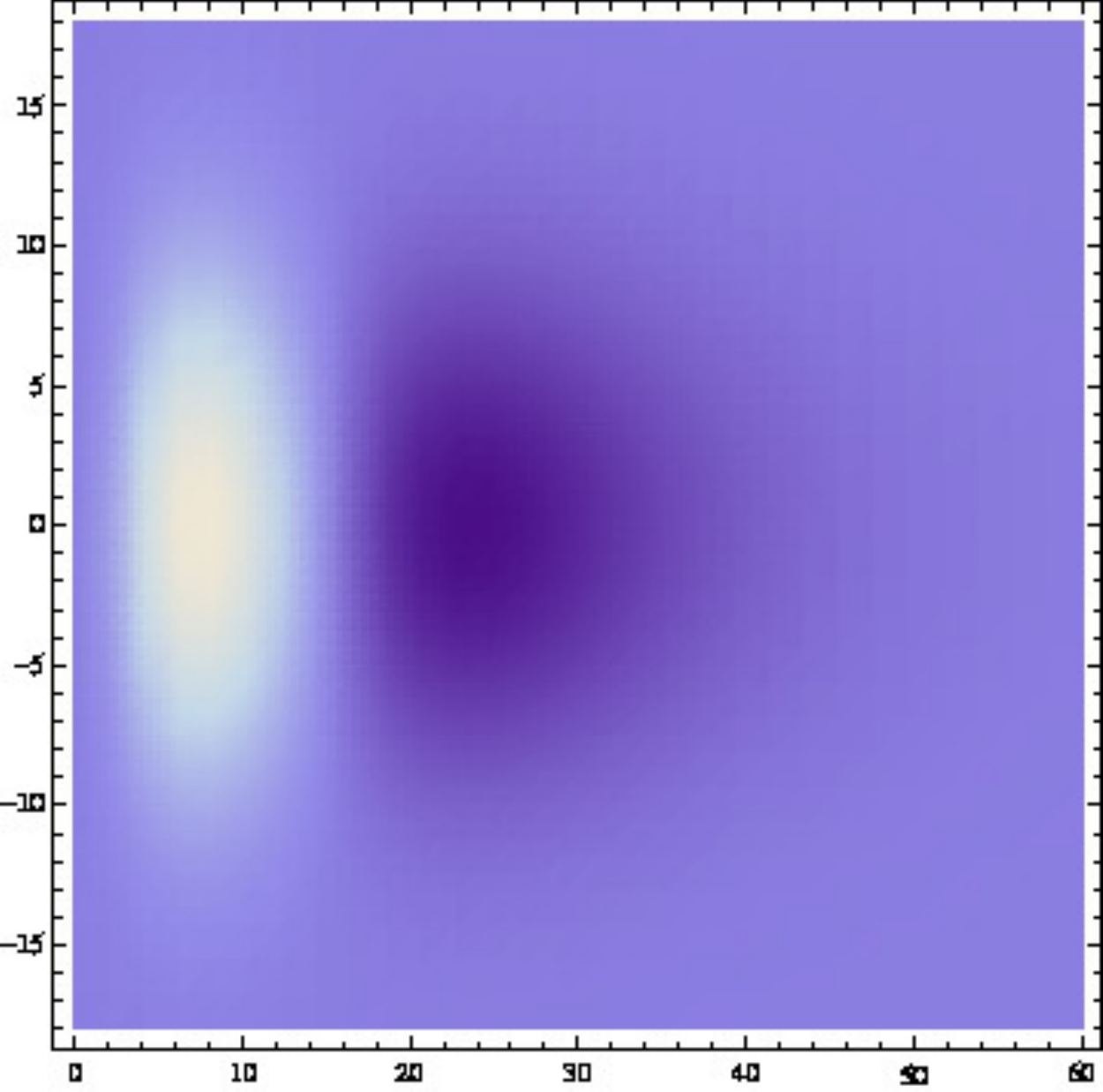} \\
    \end{tabular} 
  \end{center}
  \vspace{-5mm}
  \begin{center}
    \begin{tabular}{cc}
      {\small $\partial_{\nu\nu} T$} & {\small $\partial_{\nu\nu} T$}\\
      \includegraphics[width=0.22\textwidth]{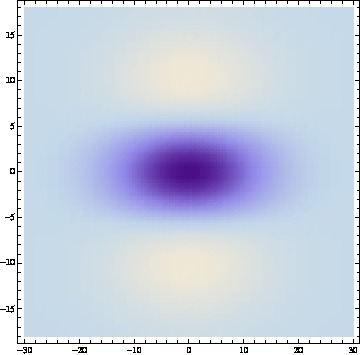} \hspace{-2mm} &
      \includegraphics[width=0.22\textwidth]{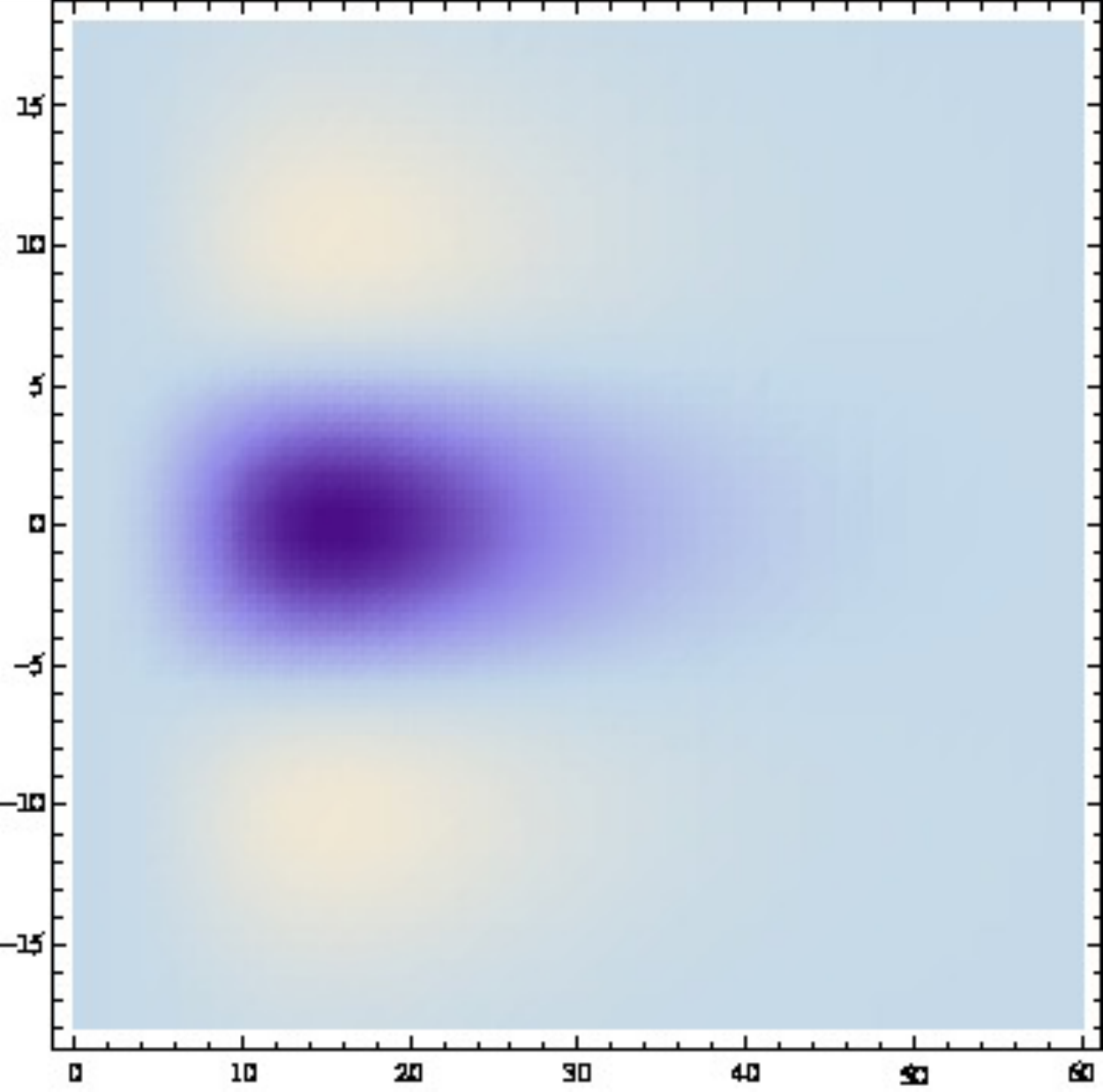} \\
    \end{tabular} 
  \end{center}
  \vspace{-5mm}
  \begin{center}
    \begin{tabular}{cc}
      {\small $\partial_{t\nu\nu} T$} & {\small $\partial_{t\nu\nu} T$}\\
      \includegraphics[width=0.22\textwidth]{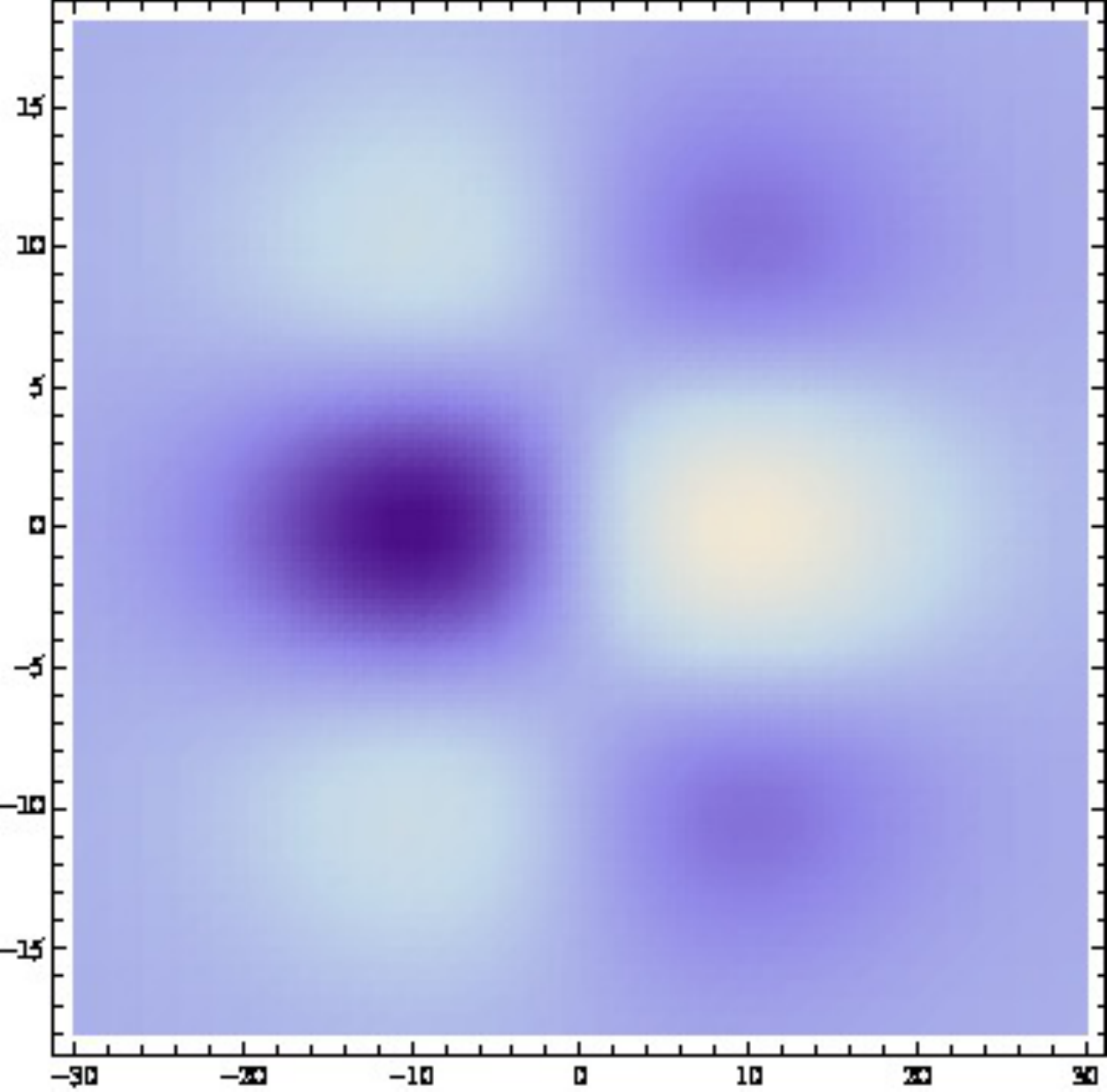} \hspace{-2mm} &
      \includegraphics[width=0.22\textwidth]{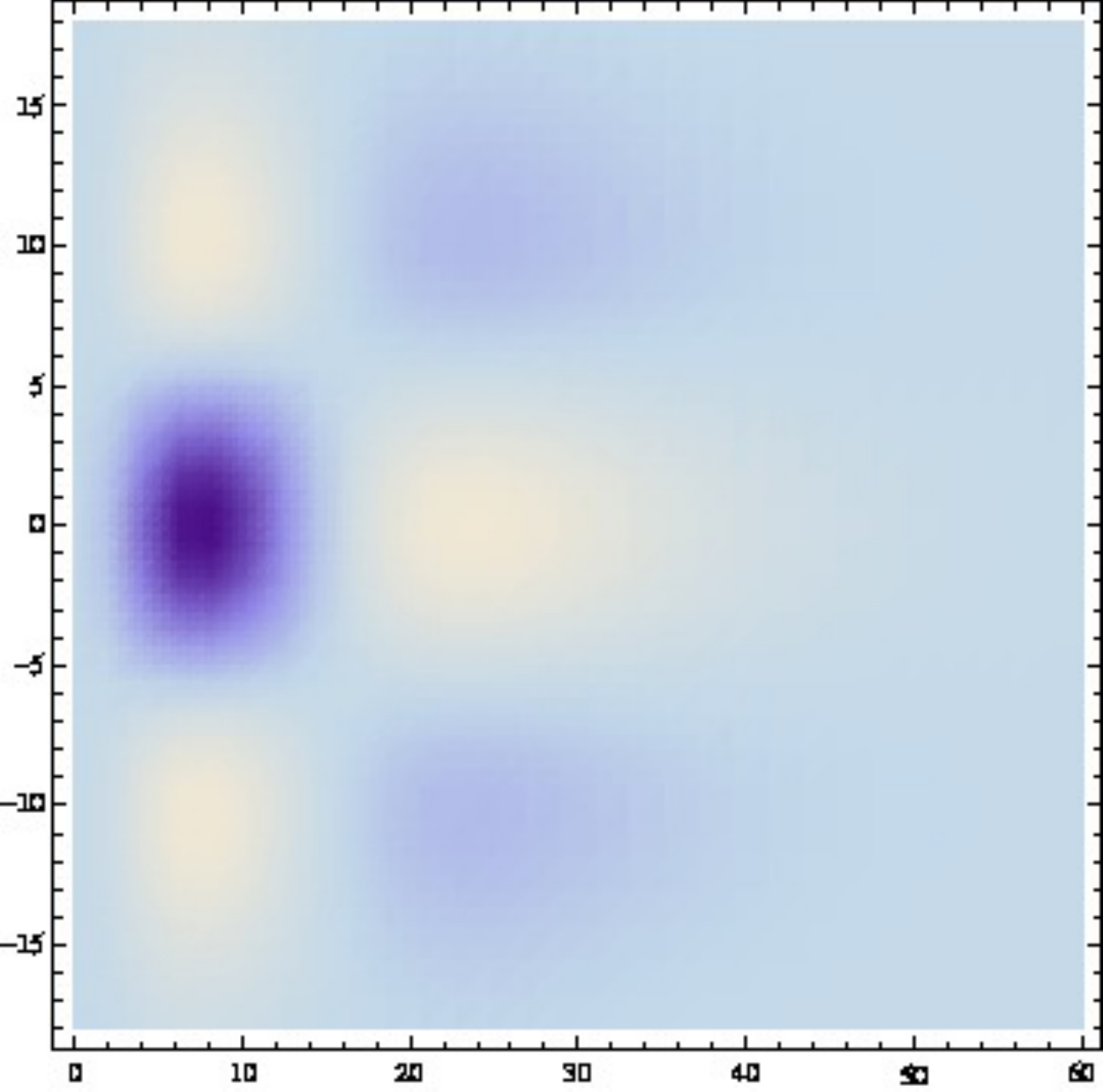} \\
    \end{tabular} 
  \end{center}
  \vspace{-5mm}
  \begin{center}
    \begin{tabular}{cc}
      {\small\em Glissando-adapted} & {\small\em Glissando-adapted} \\
      {\small $\partial_{\nu\nu} T$} & {\small $\partial_{\nu\nu} T$}\\
      \includegraphics[width=0.22\textwidth]{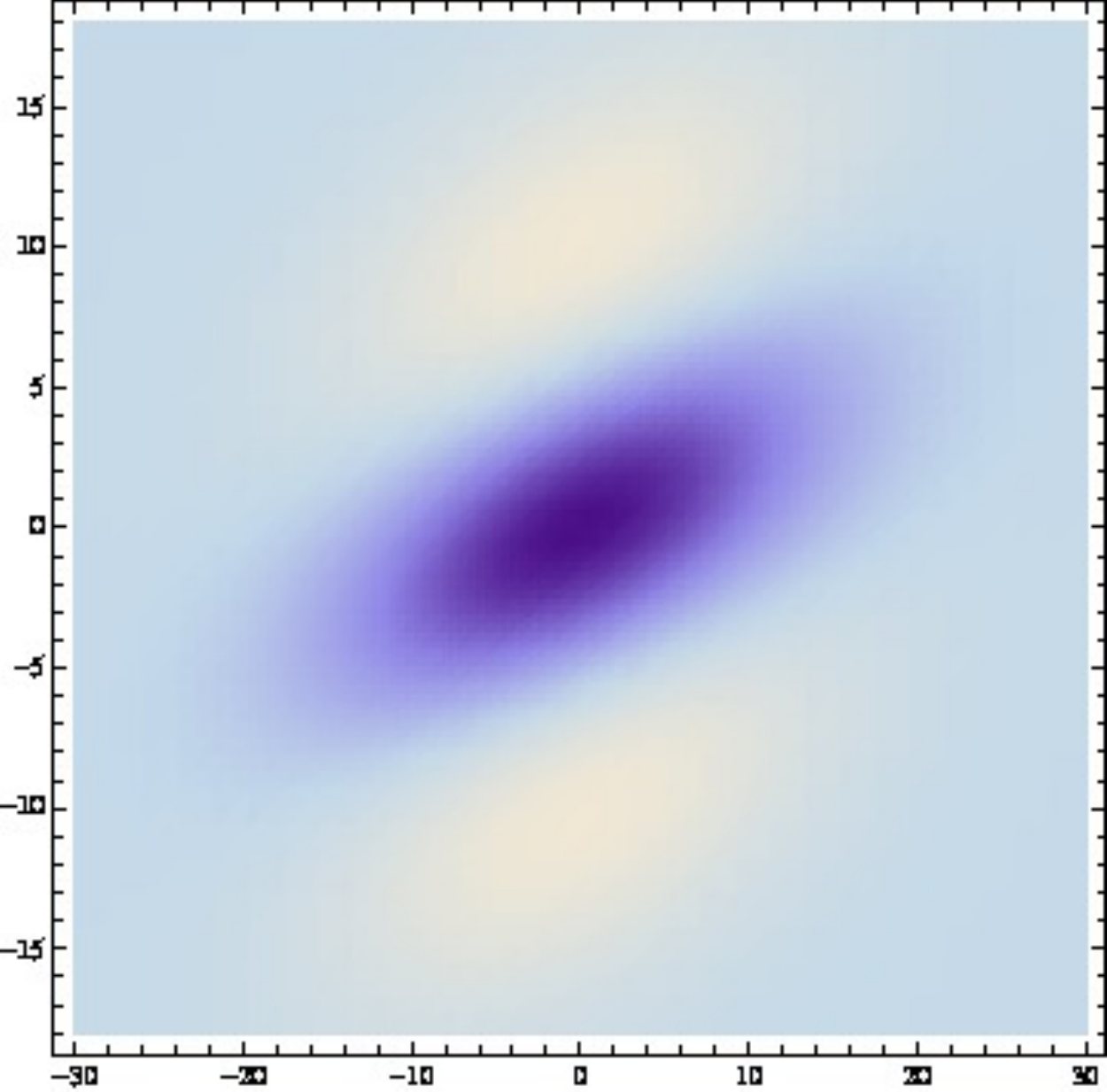} \hspace{-2mm} &
      \includegraphics[width=0.22\textwidth]{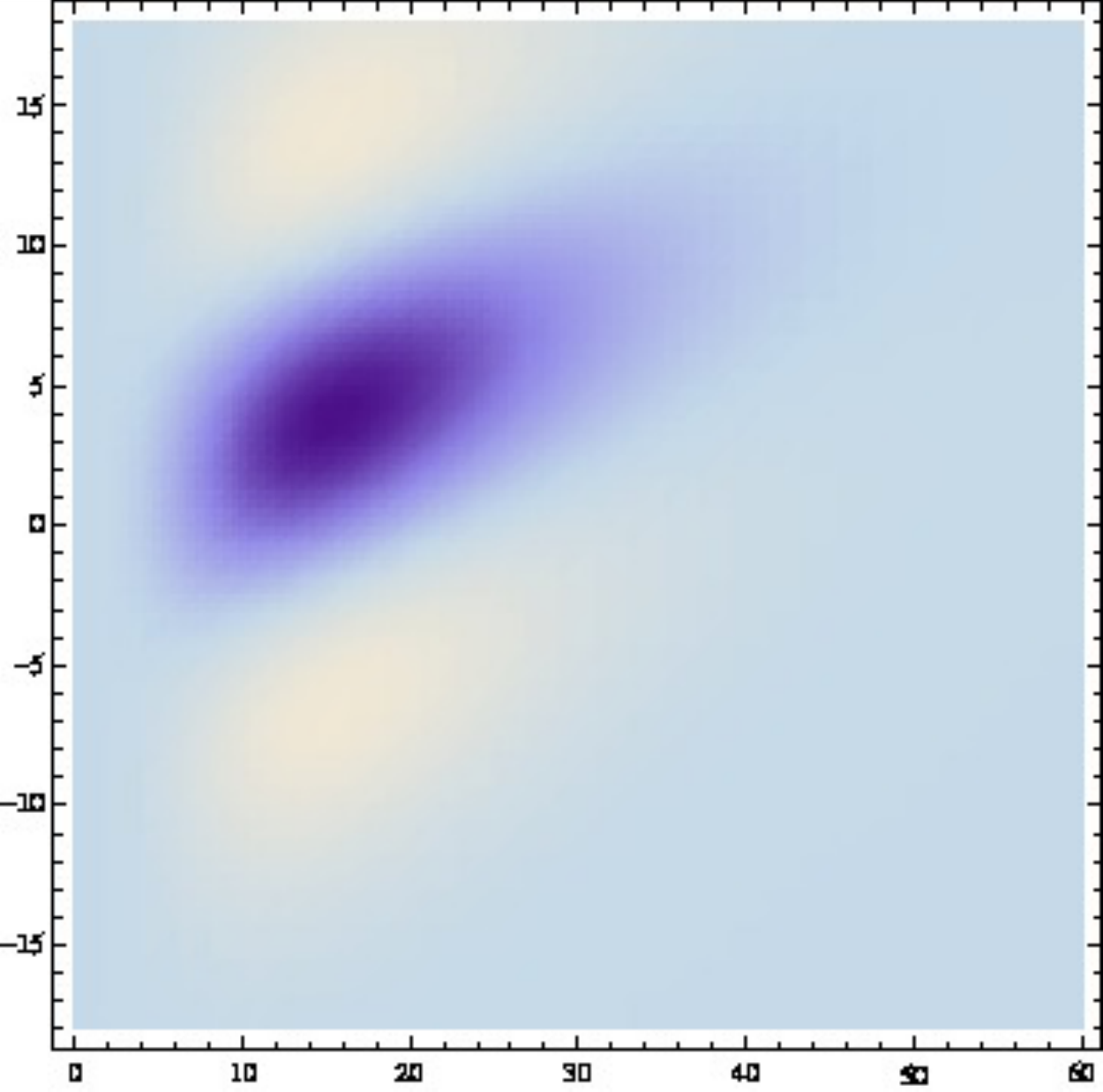} \\
    \end{tabular} 
  \end{center}
  \vspace{-3mm}
  \caption{Examples of idealized spectro-temporal receptive fields as
    obtained from spectro-temporal derivatives of spectro-temporal
    smoothing kernels based on (left column) the non-causal Gaussian
    scale-space concept and (right column) the time-causal scale-space
  concept corresponding to first-order integrators coupled in cascade
  (here using five temporal scale levels and $c = \sqrt{2}$).
  The first four columns show separable receptive fields whereas the
  bottom row shows a non-separable glissando-adapted receptive field.
  (Horizontal dimension: time $t$ in ms. Vertical dimension:
  Logarithmic frequency: $\nu$. Temporal scale: $\sigma_t = \sqrt{\tau}
= 10~\mbox{ms}$. Logspectral scale: $\sigma_{\nu} = \sqrt{s} = 6$.)}
  \label{fig-ideal-spect-temp-rec-fields}
\end{figure}

\subsection{Idealized models for spectro-temporal receptive fields}
\label{sec-ideal-rf}

Given the structural requirements above, it can from derivations%
\footnote{The proofs concerning spectro-temporal receptive fields are similar to
  those regarding spatio-temporal receptive fields over a 1+1-D
  spatio-temporal domain with the spatial dimension replaced by a
  spectral dimension.} 
similar to those that are used for
constraining visual receptive fields given structural
requirements on a visual front-end \cite{Lin13-BICY}
be shown that the second layer of auditory receptive fields should be 
based on spectro-temporal receptive fields of the form
\begin{equation}
  \label{eq-spectr-temp-recfields-gen-form}
   A(t, \nu;\; \Sigma) 
  = \partial_{t^{\alpha}} \partial_{\nu^{\beta}} 
      \left( 
         g(\nu - v t;\; s) \,
         T(t;\; \tau_a) 
      \right)
\end{equation}
where
\begin{itemize}
\item
  $\partial_{t^{\alpha}}$ represents a temporal derivative of order
  $\alpha$ with
  respect to time $t$ which could alternatively be replaced by a glissando-adapted
  temporal derivatives of the form $\partial_{\overline{t}} = \partial_t + v \, \partial_{\nu}$,
\item
  $\partial_{\nu^{\beta}}$ represents a derivative operator of order
  $\beta$ with
  respect to logarithmic frequency $\nu$,
\item
  $T(t;\; \tau_a)$ represents a temporal smoothing kernel with
  temporal scale parameter $\tau_a$ which should either be
  (i)~a temporal Gaussian
  kernel $g(t;\; \tau_a)$ of the form
  (\ref{eq-gauss-gen-spattemp-2+1-D}) or
  (ii)~the equivalent kernel $h_{composed}(t\; \mu)$ according to
  (\ref{eq-expr-comp-kern-trunc-exp-filters})
  corresponding to a set of truncated exponential filters 
  coupled in cascade,
\item
  $g(\nu - v t;\; s)$ represents a Gaussian smoothing kernel over logarithmic
  frequencies $v$ with spectral scale parameter $s$ and $v$
  representing a glissando parameter that makes it possible to adapt the
  receptive fields to variations in frequency $\nu' = \nu + v t$ over time.
\end{itemize}
Thereby, the spectro-temporal receptive fields according to
(\ref{eq-spectr-temp-recfields-gen-form}) constitute a combination of
the purely temporal receptive fields according to the theory in
section~\ref{sec-struct-req-temp-rec-fields} and
section~\ref{sec-spat-temp-scsp-concepts} with a Gaussian scale-space
concept over the log-spectral dimension.

Figure~\ref{fig-ideal-spect-temp-rec-fields} shows examples of spectro-temporal receptive fields
obtained in this way for the two different types of underlying
temporal scale-space concepts. 
For $\nu = 0$, the resulting receptive fields are separable over the
spectro-temporal domain, whereas $\nu \neq 0$ leads to non-separable
glissando-adapted spectro-temporal receptive fields.

\paragraph{Filter parameters of auditory receptive fields.}

The auditory features that are computed from these types of
receptive fields depend on three different scale parameters:
\begin{itemize}
\item
   a temporal window scale parameter $\tau_f$ defining the temporal
   extent of the windows over which the windowed Fourier transforms in
   the spectrograms are defined,
\item
  a secondary temporal integration scale parameter $\tau_a$ defining the temporal
  extent over which the
  magnitude values in the spectogram are integrated over time and
\item
  a spectral scale parameter $s$ defining the extent over
  which smoothing is performed over logarithmic frequencies $\nu$.
\end{itemize}
In addition, this class of spectro-temporal receptive fields comprises:
\begin{itemize}
\item
  a glissando parameter $v$ that makes it possible to adapt the
  receptive fields to variations on the logarithmic frequencies $\nu$
  over time $t$,
\end{itemize}
  and each parameterized spectro-temporal receptive field may occur for different orders
  of differentiation $\alpha$ and $\beta$ with respect to time and
  logarithmic frequencies, respectively.

\begin{figure}[hbtp]
 \begin{center}
    \begin{tabular}{cc}
      {\small\em Spectrogram: $\log |S|$} 
      & {\small\em Spectral smoothing: ${\cal T} \log |S|$} \\
      \includegraphics[width=0.48\textwidth]{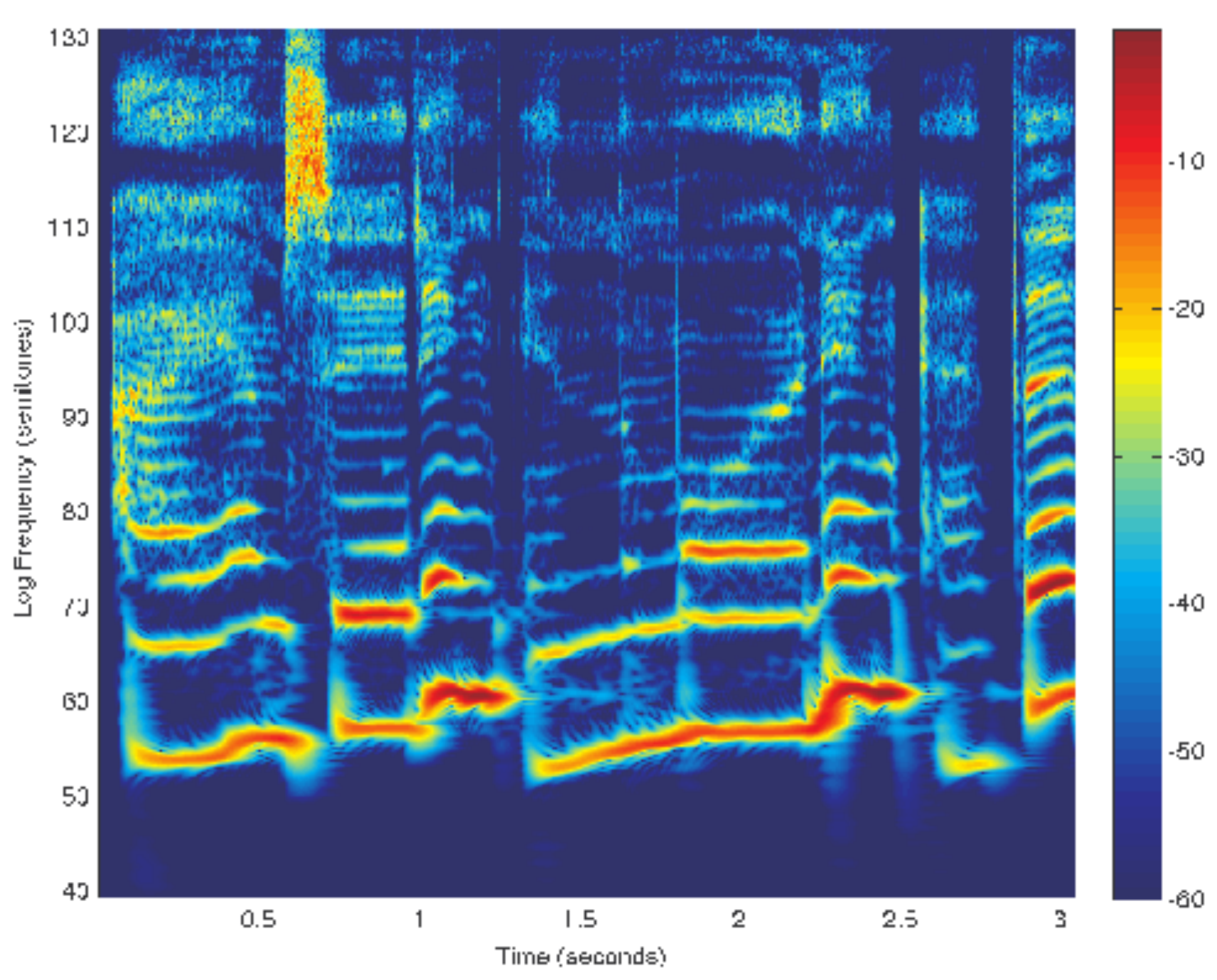} \hspace{-2mm} 
      & \includegraphics[width=0.48\textwidth]{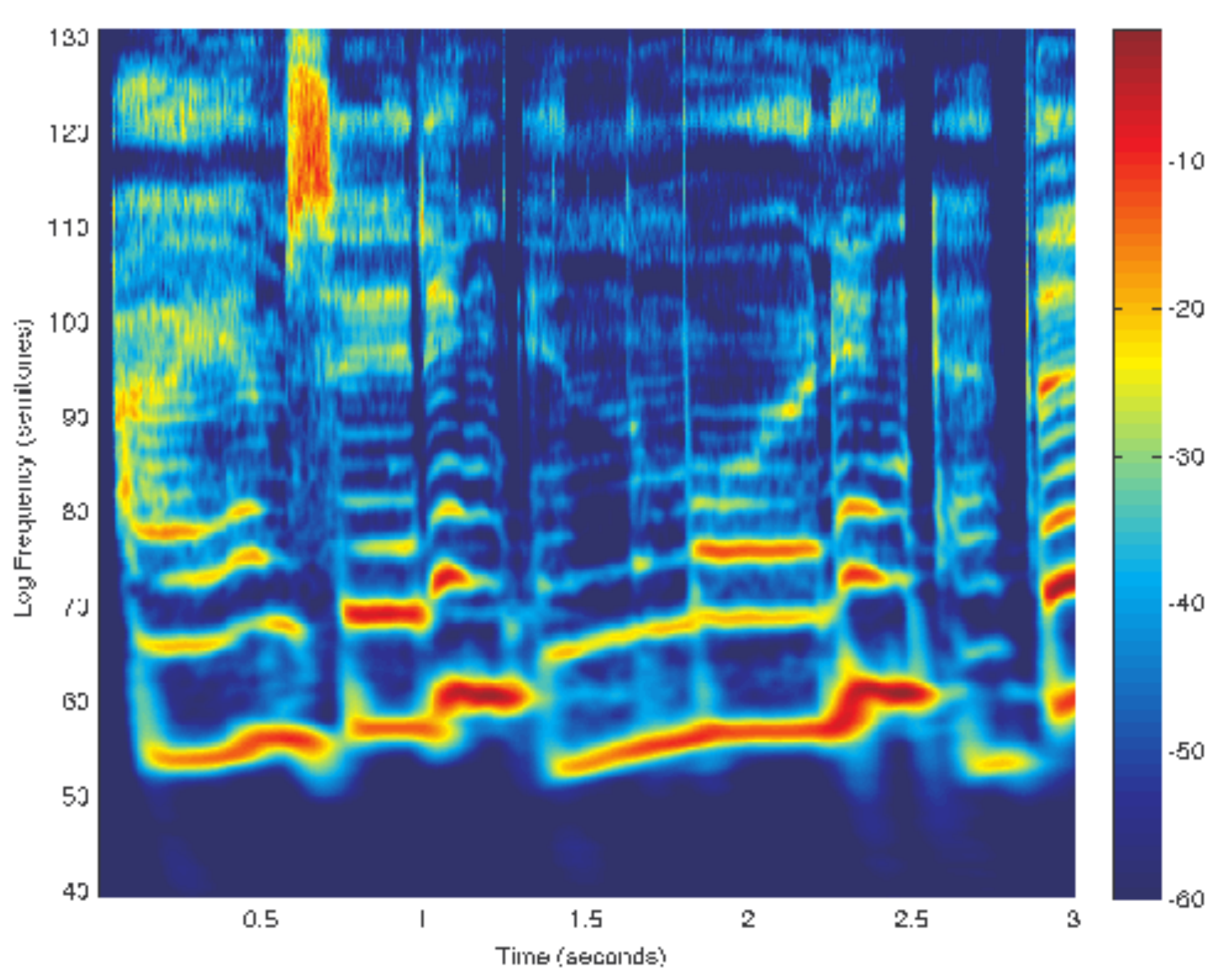}
      \\
      {\small\em Transient enhancement: $\partial_{t} {\cal T} \log |S|$} 
      & {\small\em Onset detection: $\partial_{t} {\cal T} \log |S| > 0$} \\
      \includegraphics[width=0.48\textwidth]{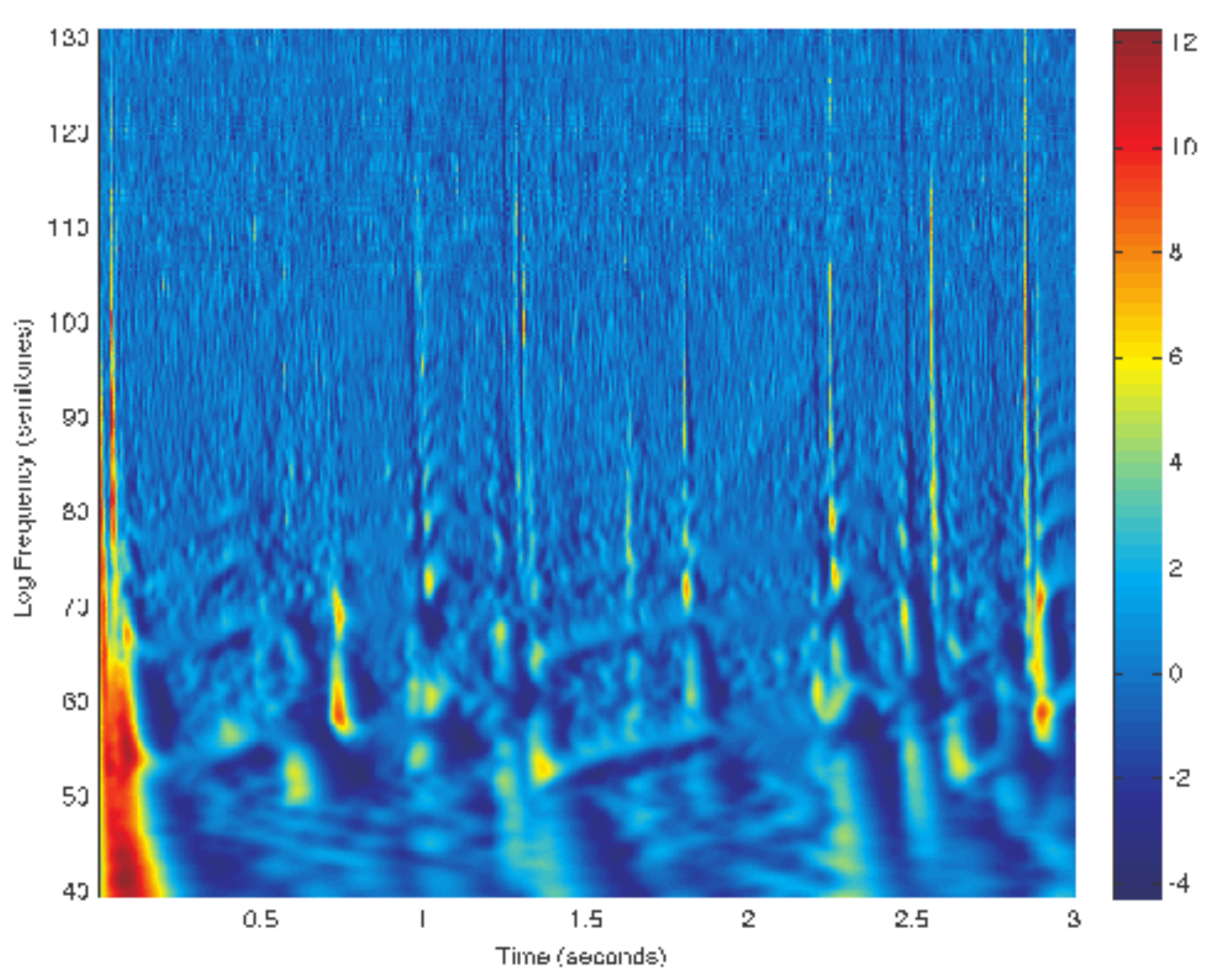} \hspace{-2mm} 
      & \includegraphics[width=0.48\textwidth]{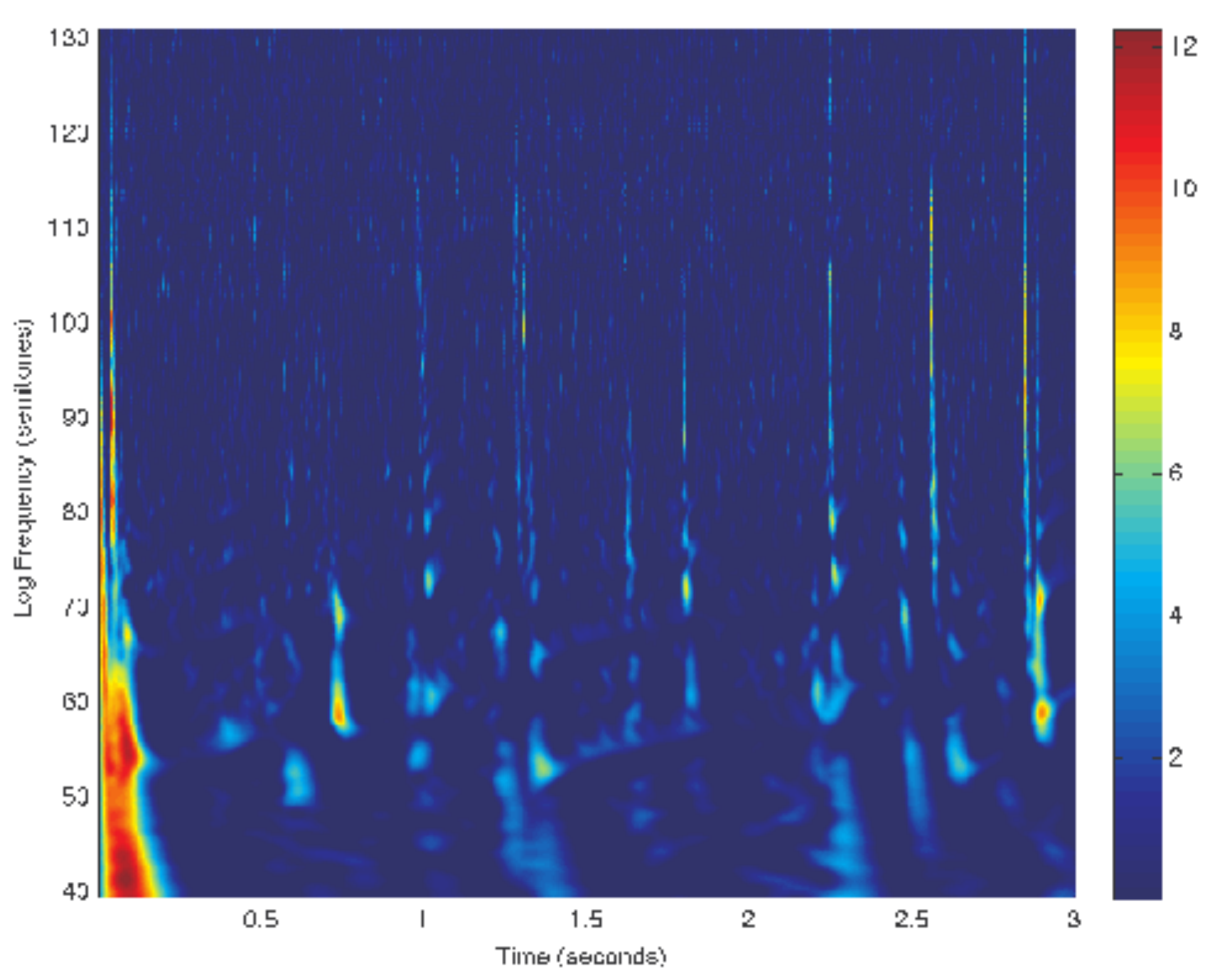} \\
      {\small\em Spectral band detection: $-\partial_{\nu\nu} {\cal T} \log |S|$} 
      & {\small\em Spectral band detection: $-\partial_{\nu\nu} {\cal T} \log |S| > 0$} \\
      \includegraphics[width=0.48\textwidth]{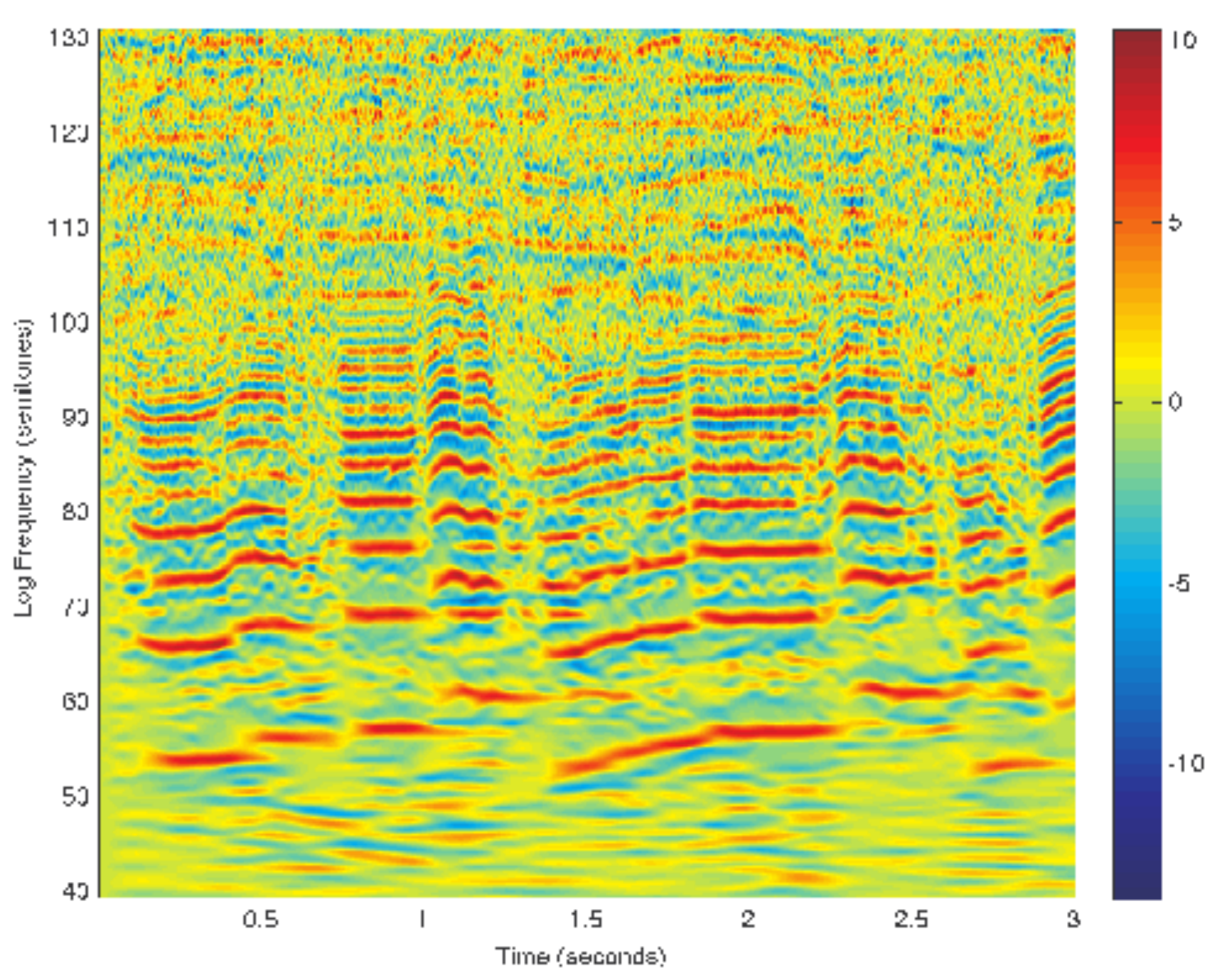} \hspace{-2mm} 
      & \includegraphics[width=0.48\textwidth]{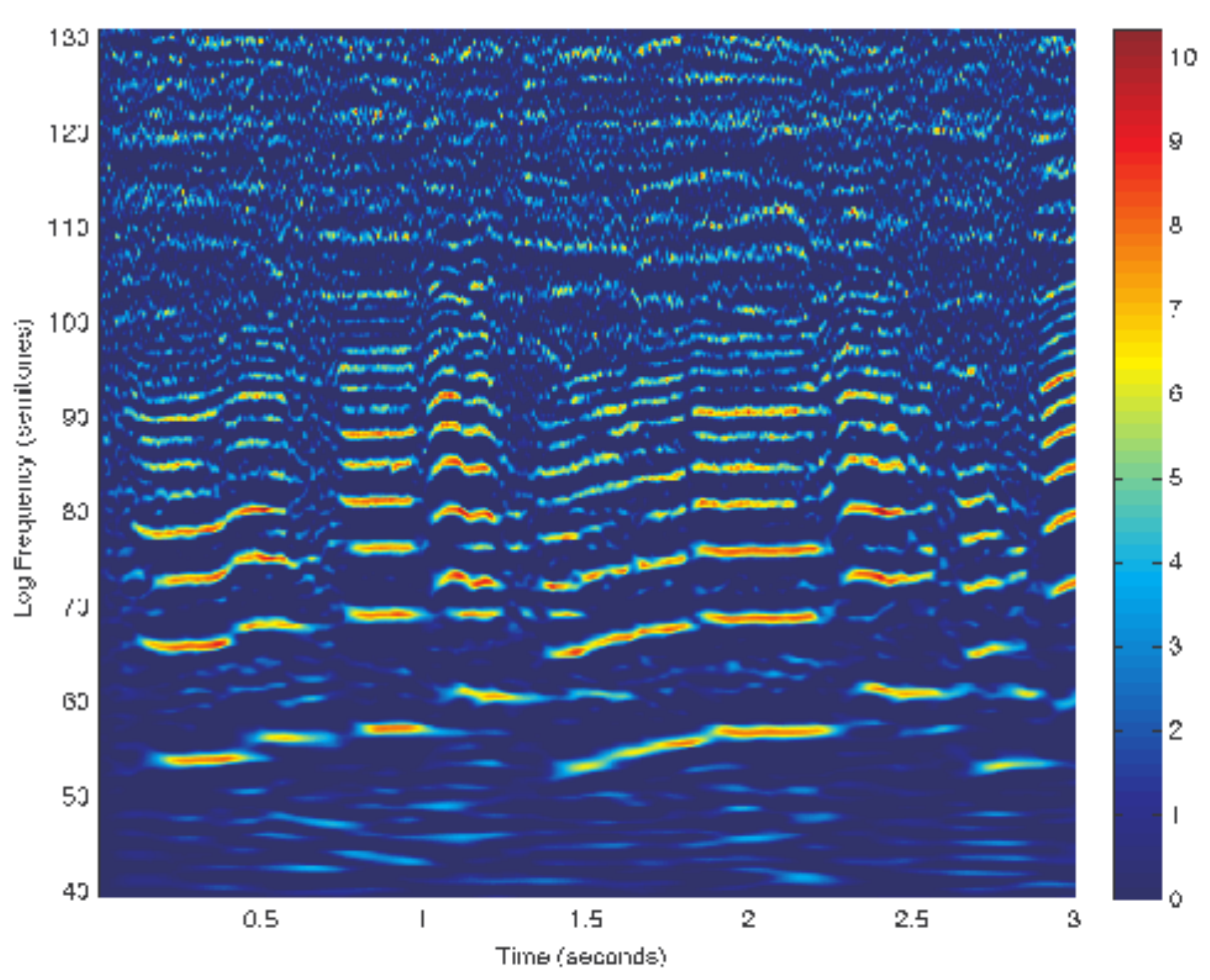} \\
    \end{tabular} 
  \end{center}
 
  \vspace{-3mm}
  \caption{Second layer receptive field responses obtained by applying
    spectro-temporal scale-space derivatives to the logarithmic 
    spectrogram of the first 3 seconds of ``Tom's diner''  by Suzanne Vega
    with the lyrics ``I am sitting in the morning at the ...''. 
    (top left) original spectrogram: (top right) 
    {\em Spectrotemporal smoothing\/}  ${\cal T} \log |S|$ using a
    cascade of four time-recursive filters with
    temporal scale proportional to the temporal window scale
    $\sigma_t = \sqrt{\tau} = 0.75 \, \sigma_w$ 
    and with logspectral smoothing scale $\sigma_{\nu} = 0.5~\mbox{semitones}$.
   (middle left) {\em Onset detection\/} from first-order temporal
   derivatives $\partial_{t} {\cal T} \log |S|$ or (middle right) the
   positive part of the first-order temporal
   derivatives $\partial_{t} {\cal T} \log |S| > 0$.
   (bottom left) {\em Spectral band detection\/} from second-order spectral
   derivatives $\partial_{\nu\nu} {\cal T} \log |S|$ or (bottom right) the
   negative part of the second-order temporal
   derivatives $\partial_{\nu\nu} {\cal T} \log |S|< 0$.
  The vertical axis shows the logarithmic frequency expressed in semitones with 69 corresponding to the tone A4 (440 Hz).
  }
  \label{fig-RF-layer2-toms-diner}
\end{figure}

\begin{figure}[hbtp]
 \begin{center}
    \begin{tabular}{cc}
      {\small\em Spectrogram: $\log |S|$} 
      & {\small\em Spectral smoothing: ${\cal T} \log |S|$} \\
      \includegraphics[width=0.48\textwidth]{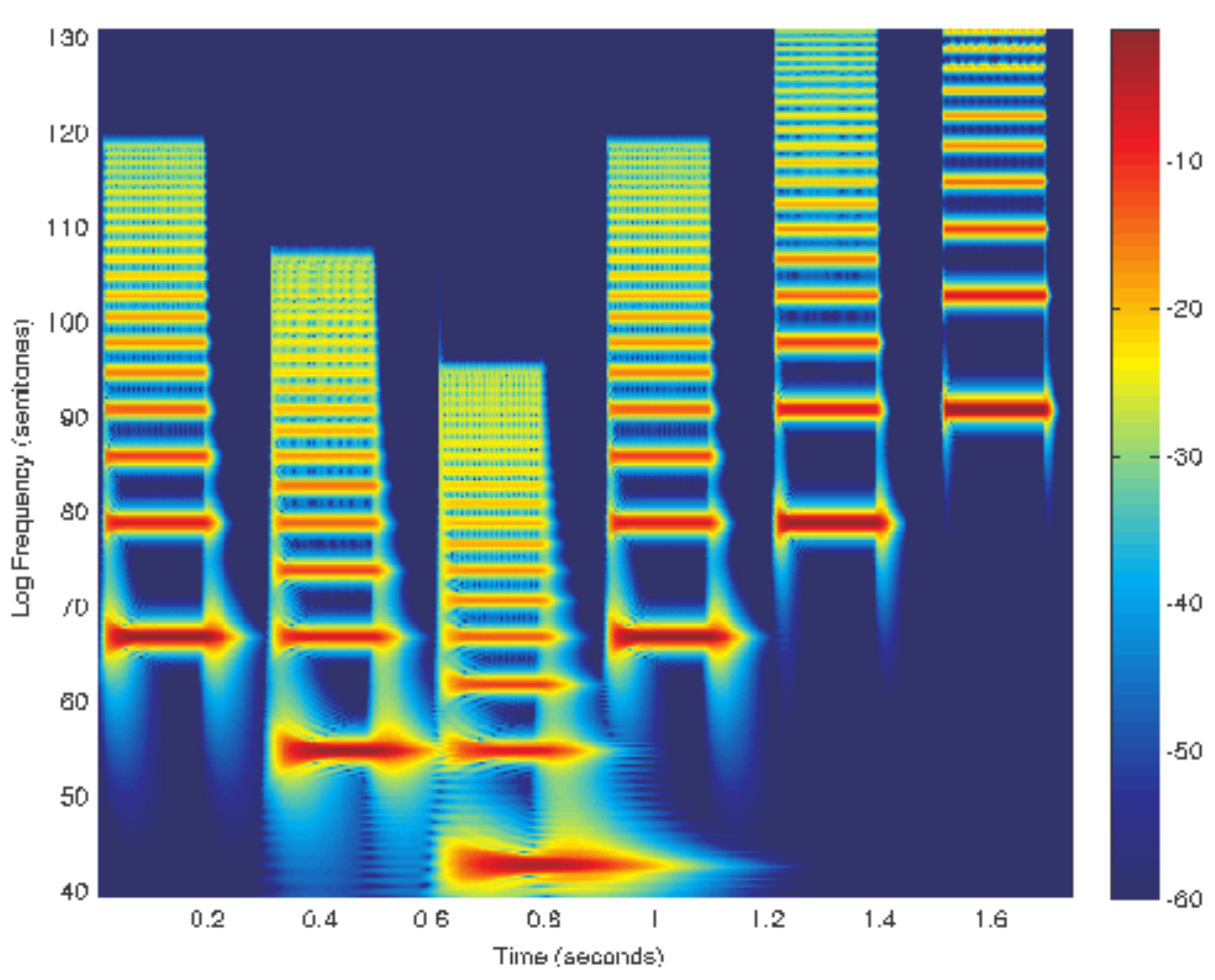} \hspace{-2mm} 
      & \includegraphics[width=0.48\textwidth]{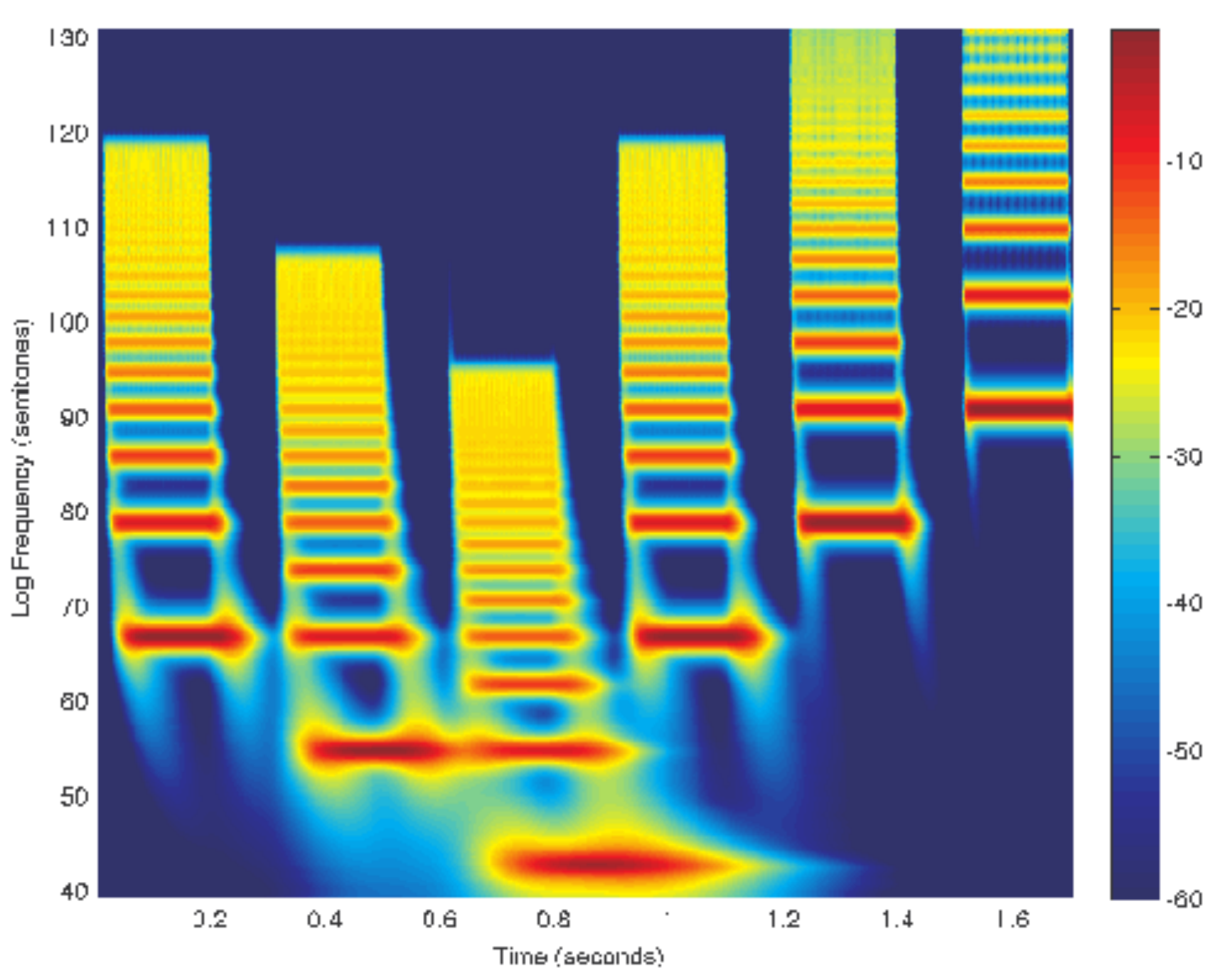} \\
      {\small\em Transient enhancement: $\partial_{t} {\cal T} \log |S|$} 
      & {\small\em Onset detection: $\partial_{t} {\cal T} \log |S| > 0$} \\
      \includegraphics[width=0.48\textwidth]{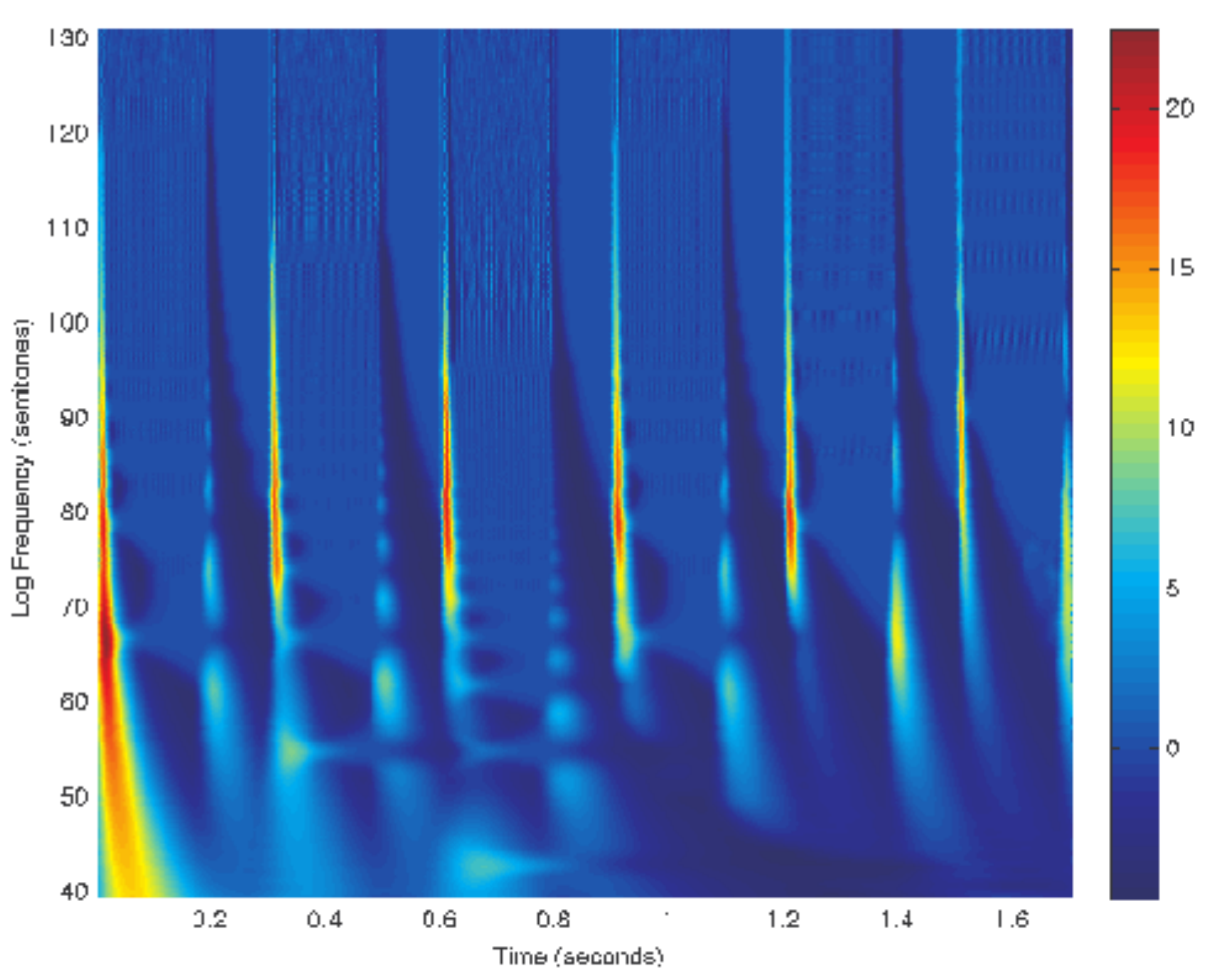} \hspace{-2mm} 
      & \includegraphics[width=0.48\textwidth]{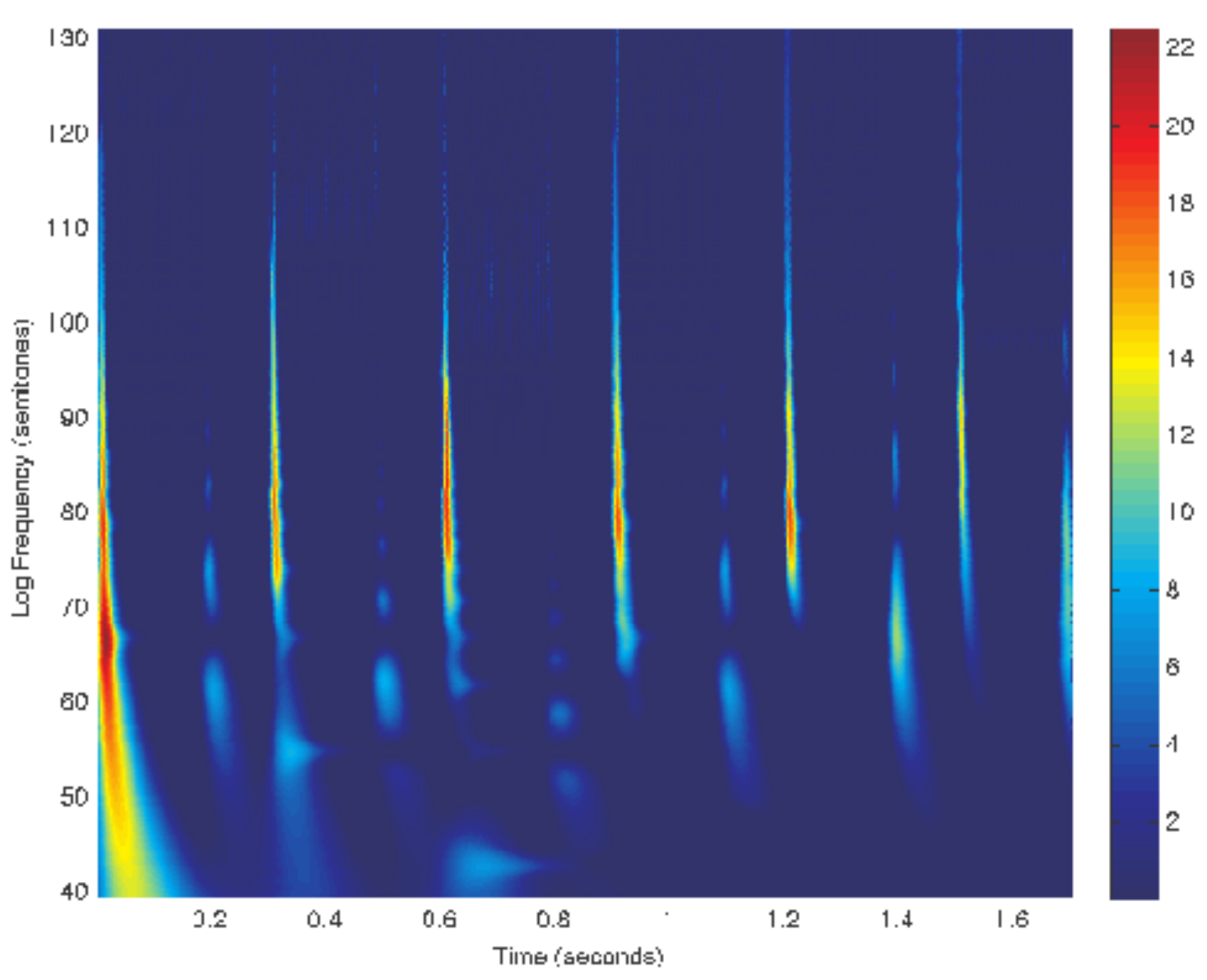} \\
      {\small\em Spectral band detection:  $-\partial_{\nu\nu} {\cal T} \log |S|$} 
      & {\small\em Spectral band detection: $-\partial_{\nu\nu} {\cal T} \log |S| > 0$} \\
      \includegraphics[width=0.48\textwidth]{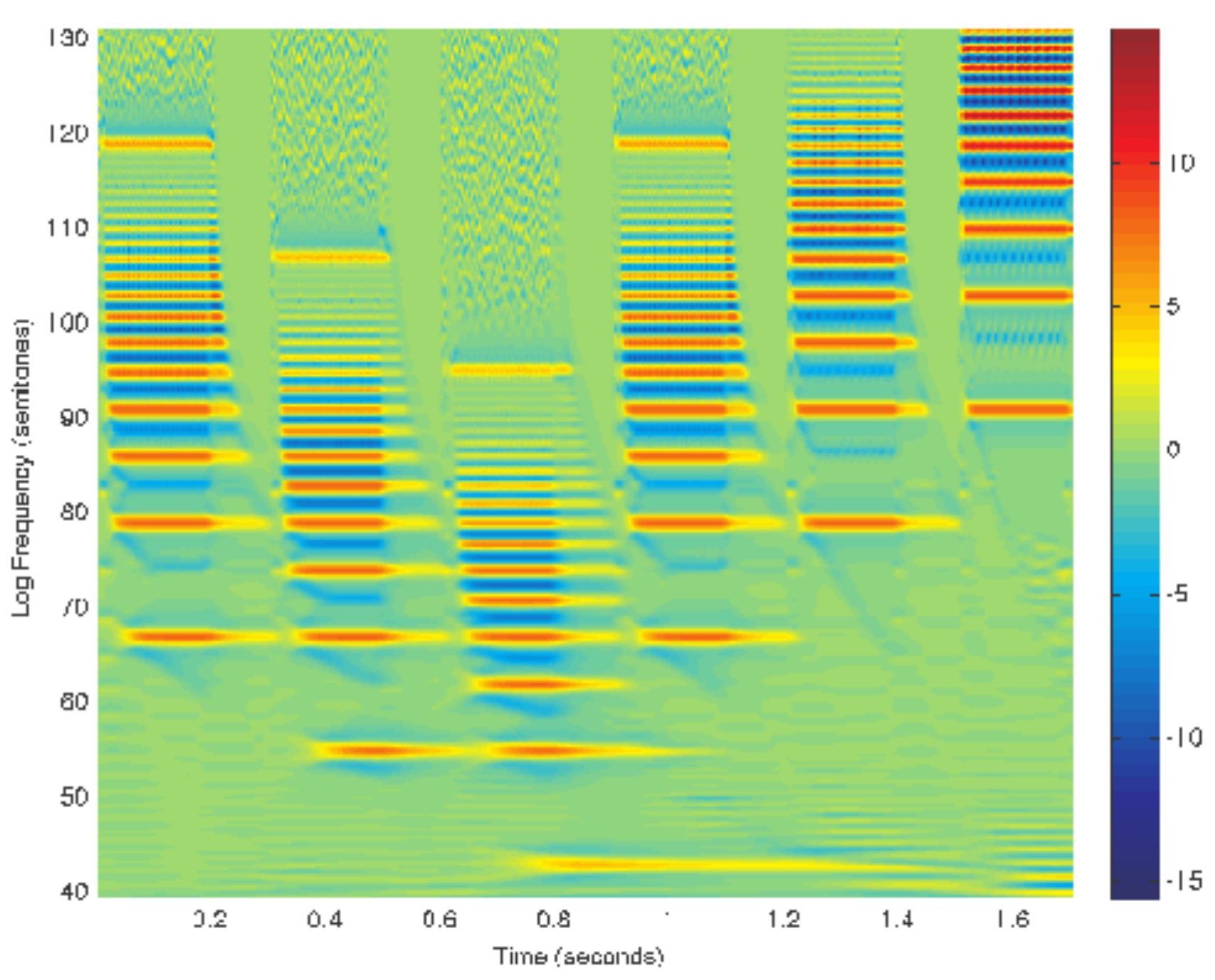} \hspace{-2mm} 
      & \includegraphics[width=0.48\textwidth]{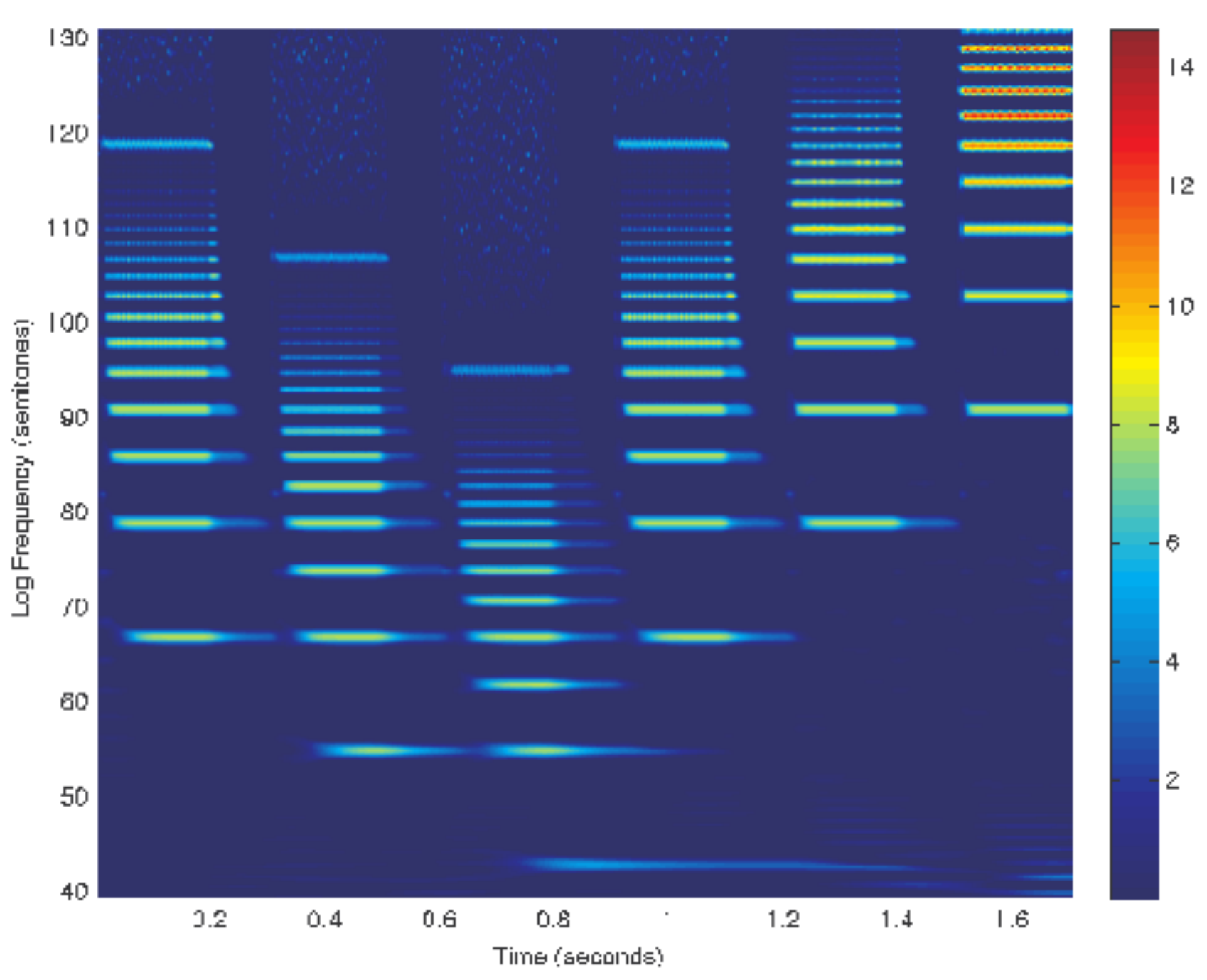} \\
    \end{tabular} 
  \end{center}
 
  \vspace{-3mm}
  \caption{Second layer receptive field responses obtained by applying
    spectro-temporal scale-space derivatives to a synthetic signal
    containing of harmonic spectra with 20 partials and a spectral 
   slope of 6 dB/octave at different fundamental frequencies:
    (top left) original spectrogram, (top right) 
    {\em Spectrotemporal smoothing\/}  ${\cal T} \log|S|$ using a
    cascade of four time-recursive filters with
    temporal scale proportional to the temporal window scale
    $\sigma_t = \sqrt{\tau} = 0.75 \, \sigma_w$ 
    and with logspectral smoothing scale $\sigma_{\nu} = 0.5~\mbox{semitones}$.
   (middle left) {\em Onset detection\/} from first-order temporal
   derivatives $\partial_{t} {\cal T} \log |S|$ or (middle right) the
   positive part of the first-order temporal
   derivatives $\partial_{t} {\cal T} \log |S| > 0$.
   (bottom left) {\em Spectral band detection\/} from second-order spectral
   derivatives $\partial_{\nu\nu} {\cal T} \log |S|$ or (bottom right) the
   negative part of the second-order temporal
   derivatives $\partial_{\nu\nu} {\cal T} \log |S| < 0$. 
  The vertical axis shows the logarithmic frequency expressed in semitones with 69 corresponding to the tone A4 (440 Hz).
}
  \label{fig-RF-layer2-meloctaves-diner}
\end{figure}

\begin{figure}[hbtp]
 \begin{center}
    \begin{tabular}{cc}
   {\small $-\partial_{\nu\nu} {\cal T} \log |S|$} 
      & {\small $-\partial_{\nu\nu} {\cal T} \log |S| > 0$} \\
      \includegraphics[width=0.48\textwidth]{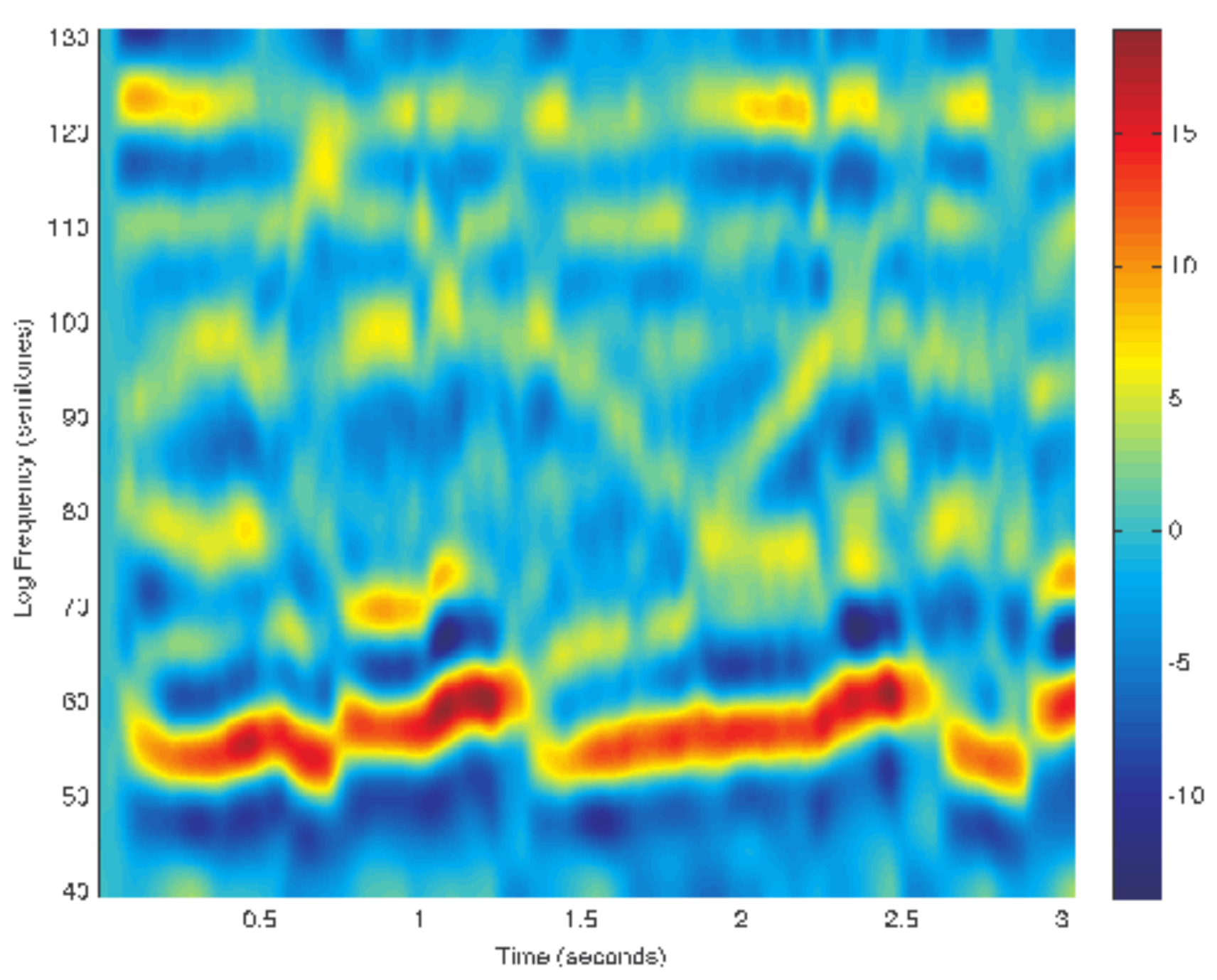} \hspace{-2mm} 
      & \includegraphics[width=0.48\textwidth]{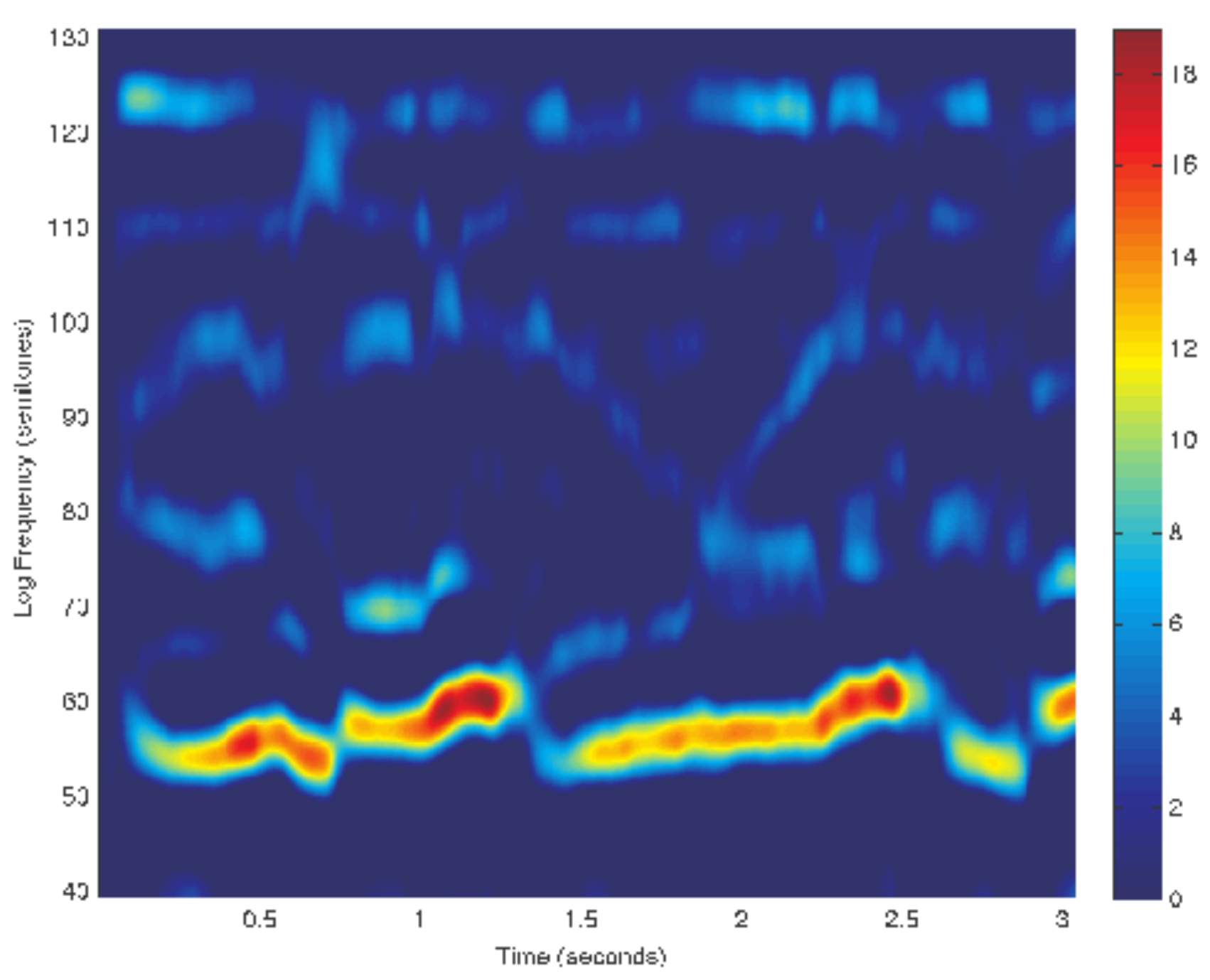} \\
    \end{tabular} 
  \end{center}
 
  \vspace{-3mm}
  \caption{Spectral sharpening at a coarser log-spectral scale
    ($\sigma_{\nu} = 4~\mbox{semitones})$ applied
    to the spectrogram of the first 3 seconds of ``Tom's diner'' 
    with the lyrics ``I am sitting in the morning at the ...'' and
    using recursive filters at composed temporal scale 20~ms.
   Note how this operation reveals the formants of the vowels in the
   frequency range between roughly MIDI 70 and MIDI 110,
   corresponding to a frequencies beween roughly 450 Hz and 4.7 kHz.
  The vertical axis shows the logarithmic frequency expressed in semitones with 69 corresponding to the tone A4 (440 Hz).}
  \label{fig-formant-enhance-toms-diner}
\end{figure}

\begin{figure}[hbtp]
 \begin{center}
    \begin{tabular}{c}
       \includegraphics[width=0.96\textwidth]{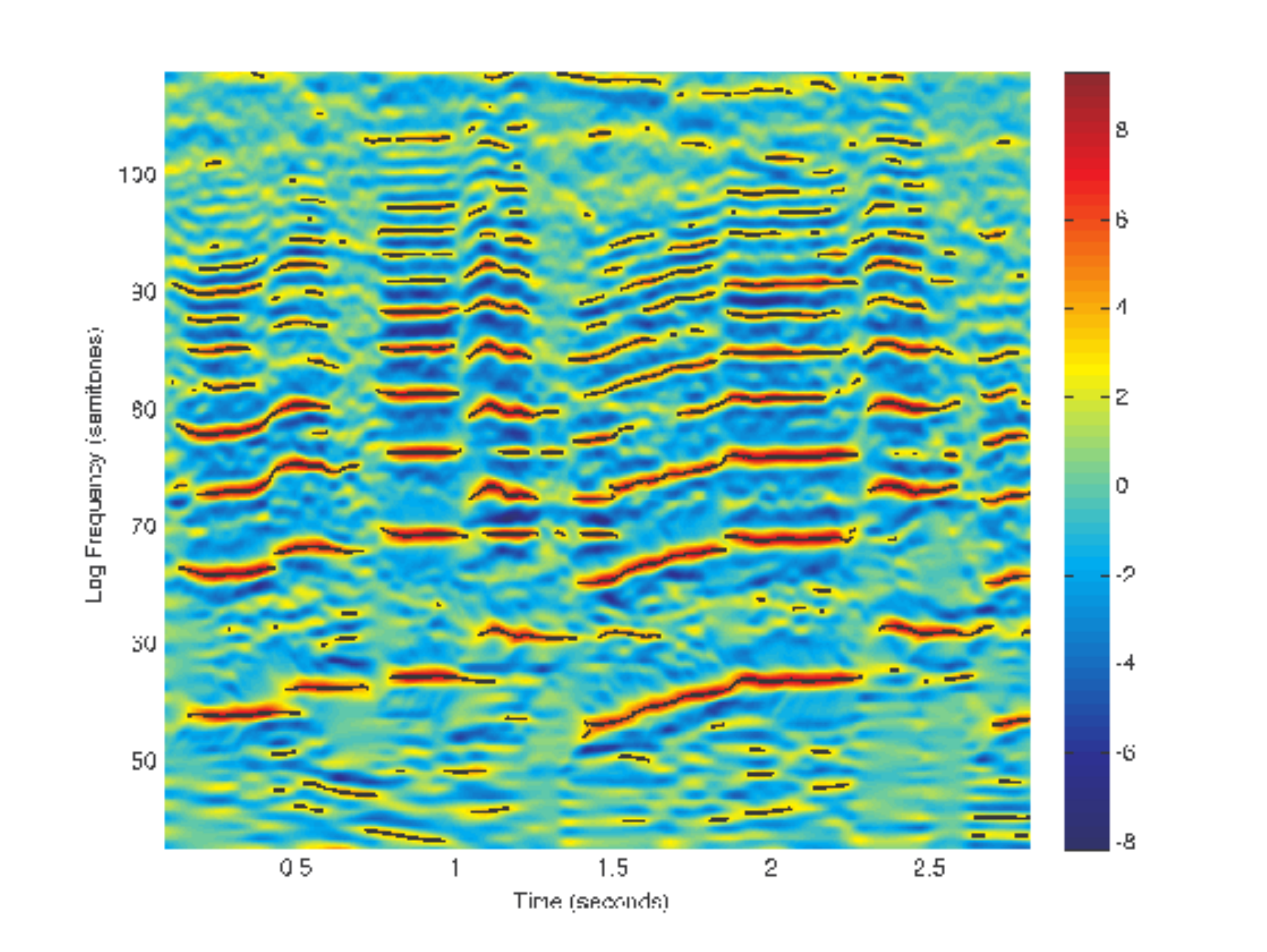}
    \end{tabular} 
  \end{center}
 
  \vspace{-3mm}
  \caption{Spectro-temporal curves that make explicit how the
    frequencies of partial tones vary over time, computed as the
    zero-crossing curves of the spectral derivative
    $\partial_{\nu} (-{\cal D}_{\nu\nu} \, \log |S|) = 0$ that satisfy
    $\partial_{\nu\nu} (-{\cal D}_{\nu\nu} \, \log |S|) < 0$ and which
    thereby become continuous curves over time (drawn in black,
    thresholded on $-{\cal D}_{\nu\nu} \, \log |S| \geq C$ for $C = 3$ and
    overlayed on $-{\cal D}_{\nu\nu} \, \log |S|$). 
    (Horizontal dimension: time $t$, Vertical dimension: logarithmic
    frequency $\nu$.) }
  \label{fig-freq-curves}     
\end{figure}

\begin{figure}[hbtp]
 \begin{center}
    \begin{tabular}{cc}
      {\small\em Glissando $v = +80~\mbox{semitones/s}$} 
      & {\small\em Glissando $v = -20~\mbox{semitones/s}$} \\
      \includegraphics[width=0.48\textwidth]{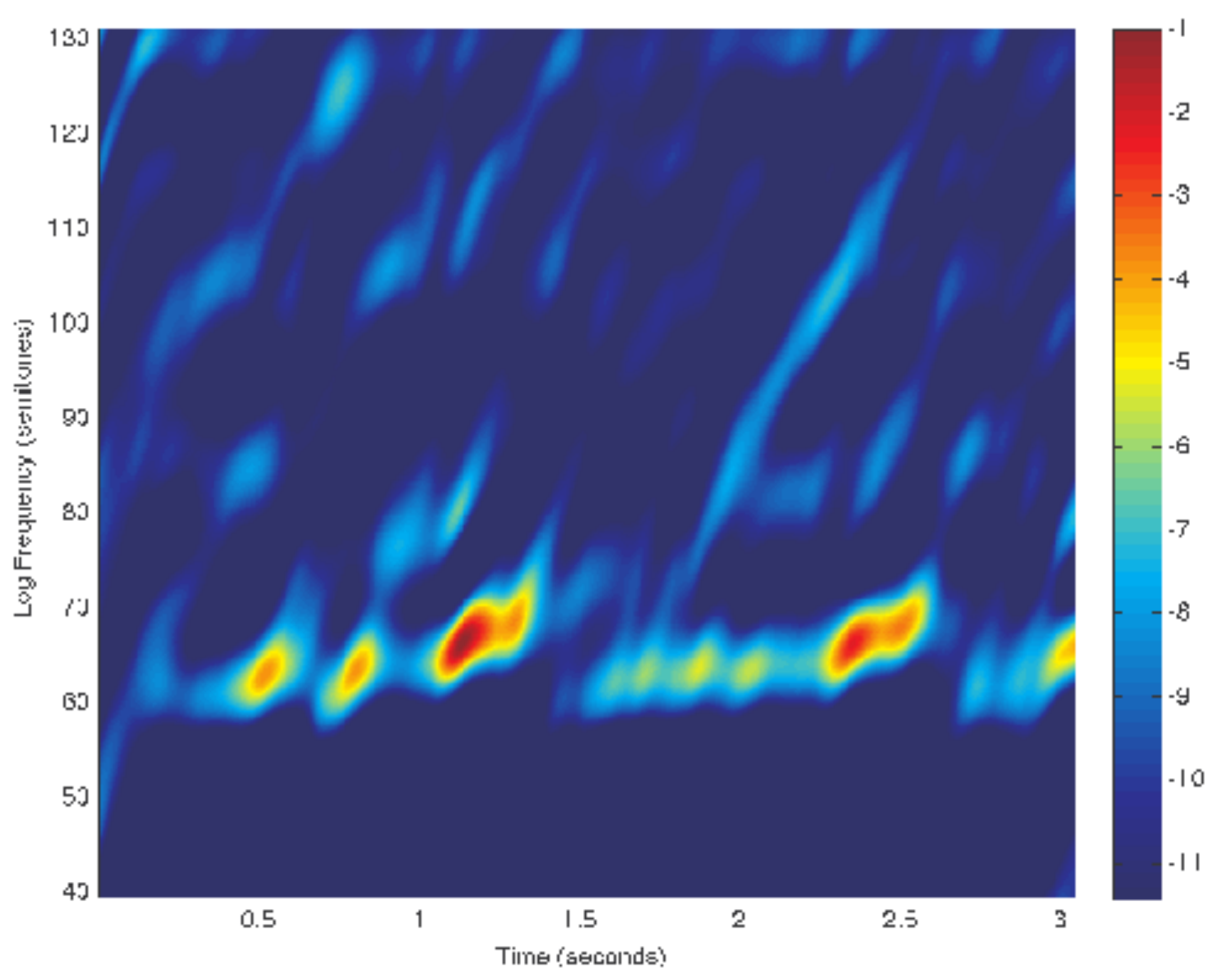} \hspace{-2mm} 
      & \includegraphics[width=0.48\textwidth]{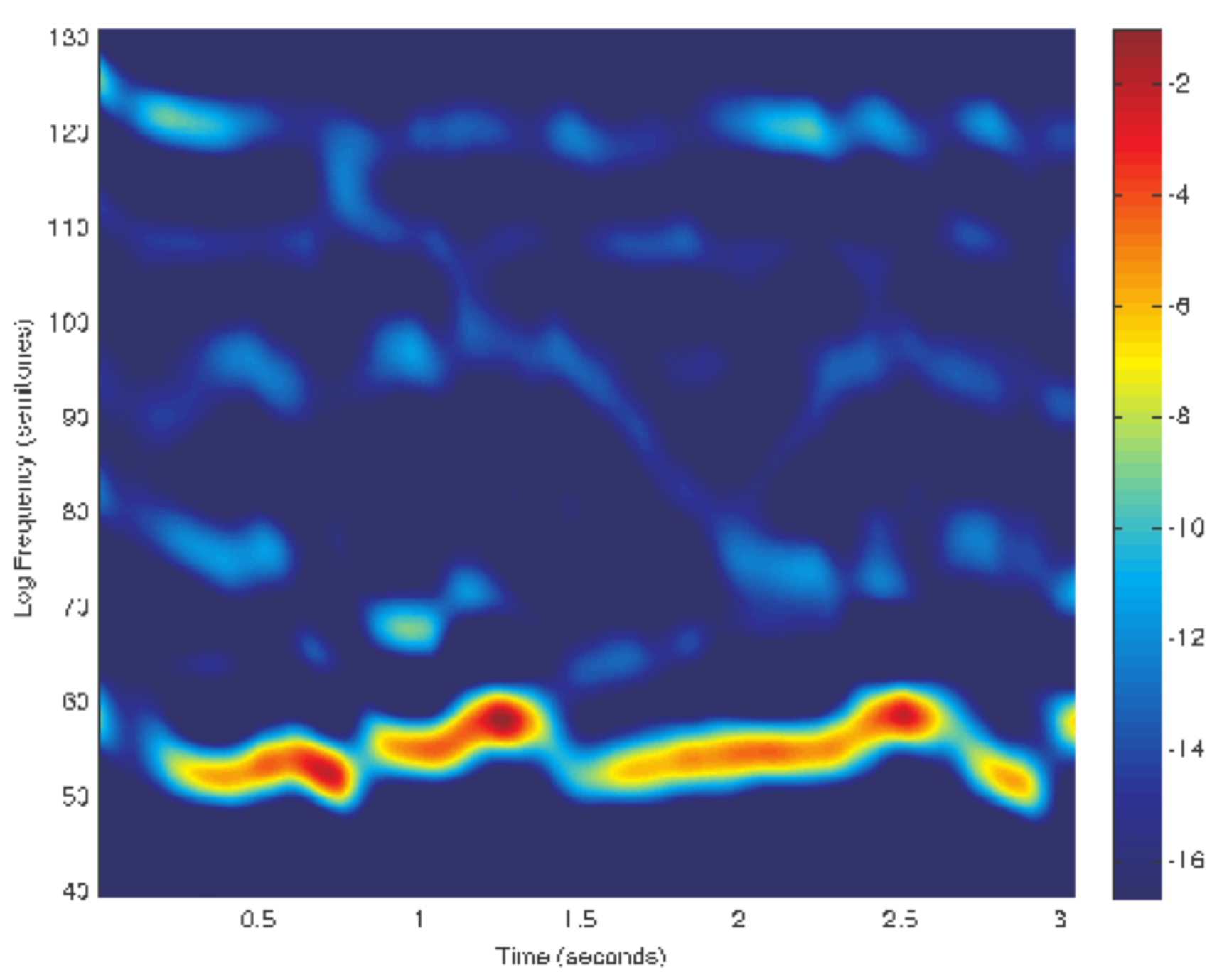} \\
      {\small\em Glissando $v = +40~\mbox{semitones/s}$} 
      & {\small\em Glissando $v = -40~\mbox{semitones/s}$} \\
      \includegraphics[width=0.48\textwidth]{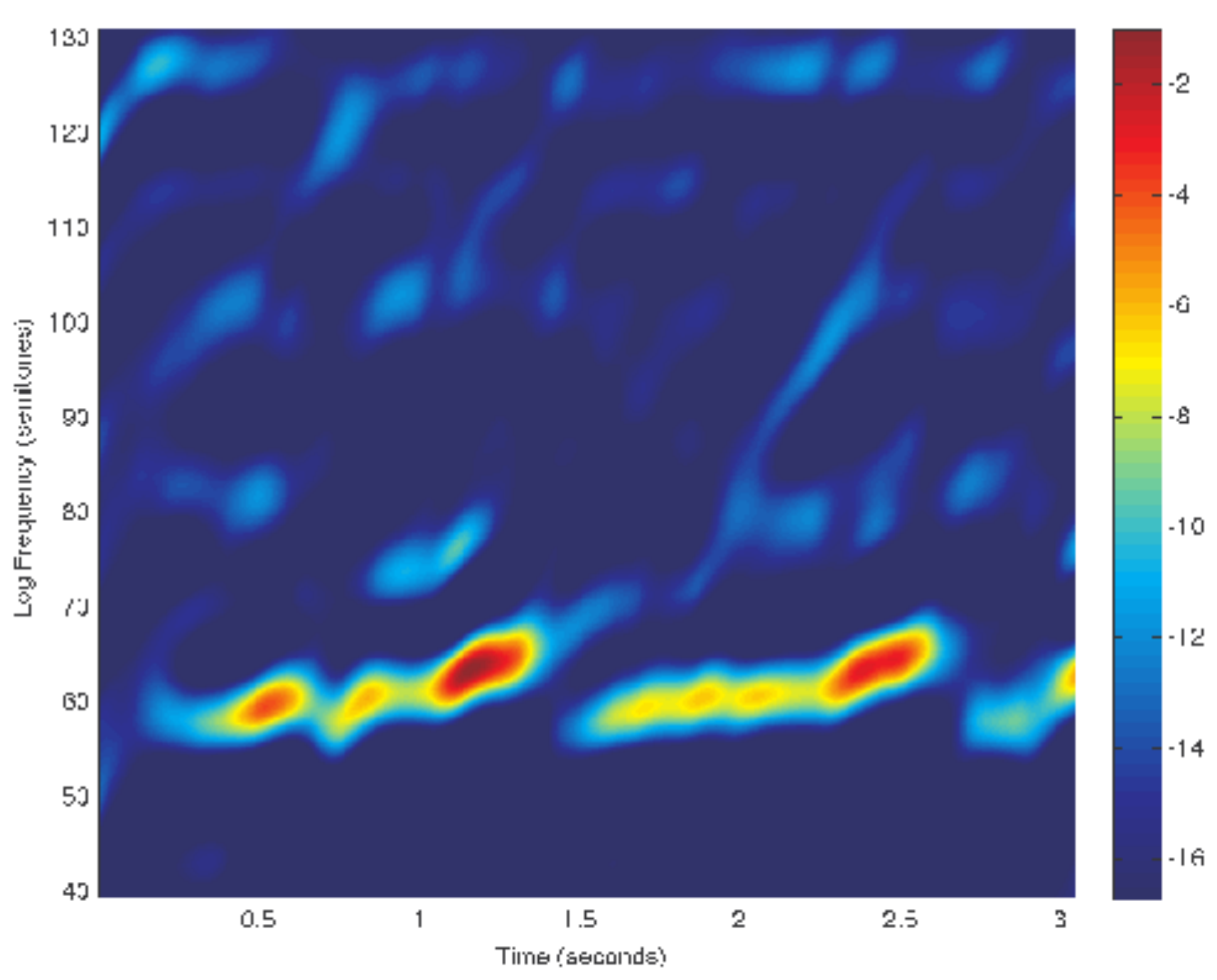} \hspace{-2mm} 
      & \includegraphics[width=0.48\textwidth]{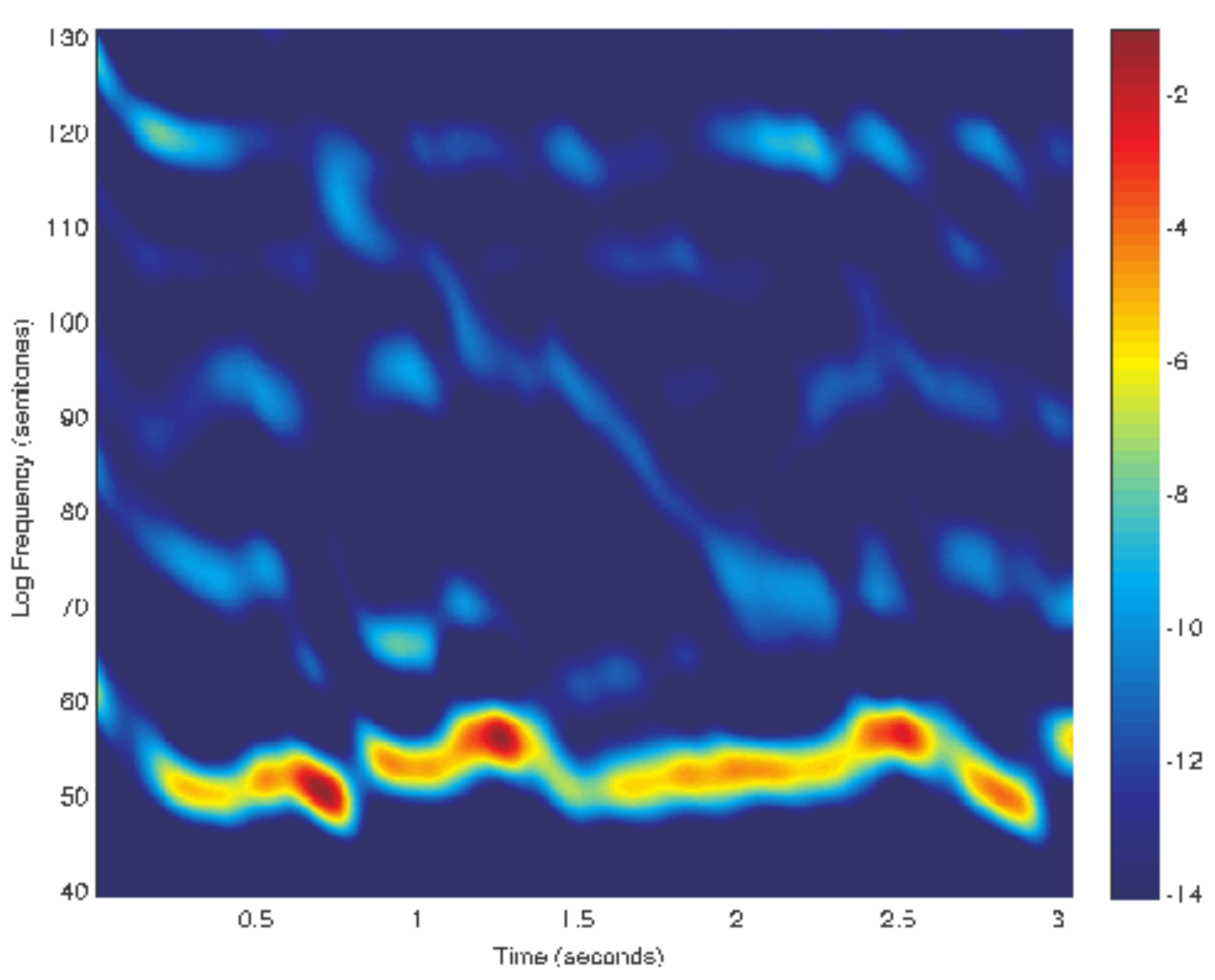} \\
      {\small\em Glissando $v = +20~\mbox{semitones/s}$} 
      & {\small\em Glissando $v = -80~\mbox{semitones/s}$} \\
      \includegraphics[width=0.48\textwidth]{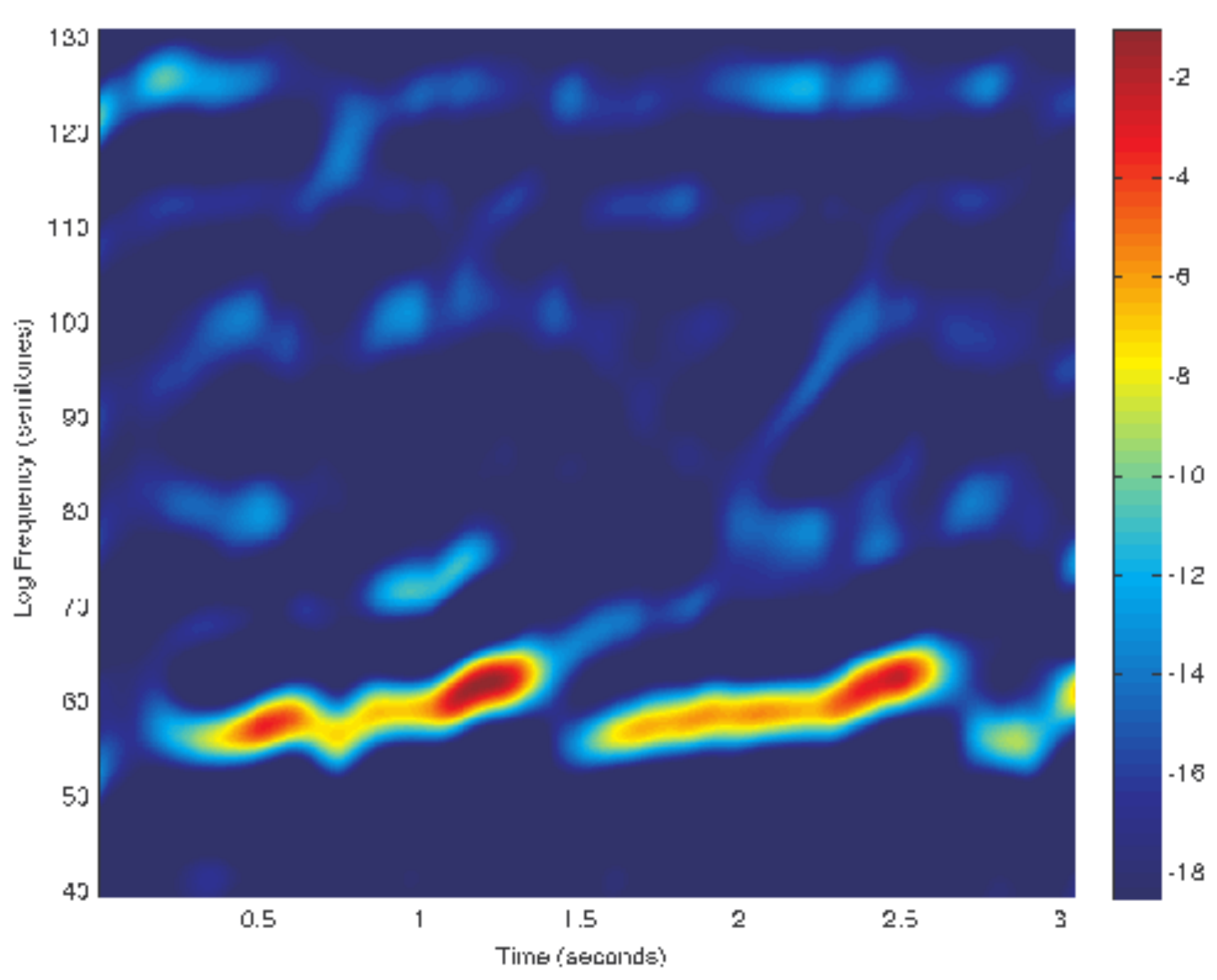} \hspace{-2mm} 
      & \includegraphics[width=0.48\textwidth]{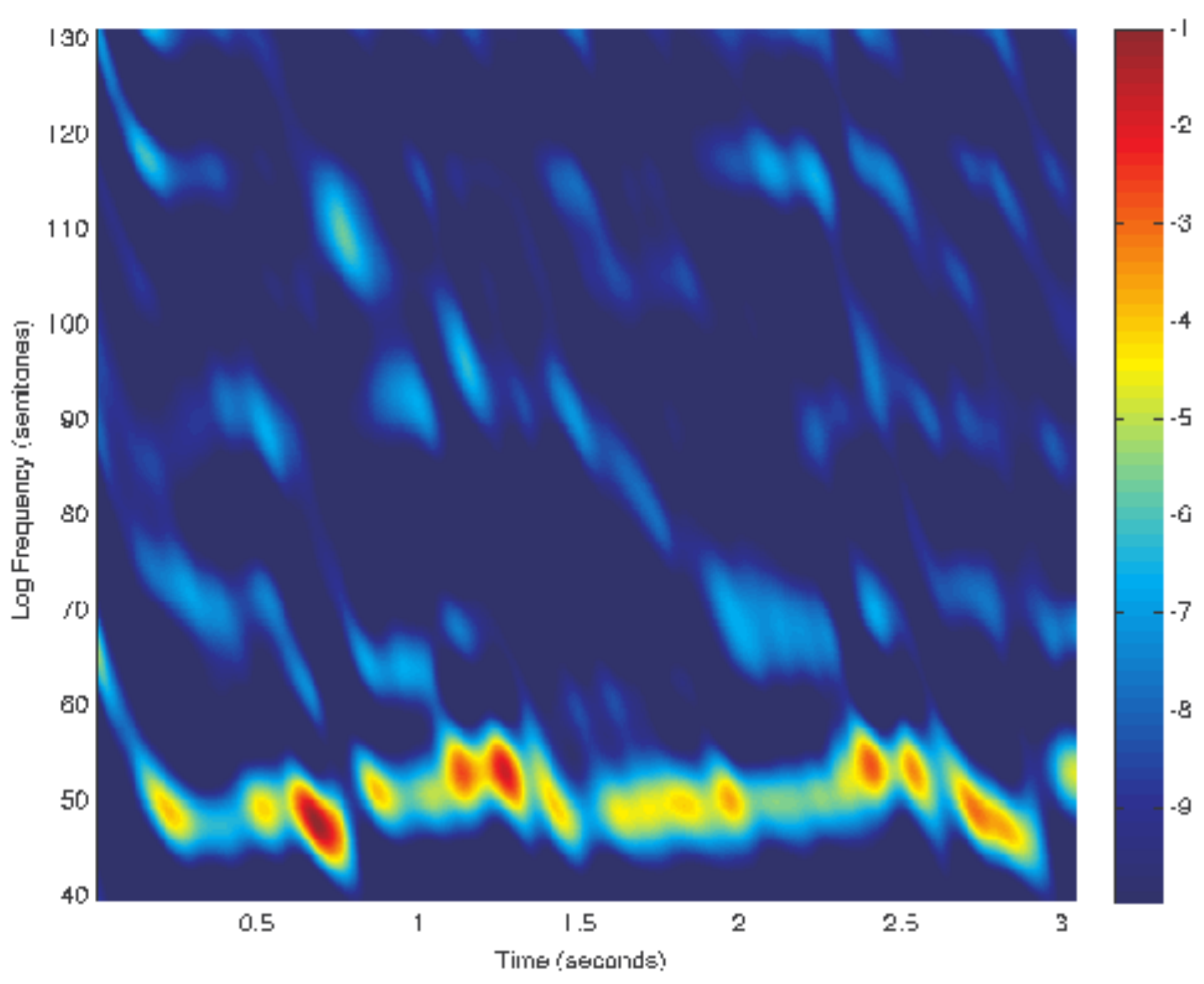} \\
    \end{tabular} 
  \end{center}
 
  \vspace{-3mm}
  \caption{Enhancement of the formants using glissando-adapted receptive
    fields corresponding to second-order derivatives with respect to
    logarithmic frequency $\nu$ for different glissando values $v =
    +80$, +40, 20, -20, -40 and -80 and
    applied to the first 3 seconds of ``Tom's diner'' computed with
    time-causal receptive fields at temporal
    scale $\sigma_t = 20$~ms and logspectral scale  $\sigma_{\nu} =
    4$~semitones (compare with
    figure~\protect\ref{fig-formant-enhance-toms-diner} that shows
    corresponding results for non-adapted separable receptive fields). Note how the formant variations for
    different amounts of glissando are enhanced by glissando-adapted
    receptive fields for corresponding values of the glissando
    parameter.
    Such a set of glissando-adapted receptive fields for a
    logarithmic distribution of the glissando values $v$ can serve as a
    filter bank for algorithms that operate on these receptive field
    responses as input.
    (Horizontal dimension: time $t$, Vertical dimension: logarithmic
    frequency $\nu$.).}
  \label{fig-RF2-toms-formants-glissando-var}
\end{figure}

\subsection{Auditory features from second layer
  spectro-temporal receptive fields}

In the following, we will show examples of auditory features
that can be defined from a second layer of auditory receptive
fields of this form. 

\paragraph{Spectro-temporal smoothing.} Auditory receptive fields $A$
that correspond to convolution with a
spectro-temporal smoothing kernel $T$ over the spectro-temporal domain
(see figure~\ref{fig-RF-layer2-toms-diner}, top right)
\begin{equation}
   A(t, \nu;\; \tau_a, s, v) = T(t, \nu;\; \tau_a, s, v)
\end{equation}

\paragraph{Onset and offset detection.} Computation of first-order temporal
derivatives
(see figure~\ref{fig-RF-layer2-toms-diner}, middle left)
\begin{equation}
   {\cal D}_t (t, \nu;\; \tau_a, s) = \sqrt{\tau_a} \, \partial_{t} T(t, \nu;\; \tau_a, s)
\end{equation}
where $\sqrt{\tau_a}$ is a scale normalization factor to approximate
scale-normalized derivatives \cite{Lin97-IJCV} by variance
normalization \cite{LinBre03-ScSp}.

This operation is similar to edge detection in image processing and
computer vision \cite{Can86-PAMI,Lin98-IJCV}
with the differences that (i)~the underlying derivatives are computed in
a fixed direction and that (ii)~in the case of a time-causal treatment of
time, the onset detection will also be associated with a temporal
delay.
The signed derivative operator responds to an increase in the
magnitude of the signal by a positive response and to a decrease in
the magnitude by a negative response.
To select receptive field responses that correspond to onsets only,
this operation is naturally combined with the (non-linear) logical operation:
$D_t > 0$ such that (see figure~\ref{fig-RF-layer2-toms-diner}, middle right)
\begin{equation}
   {\cal A}_{onset} \, S_{dB} = 
  \left\{ 
    \begin{array}{ll} 
      {\cal D}_t \, S_{dB}
      & \mbox{if ${\cal D}_t \, S_{dB} > 0$} \\
      0
      & \mbox{otherwise}
    \end{array}
 \right.
\end{equation}
In a corresponding manner, offset detection can be performed using
\begin{equation}
   {\cal A}_{offset} \, S_{dB} = 
  \left\{ 
    \begin{array}{ll} 
      -{\cal D}_t \, S_{dB}
      & \mbox{if ${\cal D}_t \, S_{dB} < 0$} \\
      0
      & \mbox{otherwise}
    \end{array}
 \right.
\end{equation}

\paragraph{Spectral sharpening.} Computation of second-order Gaussian
derivatives over the log-spectral domain
(see figure~\ref{fig-RF-layer2-toms-diner}, bottom left)
\begin{equation}
  \label{eq-spectr-sharp}
   {\cal D}_{\nu\nu}(t, \nu;\; \tau_a, s) = s \, \partial_{\nu\nu} T(t, \nu;\; \tau_a, s)
\end{equation}
where the factor $s$ is a scale normalization factor for
scale-normalized derivatives based on the Gaussian scale-space concept
\cite{Lin97-IJCV}.
Depending on the value of the log-spectral scale parameter, this
operation may either enhance partial tones or formants.
This operation is naturally combined with the (non-linear) logical operation 
${\cal D}_{\nu\nu} < 0$ such that
\begin{equation}
   {\cal A}_{band} \, S_{dB} = 
  \left\{ 
    \begin{array}{ll} 
      -{\cal D}_{\nu\nu} \, S_{dB}
      & \mbox{if ${\cal D}_{\nu\nu} \, S_{dB} < 0$} \\
      0
      & \mbox{otherwise}
    \end{array}
 \right.
\end{equation}
When applied at a fine log-spectral scale, this operation can be used
for enhancing spectral bands corresponding to the fundamental frequency and its overtones
(see figure~\ref{fig-RF-layer2-toms-diner}, bottom right).
 When applied at a coarser log-spectral scale, corresponding spectral
sharpening can be used for enhancing the formants of vowels (see
figure~\ref{fig-formant-enhance-toms-diner}).
A similar approach involving a combination of Gaussian functions 
was used by \cite{BaeMooGat93-RehabResDevel} for enhancing spectral
contrast  for listeners with hearing impairment and by 
\cite{HecDomJouGoe11-SpeechComm} as a part of feature extraction for
automatic speech recognition.

By comparing the responses to the partial tones in the second-order log-spectral
derivatives to the partial tones in the raw logarithmic spectrogram,
we can note that the responses to the partial tones are far more
similar between different partial tones in the log-spectral
derivatives compared to the raw spectrogram.
This property can be understood from the invariance of
spectro-temporal derivatives to local multiplications of the magnitude
of a signal pointed out in section~\ref{sec-log-trans-spectr}.
If we model the partial tones as self-similar copies of each other at
different frequencies while having different relative strength
(sound pressure),
then by combination of the invariance under multiplications of the
magnitude in section~\ref{sec-log-trans-spectr} with the invariance of the
relative bandwidth under multiplicative frequency transformations in
appendix~\ref{app-freq-anal}, it follows that the spectro-temporal
derivative responses to different overtones can be expected to have a
similar appearance.

Figure~\ref{fig-RF2-toms-formants-glissando-var} shows an extension of
this approach, where formant enhancement is performed using
glissando-adapted receptive fields, demonstrating how
how formant variations for
    different amounts of glissando are enhanced by glissando-adapted
    receptive fields for corresponding values of the glissando
    parameter.

\paragraph{Capturing frequency variations over time.}

Given that local spectral bands have been enhanced by second-order derivatives
over logarithmic frequencies (\ref{eq-spectr-sharp}),
we can compute local extrema over frequencies by differentiating this
response
\begin{align}
  \begin{split}
     \label{eq-DvDvv-zero}
     \partial_{\nu} (-{\cal D}_{\nu\nu} \, S_{dB}) = 0,
  \end{split}\\
  \begin{split}
     \label{eq-DvvDvv-neg}
     \partial_{\nu\nu} (-{\cal D}_{\nu\nu} \, S_{dB}) < 0.
  \end{split}
\end{align}
By interpolating for the zero-crossings of (\ref{eq-DvDvv-zero}) that
satisfy the sign constraint (\ref{eq-DvvDvv-neg}) we can obtain
subresolution curves of how the frequency of partials vary over time
(see figure~\ref{fig-freq-curves}).

\paragraph{Glissando estimation.}

One way of estimating explicitly how the frequencies vary over time, is by
estimating the temporal variation in the above curves, corresponding
to feature tracking in the area of computer vision.

An alternative more receptive field based approach is by computing 
the receptive field responses for a filter bank of different
glissando-receptive fields ${\cal D}(v)$ for different amounts of glissando
$v$ ({\em e.g.\/} using second-order spectral derivatives
${\cal D}(v) = {\cal D}_{\nu\nu}(v)$) analogous to the way ridge
detection methods in computer vision can be expressed in terms of
second-order derivatives of image intensity \cite{Lin97-IJCV}) and 
then selecting the maximum response over this filter bank as the
glissando estimate
\begin{equation}
  \label{eq-gliss-est-max-filt-bank}
   \hat{v} = \argmax_v {\cal D}(v) \, S_{db}
\end{equation}
preferably complemented by interpolation to estimate the amount of
glissando by higher accuracy than the actual sampling, 
compare with \cite{LapLin03-IVC,Lin13-PONE} for corresponding filter-based
approaches for estimating image velocities in computer vision using a
filter bank approach over different Galilean transformations.

Yet a more direct approach can be obtained by computing a
spectro-temporal second-moment matrix
\begin{align}
  \begin{split}
  \label{def-mu}
  \Upsilon(x, y;\; t, s) 
  & = \left(
       \begin{array}{cc}
        \Upsilon_{tt}  & \Upsilon_{t\nu} \\
         \Upsilon_{t\nu} & \Upsilon_{\nu\nu}
      \end{array}
   \right)
  \end{split}\nonumber\\
  \begin{split}
  & = \int_{(\xi, \eta) \in \bbbr^2}
     \left(
       \begin{array}{cc}
         L_{t}^2(\xi, \eta;\; t)                              & L_{t}(\xi, \eta;\; t) \, L_{\nu}(\xi, \eta;\; t) \\
         L_{t}(\xi, \eta;\; t) \, L_{\nu}(\xi, \eta;\; t) & L_{\nu}^2(\xi, \eta;\; t)
      \end{array}
   \right)
  T(t-\xi, \nu-\eta;\; \tau, s) \, d\xi \, d\eta
  \end{split}
\end{align}
by a third layer of spectro-temporal smoothing applied to the products
$L_{t}^2$, $L_{t}  L_{\nu}$ and $L_{\nu}^2$ of the spectro-temporal derivatives 
$L_{t} = \partial_t {\cal T}_{\Sigma} S_{db}$ and
$L_{\nu} = \partial_{\nu} {\cal T}_{\Sigma} S_{db}$ and then computing
the glissando estimate as the value
\begin{equation}
  \label{eq-gliss-est-2nd-mom}
  v = - \frac{\Upsilon_{t\nu}}{\Upsilon_{\nu\nu}}
\end{equation}
that transforms the spectro-temporal moment matrix to diagonal form
with the mixed $\Upsilon'_{t\nu}$ being zero
and corresponding to the estimation of optic flow and Galilean
invariant image descriptors in the area of computer vision
\cite{LukKan81-IU,LapCapSchLin07-CVIU,Lin13-PONE}.
Specifically, by computing receptive field responses using a glissando
estimate according to (\ref{eq-gliss-est-max-filt-bank}) alternatively
for a glissando value that corresponds to a fixed-point of 
(\ref{eq-gliss-est-2nd-mom}), it can be shown that the resulting
spectro-temporal receptive field responses will be invariant under
glissando transformations, which would not be fully possibly based on
separable spectro-temporal receptive fields only
(see also \cite{Lin13-BICY,Lin13-PONE} for analogous results regarding
Galilean invariance in vision).

\begin{figure}[hbtp]
 \begin{center}
    \begin{tabular}{c}
        \includegraphics[width=0.80\textwidth]{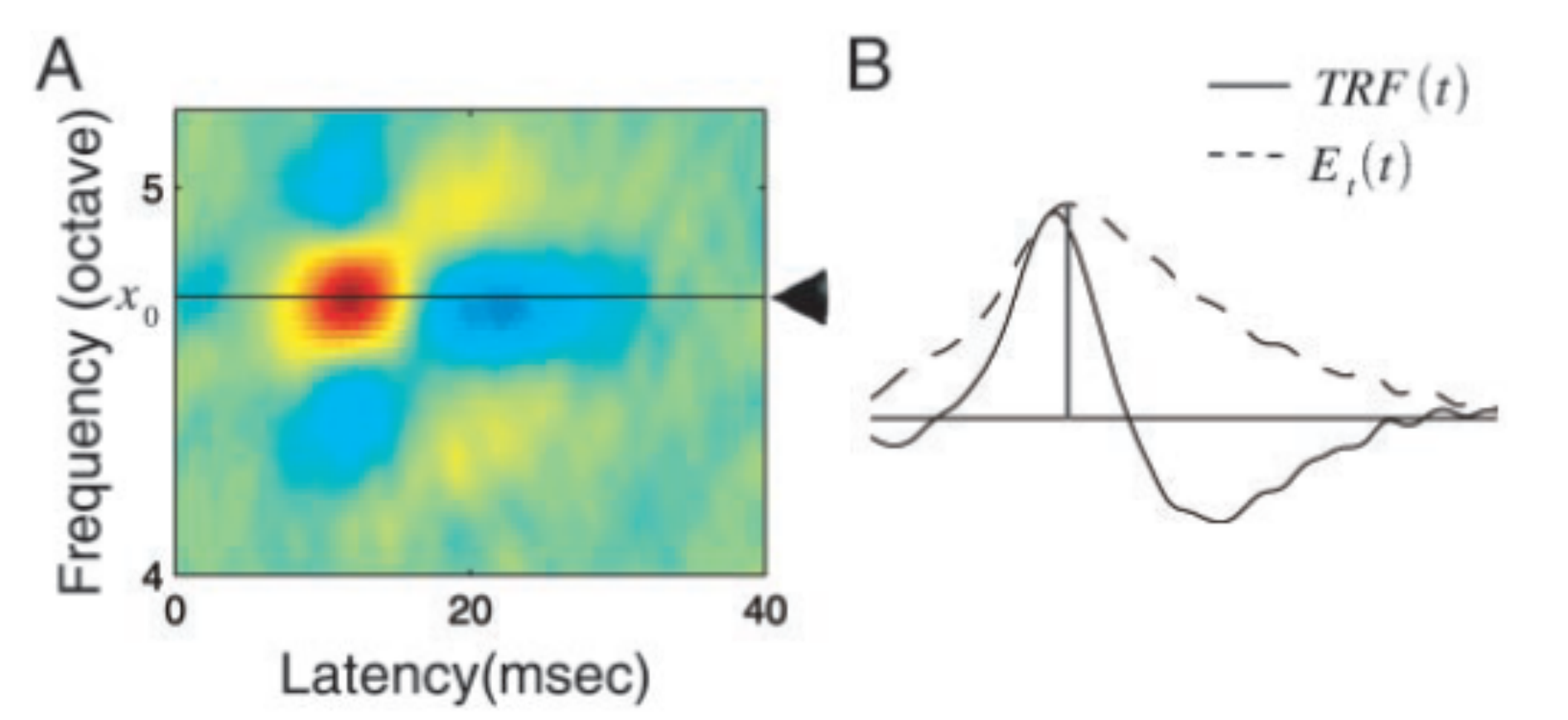}
    \end{tabular} 
    \begin{tabular}{cc}
        \hspace{3mm} {\em\small Time-causal model\/} 
& {\em\small Gaussian model\/} \\
        \hspace{3mm} \includegraphics[width=0.32\textwidth]{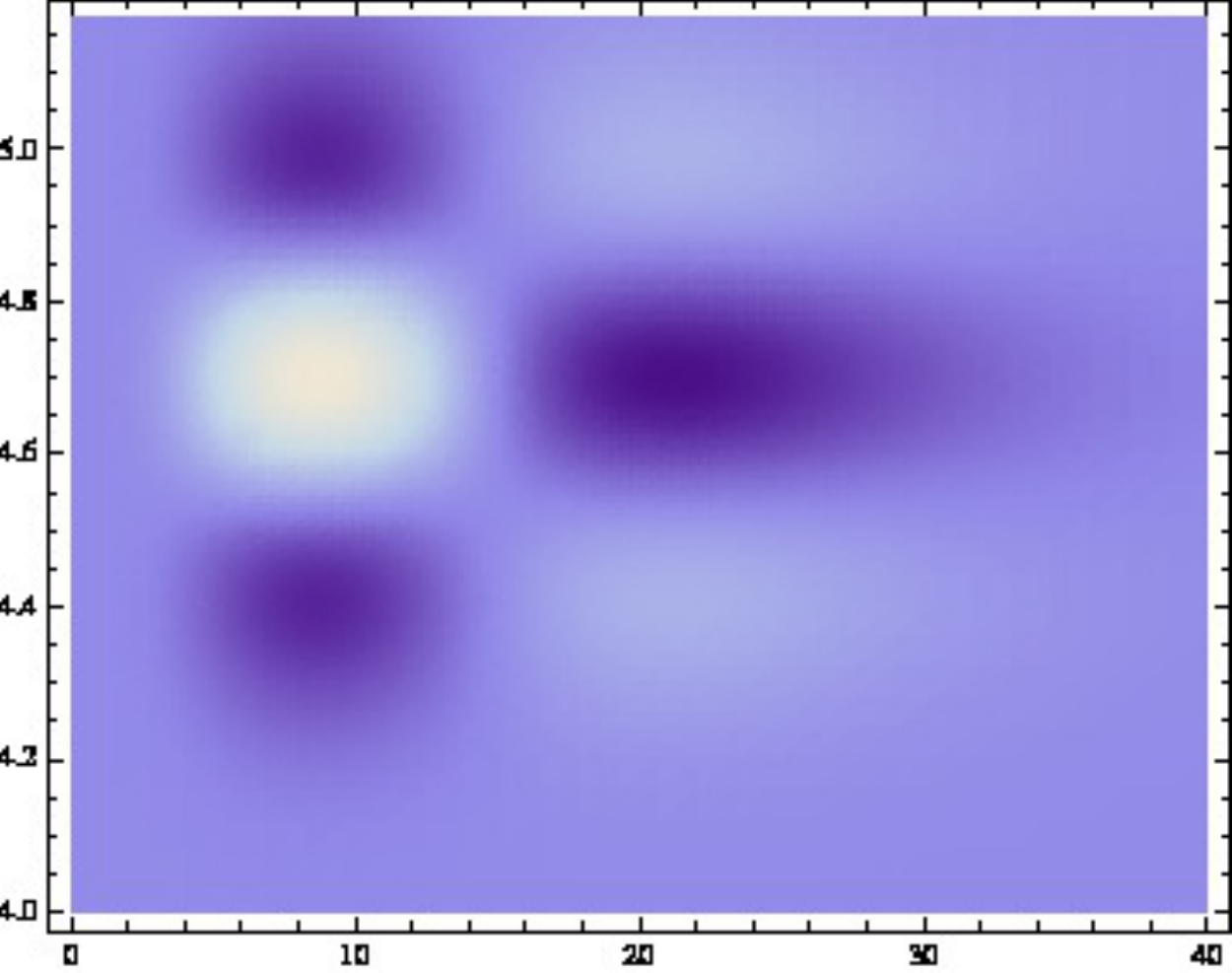}
        & \includegraphics[width=0.32\textwidth]{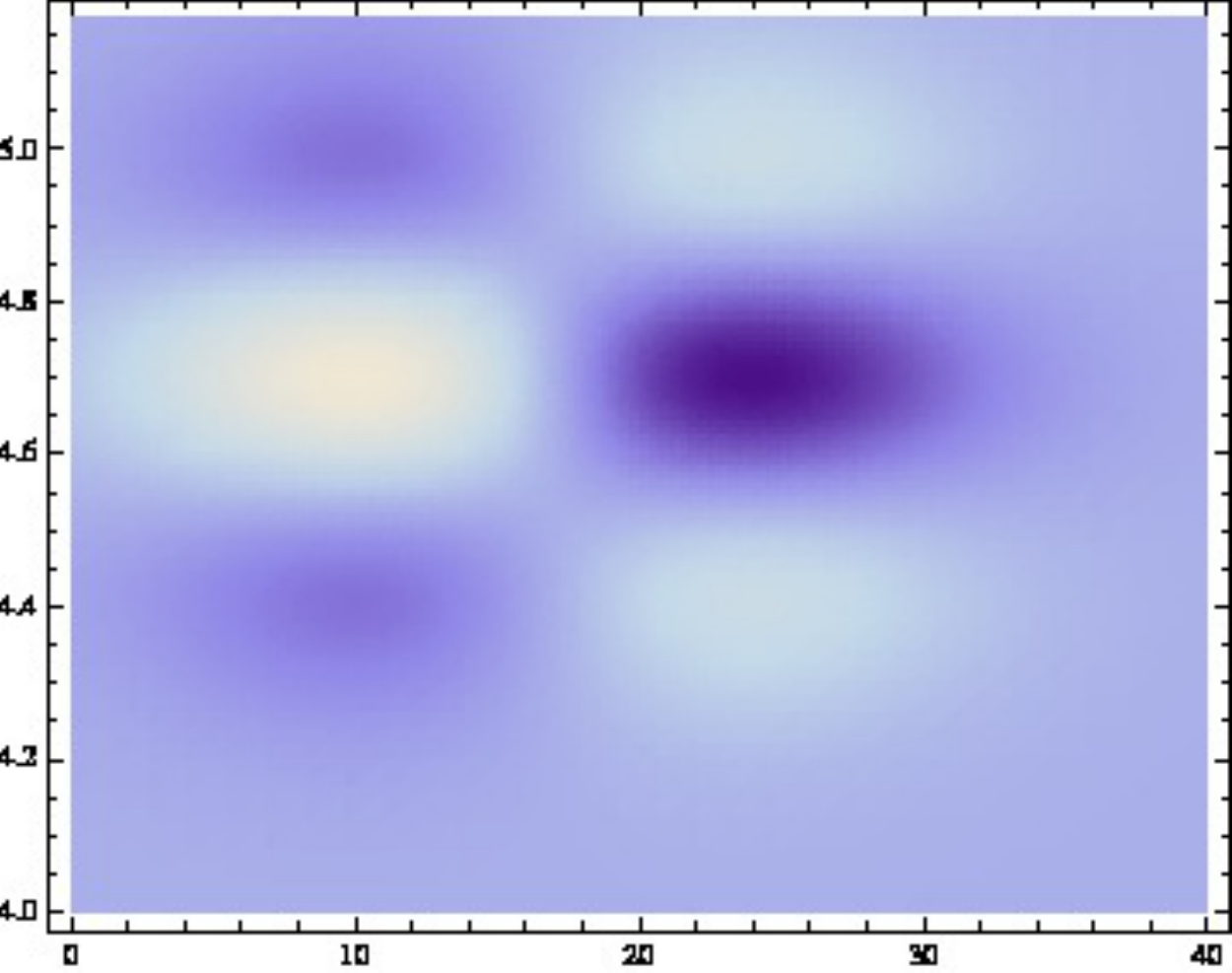}
    \end{tabular} 
  \end{center}
 
  \vspace{-3mm}
  \caption{(top row) A separable monaural spectro-temporal receptive field in the
    central nucleus of the inferior colliculus (ICC) of cat as reported
    by \protect\cite{QiuSchEsc03-JNeuroPhys}.
    (bottom row) Idealized receptive fields models 
    (\ref{eq-spectr-temp-recfields-gen-form}) corresponding to a
    first-order derivative with respect to time and a second-order
    derivative with respect to logarithmic frequency centered at $\nu
    = 4.7$~octave, temporal scale $\sigma_t = 7$~ms, logspectral scale
  $\sigma_{\nu} = 0.17$~octave for both models and additionally
  temporal delay $\delta = 17$~ms for the Gaussian model.}
  \label{fig-QiuSchEsc03-sep}
\end{figure}

\begin{figure}[hbtp]
 \begin{center}
    \begin{tabular}{ccc}
        \hspace{-2mm} &\hspace{-3mm}  {\em\small Time-causal model\/}  & {\em\small Gaussian model\/} \vspace{-6.2mm} \\ 
       \hspace{-2mm} \includegraphics[width=0.33\textwidth]{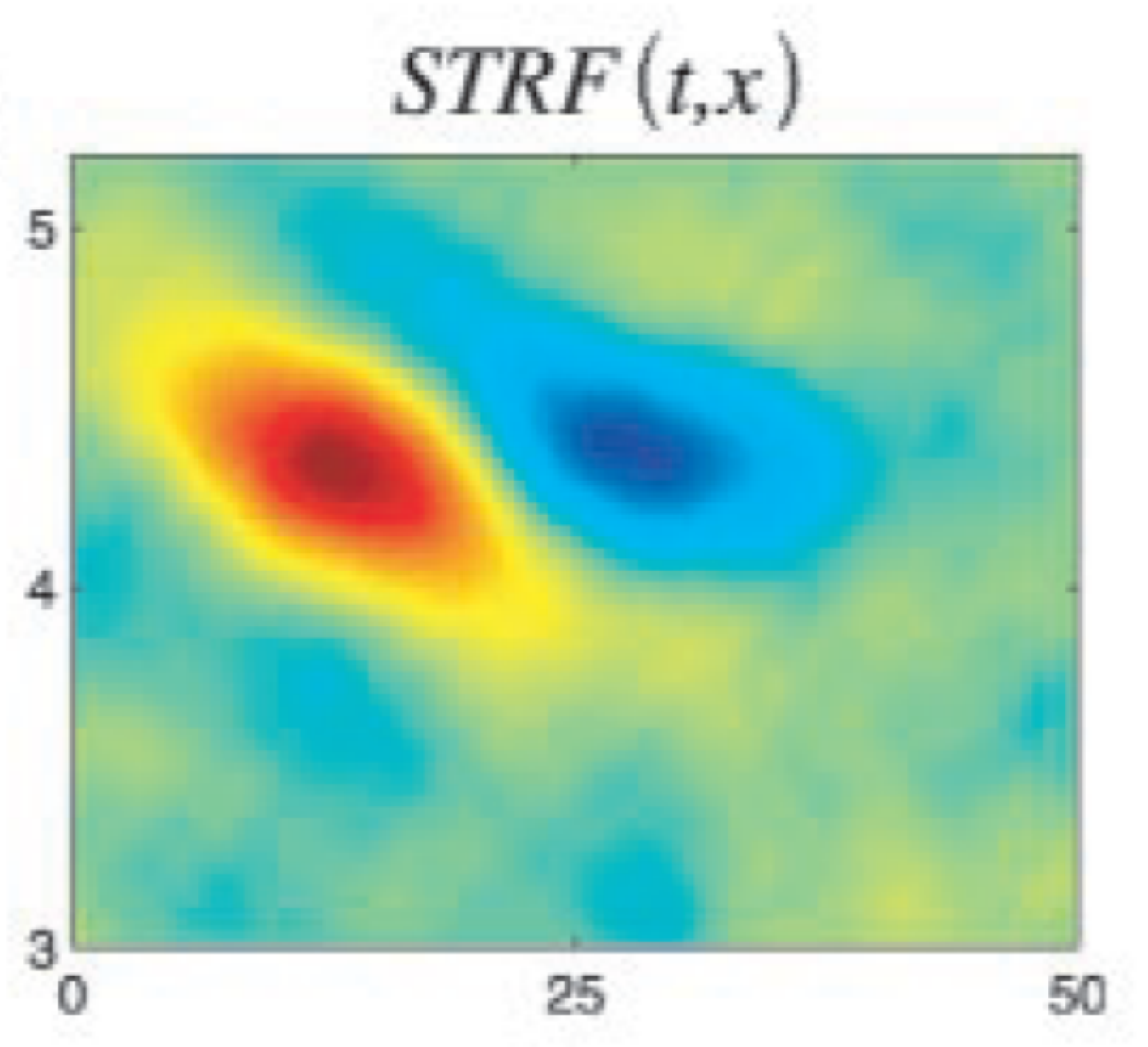}
        & \hspace{-3mm} \includegraphics[width=0.32\textwidth]{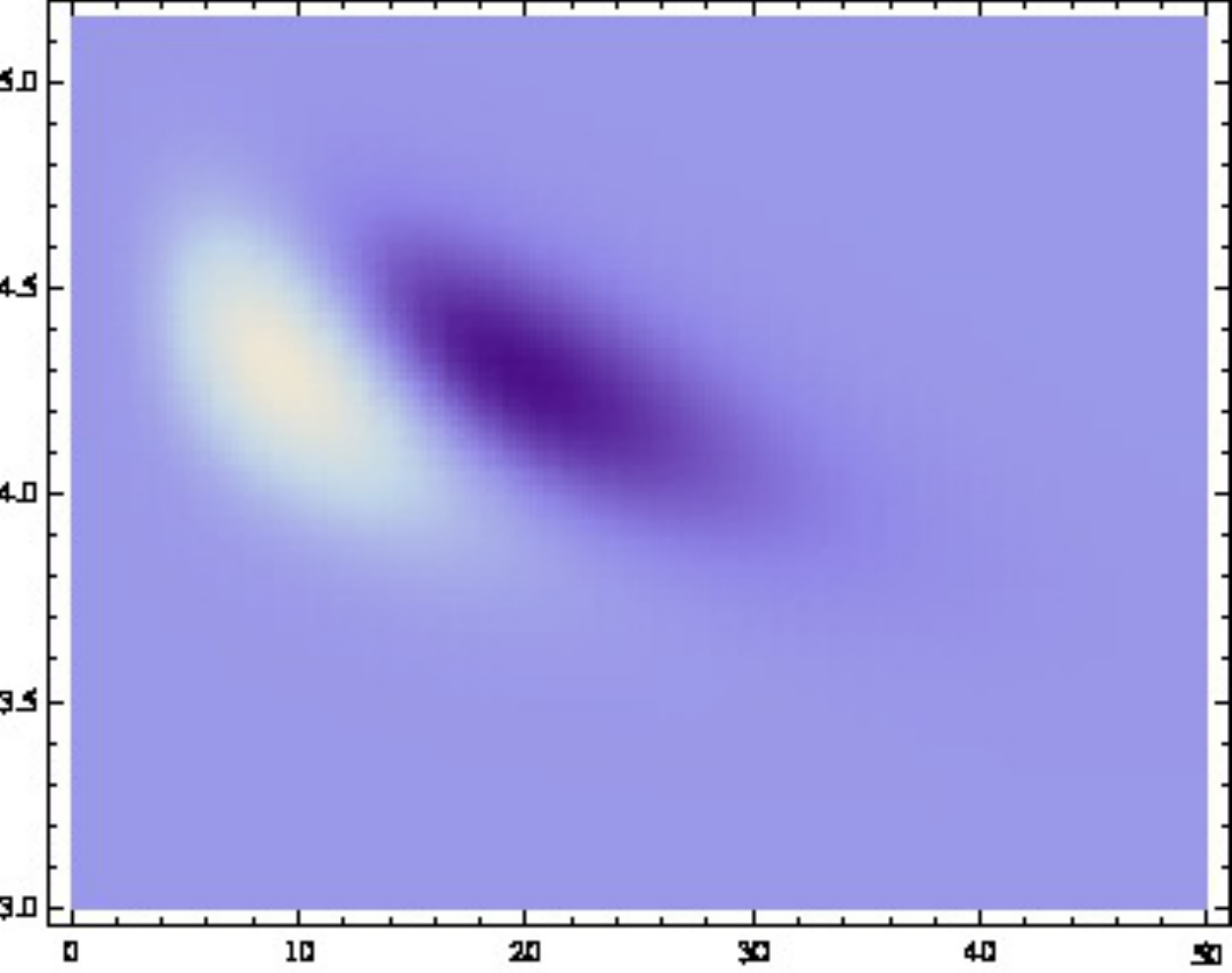}
        & \includegraphics[width=0.32\textwidth]{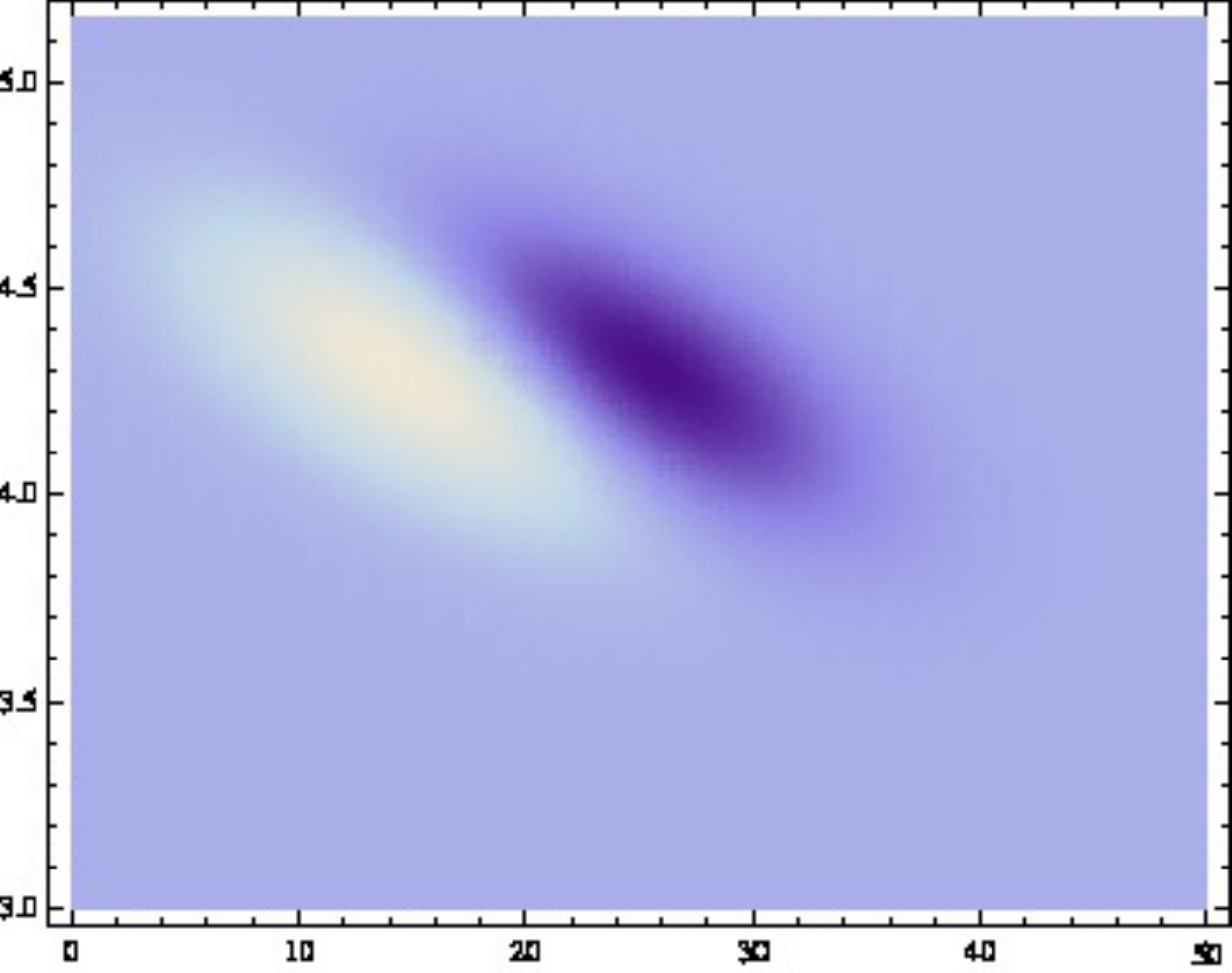}
    \end{tabular} 
  \end{center}
 
  \vspace{-3mm}
  \caption{(top row) A non-separable spectro-temporal receptive fields in the
    central nucleus of the inferior colliculus (ICC) of cat as reported
    by \protect\cite{QiuSchEsc03-JNeuroPhys}.
    (bottom row) First-order temporal derivative of idealized
    glissando-adapted receptive fields models 
    (\ref{eq-spectr-temp-recfields-gen-form}) centered at $\nu
    = 4.3$~octave, temporal scale $\sigma_t = 7$~ms, logspectral scale
  $\sigma_{\nu} = 0.20$~octave and glissando $v = -0.02$~octave/ms for both models and additionally
  temporal delay $\delta = 23$~ms for the Gaussian model.}
  \label{fig-QiuSchEsc03-nonsep}
\end{figure}

\begin{figure}[hbtp]
 \begin{center}
    \begin{tabular}{ccc}
        \hspace{-2mm} &\hspace{-3mm}  {\em\small Time-causal model\/}  & {\em\small Gaussian model\/} \vspace{-0mm} \\ 
        \hspace{-2mm}
       \includegraphics[width=0.35\textwidth]{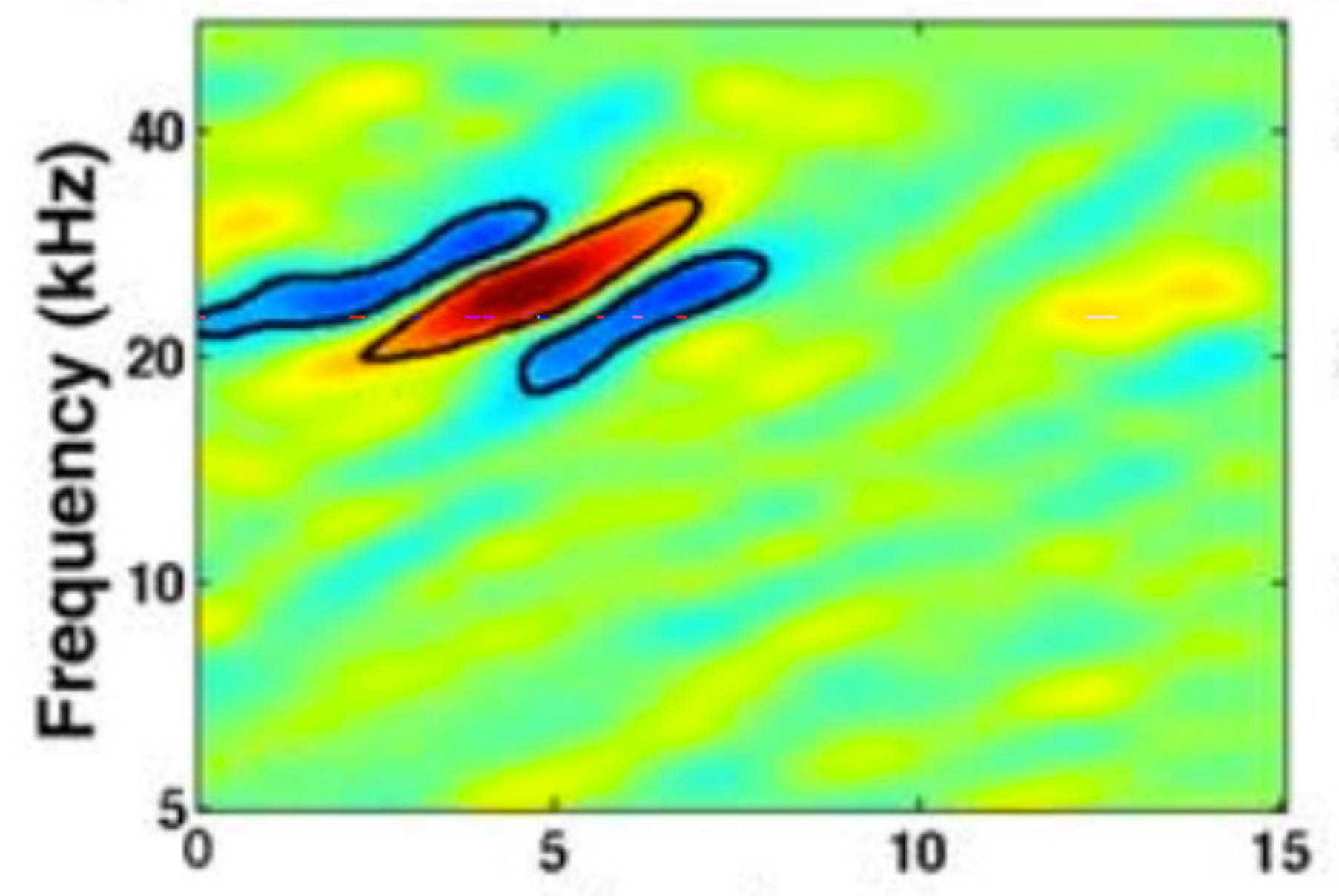}
         & \hspace{-3mm} \includegraphics[width=0.30\textwidth]{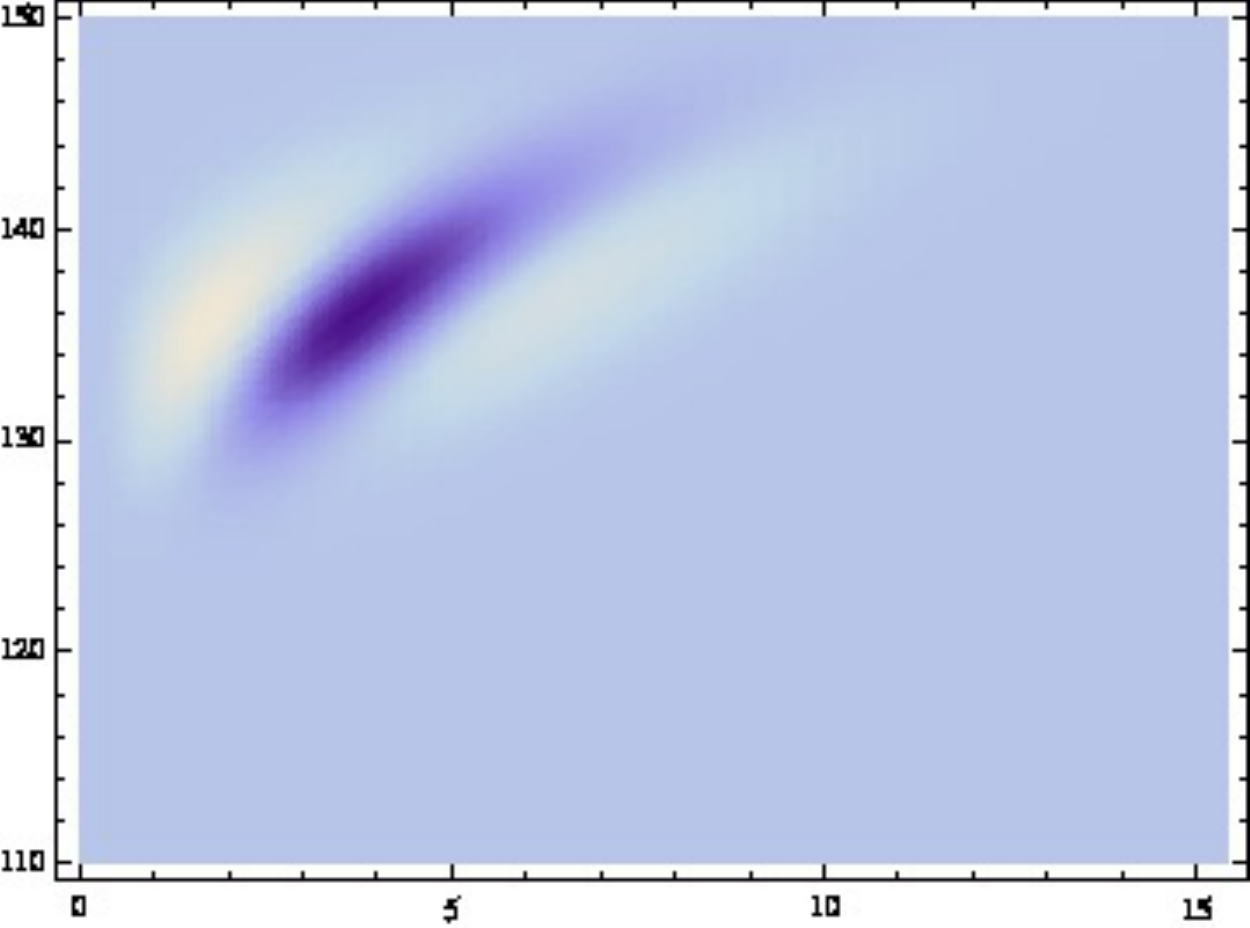}
        & \includegraphics[width=0.30\textwidth]{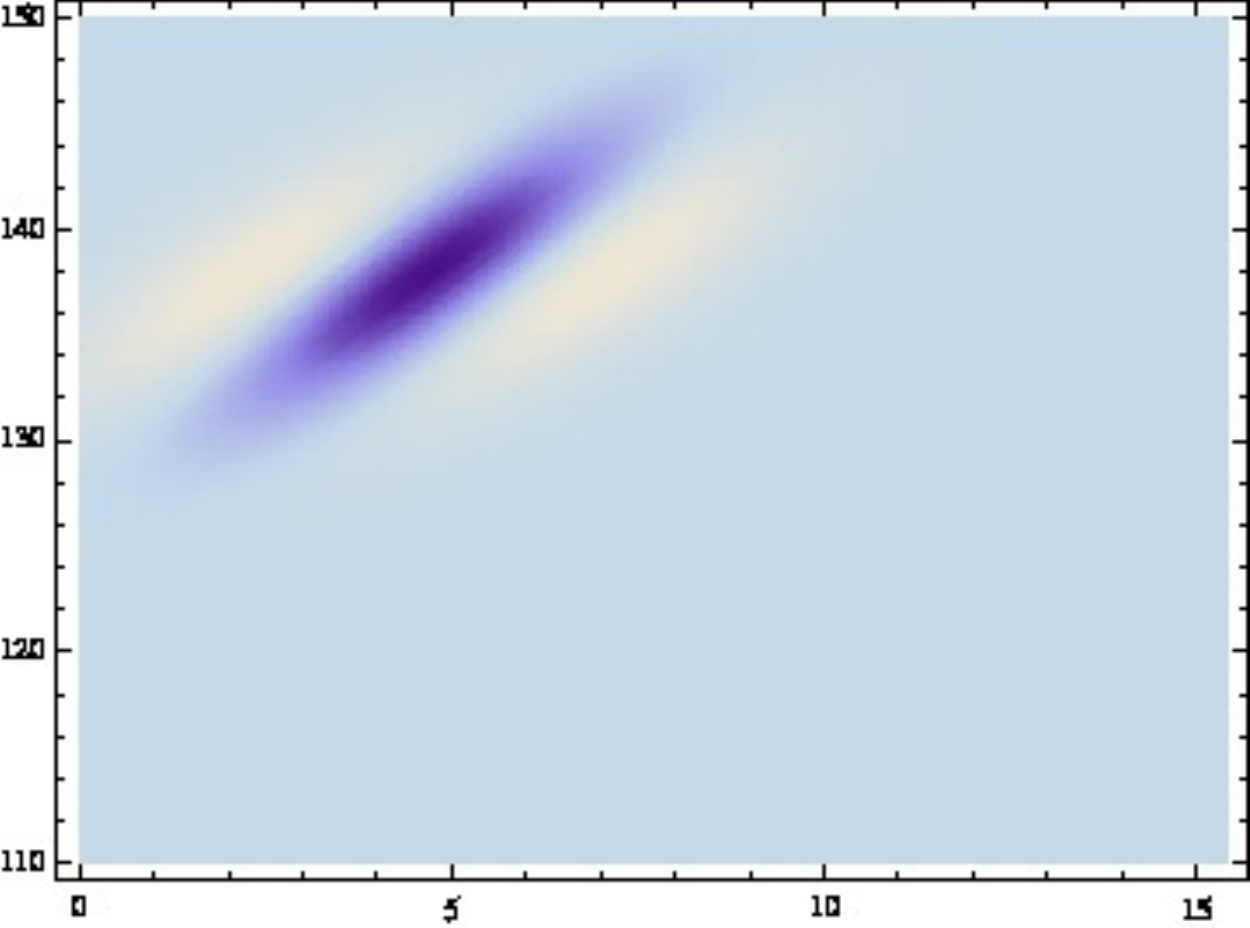}
    \end{tabular} 
  \end{center}
 
  \vspace{-3mm}
  \caption{(top row) A non-separable spectro-temporal receptive fields in the
    inferior colliculus (ICC) of Mexican
    free-tailed bat as reported
    by \protect\cite{AndLiPol07-JNeuroSci}.
(bottom row) Second-order temporal derivative of idealized
    glissando-adapted receptive fields models 
    (\ref{eq-spectr-temp-recfields-gen-form}) centered at semitone $\nu
    = 138$ , temporal scale $\sigma_t = 2$~ms, logspectral scale
  $\sigma_{\nu} = 3$~semitones and glissando $v = 1.5$~semitones/ms for both models and additionally
  temporal delay $\delta = 4.7$~ms for the Gaussian model.}
  \label{fig-AndLiPol07-nonsep}
\end{figure}

\begin{figure}[hbtp]
 \begin{center}
    \begin{tabular}{ccc}
        \hspace{-3mm} & {\em\small Time-causal model\/}  & {\em\small Gaussian model\/} \vspace{-8mm} \\ 
        \hspace{-3mm}
        \includegraphics[width=0.37\textwidth]{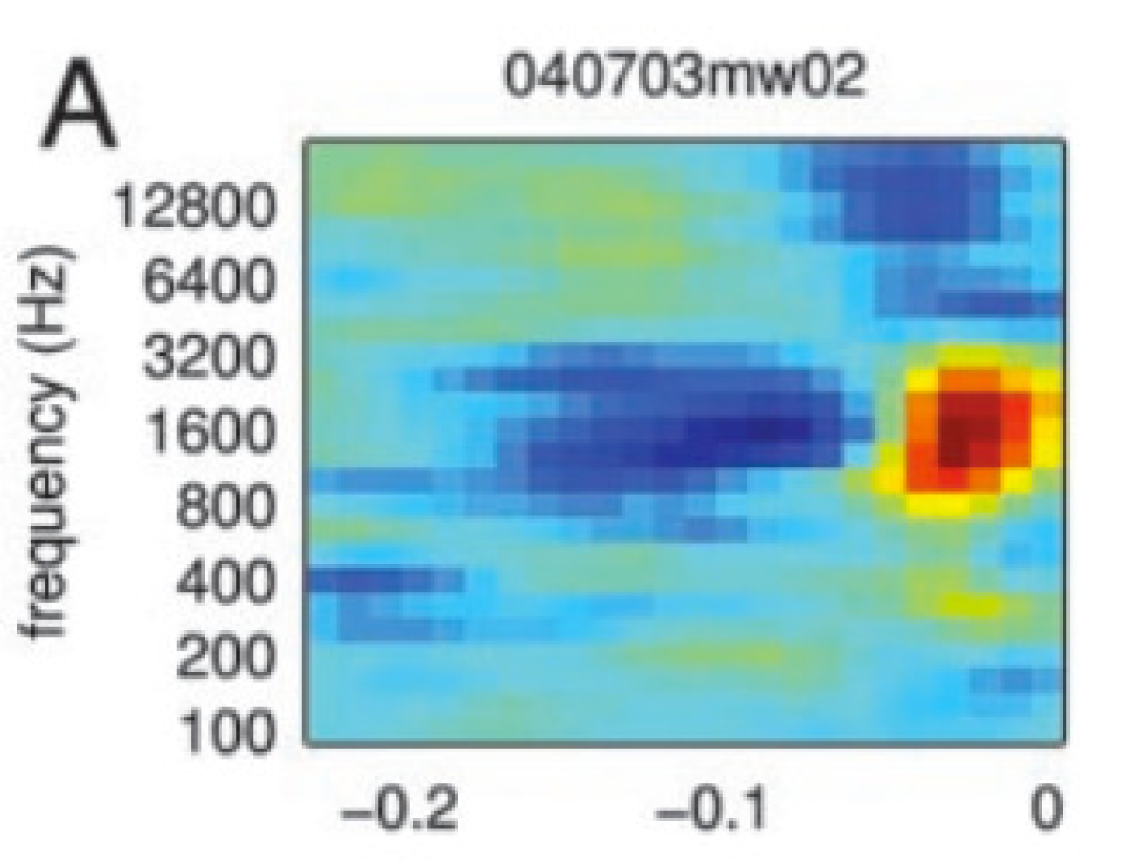}
        & \includegraphics[width=0.28\textwidth]{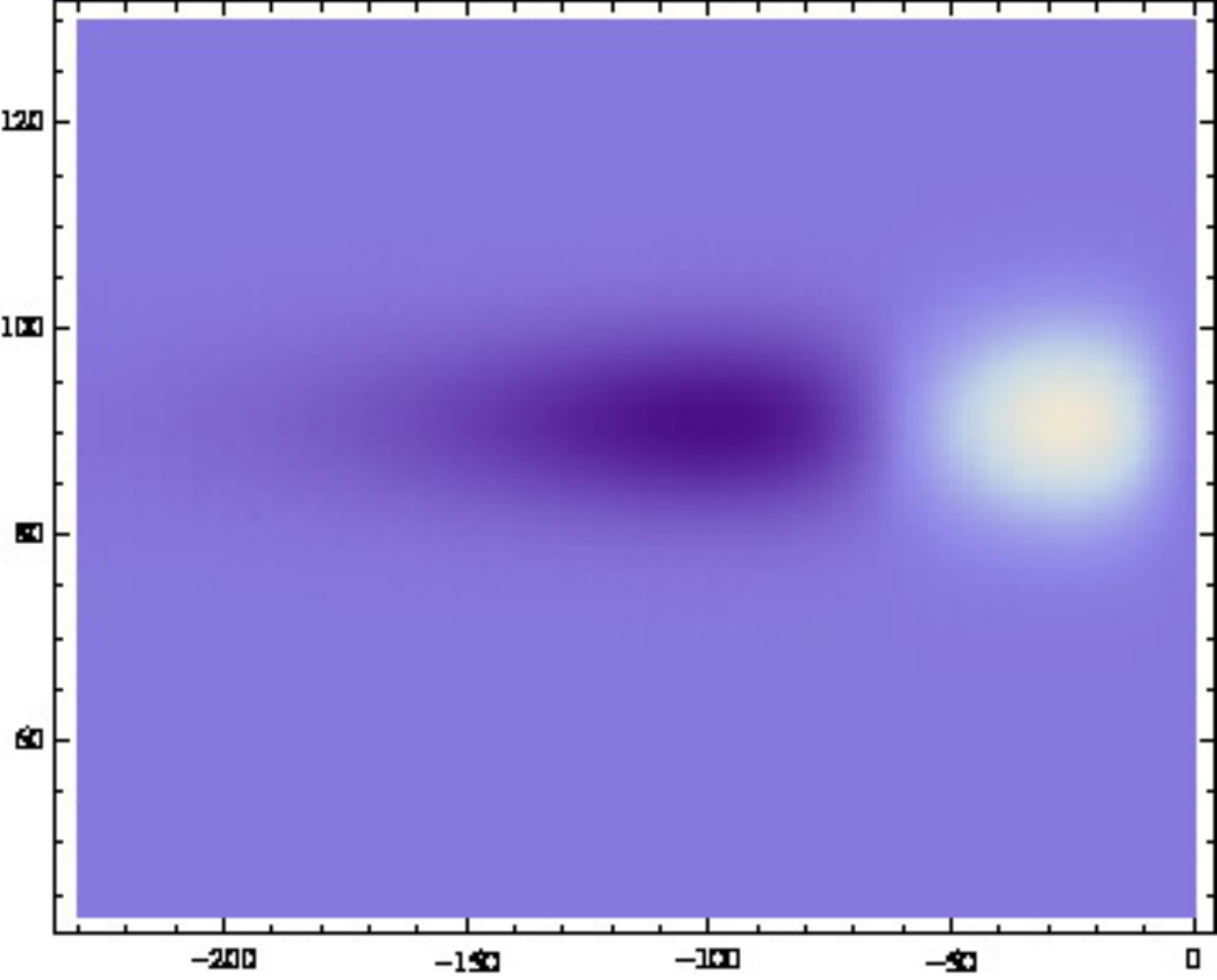}
        & \includegraphics[width=0.28\textwidth]{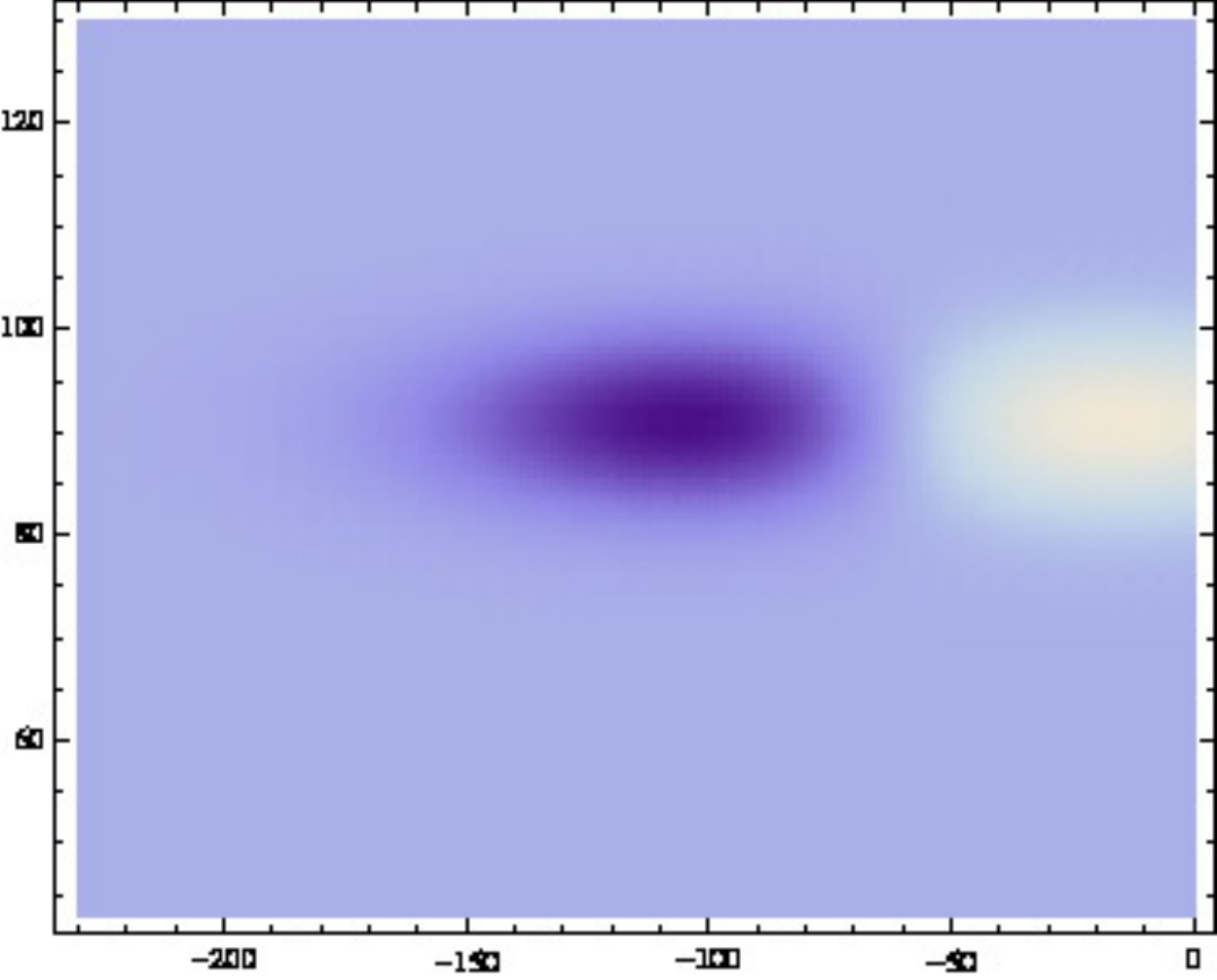}\\
        \hspace{-3mm} \includegraphics[width=0.37\textwidth]{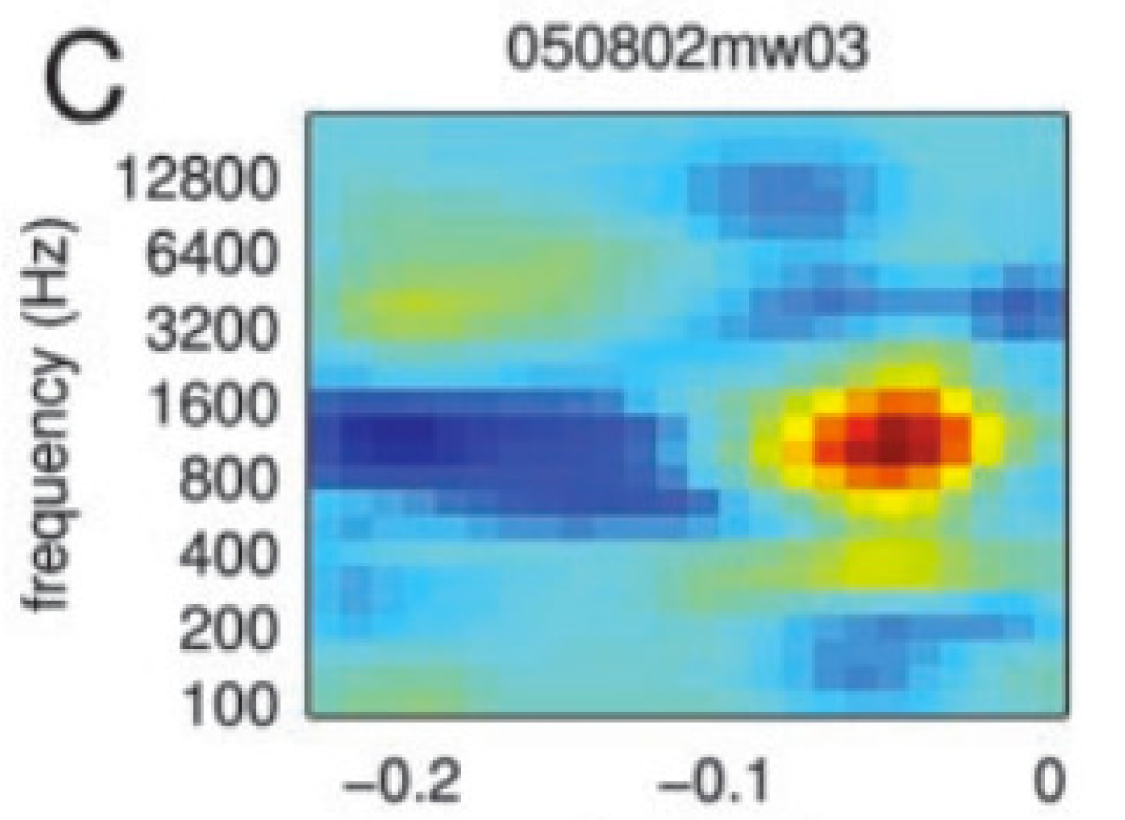}
        & \includegraphics[width=0.28\textwidth]{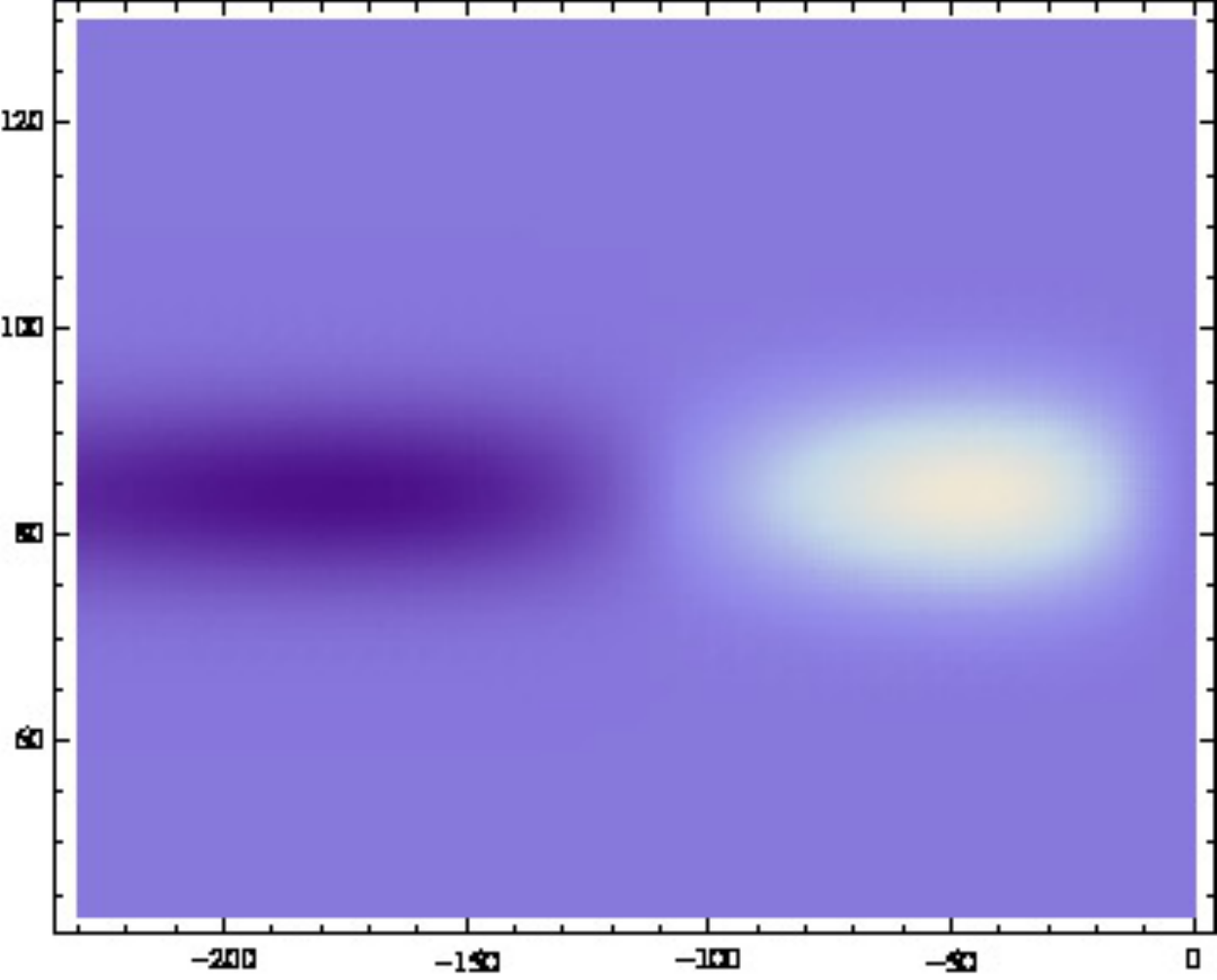}
        & \includegraphics[width=0.28\textwidth]{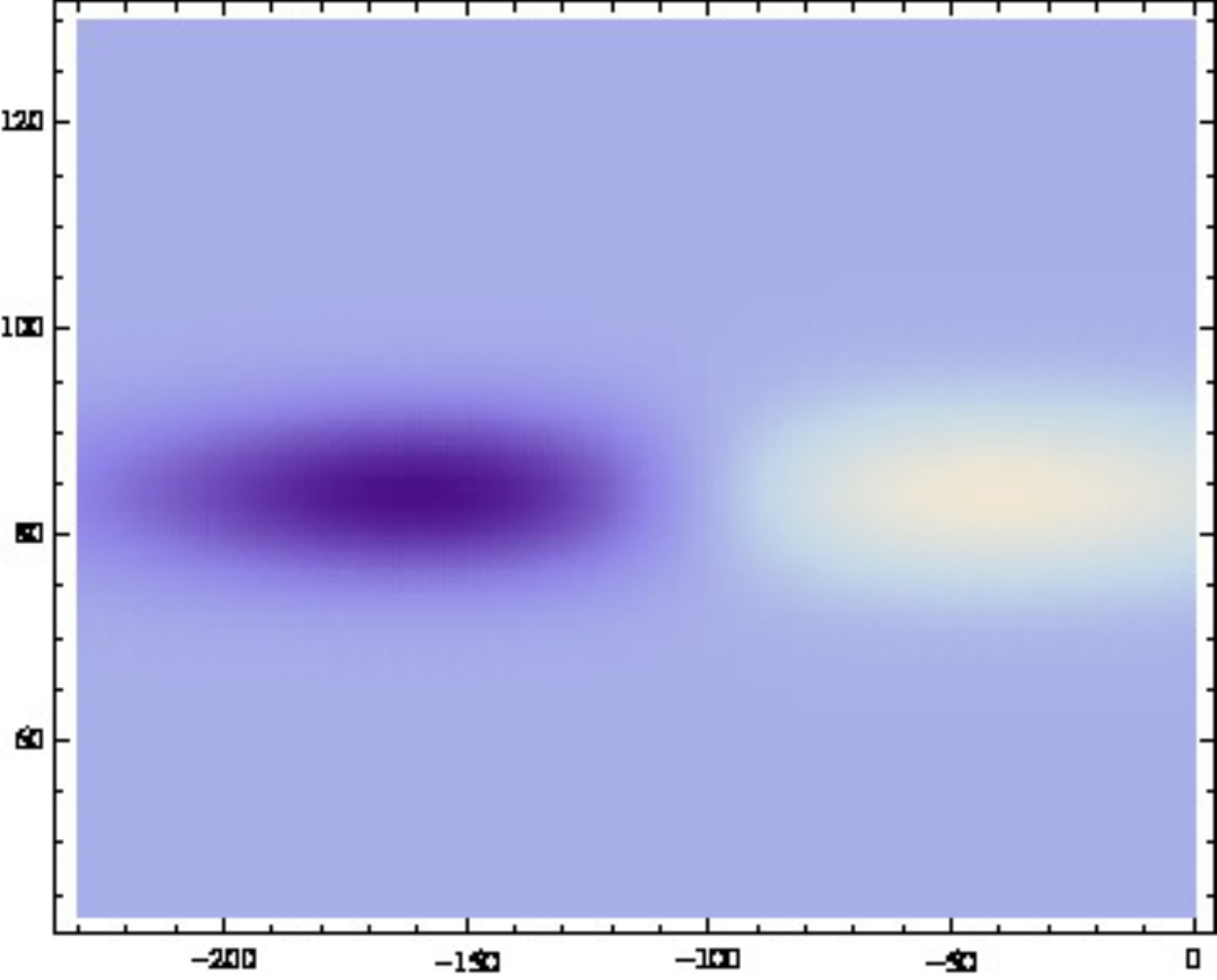}
    \end{tabular} 
  \end{center}
 
  \vspace{-3mm}
  \caption{(left column) Separable spectro-temporal receptive fields in the
    primary auditory cortex (A1) of Sprague Dawley rat as reported
    by \protect\cite{MacWehZad04-JNeuroSci}.
    (middle and right columns) Idealized receptive fields models 
    (\ref{eq-spectr-temp-recfields-gen-form}) corresponding to
    first-order derivatives with respect to time, in the top row centered at semitone $\nu
    = 91$ , temporal scale $\sigma_t = 45$~ms, logspectral scale
  $\sigma_{\nu} = 6$~semitones for both models and additionally
  temporal delay $\delta = 60$~ms for the Gaussian model and in the
  bottom row centered at semitone $\nu
    = 84$ , temporal scale $\sigma_t = 60$~ms, logspectral scale
  $\sigma_{\nu} = 6$~semitones for both models and additionally
  temporal delay $\delta = 100$~ms for the Gaussian model}
  \label{fig-MacWehZad04-A1}
\end{figure}

\begin{figure}[hbtp]
 \begin{center}
    \begin{tabular}{ccc}
        \hspace{-3mm} & {\em\small Time-causal model\/}  & {\em\small Gaussian model\/} \vspace{-9mm} \\ 
        \hspace{-3mm}
        \includegraphics[width=0.28\textwidth]{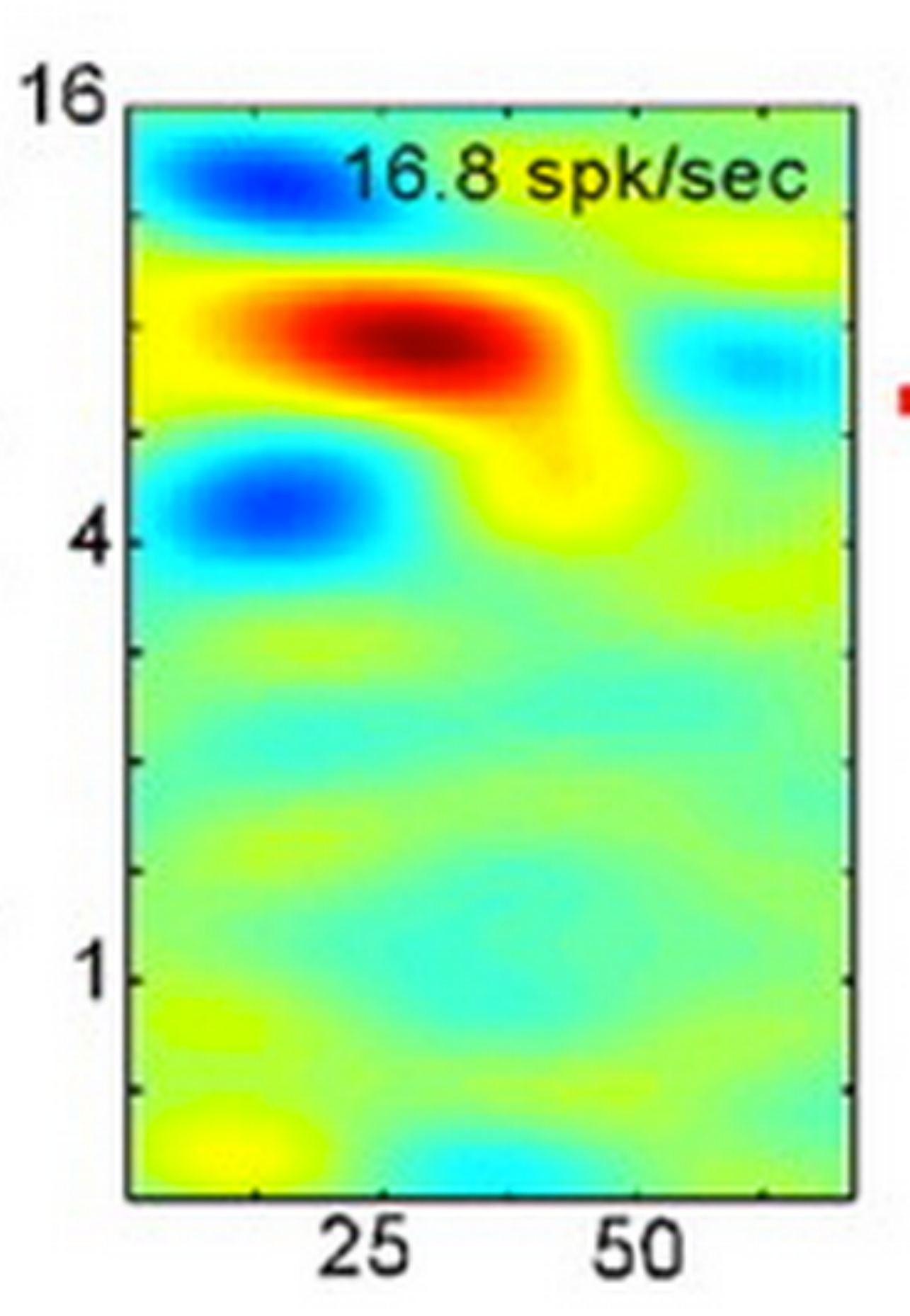}
        & \includegraphics[width=0.23\textwidth]{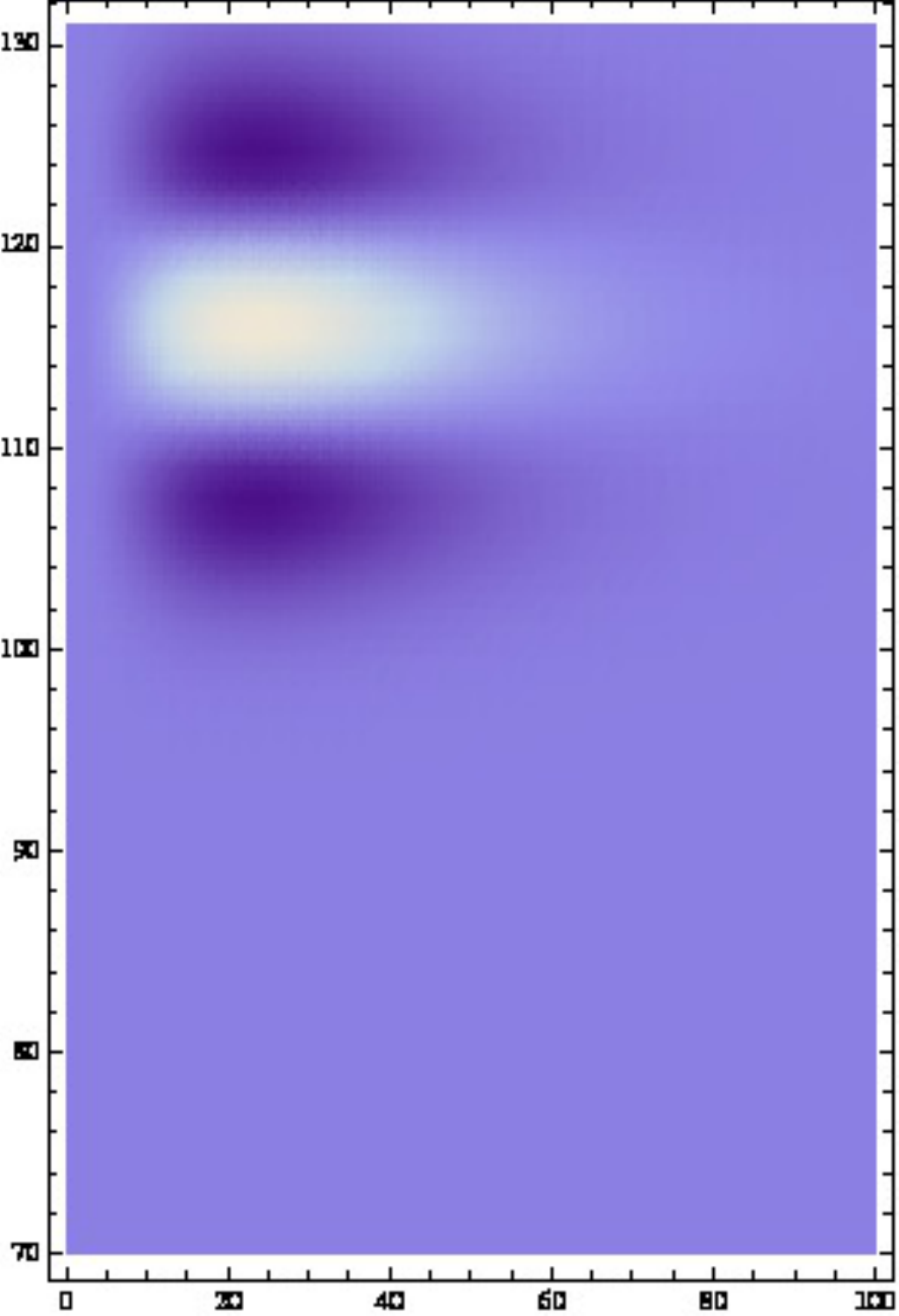}
        & \includegraphics[width=0.23\textwidth]{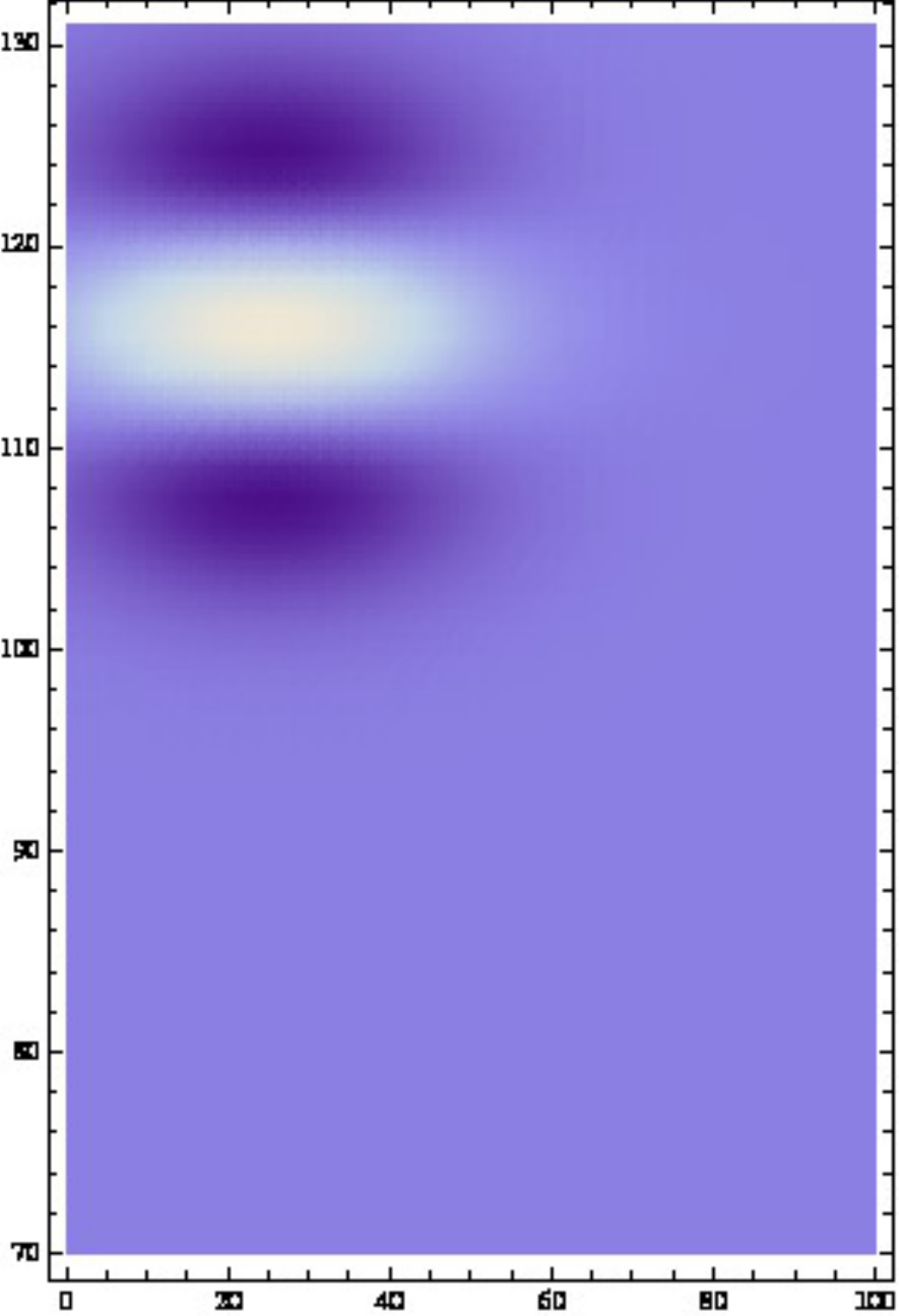}
    \end{tabular} 
  \end{center}
 
  \vspace{-3mm}
  \caption{(left) A separable spectro-temporal receptive fields in the
    primary auditory cortex (A1) of ferret as reported
    by \protect\cite{ElhFriChiSha07-JNeuroSci}. 
   (middle and right) Idealized receptive fields models 
    (\ref{eq-spectr-temp-recfields-gen-form}) corresponding to
    second-order derivatives with respect to logarithmic frequency centered at semitone $\nu
    = 116$ , temporal scale $\sigma_t = 17$~ms, logspectral scale
  $\sigma_{\nu} = 5$~semitones for both models and additionally
  temporal delay $\delta = 25$~ms for the Gaussian model.}
  \label{fig-ElhFriChiSha07-A1}

  \bigskip

 \begin{center}
    \begin{tabular}{c}
        \includegraphics[width=0.80\textwidth]{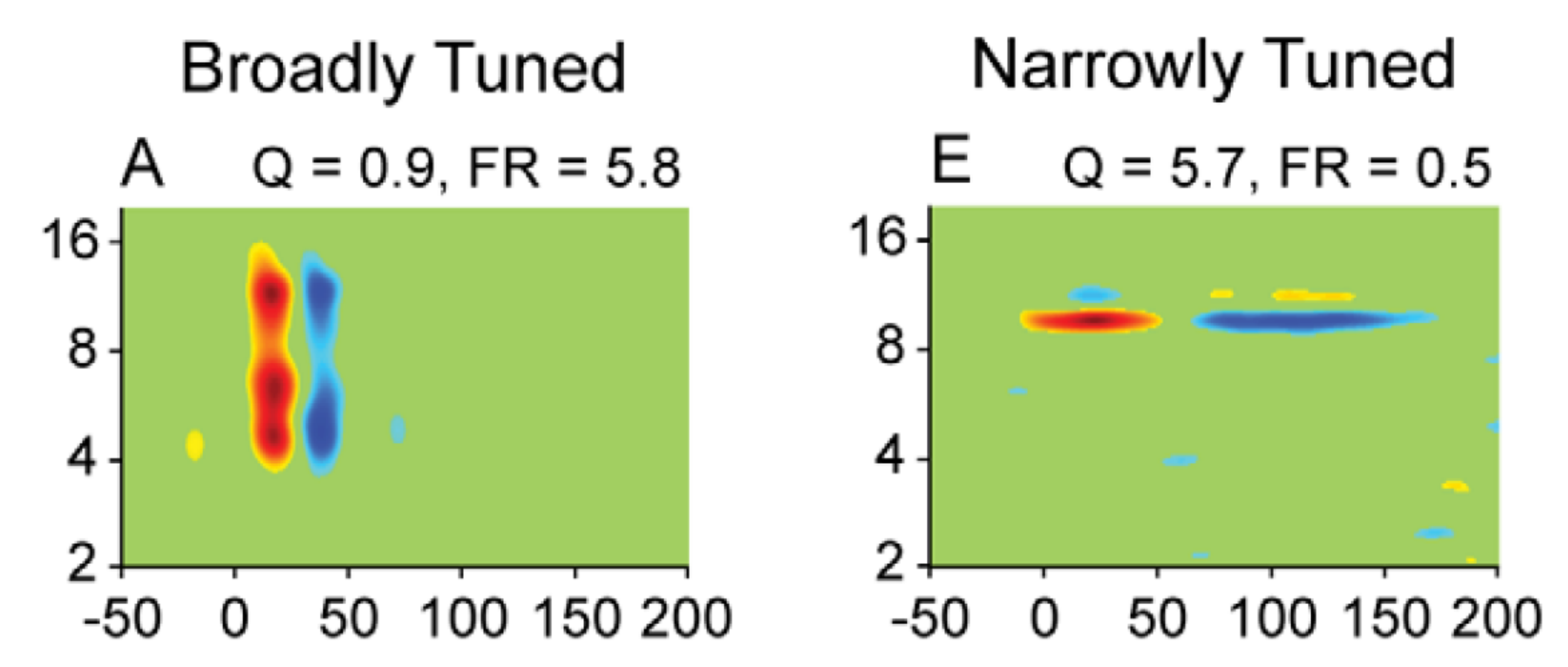}
    \end{tabular} 
    \begin{tabular}{cc}
       \hspace{-0mm}  {\em\small Time-causal model\/} \hspace{3mm}  
        & \hspace{6mm} {\em\small Time-causal model\/} \hspace{0mm}  \vspace{0mm} \\ 
        \includegraphics[width=0.31\textwidth]{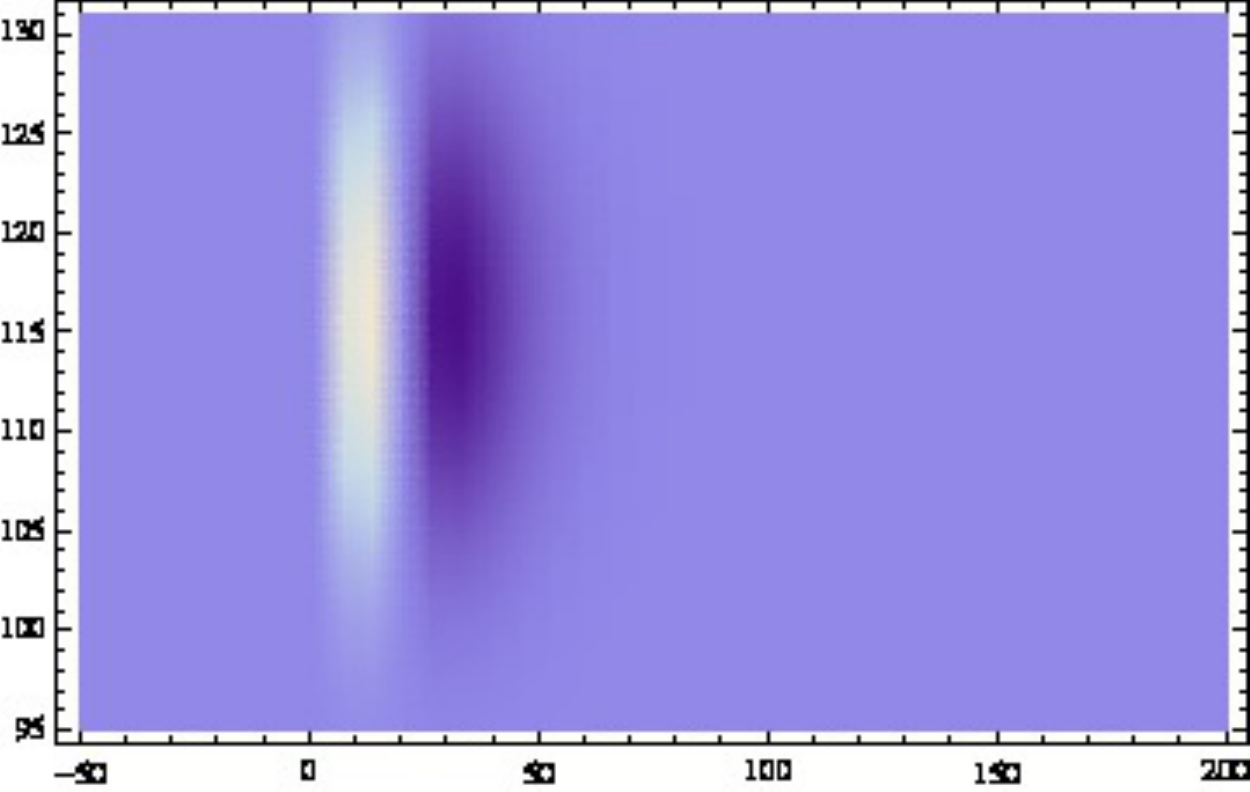} \hspace{3mm}  
        &  \hspace{6mm} \includegraphics[width=0.31\textwidth]{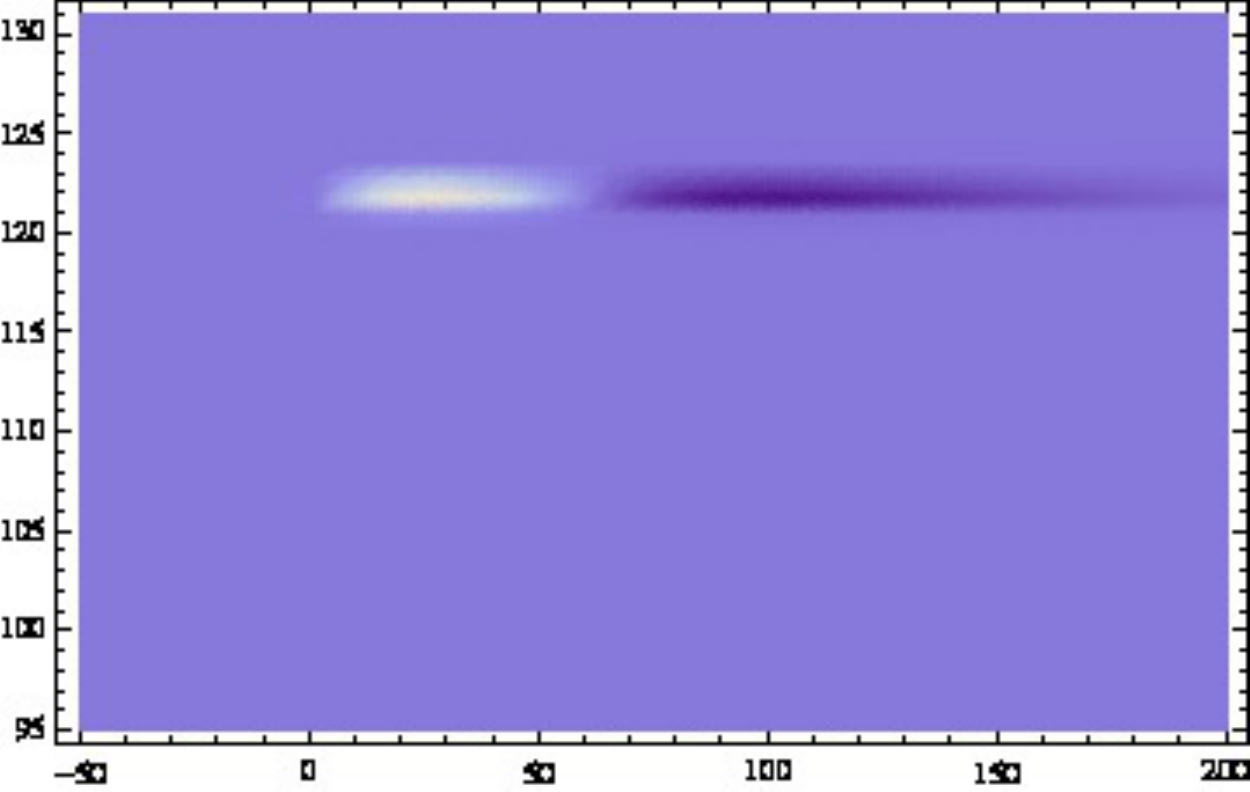}  \\
       \hspace{0mm} {\em\small Gaussian model\/}  \hspace{0mm}  
        & \hspace{6mm} {\em\small Gaussian model\/} \hspace{0mm}  \vspace{0mm} \\ 
        \includegraphics[width=0.31\textwidth]{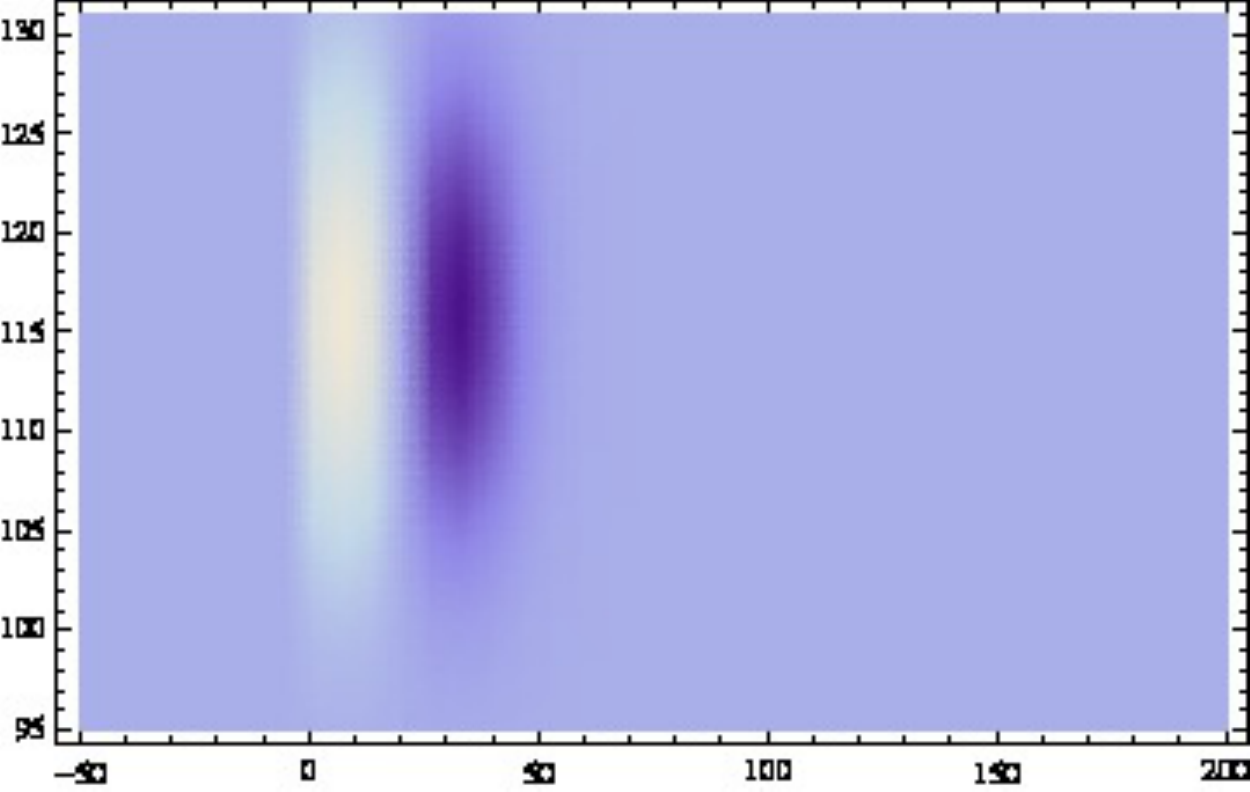} \hspace{3mm}  
        &  \hspace{4mm} \includegraphics[width=0.31\textwidth]{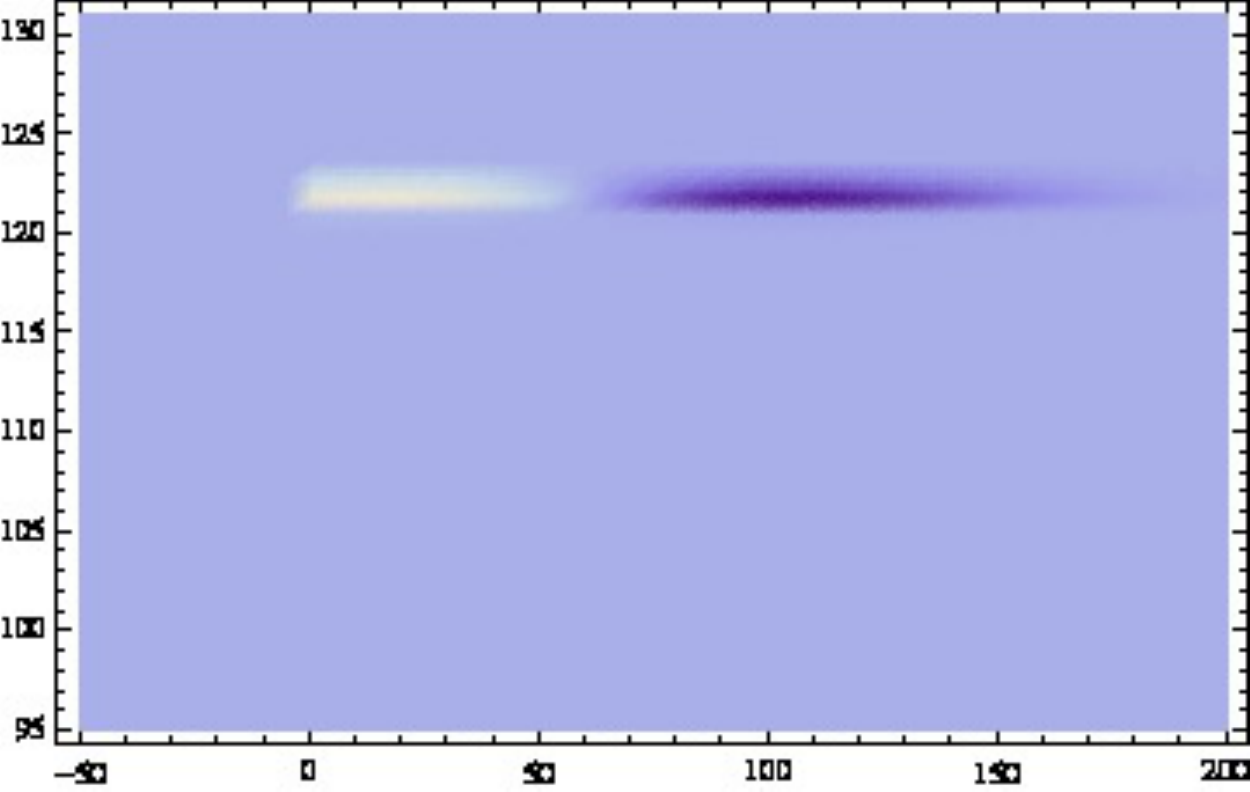} \\
    \end{tabular} 
    \end{center}
 
  \vspace{-3mm}
  \caption{(top row) Spectro-temporal receptive fields of broadly and narrowly
    tuned neurons in the primary auditory cortex of cats as reported
    by \protect\cite{AteSch12-PONE}.
    In terms of the idealized models of receptive fields, such broad
    or narrowly tuned receptive fields correspond to different values
  of the spectral scale parameter $s$.
  (middle and bottom rows) Idealized receptive fields models 
    (\ref{eq-spectr-temp-recfields-gen-form}) corresponding to
    first-order derivatives with respect to time, in the left column centered at semitone $\nu
    = 119$ , temporal scale $\sigma_t = 12$~ms, logspectral scale
  $\sigma_{\nu} = 8$~semitones for both models and additionally
  temporal delay $\delta = 20$~ms for the Gaussian model and in the
  bottom column centered at semitone $\nu
    = 122$ , temporal scale $\sigma_t = 45$~ms, logspectral scale
  $\sigma_{\nu} = 0.5$~semitones for both models and additionally
  temporal delay $\delta = 60$~ms for the Gaussian model. }
  \label{fig-AteSch-PONE-broad-narrow-tuning}
\end{figure}

\section{Relations to biological receptive fields}
\label{sec-biol-aud-rec-field}

In the central nucleus of the inferior colliculus of cats,
\cite{QiuSchEsc03-JNeuroPhys} report that about 60~\% of the neurons
can be described as separable in the time-frequency domain (see figure~\ref{fig-QiuSchEsc03-sep}),
whereas the remaining neurons are either obliquely oriented (see figure~\ref{fig-QiuSchEsc03-nonsep})
or contain multiple excitatory/inhibitory subfields.
This overall structure is nicely compatible with the treatment
in section~\ref{sec-ideal-rf},
where the second-layer receptive fields are expressed in terms of
spectro-temporal derivatives of either time-frequency separable spectro-temporal
smoothing operations or corresponding glissando-adapted features as
motivated by the structural requirements in
section~\ref{sec-struct-req-2nd-layer-RF}.

Qualitiatively similar shapes of receptive fields can also be
measured from neurons in
the primary auditory cortex (see figure~\ref{fig-ElhFriChiSha07-A1}
and figure~\ref{fig-MacWehZad04-A1}) 
and for binaural receptive fields \cite{MilEscReaSch01-JNeuroPhys}.
Specifically, the use of multiple temporal and spectral scales as a
main component in the theoretical model is
good agreement with biological receptive fields having different
degrees of spectral tuning ranging from narrow to broad
(see figure~\ref{fig-AteSch-PONE-broad-narrow-tuning})
and different temporal extent (see figure~\ref{fig-MacWehZad04-A1}). 
Corresponding tradeoffs between spectral and temporal tuning occur 
in the inferior colliculus \cite{RodReaEsc10-JNeuroPhys}.
The distribution of latencies is, however, towards somewhat shorter
latencies in the thalamus than in the auditory cortex
\cite{MilEscReaSch01-JNeuroPhys}, consistent with larger temporal
scales in the auditory cortex than in the inferior colliculus.

Whereas spatio-temporal receptive fields 
estimated from neurons in the auditory cortex have been reported to
reasonably well predict the neural responses for subsets of natural stimuli,
\cite{MacWehZad04-JNeuroSci} report that for many natural stimuli
the responses of the auditory neurons in the primary auditory
cortex cannot be predicted by the estimated linear receptive fields.
\cite{AteShaSch12-JNeuroPhys} have also reported that the
dimensionality and thereby the variability of the receptive fields
in the auditory cortex is significantly richer compared to the receptive 
fields in the inferior colliculus in the midbrain.
Thus, the neurons in the primary auditory cortex appear to contain
non-linearities whose functionality remains to be understood.
In the inferior colliculus in the midbrain,
\cite{EscSch02-JNeuroSci} do on the other hand report that 
about 60~\% of the receptive fields can be well described in terms of linearily integrating neurons.

In the work by \cite{QiuSchEsc03-JNeuroPhys}, the measured biological receptive
fields were fitted to Gabor functions as motivated by previous
use of Gabor functions for modeling visual receptive fields 
\cite{Mar80-JOSA,JonPal87a,JonPal87b}.
In vision, 
the use of Gabor functions for modelling visual receptive fields can,
however, be questioned both on theoretical and empirical grounds
\cite{StoWil90-JOSA,Lin13-BICY,Lin13-PONE}.%
\footnote{
\cite{StoWil90-JOSA} argue that (i)~only complex-valued Gabor
functions that cannot describe single receptive field minimize the
uncertainty relation, (ii)~the real functions that minimize this
relation are Gaussian derivatives rather than Gabor functions and
(iii)~comparisons among Gabor and alternative fits to both
psychophysical and physiological data have shown that in many cases
other functions (including Gaussian derivatives) provide better fits
than Gabor functions do.
\cite{Lin13-BICY,Lin13-PONE} argues that in relation to invariance properties, 
the family of affine Gaussian
kernels is closed under affine image deformations, whereas the family
of Gabor functions obtained by multiplying rotationally symmetric
Gaussians with sine and cosine waves is not closed under affine image
deformations. This means that it is not possibly to compute truly
affine invariant image representations from such Gabor functions.
Instead, given a pair of images that are related by a non-uniform
image deformation, the lack of affine covariance implies that there 
will be a systematic bias in image representations derived from such
Gabor functions, corresponding to the difference between the
backprojected Gabor functions in the two image domains.
If using receptive profiles defined from directional derivatives of
affine Gaussian kernels, it will on the other hand be possible to
compute affine invariant image representations.
Similar arguments about Galilean invariance hold regarding theoretical
modelling of spatio-temporal receptive fields.
In addition, the zero-order Gaussian receptive as well as the derivative based
receptive fields can be modelled by diffusion equations, 
and can therefore be implemented by computations between 
neighbouring computational units.}
Specifically, biological spectro-temporal receptive fields show 
a marked temporal asymmetry that cannot be captured by Gabor functions
for which the locations of excitatory and inhibitory subregions are uniformly spaced.
Therefore, \cite{QiuSchEsc03-JNeuroPhys} performed additional non-linear time warping
to be able to fit the model to the data.
Then, they modelled oblique receptive fields over the time-frequency
domain using singular value decomposition to express
any oblique receptive field as a sum of separable receptive fields
defined from Gabor functions.

By modeling the spatio-temporal receptive fields by a combination of time-causal
scale-space kernels over time and Gaussian receptive fields over
logarithmic frequencies according to
(\ref{eq-spectr-temp-recfields-gen-form}), 
the temporal asymmetry of the kernels
constitutes an integrated part of the theory.
Furthermore, oblique receptive fields in the time-frequency domain
do also constitute an integrated part of the theory in terms of glissando
transformations, and there is no need to decompose a
glissando-receptive field as a sum of a possibly rather large number
of spectro-temporal receptive fields, to be able to model the
spectro-temporal receptive in a quantitiative manner.
In addition, the model has been derived in a mathematically principled
way from a set of structural requirements and the idealized receptive fields can be computed by a 
combination of diffusion equations and first-order integrators,
and therefore by a biologically plausible neural architecture.

\section{Relations to previous work in audio processing}
\label{sec-rel-audio-proc}

Previous auditory models of human hearing have to a large extent focused on the first stages of processing including the acoustical response of the cochlea and the following neurological responses in the auditory nerve. The auditory periphery is also the stage at which it was possible to collect the first physiological and neurological data. Recently, partly due to the fast technological progress in measurement techniques, models of more high-level functions in the auditory cortex begin to emerge \cite{MedLopFayPop10-book}. Thus the purpose has been to convert the incoming audio into a frequency-time representation in a similar way as is done in the physical-neural system in the cochlea. This is typically done in several stages taking into account both biological measurements and psychoacoustic listening test data.

The main stage is to simulate the physical resonance system in the
cochlea. It is often implemented as filter bank in which the bandwidth
and frequency position of each band is separated in a similar way as
the cochlear nerves. A gammatone filter is often used since it has
been show to be a reasonable approximation of the acoustic properties
in the cochlea leading to the first neural input
\cite{PatMoo86-FreqSel,PatRobHolMcKeoZhaAll92-AudPhysPerc}.
Many other types of filters have been proposed, such as the
gammachirp providing better fit to non-linearities with regard to the
asymmetric frequency response at loud sound pressures \cite{IriPat97-JASA} (see also \cite{CheHuGlaMoo11-HearRes,LopMedd01-JASA}). In a secondary stage the response of the auditory nerves arriving from the inner hair cells in the cochlea are modeled. A common version is to make a half-wave rectification, compression (square-root or logarithmic) and low-pass filtering \cite{PatAllGig95-JASA}. In addition the local contrast can be enhanced both in time and frequency using an adaptive procedure \cite{PatHol96-AdvSpeechHear}.

Previous computational toolboxes include the auditory image model
(Aim-mat) by \cite{BleIvePat04-ActaAcust} see also
\cite{PatUnoIri03-JASA}, which include all of the different parts above as well as some additional parts, and the auditory model by \cite{Sla98-TR}.

Thus, for example the auditory image model above is quite advanced and takes into account a number of biological and perceptual phenomena. However, the biological data supporting these stages seems rather scarce in particular for complex sound signals such as music. For example, the nerve responses have often been measured in animals rather than humans and with rather simple stimuli such as stationary sinusoids \cite{Rug92-CurrOpNeuroBiol}. Many of the parts are modeled after psychoacoustic data again with simple stimuli and thus involving the whole auditory cortex and brain. A limitation is that advanced perceptual models work can be adapted to closely model a certain type of perceptual data but will typically not extrapolate to other perceptual conditions. For example, the recent loudness model by \cite{CheHuGlaMoo11-HearRes} closely approximates perceptual data concerning several aspects of loudness but applies only to steady-state sounds. This is an indication that the formulation of the underlying model(s) is still potentially open to alternative solutions. This is not surprising giving the complexity of the task and difficulties in obtaining biological data.

The traditional auditory models are not necessarily the best choice as a front-end for modeling more high-level perceptual aspects of music and speech. These models are often very demanding in terms of computer power and memory. In addition, the resulting data may not be suitable for further processing. Therefore, in practical applications, a front-end used for analyzing perceptual phenomena is often a simplification of the complete model. Recently, alternative models have been suggested which apply general principles of auditory perception but leaving out the detailed aspects. Such a model can be a good compromise between biological/perceptual reality and computational clarity and efficiency. For example, \cite{ChiGaoCuyMatRuSha99-JASA} used for the first stage 128 overlapping constant-Q bandpass filters with 24 filters/octave, an hair cell model with a high-pass filter, a non-linear compression, a membrane leakage low-pass filter, and a simplification of a lateral inhibitory network. In addition, they also modeled a second stage of cortical processing using spectro-temporal receptive fields (STRFs) applied on the spectrogram derived in the first stage \cite{ChiGaoCuyMatRuSha99-JASA}. Another example of such a simplified two-stage model including an auditory spectrogram with STRFs applied to speech recognition was presented by \cite{HecDomJouGoe11-SpeechComm}.

Traditionally, the short-time Fourier transform (STFT) has been used extensively for converting the signal to a time-frequency representation, presumably due to its efficient computer implementation, the fast Fourier transform (FFT). One of the major drawbacks with the STFT is the frequency resolution, which is constant in terms of Hertz. Since the ear is approximately logarithmic with regard to the frequency band distribution, a major part of the frequency data in the upper treble region is less relevant for a perceptual analysis. Similarily the resolution in the bass range is not enough for example for determining the musical pitch using a time window that can capture the onsets of fast notes \cite{MulEllKlaRic-SelTopSignProc}.

One possibility to achieve a log-frequency spectrum is to use a bank
of individual bandpass filters as discussed above and also in line
with the currently proposed model.  This can, however, be rather time consuming. An interesting computationally efficient compromise is therefore the constant-Q transform which uses traditional STFTs applied in a combination of downsampling and different time resolutions for different octaves \cite{BroPuc92-JASA}. A computational toolbox was recently presented by \cite{SchKla10-SoundMusicComp}. Using the constant-Q method, the frequency resolution will be the same across the spectrum. The time resolution will, however, vary significantly across the spectrum and still exhibit poor time resolution in the bass region.

Similar approaches using a second layer of receptive fields applied on
the spectrogram have been used in particular in speech research using
Gabor functions \cite{Kle02-ActAcust,EzzBouPog07-InterSpeech,MeyKol08-InterSpeech,HecDomJouGoe11-SpeechComm,WuZhaShi11-ASLP}.
For example, \cite{HecDomJouGoe11-SpeechComm}
used Gabor-based receptive fields of different orientations in the
time-frequency plane of the spectrogram in combination with different
transformations inspired by visual object recognition in order to
capture the formant trajectories over time. The resulting features
were shown to improve the performance of a speech recognition system
in combination with traditional features such as mel frequency
cepstral components (MFCC).
In this article, we show how such and related auditory operations can be derived in
principled manner.

\section{Summary and discussion}
\label{sec-summ-disc}

We have presented a theory for how idealized models of auditory receptive fields 
can be formulated based on structural constraints on the first stages
of auditory processing.
The theory includes (i)~the definition of multi-scale spectrograms at
different temporal scales in such a way that a spectrogram at any
coarser temporal scale can be related to a corresponding 
spectrogram any finer temporal scale using theoretically 
well-defined scale-space operations, and
additionally (ii)~how a second-layer of spectro-temporal receptive fields
can be defined over a logarithmically transformed spectrogram in
such a way that the resulting spectro-temporal receptive fields obey
invariance or covariance properties under natural
sound transformations including temporal shifts,
variations in the sound pressure, the distance between the sound source
and the observer, or a shift in the frequencies of auditory stimuli.
Specifically, theoretical arguments have been presented showing how
these idealized receptive fields are constrained to the presented forms from
symmetry properties of the environment in combination with assumptions
about the internal structure of auditory operations as motivated from
requirements of handling different temporal and spectral scales in a
theoretically well-founded manner.

By combining the scale-space approach with a local frequency analysis,
we obtain a new way of deriving the Gabor filters as a complex-valued
scale-space transform resulting from the Gaussian scale-space concept being
applied to a temporal signal multiplied by a complex sine wave.
We can also derive the Gamma-tone filters in a corresponding manner,
as a time-causal complex scale-space transform obtained by
applying a set of time-causal scale-space kernels based on
first-order integrators with equal time constants coupled in cascade
and applied to a temporal signal multiplied by a complex sine wave.
In addition, the scale-space approach to multi-scale spectrograms
leads to a new family of generalized Gamma-tone filters obtained by
instead using a logarithmic distribution of the intermediate temporal
scales, and which allow for different trade-offs between filter
characteristics such as frequency selectivity and temporal delay.

Then, given that a multi-scale spectrogram has been defined and
been transformed by taking the logarithm of the magnitude values and
expressing the frequencies on a logarithmic frequency scale,
to ensure natural covariance properties under variations of the sound pressure
or a frequency shift in the stimulus, the theory provides a second
layer of receptive fields applied to the spectrogram, based on
spectro-temporal derivatives of spectro-temporal scale-space kernels.
We have shown how that the derived models of idealized spectro-temporal
receptive fields are uniquely determined given natural symmetry
properties (scale-space axioms) and we have shown examples of how
basic auditory features can be computed in this way.

Thus, the presented scale-space theory for auditory signals
can be both related to existing models for auditory analysis 
and additionally leads to the formulation of a set of new models.
Specifically, the presented theory provides a coherent framework by which
auditory receptive fields at the first levels of processing
in the auditory hierarchy can be expressed within the same theoretical framework.
Moreover, the theory allows for provable invariance properties under temporal
shifts, variations in sound pressure and logarithmic frequency shifts.

Concerning limitations of the approach, we have in the present
treatment defined the second layer of receptive fields from the
magnitude values of the spectrogram only, thus ignoring the 
local phase information. 
A natural extension would be to extend the formulation of the second layer
of receptive fields to include the local phase of the spectrogram,
which for example may provide important cues to judge if partial tones
may constitute components of a harmonic spectrum belonging to the same
physical source, and to formulate binaural receptive fields that are
sensitive to the volumes in auditory space where a stimulus occurs.

It should also be stressed that the present approach constitutes a
linear and pure feed-forward model for local receptive fields, corresponding to
constant values of the filter parameters in the local diffusion equations
and recurrence relations that determine the formation of the receptive
fields.
An interesting extension would be to adapt these filter parameters to the
local input data or using top-down information, which could then provide
computational mechanisms to express stimulus- and/or task-dependent receptive fields as
reported by \cite{FriShaElhKle03-NatureNeuSci,MacWehZad04-JNeuroSci,ElhFriChiSha07-JNeuroSci,Egg11-HearRes,DavFriSha12-PNAS,LauEdeHue12-PONE}
and furthermore to extend the use of local receptive fields that are centered around a
single frequency to multi-local operations
that combine information from several distinct frequencies \cite{PieHar05-JNeuroPhys}.

In relation to such more complex non-linear mechanisms, the presented linear theory
can be seen as a first principled starting point that 
(i)~enables the
computation of basic auditory features for audio processing
and 
(ii)~generates predictions about
basic receptive field profiles that are qualitatively similar to biological
receptive fields as measured by cell recordings in the inferior
colliculus (ICC) and the primary auditory cortex (A1).

\appendix

\section{Frequency selectivity of the spectrograms}
\label{app-freq-anal}

Consider a sine wave signal with angular frequency $\omega_0$:
\begin{equation}
  f(t) = \sin \omega_0 t
\end{equation}
When computing the windowed spectrogram, we multiply this signal by sine and
cosine waves of different angular frequencies $\omega$ and integrate
by a window function $h(t;\; \tau)$ with temporal extent $\tau$:
\begin{align}
  \begin{split}
     c(t) = h(t;\; \tau) * \left( f(t) \cos \omega t) \right)
  \end{split}\\
  \begin{split}
     s(t) = h(t;\; \tau) * \left( f(t) \sin \omega t) \right)
  \end{split}
\end{align}
By the use of basic rules for trigonometric functions, we have
\begin{align}
  \begin{split}
     f(t) \cos \omega t 
     = \sin \omega_0 t \, \cos \omega t 
     = \frac{1}{2} 
         \left( 
            -\sin (\omega - \omega_0) t + \sin (\omega + \omega_0) t
         \right)
  \end{split}\\
  \begin{split}
     f(t) \sin \omega t 
     = \sin \omega_0 t \, \sin \omega t 
     = \frac{1}{2} 
         \left( 
            \cos (\omega - \omega_0) t - \cos (\omega + \omega_0) t
         \right)
  \end{split}
\end{align}
and the result of convolving these components with the window
function $h(t;\; \tau)$ can be expressed by multiplication with the Fourier transform $\hat{h}(t;\; \tau)$:
\begin{align}
  \begin{split}
     c(t) = \frac{1}{2} 
         \left( 
            -\hat{h}(\omega - \omega_0;\; \tau) \, \sin (\omega - \omega_0) t 
             + \hat{h}(\omega + \omega_0;\; \tau)  \, \sin (\omega + \omega_0) t
         \right)
  \end{split}\\
  \begin{split}
     s(t) = \frac{1}{2} 
         \left( 
            \hat{h}(\omega - \omega_0;\; \tau) \, \cos (\omega - \omega_0) t 
            - \hat{h}(\omega + \omega_0;\; \tau)  \, \cos (\omega + \omega_0) t
         \right)
  \end{split}
\end{align}
Concerning the magnitude of the spectrogram
\begin{equation}
   S(t) = \sqrt{c(t)^2 + s(t)^2}
\end{equation}
it follows that
\begin{equation}
  S(t)^2 
  = \frac{1}{4} 
      \left(
         \hat{h}(\omega - \omega_0;\; \tau)^2  
         + \hat{h}(\omega + \omega_0;\; \tau)^2
         - 2 \cos(2 \omega_0 t) \, \hat{h}(\omega - \omega_0;\; \tau) \, \hat{h}(\omega + \omega_0;\; \tau)
      \right)
\end{equation}
If we assume that the window function $h$ should be a low-pass filter,
then for $\omega$ close to $\omega_0$ it seems reasonable to assume
that 
\begin{equation}
  |\hat{h}(\omega - \omega_0;\; \tau)| \gg |\hat{h}(\omega + \omega_0;\; \tau)| 
\end{equation}
Thereby, the dominant component of the spectrogram near $\omega_0$
will be given by
\begin{equation}
   S_{magn}(\omega;\; \tau) \approx \frac{|\hat{h}(\omega - \omega_0;\; \tau)|}{2}
\end{equation}
By normalizing this entity such that the maximum value at 
$\omega = \omega_0$ is equal to one, we can quantify the frequency selectivity for
a frequency dependent window scale $\tau(\omega)$ as
\begin{equation}
  R(\omega) = |\hat{h}(\omega - \omega_0;\; \tau(\omega))|
\end{equation}
which on a logarithmic dB scale assumes the form
\begin{equation}
  R_{dB}(\omega) = 20 \log_{10} |\hat{h}(\omega - \omega_0;\; \tau(\omega))|
\end{equation}
where we would ideally choose the temporal extent of the kernel in units of
$\sigma = \sqrt{\tau}$ proportional to the wavelength 
$\lambda = 2 \pi/\omega$ for any angular frequency $\omega$:
\begin{equation}
  \label{eq-temp-sc-prop-wavelength}
   \tau(\omega) = \sigma^2 = (n * \lambda)^2 = \left( \frac{2 \pi
       n}{\omega} \right)^2
\end{equation}

\paragraph{Gaussian window functions.}

For a Gaussian window function we have
\begin{equation}
  \hat{g}(\omega;\; \tau) 
  = \int_{t = -\infty}^{\infty} g(t;\; \tau) \, e^{-i \omega t} dt
  = e^{-\omega^2 \tau/2}
\end{equation}
With the temporal extent of the window function proportional to the
wavelength for any frequency according to
(\ref{eq-temp-sc-prop-wavelength}),
the frequency selectivity is given by
\begin{equation}
  \label{eq-freq-sel-gauss}
  R_{gauss}(\omega) = e^{- \frac{2\pi^2 n^2  (\omega-\omega_0)^2}{\omega^2}} 
\end{equation}
or in dB
\begin{equation}
  R_{dB,gauss}(\omega) 
  = - \frac{40 \pi^2 n^2 (\omega-\omega_0)^2}{\log 10 \, \omega^2} 
\end{equation}

\paragraph{Window functions defined from cascade of truncated
  exponential functions.}

For the truncated exponential filters coupled in cascade,
the Laplace transform is
\begin{equation}
    H_{composed}(q;\; \mu) 
    = \int_{t = - \infty}^{\infty} (*_{i=1}^{k} h_{exp}(t;\; \mu_i)) \, e^{-qt} \, dt
    =  \prod_{i=1}^{k} \frac{1}{1 + \mu_i q}
\end{equation}
implying that the Fourier transform is given by
\begin{equation}
 \hat{h}_{composed}(\omega;\; \mu) 
  = H_{composed}(i \omega;\; \mu) 
  = \prod_{k=1}^{K} \frac{1}{1 + i \, \mu_k \, \omega}
\end{equation}
In the special case when all the time constants $\mu_k$ are equal, we have
\begin{equation}
  \mu_k = \sqrt{\frac{\tau}{K}}
\end{equation}
With the temporal extent of the window function proportional to the
wavelength according to (\ref{eq-temp-sc-prop-wavelength}),
the frequency selectivity is given by
\begin{equation}
  \label{eq-freq-sel-rec-uni}
  R_{rec-uni}(\omega) =
   |\hat{h}_{composed}(\omega-\omega_0;\; \tau(\omega), K)|
   = \frac{1}
              {\left(
                  1 
                  + \frac{4 \pi^2 n^2 (\omega-\omega_0)^2}
                             {K \omega^2} 
                  \right)^{K/2}}
\end{equation}
or in dB
\begin{equation}
  R_{dB,rec-uni}(\omega) 
  = - \frac{K}{2 \log 10}
          \log 
             \left(
                  1 
                  + \frac{4 \pi^2 n^2 (\omega-\omega_0)^2}
                             {K \omega^2} 
            \right)
\end{equation}
In the special case when all the intermediate temporal scale levels 
$\tau_k$ are instead distributed according to a logarithmic
distribution with $\tau_k = c^{2(k-K)} \tau$ and $\mu_k$ according to
(\ref{eq-mu1-log-distr}) and (\ref{eq-muk-log-distr}),
and with the temporal extent of the window function proportional to
the wavelength according to (\ref{eq-temp-sc-prop-wavelength}),
we obtain
\begin{equation}
  \label{eq-freq-sel-rec-log}
  R_{rec-log}(\omega) 
      = \frac{1}
              {\sqrt{
                  1 
                  + \frac{4 \pi^2 c^{2(1-K)} n^2 (\omega-\omega_0)^2}{\omega^2}}
               \, 
               \prod_{k=2}^K
                  \sqrt{
                  1 
                  + \frac{4 \pi^2 c^{2(k-K-1)} (c^2-1) n^2 (\omega-\omega_0)^2}{\omega^2}}}
\end{equation}
or in dB
\begin{align}
  \begin{split}
  R_{dB,rec-log}(\omega) 
  = & - \frac{10}{\log 10}
          \log 
            \left(
                    1 
                    + \frac{4 \pi^2 \, c^{2(1-K)} \, n^2 (\omega-\omega_0)^2}{\omega^2}
            \right)
   \end{split}\\
  \begin{split}
       - \frac{10}{\log 10}
              \sum_{k=2}^K
                 \log 
                    \left(
                            1 
                            + \frac{4 \pi^2 c^{2(k-K-1)} \, (c^2-1) \, n^2 (\omega-\omega_0)^2}{\omega^2}
                    \right)
  \end{split}
\end{align}

\begin{figure}[!hbt]
  \begin{center}
    \begin{tabular}{cccc}
      {\footnotesize $R_{dB,gauss}(\omega)$}
      & {\footnotesize $R_{dB,rec-uni}(\omega)$}
      & {\footnotesize $R_{dB,rec-log}(\omega)$} 
      & {\footnotesize $R_{dB,rec-log}(\omega)$} \\
    $\,$ 
      & {\footnotesize ($K=4$)} 
      & {\footnotesize ($K=4$, $c = \sqrt{2}$)} 
      & {\footnotesize ($K=4$, $c = 2$)} \\ 
      \includegraphics[width=0.23\textwidth]{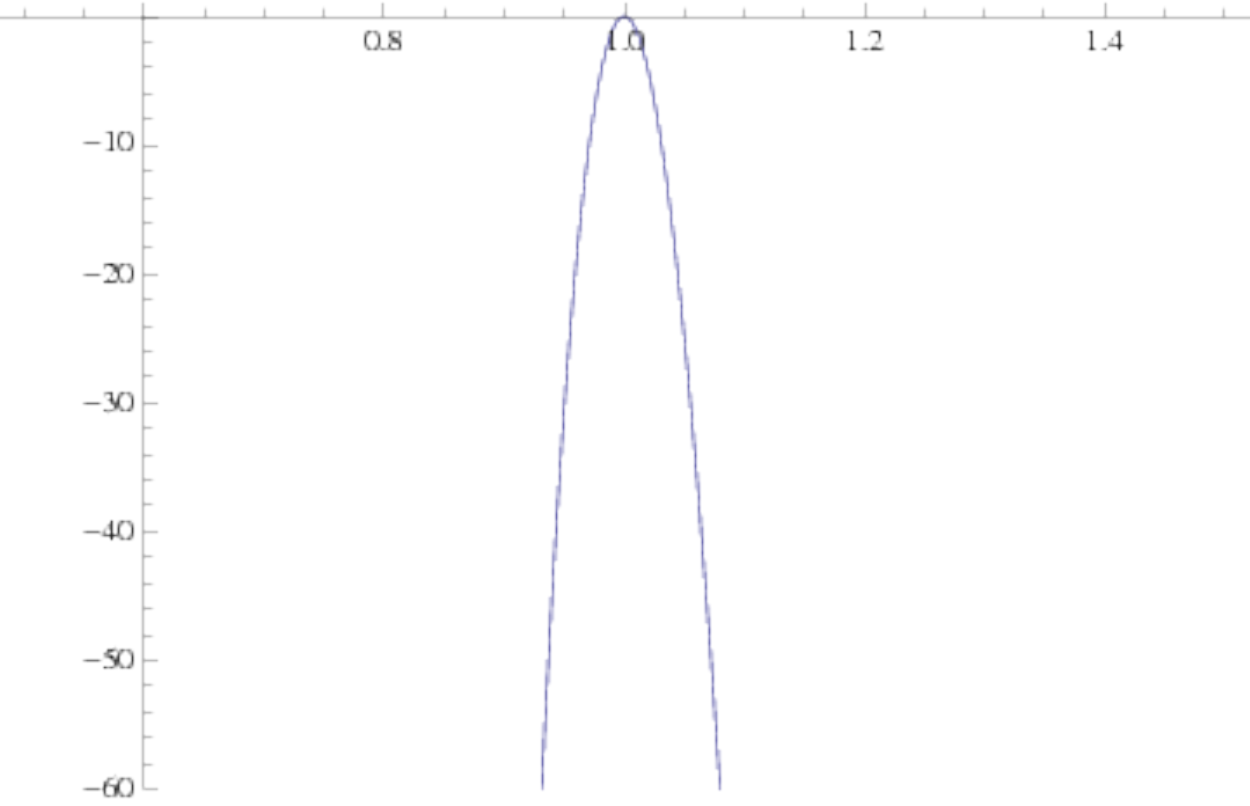} \hspace{-2mm} &
      \includegraphics[width=0.23\textwidth]{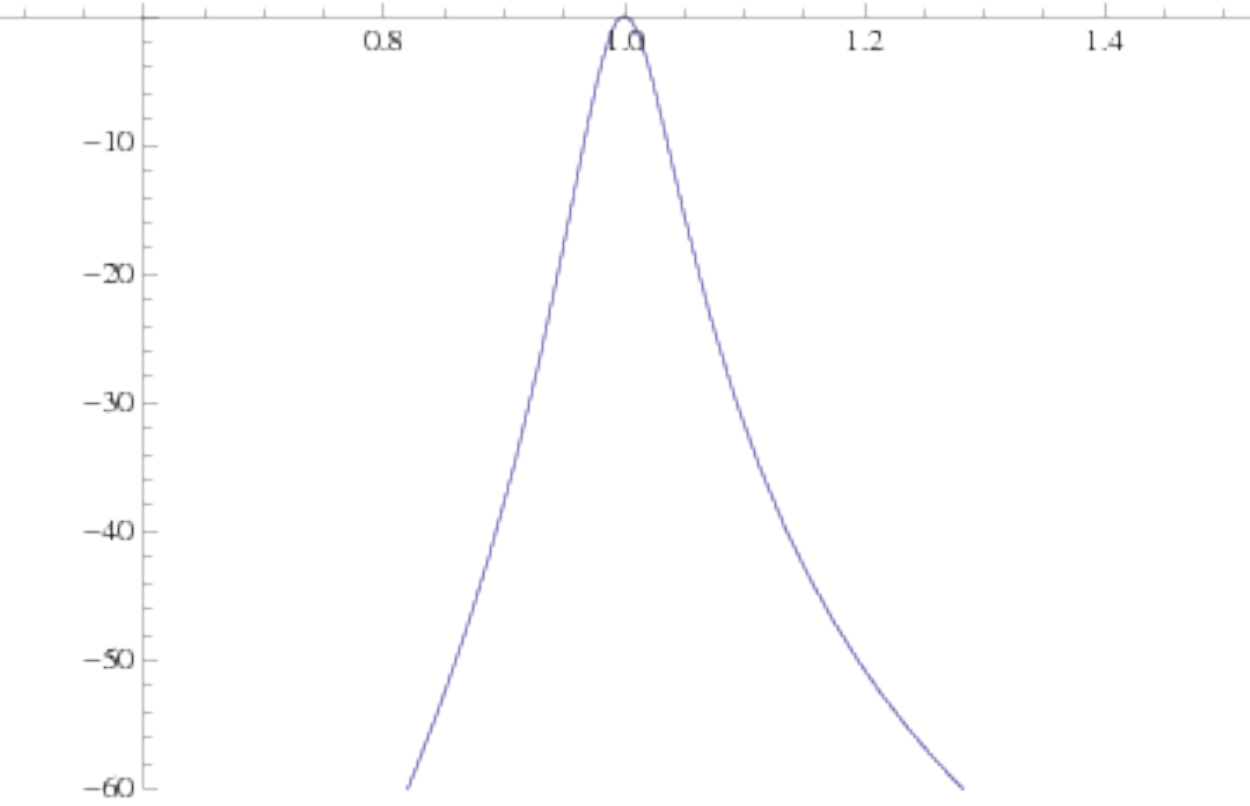} \hspace{-2mm} &
      \includegraphics[width=0.23\textwidth]{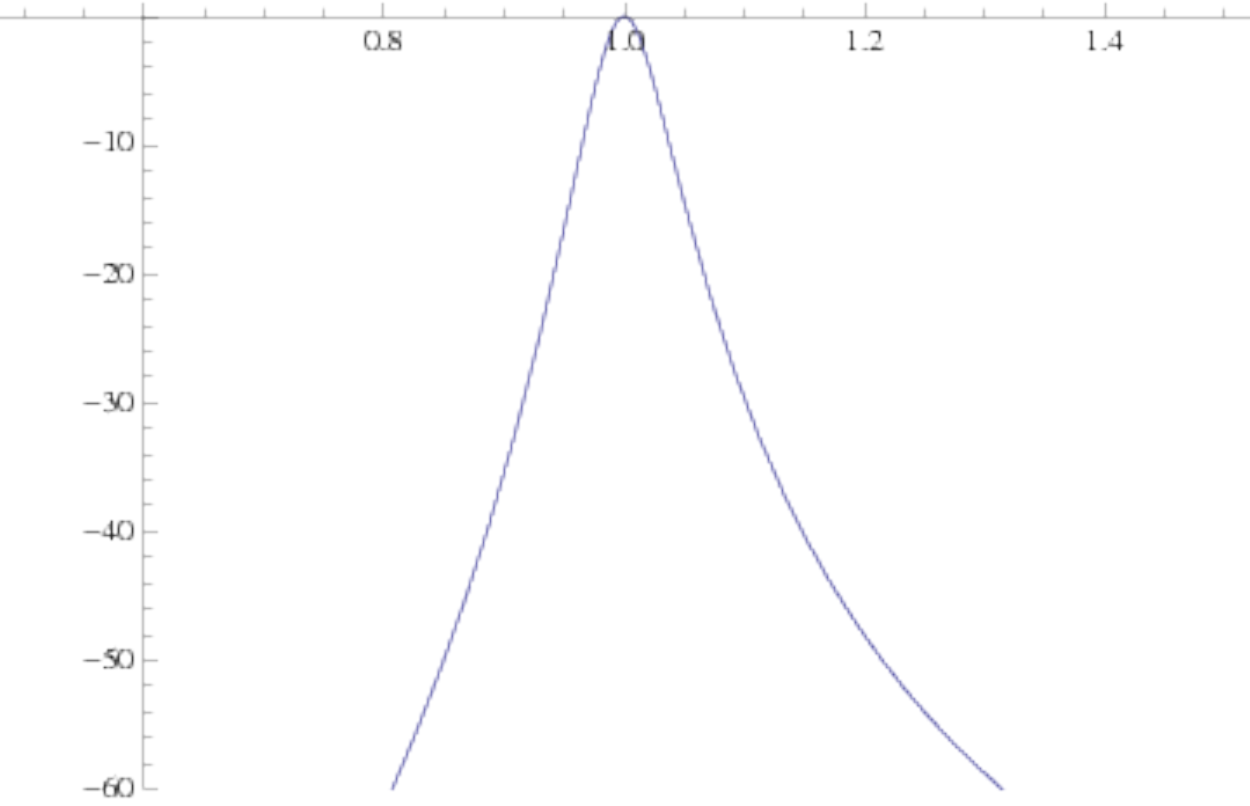} &
      \includegraphics[width=0.23\textwidth]{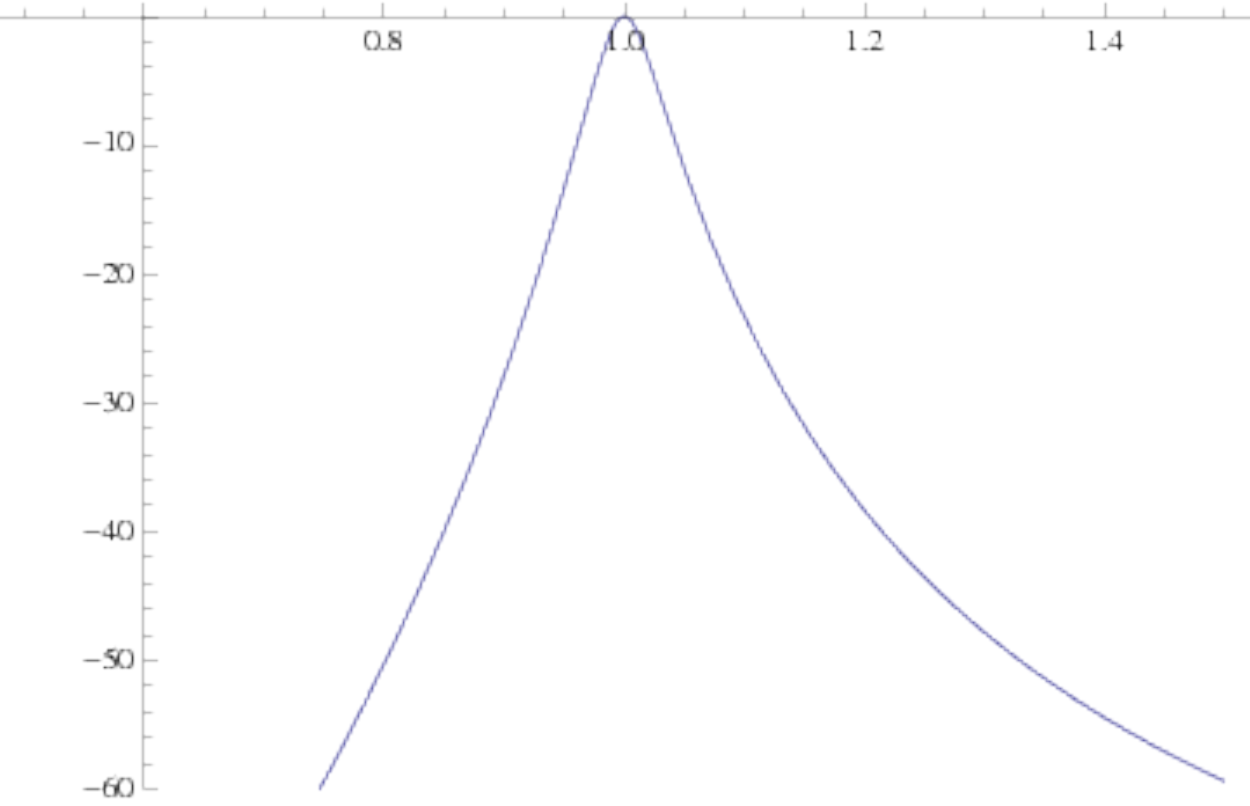} \\
      $\,$ \\
      & {\footnotesize $R_{dB,rec-uni}(\omega)$} 
      & {\footnotesize $R_{dB,rec-log}(\omega)$} 
      & {\footnotesize $R_{dB,rec-log}(\omega)$} \\
    $\,$ 
      & {\footnotesize ($K=7$)} 
      & {\footnotesize ($K=7$, $c = \sqrt{2}$)} 
      & {\footnotesize ($K=7$, $c = 2$)}\\ 
      & \includegraphics[width=0.23\textwidth]{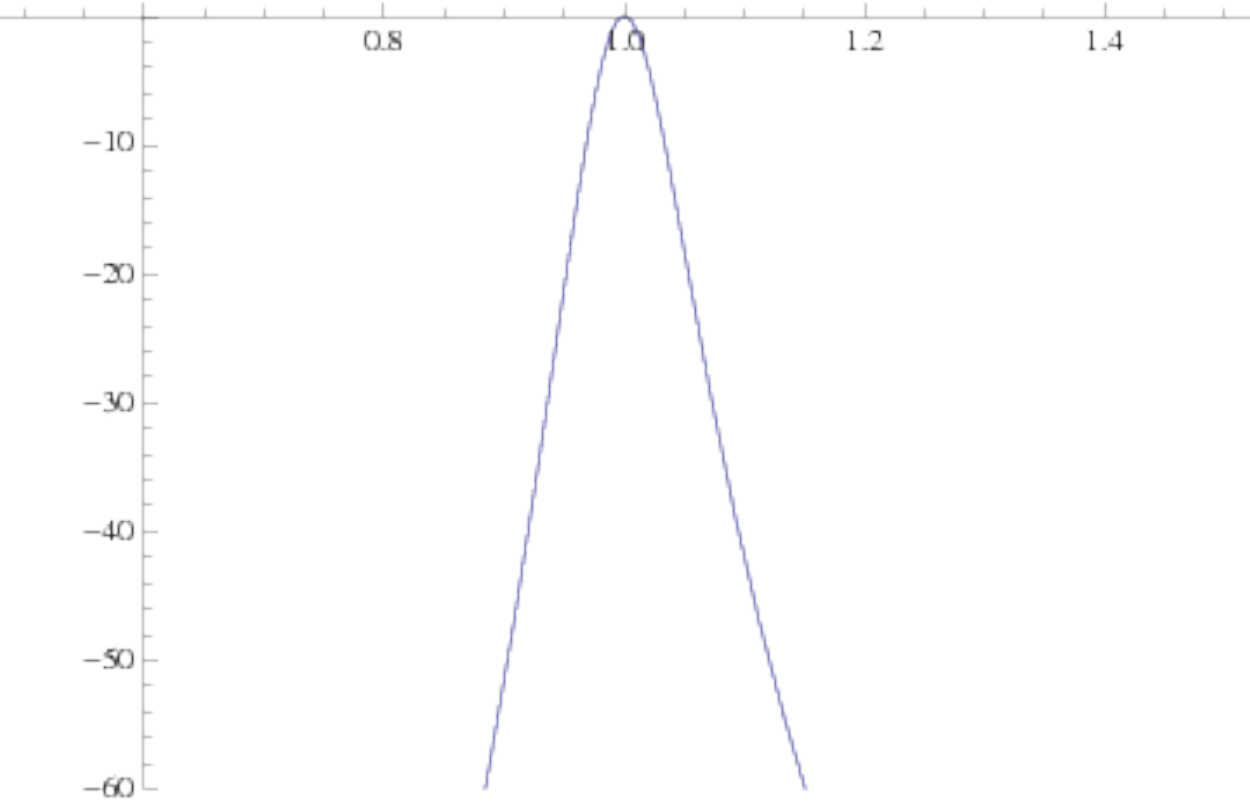} \hspace{-2mm} &
      \includegraphics[width=0.23\textwidth]{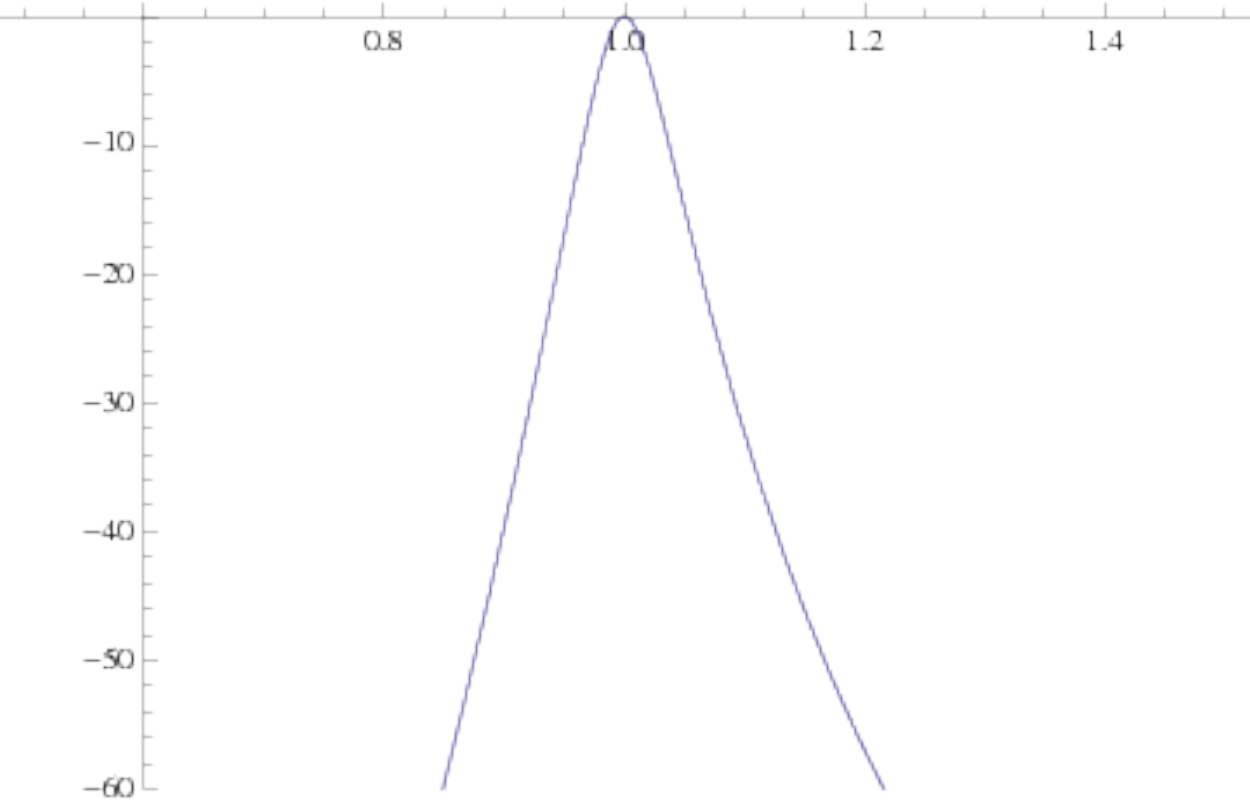} &
      \includegraphics[width=0.23\textwidth]{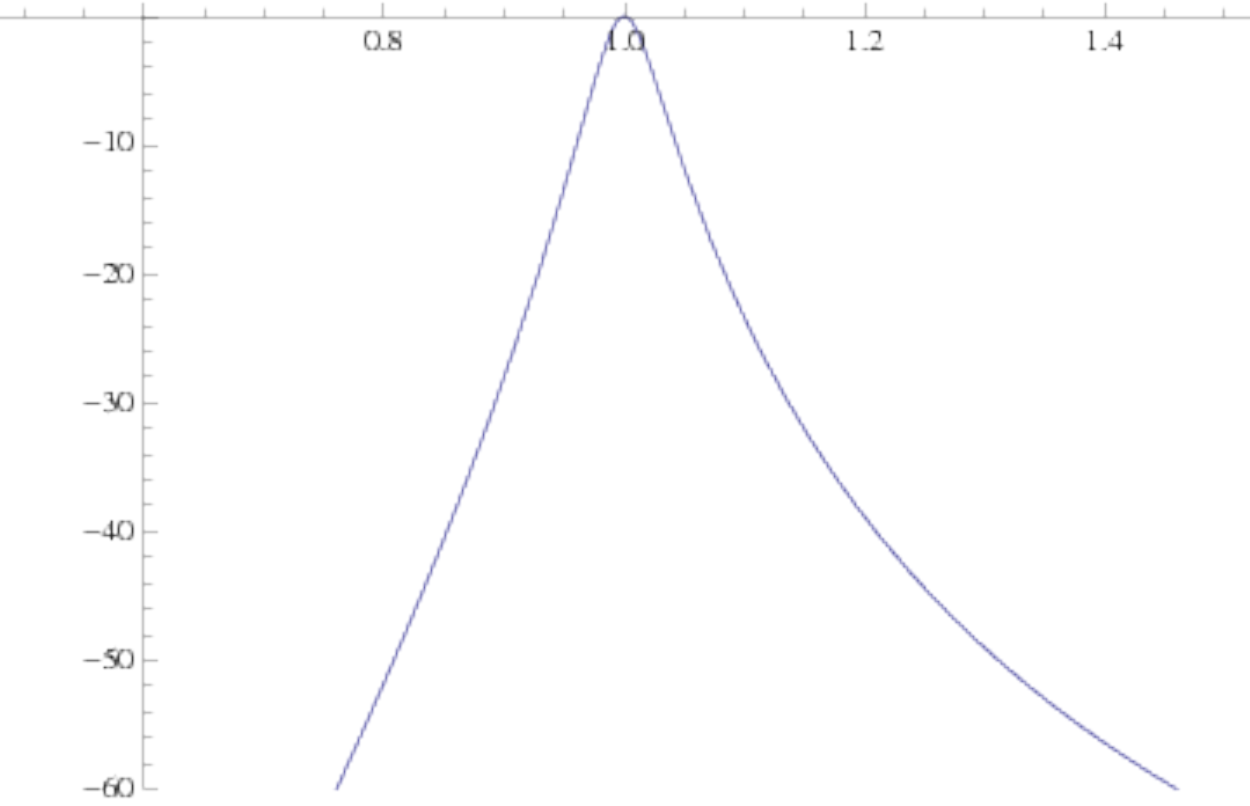} \\
     \end{tabular} 
  \end{center}
  \vspace{-3mm}
  \caption{Graphs of the frequency selectivities of Gaussian and
    time-causal window functions with the temporal extent of the
    window function proportional to the corresponding wavelength of
    the spectrogram with $n = 8$. For the time-causal filters, 
    $K = 4$ or $K = 7$ filters have been coupled in cascade.
   For the logarithmic distribution, the ratio between successive
  temporal scale levels $\tau_{k} = c^{2(k-K)} \tau$ has been
  determined from $c = \sqrt{2}$ or $c = 2$. (Horizontal axis: Angular frequency
  $\omega$ in units of $\omega_0$. Vertical axis: dB values down to -60 dB.)}
  \label{fig-freq-sel-graphs}
 \end{figure}

\noindent
Figure~\ref{fig-freq-sel-graphs} shows graphs of the frequency
selectivity of the different types of temporal window functions for a
few combinations of the underlying filter parameters.
As can be seen from these graphs, the non-causal
Gaussian kernel has sharper frequency selectivity compared to the
time-causal kernels.
Within the class of time-causal kernels, the frequency selectivity
increases with the number of truncated exponential kernels that are
coupled in cascade.
For the logarithmic distribution of the intermediate temporal scale
levels, the frequency selectivity also increases with decreasing
values of the distribution parameter $c$.

\paragraph{Dependency of the relative bandwidth on $n$.}

Notably all these expressions are functions of the ratio
\begin{equation}
  \frac{n^2 (\omega-\omega_0)^2}{\omega^2} = C^2
\end{equation}
By solving for $\omega$ and assuming $C > 0$
\begin{equation}
  \omega_1= \frac{\omega_0}{1 + \frac{C}{n}}
  \quad
  \omega_2= \frac{\omega_0}{1 - \frac{C}{n}}
\end{equation}
we get explicit expressions for how the relative 
bandwidth of the spectrogram 
\begin{equation}
  \label{eq-rel-bandwidth-general}
  \frac{\omega_2 - \omega_1}{\omega_0}
  = 
     \left( \frac{1}{1 - \frac{C}{n}} - \frac{1}{1 + \frac{C}{n}}\right)
 = \frac{\frac{2C}{n}}{1- \left( \frac{C}{n} \right)^2}
  \approx \frac{2C}{n} +{\cal O} \left( \left( \frac{C}{n} \right)^3 \right)
\end{equation}
alternatively in logarithmic MIDI units
\begin{equation}
  12 \log_2 \left( \frac{\omega_2}{\omega_1} \right)
  =  12 \log_2 \left( \frac{1 + \frac{C}{n}}{1 - \frac{C}{n}} \right)
  \approx \frac{24}{\log 2} \frac{C}{n} +{\cal O} \left( \left( \frac{C}{n} \right)^3 \right)
\end{equation}
depends on $n$ for any $C$,
implying that the relative bandwidth
increases approximately linearily with the proportionality constant
$n$, where $C$ is related to the dB level $R_{dB} < 0$ according to
\begin{equation}
  C_{gauss} = \frac{\sqrt{\log 10}}{2\pi} \sqrt{\frac{-R_{dB}}{10}}
\end{equation}
for the Gaussian window functions and according to
\begin{equation}
  C_{rec-uni} 
  = \frac{\sqrt{K}}{2 \pi} \sqrt{10^{-\frac{R_{dB}}{10 K}}-1}
\end{equation}
for the time-causal kernels having a uniform distribution of the
intermediate temporal scale levels.
For the time-causal kernels having a logarithmic distribution of the
intermediate scale levels, the parameter $C$ can be determined
by solving the following equation
\begin{equation}
R_{dB}= 
-\frac{10 \log \left(4 \pi ^2 C^2 c^{2-2 K}+1\right)}{\log (10)} 
- \frac{20 \sum _{k=2}^K \frac{1}{2} \log \left(4 \pi^2 \left(c^2-1\right) C^2 c^{2 k-2 K-2}+1\right)}{\log (10)}
\end{equation}
numerically given specific values of $K$ and $c$.
Table~\ref{tab-C-uni-log-compare} shows such values for $K = 4$
and $K = 7$ for $c = \sqrt{2}$ and $c = 2$ as well as corresponding
values for a uniform distribution of the intermediate scale levels 
and for a Gaussian window function.

\begin{table}[!hbt]
  \begin{center}
   \footnotesize
  \begin{tabular}{lcccc}
  \hline
   \multicolumn{5}{c}{Relative bandwidth of temporal window functions} \\
  \hline
                      & -3 dB  & -10 dB  & -20 dB & -30 dB \\ % & -40 dB \\
  \hline
    $C_{gauss}$  & 0.132 & 0.242 & 0.342  & 0.418 \\ % & 0.483 \\
  $C_{rec-uni}$ ($K = 4$)
                      & 0.138 & 0.281 & 0.468 & 0.684 \\ % & 0.955 \\
  $C_{rec-log}$ ($K = 4, c = \sqrt{2}$)
                      & 0.140 & 0.292 & 0.498 & 0.736 \\ % & 1.033 \\
  $C_{rec-log}$ ($K = 4, c = 2^{3/4}$)
                      & 0.143 & 0.312 & 0.553 & 0.838 \\ % & 1.033 \\
 $C_{rec-log}$ ($K = 4, c = 2$)
                       & 0.146 & 0.332 & 0.619 & 0.971 \\ % & 1.414 \\
  $C_{rec-uni}$ ($K = 7$)
                      & 0.136 & 0.263 & 0.406 & 0.546 \\ % & 0.695 \\
  $C_{rec-log}$ ($K = 7, c = \sqrt{2}$)
                       & 0.140 & 0.289 & 0.478 & 0.678 \\ % & 0.897 \\
  $C_{rec-log}$ ($K = 7, c = 2^{3/4}$)
                      & 0.143 & 0.311 & 0.547 & 0.816 \\ % & 1.033 \\
 $C_{rec-log}$ ($K = 7, c = 2$)
                        & 0.146 & 0.332 & 0.617 & 0.963 \\ % & 1.387 \\
  \hline
  \end{tabular}
\end{center}
\caption{Numerical values of the parameter $C$ that determines the
  influence of the type of window function on the relative bandwidth
  $\frac{2C}{n}$ 
  of the spectrogram according to
  (\protect\ref{eq-rel-bandwidth-general}), 
  for a Gaussian function or $K$ truncated
    exponential kernels in cascade in the cases of a uniform
    distribution of the intermediate temporal scale levels $\tau_k = \tau/K$ or a
    logarithmic distribution $\tau_k = c^{2(k-K)} \tau$ with $c > 1$.}
  \label{tab-C-uni-log-compare}
\end{table}

\paragraph{Frequency invariance.}

From the invariance of the expressions (\ref{eq-freq-sel-gauss}),
(\ref{eq-freq-sel-rec-uni}) and (\ref{eq-freq-sel-rec-log})
under frequency transformations of the form
\begin{align}
   \begin{split}
      \omega \mapsto \alpha \, \omega 
   \end{split}\\
  \begin{split}
      \omega_0 \mapsto \alpha \, \omega_0
   \end{split}
\end{align}
for any $\alpha > 0$, it follows that the spectral sensitivity of the spectrogram will be
independent of the angular frequency $\omega$.
Thereby, over the range of frequencies for which the temporal extent
of the window function is proportional to the wavelength, it follows
that the spectral sensitivity is invariant under a shift in frequency of the form
$\omega \mapsto \alpha \, \omega $, thus
providing a foundation frequency covariant receptive fields 
at higher levels in the auditory hierachy.

\section{Temporal dynamics of the time-causal scale-space kernels}
\label{app-temp-dyn}

For the time-causal filters obtained by coupling truncated exponential
kernels in cascade, there will be an inevitable temporal delay 
depending on the time constants $\mu_k$ of the individual filters.

A most straightforward way of estimating this delay is by using the
additive property of mean values under convolution
\begin{equation}
  m = \sum_{k=1}^K \mu_k
\end{equation}
In the special case of all the time constants being equal $\mu_k =
\sqrt{\tau/K}$, this measure is given by
\begin{equation}
  \label{eq-delta-recfilt-uni}
  m_{uni} = \sqrt{K \tau} = \frac{2 \pi  \sqrt{K} n}{\omega }
\end{equation}
showing that the temporal increases if the temporal smoothing
operation is divided into a lower number individual smoothing steps.

In the special case when the intermediate temporal scale levels are
instead distributed logarithmically according to (\ref{eq-distr-tau-values}), 
with the corresponding time constants given by
(\ref{eq-mu1-log-distr}) and (\ref{eq-muk-log-distr}),
this measure for the temporal delay does instead assume the form
\begin{align}
  \begin{split}
  m_{log} 
   & = \frac{c^{-K} \left(c^2-\left(\sqrt{c^2-1}+1\right) c+\sqrt{c^2-1} c^K\right)}{c-1} \, \sqrt{\tau }
  \end{split}\nonumber\\
  \begin{split}
  \label{eq-delta-recfilt-log}
   & = \frac{2 \pi  n \, c^{-K} \left(c^2-\left(\sqrt{c^2-1}+1\right) c+\sqrt{c^2-1} c^K\right)}{(c-1) \, \omega }
  \end{split}
\end{align}
with the limit value
\begin{equation}
  \label{eq-delta-recfilt-log-limit}
  m_{log-limit} = \lim_{K \rightarrow \infty} m_{log} 
   = \frac{\sqrt{c^2-1}}{c-1} \sqrt{\tau}
   = \frac{\sqrt{c^2-1}}{c-1} \frac{2 \pi n}{\omega}
\end{equation}
when the number of filters tends to infinity.

By comparing equations (\ref{eq-delta-recfilt-uni}),
(\ref{eq-delta-recfilt-log}) and (\ref{eq-delta-recfilt-log-limit}),
we can specifically note that with increasing number of intermediate
temporal scale levels, a logarithmic distribution of the intermediate
scale levels implies shorter temporal delays than a uniform
distribution of the intermediate scale levels.

Table~\ref{tab-tmean-uni-log-compare} shows numerical values
of these measures for different values of $K$ and three values of $c$.
Notably, the logarithmic
distribution of the intermediate scale levels allows for significantly
faster temporal dynamics than a uniform distribution.

\begin{table}[!hbt]
  \begin{center}
   \footnotesize
  \begin{tabular}{ccccc}
  \hline
   \multicolumn{5}{c}{Temporal mean values of time-causal kernels} \\
  \hline
    $K$ & $m_{uni}$  & $m_{log}$  ($c = \sqrt{2}$) & $m_{log}$  ($c = 2^{3/4}$) & $m_{log}$ ($c = 2$) \\
  \hline
    2 & 1.414 & 1.414 & 1.399 & 1.366 \\
    3 & 1.732 & 1.707 & 1.636 & 1.549 \\
    4 & 2.000 & 1.914 & 1.777 & 1.641 \\
    5 & 2.236 & 2.061 & 1.860 & 1.686 \\
    6 & 2.449 & 2.164 & 1.910 & 1.709 \\
    7 & 2.646 & 2.237 & 1.940 & 1.721 \\
    8 & 2.828 & 2.289 & 1.957 & 1.732 \\
  \hline
  \end{tabular}
\end{center}
\caption{Numerical values of the temporal delay 
    in terms of the temporal mean $m = \sum_{k=1}^K \mu_k$
    in units of $\sqrt{\tau}$ for time-causal kernels obtained by coupling $K$ truncated
    exponential kernels in cascade in the cases of a uniform
    distribution of the intermediate temporal scale levels $\tau_k = k \tau/K$ or a
    logarithmic distribution $\tau_k = c^{2(k-K)} \tau$ with $c > 1$.}
  \label{tab-tmean-uni-log-compare}
\end{table}

\paragraph{Additional temporal characteristics.}

Because of the asymmetric tails of the time-causal temporal smoothing
kernels, temporal delay estimation by the mean value may however lead
to substantial overestimates compared to {\em e.g.\/} the position of the local maximum.
To provide more precise characteristics in the case of a uniform
distribution of the intermediate temporal scale levels, for which
a compact closed form expression is available for the composed kernel
\begin{equation}
  h_{composed}(t;\; \mu, K) = \frac{t^{K-1} \, e^{-t/\mu}}{\mu^K \, \Gamma(K)}
\end{equation}
let us differentiate this function
\begin{align}
  \begin{split}
    \partial_t \left( h_{composed}(t;\; \mu, K) \right)
    = \frac{e^{-\frac{t}{\mu }} ((K-1) \mu -t) \left(\frac{t}{\mu
        }\right)^{K+1}}{t^3 \, \Gamma(K)}
  \end{split}\\
  \begin{split}
    \partial_t \left( h_{composed}(t;\; \mu, K) \right)
    = \frac{e^{-\frac{t}{\mu }} \left(\frac{t}{\mu }\right)^K \left(\left(K^2-3
   K+2\right) \mu ^2-2 (K-1) \mu  t+t^2\right)}{\mu^2 \, t^3 \, \Gamma (K)}
  \end{split}
\end{align}
and solve for the positions of the local maximum and the 
inflection points
\begin{align}
  \begin{split}
     \label{eq-tmax-recfilt-uni}
     t_{max,uni} = (K-1) \, \mu 
                = \frac{(K-1) }{\sqrt{K}} \sqrt{\tau} 
                = \frac{2 \pi  (K-1) \, n}{\sqrt{K} \, \omega }
  \end{split}\\
  \begin{split}
     \label{eq-tinflect-recfilt-uni-1}
     t_{infl1,uni} = \left(K-\sqrt{K-1}-1\right) \mu 
                  = \frac{\left(K-\sqrt{K-1}-1\right) \sqrt{\tau }}{\sqrt{K}}
                  = \frac{2 \pi  \left(K-\sqrt{K-1}-1\right) n}{\sqrt{K} \, \omega }
  \end{split}\\
  \begin{split}
     \label{eq-tinflect-recfilt-uni-2}
     t_{infl2,uni} = \left(K+\sqrt{K-1}-1\right) \mu 
                        = \frac{\left(K+\sqrt{K-1}-1\right) \sqrt{\tau }}{\sqrt{K}}
                        = \frac{2 \pi  \left(K+\sqrt{K-1}-1\right) n}{\sqrt{K} \, \omega }
  \end{split}
\end{align}
Table~\ref{tab-tmax-uni-log-compare} shows numerical values for the
position of the local maximum for both types of time-causal kernels.
As can be seen from the table, the temporal response properties are
significantly faster for a logarithmic distribution of the
intermediate scale levels compared to a uniform distribution,
and the difference increases rapidly with $K$.
These temporal delay estimates are also significantly shorter than the
temporal mean values, in particular for the logarithmic distribution
of the intermediate scale levels.

\begin{table}[!hbt]
  \begin{center}
   \footnotesize
  \begin{tabular}{ccccc}
  \hline
   \multicolumn{5}{c}{Temporal delays from the maxima of time-causal kernels} \\
  \hline
    $K$ & $t_{max,uni}$  & $t_{max,log}$ ($c = \sqrt{2}$) & $t_{max,log}$ ($c = 2^{3/4}$) & $t_{max,log}$ ($c = 2$) \\
  \hline
    2 & 0.707 & 0.707 & 0.688 & 0.640 \\
    3 & 1.154 & 1.122 & 1.027 & 0.909 \\
    4 & 1.500 & 1.385 & 1.199 & 1.014 \\
    5 & 1.789 & 1.556 & 1.289 & 1.060 \\
    6 & 2.041 & 1.669 & 1.340 & 1.083 \\
    7 & 2.268 & 1.745 & 1.370 & 1.095 \\
    8 & 2.475 & 1.797 & 1.388 & 1.100 \\
  \hline
  \end{tabular}
\end{center}
\caption{Numerical values for the time delay of the local maximum in
  units of $\sqrt{\tau}$
    for time-causal kernels obtained by coupling $K$ truncated
    exponential kernels in cascade in the cases of a uniform
    distribution of the intermediate temporal scale levels $\tau_k = k \tau/K$ or a
    logarithmic distribution $\tau_k = c^{2(k-K)} \tau$ with $c > 1$.}
  \label{tab-tmax-uni-log-compare}
\end{table}

If we consider a temporal event that occurs as a step function over
time ({\em e.g.\/} as an onset in the magnitude of the spectrogram
which is then processed by a second layer of spectro-temporal
receptive fields)
and if the temporal position of this onset is estimated from a the local maximum over
time in the first-order temporal derivative response, 
then the temporal variation in the response over time will be given by shape of the temporal
smoothing kernel and the local maximum over time will occur at a time
delay equal to the time at which the temporal kernel has its maximum
over time. Thus, the position over time of the local maximum of the
temporal smoothing kernel is highly relevant for quantifying the
temporal responses characteristics of time-causal filtering operations.

\section{Computational implementation}
\label{sec-comp-impl}

The computational model for auditory receptive fields presented in
this paper is based on auditory signals that are assumed to be
continuous over time and with frequencies that are also assumed to
take values over a continuous frequency domain.
When implementing this model computationally on sampled sound signals,
the continuous theory must be transferred to a discrete temporal domain
where also a finite set of discrete frequencies are being used.

In this appendix we describe how the temporal and spectro-temporal
receptive fields can be implemented in terms of corresponding discrete
scale-space kernels that possess scale-space properties over discrete
temporal and spectro-temporal domains.

\subsection{Discrete temporal scale-space kernels based on
  recursive filters}
\label{app-disc-temp-smooth}

Given a temporal signal that has been sampled for some temporal
sampling density $\phi_0$, the temporal
scale $\tau$ in the continous model in units of seconds is first 
transferred to a temporal scale relative to a unit time sampling according to
\begin{equation}
  \label{eq-transf-tau-sampl}
  \tau_{sampl} = \phi_0^2 \, \tau
\end{equation}
where we have usually used sound signals
with $\phi_0 = 44.1~\mbox{kHz}$ in the experiments.
Then, a discrete set of intermediate temporal scale levels is defined
according to (\ref{eq-distr-tau-values})
\begin{equation}
  \tau_k = c^{2(k-K)} \tau_{sampl} \quad\quad (1 \leq k \leq K)
\end{equation}
or (\ref{eq-distr-tau-values-uni})
\begin{equation}
  \tau_k = \frac{k}{K} \, \tau_{sampl}
\end{equation}
with the difference between successive scale levels according to (with $\tau_0 = 0$)
\begin{equation}
  \Delta \tau_k = \tau_k - \tau_{k-1}
\end{equation}
For implementing the temporal smoothing operation between two such
adjacent scale levels, we make use of a first-order recursive filter
\begin{equation}
  \label{eq-norm-update}
  f_{out}(t) - f_{out}(t-1)
  = \frac{1}{1 + \mu_k} \,
    (f_{in}(t) - f_{out}(t-1)).
\end{equation}
with generating function
\begin{equation}
  \label{eq-gen-fcn-prim-norm-rec-filt}
  \htransf_{geom}(z) = \frac{1}{1 - \mu \, (z - 1)},
\end{equation}
which is a time-causal kernel and can be shown to satisfy discrete
scale-space properties in the sense of guaranteeing that the number of local extrema
or zero-crossings in the signal will not increase \cite{Lin90-PAMI,LF96-ECCV}.
Each such filter has temporal mean value $m_k = \mu_k$ and temporal variance
$\Delta \tau_k = \mu_k^2 + \mu_k$, and we compute $\mu_k$ from 
$\Delta \tau_k$ according to
\begin{equation}
  \mu_k = \frac{\sqrt{1 + 4 \Delta \tau_k}-1}{2}
\end{equation}
By the additive property of variances under convolution with a
positive kernel it follows that the discrete variances of
the discrete temporal scale-space kernels will perfectly match those
of the continuous model, whereas the mean values and the temporal
delays will be somewhat different. 
If the temporal scale $\tau_k$ is large relative to the
temporal sampling density, the discrete model can however also
be seen as a good approximation in this respect.

By the time-recursive formulation of this temporal scale-space
concept, it follows that the computations can be performed based on a
compact temporal buffer over time which contains the temporal
scale-space representations at temporal scales $\tau_k$, and there is
therefore no need for storing any additional temporal buffer of what
has occured in the past to perform the corresponding temporal operations.

\subsection{Discrete implementation of Gaussian smoothing}
\label{app-disc-gauss-smooth}

In our model, Gaussian smoothing is used both for smoothing over the
spectral domain and as a non-causal model for smoothing over the temporal domain.
To implement this operation on discrete sampled data, we do first
(i)~in the case of purely temporal smoothing transform a temporal 
variance $\tau$ in units of seconds to a temporal variance relative to 
a unit sampling density $s_{sampl}$ according to
\begin{equation}
  \label{eq-transf-tau-sampl-s}
  s_{sampl} = \phi_0^2 \, \tau
\end{equation}
or (ii)~in the case of purely spectral smoothing transform a spectral
smoothing scale $\sigma$ in units of semitones to a spectral smoothing scale
relative to the logspectral sampling distance $\Delta \nu$ and in
units of variance according to
\begin{equation}
  \label{eq-transf-s-sampl-midi}
  s_{sampl} = \left( \frac{\sigma}{\Delta \nu} \right)^2
\end{equation}
Then, we perform convolution with the discrete analogue of the
Gaussian kernel \cite{Lin90-PAMI}
\begin{equation}
  T(n;\; s_{sampl}) = e^{-s_{sampl}} I_n(s_{sampl})
\end{equation}
where $I_n$ denotes the modified Bessel functions of integer order
and which corresponds to the solution of the semi-discrete diffusion
equation
\begin{equation}
  \partial_s L(n;\; s) 
  \frac{1}{2} 
  \delta_{xx}
  = \frac{1}{2} 
   \left( 
     L(n-1;\; s) - 2 L(n;\; s) + L(n+1;\; s) 
   \right)
\end{equation}
where $x$ denotes the variable over the dimension of the domain,
which can either be time $t$ or logarithmic frequency $\nu$.

It can be shown that these kernels constitute the natural way to define
a scale-space concept for discrete signals corresponding to the
Gaussian scale-space over a symmetric domain in the sense of guaranteeing that the number of
local extrema or zero-crossings must not increase with scale, while
also ensuring a semi-group property
\begin{equation}
  T(\cdot;\; s_1) * T(\cdot;\; s_2) = T(\cdot;\; s_1 + s_2)
\end{equation}
over the discrete domain which implies that representations at coarser
scales can be computed from representations at finer scales using the
cascade property (\ref{eq-casc-prop-spat-temp}).

In practice, we do based on the (exact) relation 
$\sum_{n=-\infty}^{\infty} T(n;\, s) = 1$ truncate the infinite discrete kernel at the tails
such that
\begin{equation}
  \label{eq-trunc-disc-gauss-tails}
  \sum_{n=-N}^{N} T(n;\; s) > 1 - \varepsilon
\end{equation}
for some small value of $\varepsilon$ of the order $10^{-6}$ to
$10^{-4}$.
A coarse estimate of this bound can be obtained 
by estimating the corresponding tails of the continuous Gaussian
kernel
\begin{equation}
  2 \int_{x=N}^{\infty} g(x;\; s) \, dx < \varepsilon
\end{equation}
using the error function and then adjusting this estimate to match (\ref{eq-trunc-disc-gauss-tails}).

For points where some part of the kernel would stretch outside the
domain of available data, we mirror the data at the boundaries which
has the equivalent effect of solving the diffusion equation with
adiabatic boundary conditions corresponding to no heat transfer across
the boundaries of the domain where data are available.

\subsection{Discrete implementation of spectro-temporal receptive fields}

For separable spectro-temporal receptive fields, we implement the
spectro-temporal smoothing operation by separable combination of the temporal and spectral
scale-space concepts in sections~\ref{app-disc-temp-smooth} and 
\ref{app-disc-gauss-smooth}.
From this representation, separable spectro-temporal derivative
approximations are then computed from difference operators of the
following types:
\begin{align}
  \begin{split}
     \delta_t = (-1, +1)
  \end{split}\\
  \begin{split}
     \delta_{tt} = (1, -2, 1)
  \end{split}\\
 \begin{split}
     \delta_{v} = (-\frac{1}{2}, 0, -\frac{1}{2})^T
  \end{split}\\
  \begin{split}
     \delta_{vv} = (1, -2, 1)^T
  \end{split}
\end{align}
with the difference operators expressed over the appropriate
dimensions, here with the implicit convention that time corresponds to
the horizontal dimension in an auditory signal or a spectrogram and
logarithmic frequency $\nu$ to the vertical (transposed) dimension.

From the general theory in \cite{Lin93-JMIV,Lin93-Dis} it follows that
computation of discrete derivative approximation in this way
implies that the scale-space properties for the original 
zero-order signal will be transferred to the derivative
approximations, thereby implying theoretically well-founded
implementation of receptive fields in terms of derivatives.

For non-separable spectro-temporal receptive fields corresponding
to logarithmic frequencies $\nu$ that vary with time $t$ by glissando $v$,
we implement the spectro-temporal smoothing operation by first warping 
the spectro-temporal data locally
\begin{equation}
   \nu' = \nu - v \, t
\end{equation}
using spline interpolation. Then, we apply separable spectro-temporal
smoothing in the transformed domain and unwarp the result back to the
original domain.
Over a continuous domain, such an operation is equivalent to
convolution with corresponding glissando-adapted spectro-temporal
receptive fields, while being significantly faster in a discrete
implementation than corresponding explicit convolution with
non-separable receptive fields over two dimensions.

In addition to a transfer of the scale-space properties from the
continuous model to the discrete implementation, 
all the components in this discretization, the discrete Gaussian
kernel, the time-recursive filters and the discrete derivative
approximations, can be seen as mathematical approximations of the
corresponding continuous counterparts.T
Thereby, if follows that the behaviour of the discrete implementation
will approach the behaviour of the corresponding continuous model as
the temporal sampling rate and the sampling rate in the logarithmic frequency
domain increase.
Choosing appropriate sampling rates in an actual implementation is a
trade-off between computational accuracy and computational efficiency.

\bibliographystyle{agsm}

{\footnotesize
\bibliography{bib/defs,bib/tlmac}
}

\end{document}